%% file: main.tex
\documentclass[a4paper,11pt]{article}
\usepackage{jheppub} 
\usepackage[T1]{fontenc} 
\usepackage{amsmath,amsthm}
\usepackage{amssymb, amsfonts}
\usepackage{bbm}

\usepackage{graphicx}
\usepackage{dsfont}
\usepackage{relsize}
\usepackage{stmaryrd}
\usepackage{lmodern}
\usepackage{slantsc}
\usepackage{scalefnt}
\usepackage{wasysym}
\usepackage{xspace}
\usepackage{boxedminipage}
\usepackage{ifpdf}
\usepackage{multirow, hhline}
\usepackage{slashed}
\usepackage{xifthen}
\usepackage{tensor}
\usepackage{array, nicematrix}
\usepackage{tabu}
\usepackage{pbox}
\usepackage{multirow}
\usepackage{tikz}
\usepackage{overpic}
\usepackage{color}
\usepackage{diagbox}
\usepackage{makecell}
\usetikzlibrary{arrows, matrix, calc, scopes, decorations.markings}
\usepackage[mathcal]{euscript}
\usepackage{booktabs}
\usepackage{autobreak, subfig, diagbox, mathrsfs}
\usepackage{mathbbol}
\usepackage[skins,breakable]{tcolorbox}
\usepackage{hyperref}
\allowdisplaybreaks[4]

\newcommand{\ii}{{\mathbb{i}}}
\newcommand{\ee}{{\mathbb{e}}}
\newcommand{\NN}{\mathbb{N}}
\newcommand{\A}{\mathcal{A}}
\newcommand{\Hil}{\mathcal{H}}
\newcommand{\G}{\mathcal{G}}
\newcommand{\T}{\mathcal{T}}
\newcommand{\Bimod}{{\mathsf{Bimod}}}

\newcommand{\C}{{\mathbb{C}}}

\newcommand{\D}{\mathscr{D}}

\newcommand{\Fus}{\mathcal{F}}
\newcommand{\Sub}{\mathcal{S}}
\newcommand{\Cent}{\mathcal{Z}}

\newcommand{\RR}{\mathbb{R}}
\newcommand{\ZZ}{{\mathbb{Z}}}
\newcommand{\CC}{{\mathbb{C}}}
\newcommand{\idm}{{\mathds{1}}}

\newcommand{\hop}{{\mathscr{W}}}
\newcommand{\dash}{{\text{-}}}

\newcommand{\df}{{\mathrm{d}}}

\newcommand{\dk}[2][1]{{\ifthenelse{\equal{#1}{1}}{\frac{\df{#2}}{2\pi}}{\frac{\df^{#1}{#2}}{(2\pi)^{#1}}}}}

\newcommand{\Rep}{{\tt Rep}}

\renewcommand{\Vec}{{\tt Vec}}

\newcommand{\tob}{{\it 1}}

\newcommand{\eq}[1]{\begin{align*}#1\end{align*}}
\newcommand{\eqn}[2][0]{\ifthenelse{\equal{#1}{0}}{\begin{equation}\begin{aligned}#2\end{aligned}\end{equation}}{\begin{equation}\begin{aligned}#2\end{aligned}\label{#1}\end{equation}}}

\newcommand{\eqs}[1]{{\texorpdfstring{${#1}$}{Lg}}}

\newcommand{\ket}[1]{{\left\vert{#1}\right\rangle}}
\newcommand{\bra}[1]{{\left\langle{#1}\right\vert}}
\newcommand{\eval}[1]{{\left\langle{#1}\right\rangle}}
\newcommand{\inprod}[2]{{\left\langle{#1}\middle\vert{#2}\right\rangle}}

\usetikzlibrary{shapes.geometric, snakes}
\tikzset{>=latex}
\usetikzlibrary{arrows, matrix, calc, scopes, decorations.markings}
\usetikzlibrary{decorations.pathmorphing, positioning}
\tikzset{snake it/.style={decorate, decoration={snake,amplitude=0.2mm,segment length=1mm}}}
\tikzset{->-/.style={decoration={
			 markings,
			 mark=at position .5*\pgfdecoratedpathlength+2pt with {\arrow{>}}},postaction={decorate}}}

\tikzset{-<-/.style={decoration={
			 markings,
			 mark=at position .5*\pgfdecoratedpathlength+2pt with {\arrow{<}}},postaction={decorate}}}

\newtabulinestyle{dash=off 2pt}

{\null\,\vcenter\bgroup\scriptsize
\arraycolsep=.13885em
\hbox\bgroup$\array{@{}#1@{}}}
{\endarray$\egroup\egroup\,\null}

\newenvironment{example}{\begin{tcolorbox}[colback=white, colframe=teal!80!darkgray!100!, breakable=true, width=\linewidth, enhanced]}{\end{tcolorbox}}

\newenvironment{important}{\begin{tcolorbox}[colback=white, colframe=red!40!darkgray!100!, breakable=true, width=\linewidth, enhanced]}{\end{tcolorbox}}

\input{input.tex}

\begin{document}

\title{Landau-Ginzburg Paradigm of Topological Phases}

\date{\today}
\author[a]{Yu Zhao}
\author[a,b,c]{Yidun Wan\footnote{Corresponding author}}
\affiliation[a]{State Key Laboratory of Surface Physics, Center for Astronomy and Astrophysics, Department of Physics, Center for Field Theory and Particle Physics, and Institute for Nanoelectronic devices and Quantum Computing, Fudan University, 2005 Songhu Road, Shanghai 200433, China}
\affiliation[b]{Shanghai Research Center for Quantum Sciences, 99 Xiupu Road, Shanghai 201315, China}
\affiliation[c]{Hefei National Laboratory, Hefei 230088, China}
\emailAdd{yuzhao20@fudan.edu.cn, ydwan@fudan.edu.cn}

\abstract{
Topologically ordered matter phases have been regarded as beyond the Landau-Ginzburg symmetry breaking paradigm of matter phases. Recent studies of anyon condensation in topological phases, however, may fit topological phases back in the Landau-Ginzburg paradigm. To truly do so, we realized that the string-net model of topological phases is in fact an effective lattice gauge theory coupled with anyonic matter once two modifications are made: (1) We reinterpret anyons as matter fields coupled to lattice gauge fields, thus extending the HGW model to a genuine Hamiltonian lattice gauge theory. (2) By explicitly incorporating the internal degrees of freedom of anyons, we construct an enlarged Hilbert space that supports well-defined gauge transformations and covariant coupling, restoring the analogy with conventional lattice gauge field theory. In this modified string-net model, topological phase transitions induced by anyon condensation and their consequent phenomena, such as order parameter fields, coherent states, Goldstone modes, and gapping gauge degrees of freedom, can be formulated as Landau's effective theory of the Higgs mechanism. To facilitate the understanding, we also compare anyon condensation to/with the Higgs boson condensation in the electroweak theory and the Cooper pair condensation.
}

\maketitle

\section{Introduction}\label{sec:intro}

Historically, the string-net model was recognized as a Hamiltonian extension of the Turaev-Viro topological field theory \cite{Turaev1992}, achieved by matching its \emph{ground-state} wavefunction to the Turaev-Viro partition function, yet this identification remained largely formal, as the nature of the underlying gauge field was generally not specified. A full gauge-theoretic formulation must explain how anyonic excitations couple to a gauge connection and how that connection transports an anyon's internal gauge states between different positions. The original Levin-Wen construction \cite{Levin2004}, with its restricted Hilbert space, falls short of representing the full anyon spectra or capturing the internal spaces of non-Abelian anyons. By contrast, the Hu-Geer-Wu (HGW) string-net model \cite{Hu2018} extends the Levin-Wen Hilbert space to include the full anyon spectra and manifest the internal spaces of non-Abelian anyons, enabling us to explicitly study the microscopic details of topological phases, including explicit realizations of anyon-condensation-induced topological phase transitions \cite{zhao2022, zhao2024b} and their physical consequences. Nevertheless, a limitation still remains in the HGW model: anyons are essentially realized as punctures in plaquettes of specific collective lattice-field excited states determined by the Hamiltonian. Although this treatment is sufficient in manifesting (partially) the internal spaces of non-Abelian anyons, it still blurs the roles of and especially the interplay between the effective gauge field and anyons. This obstruction arises because anyon types are fully constrained by the lattice-field states and cannot be regarded as independent particles coupled to the lattice field.

Through a series of our recent studies \cite{Hu2020, Wang2020, zhao2024, zhao2024b, zhao2025}, we have realized that two crucial modifications to the HGW model are necessary to not only exhibit separate effective lattice gauge field and anyonic matter fields but also incorporate their interaction coherently, such that the model can describe topological phases in a generalized Landau-Ginzburg paradigm\footnote{Note that proposals of Landau-Ginzburg paradigm of gapped phases have been made before \cite{jia2024, bhardwaj2024, bhardwaj2025}. Nevertheless, they rely solely on the output data---modular tensor categories---of models like the string-net and quantum double models. Hence, they are not able to describe anyon condensation at a fine level as ours and thus cannot formulate order parameter fields with explicit goldstone modes. There also lacks an effective theory coupling gauge fields with anyonic matter fields, which is what a genuine Landau theory needs. Therefore, such Landau paradigms are incomplete and have limited utility.}.

\begin{table}
\centering
\begin{tabular}{|c|c|}
\hline
\textbf{Enlarged HGW Model} & \textbf{Electroweak Theory} \\ \hline
Parent Input UFC & Unbroken Gauge Group ${\tt SU(2)}_L \times {\tt U(1)}_Y$ \\ \hline
Lattice Dofs & Gauge Fields $W_{\mu}^{1,2,3},\; B_{\mu}$ \\ \hline
Anyon Fields & Matter Fields $\ell = \begin{pmatrix} e^{L} \\ \nu_e^{L} \end{pmatrix},\;
e^{R},\;
\phi=\begin{pmatrix} \phi^{1}+\ii\phi^{2} \\ \phi^{0}+\ii\phi^{3} \end{pmatrix},\ \cdots$ \\ \hline
Anyonic Excitations & Matter Particles \\ \hline
Anyons Hopping & Gauge Coupling 
$\ell^{\dagger}\gamma^\mu\left(\partial_{\mu}-\frac{\ii g}{2} W_{\mu}^{a}\sigma_{a}\right)\!\ell$ \\ \hline
Plaquette Operators & Gauge Transformation
$\ell(x)\;\rightarrow\;\ell(x)+\ii \theta^{a}(x)\sigma^{a}\,\ell$ \\ \hline
Anyon Fusion & Yukawa Coupling $\ell^{\dagger}\phi\,e^{R}$ \\ \hline
Anyon Condensation & Higgs Condensation $\langle \phi^{\dagger}\phi\rangle = v$ \\ \hline
Gapped Dofs & Gauge Bosons $W_{\mu}^\pm,\,B_{\mu}$ Acquiring Masses \\ \hline
Child Input UFC & Residual Gauge Group ${\tt U(1)}_{\mathrm{EM}}$ \\ \hline
Child Gauge Dofs & Photon $A_\mu$ \\ \hline
Order Parameter & Vacuum Expectation Values \(\eval{\phi}\) \\ \hline
Local Excitations & Higgs Field Perturbation \(h = \phi - \langle\phi\rangle\) \\ \hline
Goldstone Modes & Ground States Parameters \\ \hline
Gapped Dofs Enlargement & Mass Bosons $W_{\mu}^{\pm},\,Z_{\mu}$ Eating Goldstone Bosons \\ \hline
Anyon Splitting & Splitting of $e^{L}$ and $\nu_e^{L}$ from Isospin Doublet \\ \hline
Anyon Identification & Identified as Dirac Spinor $m_e\,\bar e^{\,L} e^{R}$ \\ \hline
Anyon Confinement & Not in Electroweak: Trivial Moduli Topology \\ \hline
\end{tabular}
\caption{Dictionary between concepts in the Enlarged HGW string-net model and those in traditional gauge field theory taking the electroweak theory as an example.}
\label{tab:rosetta}
\end{table}

The first modification is to reinterpret anyons not merely as punctures representing excited states of the lattice field, but as \emph{matter fields} that couple to the lattice field, which is now regarded as the \emph{gauge field}. With this modification, the HGW model is further extended into a \emph{Hamiltonian lattice gauge theory}, directly generalizing the conventional Hamiltonian lattice gauge theory framework with Lie group as input data \cite{kogut1975, susskind1977}. In this construction, the gauge field resides on the lattice edges and tails, while the matter fields are located in plaquettes. This separation enables an explicit distinction between the gauge field and matter fields, allowing their physical properties and interactions to be discussed independently. As will be shown, several Gauss laws will be established to recover the relationship between gauge-field states on the lattice and the distribution of anyonic matter particles in the plaquettes.

The second modification is to further enlarge the Hilbert space of the HGW model by explicitly incorporating the internal degrees of freedom (dofs) of anyons. This enlargement enables us to express the kinematic Hilbert space of the model as a tensor product of the gauge-field Hilbert space and the internal spaces of the matter particles. By imposing physical constraints---such as the Gauss law and additional conditions to be specified later---this kinematic space is projected onto the physical Hilbert space of the model. On this enlarged Hilbert space, both the gauge transformations of the gauge field and the associated transformations on the anyonic matter fields can be concretely defined. This makes it possible to explicitly formulate the covariant gauge coupling between the gauge field and the matter fields, thereby restoring all the analogies to traditional gauge field theory and rendering the model manifestly gauge invariant.

Therefore, we shall name this modified HGW model the \textbf{enlarged HGW string-net model}. We will show that this enlarged HGW string-net model is a genuine lattice gauge theory consisting of a lattice gauge field coupled to anyonic matter fields. We shall detail the matter-gauge couplings, exhibit the lattice gauge connections mediating internal anyon transformations, define the fundamental fluxes and gauge transformations, and most importantly, implement a lattice Higgs mechanism of anyon condensation, where the Goldstone modes and gapping gauge field dofs can be concretely defined and analyzed. These will construct a generalized Landau-Ginzburg paradigm of symmetry breaking that encompasses topological phases. To corroborate our statements and facilitate understanding, we will compare our formulation with traditional lattice gauge theory and the Higgs mechanism where relevant. Table \ref{tab:rosetta} serves as a Rosetta stone, mapping our Landau-Ginzburg-Higgs description onto its counterparts in traditional gauge theory.

The generalized Landau-Ginzburg paradigm in our enlarged HGW model states that certain gauge invariance of the parent topological phase is spontaneously broken to a global symmetry in the child topological phase. The resulting child phase is, in fact, a symmetry-enriched topological phase that inherits this emergent global symmetry \cite{Hung2013}. In this paper, we will briefly discuss the nature of global symmetry; a detailed construction of symmetry-enriched topological phases within the framework of our enlarged HGW model will be presented in a forthcoming work \cite{Fu2025, Fu2025b}.

Furthermore, by the topological holographic principle, in particular interpreting a bulk topological order as the symmetry topological order (symTO) of its boundary theory \cite{Bhardwaj2017, Freed2018, Ji2019, Kong2019, Kong2020, Kong2020c,Gaiotto2020, Apruzzi2021, Freed2022, Chatterjee2022a}, our generalized Landau-Ginzburg paradigm has a related and dimensionally reduced counterpart---a paradigm for (1+1)-dimensional lattice models as boundaries of the (2+1)-dimensional model of the bulk symTO. This is discussed in a related paper \cite{Hung2025}, which partially by the anyon condensation machineries developed here and in Ref. \cite{zhao2024b}, constructs a streamline for systematically producing 2D critical lattice models and (new) conformal field theories (CFTs).

\section{Enlarged HGW String-Net Model: A Lattice Effective Gauge Theory Coupled with Anyonic Matter}\label{sec:model}

In this section, we construct a lattice model that further enlarges the Hilbert space of the extended string-net model introduced by Hu, Geer, and Wu \cite{Hu2018}---to be called the HGW model---and interpret the enlarged HGW model as an effective lattice Hamiltonian gauge theory coupled with anyonic matter. We refer to the original string-net due to Levin and Wen as the LW model. The original purpose of the HGW model was to manifest the full dyonic excitation spectrum of the string-net model \footnote{The original Levin-Wen (LW) model can only predict the types of anyons and ground-state degeneracies, but cannot explicitly express the wavefunctions for each excitation state. Besides, the full anyon spectrum of the string-net model can also be revealed by attaching a tube-algebra (\(Q\)-algebra) structure to each plaquette in the LW model, where the anyons correspond to irreducible modules of the tube algebra. \cite{Lan2014b}}. Nevertheless, we realize that the full methodological and physical power of the HGW model cannot be unleashed unless we further enlarge the Hilbert space of the HGW model, such that the enlarged HGW model can be interpreted as a lattice effective gauge theory coupled with anyonic matter fields, which is fully gauge invariant. This interpretation is not possible in the LW model, where anyons are regarded as structureless punctures. 

In what follows, we shall introduce our enlarged HGW model and try to interpret it as a lattice gauge theory coupled with anyonic matter step by step, through which the necessity of enlarging the Hilbert space will be unfolded. We then present the Hamiltonian of the model. 

\begin{figure}
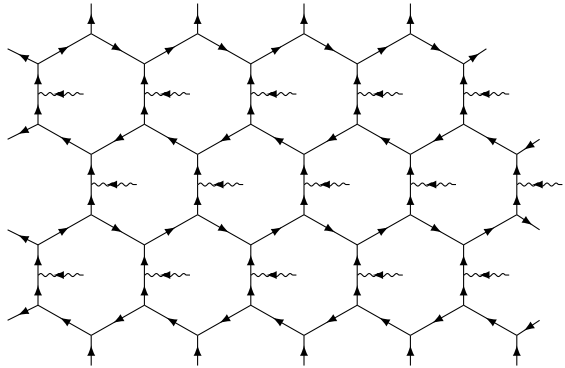

\centering
\Lattice
\caption{Lattice of the HGW model. The shapes of the plaquettes are arbitrary; however, a honeycomb lattice is usually chosen for simplicity.}
\label{fig:lattice}
\end{figure}

The HGW string-net model discretizes the \(2\)d spatial manifold using a trivalent lattice (see Figure \ref{fig:lattice}), while time remains continuous (although time evolution is not our main concern here). The HGW model differs from the traditional LW model by adding to each plaquette of the lattice a tail (wiggling lines in Fig. \ref{fig:lattice}), attached to a chosen edge of the plaquette, while the choice is topologically irrelevant (see Appendix). Each edge and tail is assigned with a direction. Each directed edge or tail \(E\) in the model carries a dof \(x_E\) valued in the simple objects of the input unitary fusion category (UFC) \(\Fus\) (to be defined explicitly in Appendix) of the model. Different choices of directions for edges and tails are physically equivalent. Specifically, if one simultaneously reverses the direction of an edge \(E \mapsto -E\) and applies the transformation
\begin{equation}
x_E \mapsto x_{-E} = x_E^\ast,
\end{equation}
then the gauge field configuration remains invariant. Here, \(x_E^\ast\) denotes the dual (or opposite) simple object of \(x_E\in L_\Fus\) in UFC \(\Fus\), where \(L_\Fus\) is the set of all simple objects in \(\Fus\). All allowed configurations are subject to \emph{fusion constraints} that for any three edges or tails \(E_1, E_2, E_3\) counterclockwise meeting at a trivalent vertex and all directed toward the vertex (with outward-pointing edges or tails reversed in both directions and labels), the fusion rule \(\delta_{x_{E_1} x_{E_2} x_{E_3}} = 1\) must be satisfied. The gauge-field Hilbert space \(\Hil_\text{GF}\) of the HGW model is defined as the span of all possible configurations obtained by assigning simple objects in \(L_\Fus\) to each edge and tail of the lattice, subject to the fusion constraints at every vertex. Our model will further enlarge this Hilbert space by introducing matter fields that couple to the HGW model.

\subsection{Gauge Field and Electric Field on The Lattice}\label{sec:gaugefield}

So as to interpret the HGW model as a lattice gauge theory coupled with anyonic matter, upon necessity, we shall bring up the ingredients of traditional lattice gauge theories that are most relevant to our purposes.

In a traditional Hamiltonian lattice gauge theory with gauge Lie group \(G\) and in temporal gauge \cite{kogut1975}, space is discretized while time remains continuous. The gauge field configuration is specified by assigning to each directed edge \(l\) a group element \(U_l \in G\), representing the gauge field configuration (the parallel transporter, in fact), and a Lie-algebra-valued element \(E_l\in\mathfrak{g}\), which is the electric field configuration. Different choices of edge directions are physically equivalent. Specifically, if one simultaneously reverses the direction of an edge \(l \mapsto -l\) and performs the transformation
\eqn{U_l \mapsto U_{-l} = U_l^{-1}, \qquad 
E_l \mapsto E_{-l} = -E_l,}
then the gauge field configuration remains invariant. The gauge curvature (magnetic field strength) associated with a plaquette \(P\) is given by the counterclockwise-ordered product of group elements on counterclockwise-directed edges and inverse group elements on clockwise-directed edges bounding \(P\). The electric Gauss law states that if the divergence of the electric field $E_l$ at a vertex \(v\), defined by
\eqn[eq:divEv]{
({\tt div} E)_v = \sum_{l\text{ pointing outward }v}E_l - \sum_{l\text{ pointing inward }v}E_l,
}
is nonzero, then there must be a matter field contributing an electric charge \(Q_v\in\mathfrak{g}\) at vertex $v$, such that
\eqn[eq:traditionalGauss]{
({\tt div} E)_v = Q_v.
}
In traditional gauge field theory, only electric charges are dynamically included, while magnetic charges (the monopoles) are absent. Consequently, the magnetic field is divergenceless, and there is no magnetic Gauss law analogous to the electric one.

Usually, the string-net model with input UFC being a finite group-graded UFC $\Vec{G}$ is regarded as the lattice dual of the quantum double model with gauge group $G$, which is a doubled lattice gauge theory that bears both charge excitations at the vertices (Gauss law) and flux excitations in the plaquettes (non-flatness). Building on this duality, although the input UFC $\Fus$ of the string-net model can be more general than $\Vec{G}$, we interpret assigning simple objects \(x_E\in L_\Fus\) in the UFC $\Fus$ on the directed edges and tails \(E\) of the lattice of the HGW model as the configuration space of a \emph{gauge field}, with the gauge curvature (field strength) is now defined at the vertices of the lattice. The condition that the fusion constraints are satisfied at all vertices implies that the corresponding gauge field is always completely flat at every \emph{trivalent vertex} of the lattice.

\begin{important}
In the HGW model with input UFC \(\Fus\), the Hilbert space \(\Hil_\text{GF}\) of the \emph{gauge field} is spanned by all admissible gauge field configurations defined by assigning to each directed edge and tail \(E\) a simple object \(x_E\) of \(\Fus\), subject to the fusion constraints (flatness condition) at every trivalent vertex.
\end{important}

\begin{figure}
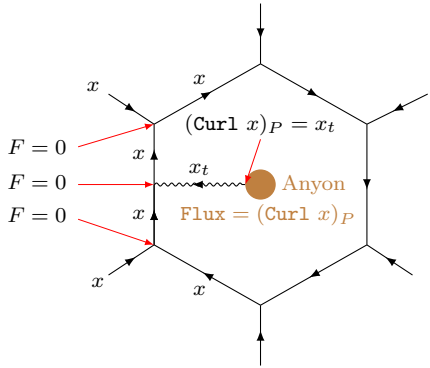
\centering
\Interpret
\caption{Gauge field and matter fields on the enlarged HGW model lattice. The gauge field is flat \(F = 0\) at every trivalent vertex. The curvature \(({\tt curl\ }x)_P\) of the gauge field in plaquette \(P\) is equal to the label \(x_t\) on tail \(t\) within plaquette \(P\), directed outward. The magnetic Gauss law implies that the curvature \(({\tt curl}\ x)_P\) of the gauge field matches the flux \(p\) carried by the anyon (the brown dot) located in \(P\).
}
\label{fig:gaugeflux}
\end{figure}

While the fusion rules are satisfied at every trivalent vertex, they are violated at the open ends of the tails (i.e., degree-one vertices); see Figure \ref{fig:gaugeflux} for an illustration. At these endpoints, the flatness condition for the topological gauge fields is also broken. As we will see in the next subsection, these endpoints host anyonic matter fields, which carry nontrivial fluxes and act as direct sources for the curvature of the gauge field.

\subsection{Coupling between the Matter Field and Gauge Field: Magnetic Gauss Law and Charge Space}\label{sec:magauss}

In traditional lattice gauge theory, a matter field carries a \(d\)-dimensional representation \(\rho\) of the gauge group \(G\). A matter particle resides at a vertex \(v\) of the lattice, with its internal gauge dof described by a \(d\)-dimensional vector \(\psi_v\in\CC^d\). There are three primary ways in which the matter field \(\psi_v\) at vertex \(v\) interacts with the gauge field. First, it contributes an electric charge \(Q_v \in \mathfrak{g}\), which acts as the source term for the electric field \(E_l \in \mathfrak{g}\) through the electric Gauss law (see Eq. \eqref{eq:traditionalGauss}). Second, the matter field couples to the gauge field via covariant derivative operators. The third interaction---the influence of the current of the matter field on the dynamics of the electric field---does not appear because by construction the gauge field is flat.

These concepts in gauge theory generalize naturally to the HGW string-net model, in which the gauge structure is generalized to a UFC \(\Fus\). In this subsection, we focus on the Hilbert spaces of anyonic matter fields, their associated gauge spaces, and how they contribute fluxes to the gauge field via the Gauss law. In the next subsection, we will introduce the hopping operators (string operators) and the classification of anyon types in the enlarged HGW model. 

\begin{itemize}
\item \textbf{Magnetic Gauss law and Anyon's Flux}: As explained earlier, in the HGW model, every trivalent vertex must satisfy the fusion rules, except at the open end of a tail, where the fusion rules may be violated, signaling the presence of a nontrivial curvature \(x_t\in L_\Fus\) at the open end, where \(x_t\) is the dof on the tail pointing toward edges. This is the divergence of the curvature, so we must place an \emph{anyon} (excitation of a \emph{matter field}) with flux \(p = x_t\) at the open end, thereby canceling the excessive flux of the gauge field via the magnetic Gauss law. Therefore, unlike the traditional gauge theory, the HGW model does have nontrivial magnetic monopoles, and formally we have the \emph{magnetic Gauss law} in each plaquette of our enlarged HGW model:
\eqn[eq:magauss]{
\text{Flux of anyon in the plaquette}\quad=\quad
\text{Curvature at tail's open end}\ .
}
This is the key distinction between our enlarged HGW string-net model and the original Levin-Wen and HGW models: In our framework, anyons are treated as physical entities (matter particles) that couple directly to the lattice gauge field, rather than merely as punctures (excitation modes) of particular excited states defined by Hamiltonian constraints on a microscopic lattice model with edge and tail dofs such as electric spins.

An anyon of type \(J\) (which will be defined explicitly later) may carry more than one flux type and correspondingly contribute more than one type of curvature. Let $L_J\subseteq L_\Fus$ be the set of all allowed flux types of anyon type \(J\).

\item \textbf{\(T\)-charges of anyons}: When the fusion rules of the input UFC \(\Fus\) is not commutative, certain anyon (matter particle) types acquire an additional internal gauge dof that we call the \(T\)-\emph{charge dof}, spanning a \(T\)-\emph{charge space}. These \(T\)-charge dofs are analogous to the colors of a quark. As we will see in Section \ref{sec:gauge}, a nontrivial \(T\)-charge space is closely related to multi-dimensional irreducible representation of the \(T\)-gauge transformations of the enlarged HGW model---a direct generalization of the gauge transformation in traditional lattice gauge theory. The dimension \(n_J^p\) of this charge space depends on not only anyon type \(J\) but also \(J\)'s flux type \(p\), beyond the context of the traditional lattice gauge theory, where the representation dimension of the matter particle is fixed merely by particle type. We have \(n_J^p = n_{J^\ast}^{p^\ast}\) for a pair of opposite anyons \(J\) and \(J^\ast\) with opposite flux types \(p\) and \(p^\ast\).

Moreover, the enlarged HGW model admits an entirely new class of gauge transformations that have no analogy in traditional group-based gauge theory, nor in the enlarged HGW model with input UFC \(\Vec(G)\) for any finite group \(G\). We refer to these as \(F\)-\emph{gauge transformations}. Notably, \(F\)-gauge transformations may require an additional multidimensional charge space, called the \(F\)-charge space, to fully characterize the gauge actions. This phenomenon is partially studied in Ref. \cite{zhao2025} and briefly discussed in Section \ref{sec:cond} and Appendix; however, a complete understanding will be presented in future work.

\item \textbf{Single-Anyon Hilbert Space}: Building on the flux types and \(T\)-charge dofs, we obtain the single-anyon Hilbert space of an anyon of type \(J\) in a plaquette \(P\):
\begin{important}
In the enlarged HGW model with input UFC \(\Fus\), an anyon of type \(J\) in plaquette \(P\) has two internal dofs: flux type $p\in L_J$ and charge $\alpha$ with $1\le\alpha\le n_J^p$. So, an orthonormal basis state of the internal space of the anyon in plaquette $P$ can be denoted as \(\ket{J, p, \alpha}_P\), such that
\eqn[eq:anyonHil]{
\Hil_J^P = {\tt span}_\CC\{\ket{J, p,\alpha}_P\mid p\in L_J, \alpha\le n_J^p\}.
}
For historical reasons \cite{Hu2018} and later convenience, we refer to this basis state \(\ket{J, p, \alpha}_P\) as a \emph{dyonic} sector (or simply a dyon). In particular, if \(n_J^p = 1\), we omit the charge dof and simply write the dyon as \((J, p)\).
\end{important}

In general, a dyonic state of an anyon is a superposition of dyonic sectors
\eqn[eq:gaugestate]{\ket{J, \psi}_P = \sum_{p\in L_J}\sum_{\alpha = 1}^{n_J^p}\psi_{p,\alpha}\ket{J,p, \alpha}_P,\qquad \psi_{p,\alpha}\in\CC,\qquad \sum_{p\in L_J}\sum_{\alpha = 1}^{n_J^p}|\psi_{p, \alpha}|^2 = 1.}
We emphasize that these sectors do not have an invariant meaning because they are subject to gauge transformations, which are to be defined soon.

For mathematical completeness, 

\begin{important}
The vacuum of the model is regarded as bearing in each plaquette a \emph{trivial anyon} \(\idm\), which carries a unique trivial flux \(\tob \in L_\Fus\) and a trivial charge space of dimension \(n_\idm^\tob = 1\).
\end{important}

The flux dof \(p\) and the charge dof \(\alpha\) of a given anyon type \(J\) together span the anyon's full internal gauge space of dimension
\[\qquad
{\tt dim}(\Hil^J_P) = \sum_{p\in L_J}n_J^p. 
\]
Note that each anyon type $J$ has a quantum dimension defined as
\[\qquad
d_J = \sum_{p\in L_J}n_J^pd_p.
\]
In the case of $\Fus=\Vec{G}$ for some finite group $G$, ${\tt dim}(\Hil_{J, P})=d_J$, but in general ${\tt dim}(\Hil_J^P)\neq d_J$.

\item \textbf{Physical observables and physical states}: The internal gauge dofs \(p, \alpha\) of anyons are unobservable. The only physical observable of an anyon state \(\ket{J, \psi}_P\) in plaquette \(P\) is its anyon type \(J\). The physical state of anyon \(J\) corresponds to the equivalence class of all gauge-equivalent states associated with the same anyon type \(J\). We will define the measurement operators of anyon types \(J\) in Section \ref{sec:Hamil}.

\end{itemize}

\begin{example}
As the first example, consider the enlarged HGW model with \({\tt Vec}(S_3)\) as its input UFC, where each edge or tail is labeled by a group element from the symmetry group
\[S_3 = \langle\ r, s\mid r^3 = s^2 = (rs)^2 = e\ \rangle,\]
and each trivalent vertex satisfies the \(S_3\) multiplication rules counterclockwise. The group \(S_3\) has three irreducible representations, denoted by \(\tob\), \(\tt sgn\), and \(\pi\), which are respectively the trivial, sign, and two-dimensional representations. 

As we will see, this \(\Vec(S_3)\) model admits eight types of anyons, denoted by \(A, B, C, D, E, F, G\), and \(H\), with the following properties:
\begin{itemize}
\item \(A\): The trivial anyon. The unique dyonic sector is \((A, e)\).

\item \(B\): Corresponding to the (\(\tt sgn\)) representation of the \(S_3\) group, so \(L_B = \{e\}\), \(n_B^e = 1\), and \(\dim(\Hil_{B, P}) = d_B = 1\). Although \(B\) has no nontrivial internal gauge space, it acquires a phase shift when hopping across the lattice. The unique dyonic sector is \((B, e)\).
\item \(C\): Corresponding to the \(2\)d irreducible representation of \(S_3\), and thus possesses a two-dimensional internal charge space:
\[
\qquad n_C^e = \dim(\pi) = 2, \qquad \dim(\Hil_C^P) = d_C = 2.
\]
Therefore, anyon \(C\) contains two dyonic sectors \((C, e, 1)\) and \((C, e, 2)\).
\item \(D\) and \(E\): Both have flux sets \(L_D = L_E = \{s, sr, rs\}\), with \(n_D^s = n_D^{sr} = n_D^{rs} = n_E^s = n_E^{sr} = n_E^{rs} = 1\), and
\eq{\dim(\Hil_D^P) = \dim(\Hil_E^P) = d_D = d_E = 3.}
Although anyons \(D\) and \(E\) share the same sets of flux types \(s, rs, sr\), they acquire different phase shifts when hopping.
\item \(F\), \(G\), and \(H\): Each has flux set \(L_F = L_G = L_H = \{r, r^2\}\), with \(n_F^r = n_F^{r^2} = n_G^r = n_G^{r^2} = n_H^r = n_H^{r^2} = 1\), and
\eq{
\dim(\Hil_F^P) = \dim(\Hil_G^P) = \dim(\Hil_H^P) = d_F = d_G = d_H = 2.}
\end{itemize}
\end{example}

\begin{example}
The simple objects of the UFC \(\Vec(G)\) for any finite group \(G\) all have quantum dimension \(1\). One of the simplest examples of a UFC with simple objects of higher quantum dimension is the \emph{Ising} UFC, which has three simple objects: \(1\), \(\sigma\), and \(\psi\), with fusion rules \(\delta_{1\psi\psi} = \delta_{1\sigma\sigma} = \delta_{\psi\sigma\sigma} = 1\).

The enlarged HGW model with input Ising UFC goes beyond traditional lattice gauge theory because its gauge structure cannot be described by a group. Instead, the enlarged HGW model has a categorical gauge structure described by the Ising UFC. It gives rise to a topological phase with nine anyon types, conventionally labeled as
\[
1\bar{1},\ \ 1\bar{\psi},\ \ 1\bar{\sigma},\ \ \psi\bar{1},\ \ \psi\bar{\psi},\ \ \psi\bar{\sigma},\ \ \sigma\bar{1},\ \ \sigma\bar{\psi},\ \ \sigma\bar{\sigma}.
\]
Since the fusion rules of the Ising UFC are commutative, no multidimensional \(T\)-charge space arises for these anyons. The properties of these anyons are as follows:
\begin{itemize}
\item \(L_{1\bar{1}} = L_{\psi\bar{\psi}} = \{1\}\), with \(n_{1\bar{1}}^1 = n_{\psi\bar{\psi}}^1 = 1\). Here, \(1\bar 1\) is the trivial anyon.
\item \(L_{1\bar{\psi}} = L_{\psi\bar{1}} = \{\psi\}\), with \(n_{1\bar{\psi}}^\psi = n_{\psi\bar{1}}^\psi = 1\).
\item \(L_{\sigma\bar{1}} = L_{1\bar{\sigma}} = L_{\sigma\bar{\psi}} = L_{\psi\bar{\sigma}} = \{\sigma\}\), with \(n_{\sigma\bar{1}}^\sigma = n_{\sigma\bar{\psi}}^\sigma = n_{1\bar{\sigma}}^\sigma = n_{\psi\bar{\sigma}}^\sigma = 1\).
\item \(L_{\sigma\bar{\sigma}} = \{1, \psi\}\), with \(n_{\sigma\bar{\sigma}}^1 = n_{\sigma\bar{\sigma}}^\psi = 1\), and
\[
\dim(\Hil_{\sigma\bar{\sigma}, P}) = d_{\sigma\bar{\sigma}} = 2.
\]
That is, anyon \(\sigma\bar\sigma\) has two types of fluxes \(1\) and \(\psi\), which span a two-dimensional internal gauge space of each \(\sigma\bar\sigma\) anyon.
\end{itemize}
\end{example}

\begin{example}
Another UFC example is the \emph{Fibonacci} UFC, which has two simple objects, \(1\) and \(\tau\), subject to the fusion rules \(\delta_{1\tau\tau} = \delta_{\tau\tau\tau} = 1\). The enlarged HGW model with Fibonacci input UFC describes four anyon types:
\begin{itemize}
\item Trivial anyon \(1\bar{1}\), where \(L_{1\bar{1}} = \{1\}\) and \(n_{1\bar{1}}^1 = 1\).
\item Anyon types \(\tau\bar{1}\) and \(1\bar{\tau}\), with \(L_{\tau\bar{1}} = L_{1\bar{\tau}} = \{\tau\}\), and \(n_{\tau\bar{1}}^\tau = n_{1\bar{\tau}}^\tau = 1\).
\item Anyon type \(\tau\bar{\tau}\), with \(L_{\tau\bar{\tau}} = \{1, \tau\}\) and \(n_{\tau\bar{\tau}}^1 = n_{\tau\bar{\tau}}^\tau = 1\). The single-anyon Hilbert space is thus two-dimensional.
\end{itemize}

\end{example}

\subsection{Kinematic and Physical Hilbert Space}\label{sec:hilbert}

Now that we have defined the gauge-field Hilbert space \(\Hil_\text{GF}\) and the single-anyon Hilbert space \(\Hil_J^P\) for each anyon type \(J\) in every plaquette \(P\), we can define the total \emph{kinematic} Hilbert space \(\Hil_\text{HGW}\) of our enlarged HGW model with input UFC \(\Fus\) as the tensor product of the gauge-field Hilbert space and the direct sum of anyons' Hilbert spaces over all plaquettes:
\eqn[eq:Hilbert]{
\Hil_\text{HGW} = \Hil_\text{GF} \otimes \bigotimes_{\text{plaquette } P} \bigoplus_{\text{anyon type } J} \Hil_J^P.
}
Here, the direct sum \(\oplus\) reflects the natural requirement that there can only be at most one anyon within a single plaquette.

Nevertheless, not all states in the kinematic Hilbert space are physically allowed in the topological phase described by the enlarged HGW model. The physical Hilbert space---a subspace of the kinematic Hilbert space, is called the \emph{dynamical Hilbert space}, defined by several fundamental constraints:
\begin{itemize}
\item \textbf{Magnetic Gauss Law:} As introduced in Eq. \eqref{eq:magauss}, an anyon's flux must match the gauge dof on the tail in the plaquette it resides.

\item \textbf{Topological Gauss Law:} The enlarged HGW model, which couples its gauge field with matter (anyonic) fields must be gauge invariant. It follows that not only the anyon's flux types but also the anyon types in plaquettes are fully reflected by the gauge-field states in \(\Hil_\text{GF}\). We will discuss this in detail in Section \ref{sec:Hamil}.

\item \textbf{Topological Superselection Rule:} To preserve a definite topological phase, in each plaquette, any superposition of different anyon types is forbidden. This superselection rule also constrains that only certain gauge field states in \(\Hil_\text{GF}\) are physically permitted. See Section \ref{sec:Hamil} for further discussion.

\item \textbf{Topological Selection Rule:} Beyond the local constraints of superselection, the topology of the underlying manifold imposes global consistency conditions on the allowed distributions of anyon types in the whole system \cite{Kitaev2003a, pachos2012}. For example, on a sphere, anyons must be created and annihilated in pairs, so physical states cannot contain a single isolated nontrivial anyon in the entire system. On a torus or more topologically nontrivial surfaces, however, single-anyon states may exist.
\end{itemize}

\subsection{Coupling between the Matter and Gauge Field: Gauge Connection and Anyon Type}\label{sec:gauge}

\begin{figure}
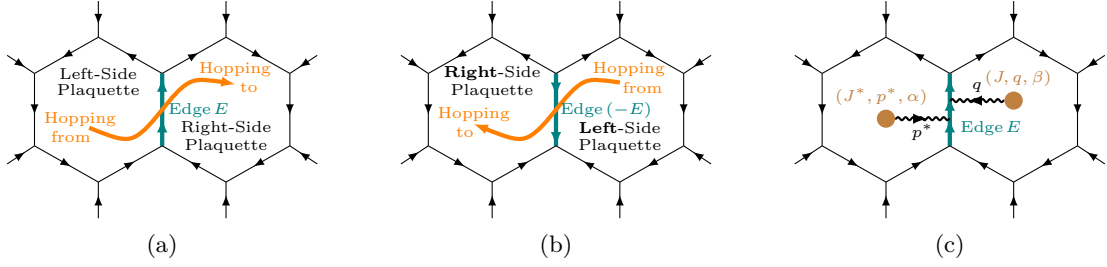
\centering
\subfloat[]{\DirectionA\label{fig:directionA}}\hspace{20pt}
\subfloat[]{\DirectionB\label{fig:directionB}}\hspace{20pt}
\subfloat[]{\DirectionC\label{fig:directionC}}
\caption{Directions of hopping and creation along a directed edge \(E\). (a) The hopping operator acting across an \emph{directed} edge \(E\) always moves an anyon from $E$'s \emph{left} plaquette to its right plaquette. Left and right is defined with respect to the \(E\)'s direction. (b) Reversing the direction of \(E\) exchanges its left and right accordingly but does not affect the configuration of the gauge field. Therefore, to hop an anyon from the right side to the left side of a directed edge $E$, one can first reverse $E$'s direction and dual the dof on this edge; then $E$'s right (left) side becomes its left (right) side. (c) When acting the creation operator \(W_E^{J; (p, \alpha)(q, \beta)}\) across a directed edge \(E\), we conventionally anchor the tail connecting to dyon \((J^\ast, p^\ast, \alpha)\) (\((J, q, \beta)\)) to \(E\) from its left (right). The anchor point of $E$'s left tail is relatively (with respect to $E$'s direction) lower than that of $E$'s right tail. This convention is rotationally invariant.}
\label{fig:direction}
\end{figure}

In traditional lattice gauge theory, besides the magnetic Gauss law \eqref{eq:magauss}, matter fields also couple to the gauge field through covariant derivative operators. \emph{Hopping operators} \(\hop_l\) on edges \(l\) play a central role in mediating this coupling. Specifically, two matter field vectors \(\psi_v\) and \(\psi_{v + l}\), located at neighboring vertices \(v\) and \(v + l\), are related by hopping operator \(\hop_l\), defined as
\eqn[eq:conventionhop]{
\hop_l\psi_{v'} = D^\rho(U_l) \psi_v\delta_{v', v+l},
}
where \(\psi_{v'}\) is an arbitrary matter state at an arbitrary vertex \(v'\) in the lattice, and \(l\) denotes the directed edge from \(v\) to \(v + l\), \(U_l\) is the group element on this edge, and \(D^\rho(U_l)\) is the representation matrix of the gauge group element \(U_l\in G\) in representation \(\rho\). 

Likewise, in the enlarged HGW model, if a dyon $(J, p, \alpha)$ in plaquette \(P\) moves across an edge \(E\) carrying simple object \(x\) to its neighbouring plaquette \(P'\), the dyon sector may morph to \((J, q, \beta)\). We now realize this process via hopping operators, which are analogous to the hopping operators in traditional lattice gauge theories. 
\begin{itemize}
\item The shortest \emph{dyon hopping operator} \(\hop_E^{J; (p, \alpha), (q, \beta)}\) moves a dyon \((J, p, \alpha)\) in $E$'s \emph{left} plaquette \(P\) to $E$'s \emph{right} plaquette \(Q\) across a directed edge \(E\) carrying dofs \(x\) and \(y\), while simultaneously transforming the dyon to \((J, q, \beta)\):
\eqn[eq:shorthop]{
\hop_E^{J;(p,\alpha),(q,\beta)}\quad\HoppingLeft =\quad
[z^J_{yx}]_{p\alpha}^{q\beta}\ \sqrt{\frac{d_p}{d_y}}\quad\HoppingRight\ ,}
where the indices of the coefficients \([z^J_{yx}]_{p\alpha}^{q\beta}\in\CC\) take values
\eq{\qquad x, y\in L_\Fus,\ \ p, q\in L_J,\ \ 1\le\alpha\le n_J^p,\ \ 1\le\beta\le n_J^q.}
The coefficients \([z^J_{yx}]_{p\alpha}^{q\beta}\) are the \emph{half-braiding tensor components} to be defined explicitly in Eq. \eqref{eq:yangbaxter}. The moving direction is defined in Fig. \ref{fig:directionA}. Here, \([z^J_{xy}]\) is a matrix indexed by pairs \((x, y) \in L_\Fus^2\): For each given pair \((x, y)\), \([z^J_{xy}]\) is a matrix with $(q\beta)$ and $(p\alpha)$ being respectively the row and column indices. This generalizes the \(G\)-indexed representation matrix \(D^\rho(U_l)\) in definition \eqref{eq:conventionhop} of gauge connection/coupling in traditional lattice gauge theory in two ways: (1) In traditional lattice gauge theory, the gauge field on an edge is specified by a single group element \(U_l\). In contrast, anyons in Eq. \eqref{eq:shorthop} carry fluxes \(p, q\) that may change the gauge configuration on edge \(E\), so the matrix \([z^J_{xy}]\) is indexed by pairs \((x, y)\) on edge \(E\). (2) The charge space dimensions of an anyon \(J\) depends on its flux types, so in general \(n_J^p \ne n_J^q\). Hence, \([z^J_{xy}]\) may be a rectangular matrix mapping between spaces of different dimensions.

Equation \eqref{eq:shorthop} that generalizes the notion of gauge connection to the enlarged HGW model can be expressed as follows, in a way analogous to Eq. \eqref{eq:conventionhop} in traditional lattice gauge theory.
\eqn{\hop_E^{J;(p,\alpha),(q,\beta)}\ket{J', \psi'}_{Q'} = \ket{\idm, \tob}_P\delta_{Q', P} +
[z^J_{yx}]_{p\alpha}^{q\beta}\sqrt{\frac{d_p}{d_y}}{}_{P}\inprod{J, p, \alpha}{J_P, \psi_P}_P\ket{J, q, \beta}_{Q}\delta_{Q, Q'},
}
where $Q'$ is an arbitrary plaquette in the lattice, and \(P, Q\) are the left and right plaquettes of directed edge $E$ carrying dofs \(x, y\).

\item In Eq. \eqref{eq:shorthop}, we impose the condition that $E$'s right plaquette---the target plaquette of hopping---must not have any anyon. This requirement is natural in topological phases, which demand that anyons be well localized and well-separated \cite{Kitaev2006, Nayak2008, pachos2012, WenTensorCat2004}. Therefore, we restrain a plaquette from hosting more than one anyon, unless we intend to explicitly alter a topological phase. We will introduce the matrix elements of hopping operators in such scenarios and show how it alters the topological phase.

\item A long hopping operator \(\hop_{E_1E_2\cdots E_n}^{J;(p_0, \alpha_0)(p_n, \alpha_n)}\) along a path crossing edges \(E_1E_2\cdots E_n\) is the composition of a sequence of shortest hopping operators:
\eqn[eq:longhop]{\hop_{E_1E_2\cdots E_n}^{J;(p_0, \alpha_0)(p_n, \alpha_n)} = \sum_{p_k\in L_J}\sum_{\alpha_k = 1}^{n_J^{p_i}}\Big[\hop_{E_n}^{J;(p_{n-1}, \alpha_{n-1})(p_n,\alpha_n)}\cdots \hop_{E_1}^{J;(p_0,\alpha_0)(p_1,\alpha_1)}\ \Big].}
For consistency, we require that all edges \(E_i\) along the path are directed in the same direction, and that \(E_{i+1}\) is always located to the right of \(E_i\). If this is not the case, we can always reverse the directions and dual the dofs on the corresponding edges.

\item The \emph{anyon hopping operators} of anyon \(J\), mediating the coupling between anyon $J$ and the gauge field, are sums of all dyon hopping operators of \(J\)
\eqn[eq:anyonhop]{\hop^J_{E_1E_2\cdots E_n} = \sum_{p,q\in L_J}\sum_{i = 1}^{n_J^p}\sum_{j = 1}^{n_J^q}\hop_{E_1E_2\cdots E_n}^{J;(p,\alpha)(q,\beta)}\ .}
\end{itemize}

Any dyonic sector of a specific anyon type shall be invariant under a loop path due to flatness of the gauge field. This fact results in for any three edges \(E_1, E_2, E_3\) incident at a vertex, the following \emph{Yang-Baxter equation}:
\eqn{\qquad\sum_{q\in L_J}\sum_{j = 1}^{n_J^q}\hop_{E_2}^{J;(q,\beta)(r,\gamma)}\hop_{E_1}^{J;(p,\alpha)(q,\beta)} = \hop_{E_3}^{J;(p,\alpha)(r,\gamma)}.}
This equation can be depicted as
\eqn[eq:halfbraidfigure]{\mathcal{T}\quad\sum_{q\in L_J}\ \sum_{j = 1}^{n_J^q}\ \ \FlatnessA\quad=\quad \quad\FlatnessB\ ,}
where ``{\color{red}\(\times\)}'' labels the plaquette to be contracted via topological moves \(\mathcal{T}\) defined in Appendix. This equation yields the defining equation of both the dimension functions \(n: L_\Fus \to \NN\) and half-braiding tensors \([z_{xy}], x, y\in L_\Fus\):
\eqn[eq:yangbaxter]{\qquad\frac{\delta_{x,t}\delta_{rst}}{d_t}[z_{tw}]_{p\alpha}^{q\beta} = \sum_{ulv\in L_\Fus}\sum_{1\le \gamma\le n^l} [z_{su}]_{p\alpha}^{l\gamma}\ [z_{rv}]_{l\gamma}^{q\beta}\ d_ud_v G^{rst}_{pwu} G^{srx}_{qwv} G^{sul}_{rvw},}
Among all possible solutions to this equation, there are minimal solutions, each of which cannot be the direct sum of any other nonzero solutions. Each such minimal solution uniquely specifies an anyon type $J$.

\begin{example}
The half-braiding tensor components of anyons in \(\Vec(S_3)\)-model, \(\Vec(\ZZ_2)\)-model, \(\Vec(\ZZ_3)\)-model, Ising model and Fibonacci model are collected in Appendix.
\end{example}

\begin{figure}
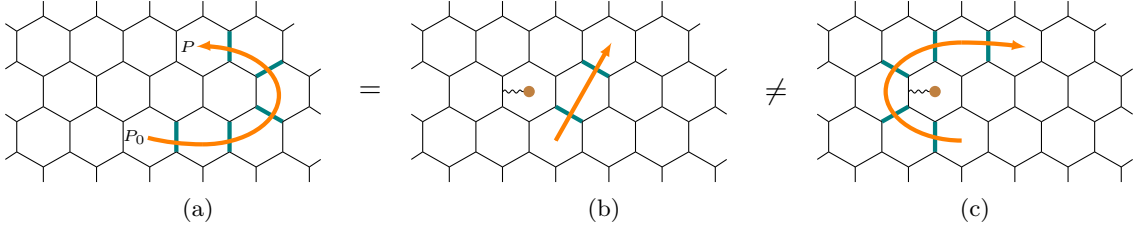
\centering
\subfloat[]{\LongHopA}
\subfloat[]{\LongHopB}
\subfloat[]{\LongHopC}
\caption{Deformation of paths of long hopping operators. The orange lines indicate the hopping paths, while the green edges mark the edges of the lattice these paths cross. Because of the flatness, the path of a long hopping operator can be arbitrarily deformed, as long as the deformation does not ``sweep'' over any plaquettes containing anyons. We omit the directions of edges in the lattice for simplicity.}
\label{fig:longhop}
\end{figure}

The Yang-Baxter equation induces that in the enlarged HGW model, whose gauge field is flat, the path of a long hopping operator \(\hop_{E_1E_2\cdots E_n}^{J;(p, \alpha)(q,\beta)}\) can be continuously deformed without affecting its matrix elements, provided the deformation does not pass over or enclose any plaquettes containing anyons. See Fig. \ref{fig:longhop}. 

Mathematically, anyon types are labeled by the simple objects in a unitary modular tensor category (UMTC)---the \emph{Drinfeld center} \(\Cent(\Fus)\) of the input UFC \(\Fus\). Each simple object \(J\) (anyon type) of \(\Cent(\Fus)\) is expressed as a pair \(({\tt Obj}(J), c_X)\), where \({\tt Obj}(J) = \bigoplus_{p \in L_J} n_J^p p\) is a (possibly composite) object of the input UFC \(\Fus\), and \(c_X: {\tt Obj}(J)\otimes X \to X\otimes {\tt Obj}(J), \forall X\in\Fus\) is the \emph{half-braiding} morphisms that describe how \({\tt Obj}(J)\) commutes with objects of \(\Fus\). The shortest hopping operator $\hop_E^{J}$, which describe how an anyon \(J\) in edge $E$'s left plaquette commutes with simple objects on \(E\) and enters $E$'s right plaquette, encode the half-braiding morphisms. The Yang-Baxter equation \eqref{eq:halfbraidfigure} expresses the mathematical fact that the half-braiding morphisms commute with the fusion of simple objects of $\Fus$. Two distinct anyon types $J$ and $J'$ may differ in their half-braiding morphisms although ${\tt Obj}(J)={\tt Obj}(J')$.

If there is no anyon excited in the two adjacent plaquettes with common edge \(E\), the \emph{shortest creation operator} $W_E^{J;(p, \alpha), (q, \beta)}$ creates a pair of dyon $(J, q, \beta)$ and its antiparticle $(J^\ast, p^\ast, \alpha)$ in these two plaquettes, defined as
\eqn[eq:shortcreation]{W_E^{J;(p,\alpha),(q,\beta)}\quad\CreationLeft\quad =\quad\sum_{u}[z^J_{xu}]_{p\alpha}^{q\beta}\sqrt{\frac{d_u}{d_x}} \quad\CreationRight\ .}

A long creation operator along a long path is defined by concatenating the shortest creation operators and a hopping operators along the path:
\[W_{E_1E_2\cdots E_n}^{J;(p, \alpha)(q, \beta)} = \sum_{r\in L_J}\sum_{\gamma = 1}^{n_J^r} \hop^{J;(r, \gamma)(q, \beta)}_{E_2E_3\cdots E_n}W_{E_1E_2}^{J;(p, \alpha)(r, \gamma)}.\]

\subsection{The \eqs{T}-Gauge Transformations}\label{secgaugetrans}

In traditional lattice gauge theory with gauge group $G$, a gauge transformation is specified by assigning to each vertex \(v\) a group element \(g_v \in G\). Different vertices can be assigned different group elements. Under such a gauge transformation, the gauge field and electric field \(U_l\), \(E_l\) transform as:
\eqn[eq:conventiongaugefield]{
U_l \ \rightarrow\ g_{v+l} U_l g_v^{-1}, \qquad
E_l \ \rightarrow\ g_{v+l} E_l g_v^{-1},}
where \(l\) denotes the directed edge pointing from vertex \(v\) to vertex \(v+l\). Simultaneously, to ensure gauge invariance of the theory, the matter field \(\psi_v\) must carry a certain representation \(\rho\) of $G$ and transform accordingly:
\eqn[eq:conventionmatterfield]{
\psi_v \ \rightarrow\ D^\rho(g_v)\, \psi_v,\qquad Q_v \ \to g_vQ_vg_v^{-1}.
}
These gauge transformations ensure that the entire lattice gauge theory remains gauge invariant---the Gauss law is exactly preserved, and the hopping operator commutes with the transformation.

The enlarged HGW model admits gauge transformations analogous to traditional gauge transformations, which transform the gauge field configuration on edges and tails, as well as transform the internal gauge states of anyons commensurately to keep the entire theory gauge invairiant. Namely,
\[
p_t\to q_t,\qquad \ket{J,p,\alpha}_P \to \sum_{\beta = 1}^{n_J^p}\cdots\ket{J,q,\beta}_P,
\]
where \(p_t, q_t\in L_\Fus\) are labels on tail \(t\) in plaquette $P$, and $\cdots$ collects the coefficients to be nailed down shortly. We shall refer to such gauge transformations as the \(T\)-\textit{gauge transformations} because they directly generalize traditional gauge transformations.

The $T$-gauge transformation consists of two parts: The first part transforms the gauge field, while the second transforms the matter fields commensurately. The first part is realized by acting on each plaquette \(P\) with a \emph{plaquette operator} $B_P^{psuq}$\cite{Lan2014b, Hu2018}:
\eqn[eq:tube]{
&B_P^{psuq}\quad\TubeGaugeE\qquad\Longrightarrow\qquad\TubeGaugeF\quad = \quad \sum_{j_0j_1j_2j_3j_4j_5j_6\in L_\Fus}\\
&\qquad\qquad \delta_{pp'}\ \prod_{k = 0}^5\Bigg[G^{sj_{k+1}^\ast i_{k+1}}_{e_ki_kj_k^\ast}\sqrt{d_{i_k}d_{j_k}}\Bigg]G^{su^\ast p^\ast}_{i_0^\ast i_6 j_6^\ast}
G^{sj_0^\ast i_0}_{j_6u^\ast q^\ast}\sqrt{d_pd_sd_ud_{i_6}d_{j_6}}\quad \TubeGaugeG\quad,}
where ``{\color{red} $\times$}'' in the second figure labels the plaquette to be contracted via topological moves \(\mathcal{T}\). Note that \(B_P^{psuq}\) acts only on gauge field configurations but not on anyons' internal gauge spaces, in this equation, we omit the anyon at the tail's open end in plaquette \(P\).

\begin{example}
To compare to gauge transformations in traditional lattice gauge theory, let's consider a concrete example of Eq. \eqref{eq:tube} in the enlarged HGW model with input UFC \(\Vec S_3\). In this case, operator $B_P^{psuq}$ is nonzero only if the group multiplications
\[u = p^\ast s,\qquad q = u^\ast s = s^\ast ps\]
are satisfied. Then the action of \(B_P^{psuq}\) becomes
\[
B_P^{psuq}\ \TubeGaugeHLeft\ :=\ \TubeGaugeHRight.
\]
This shows that when the input UFC is \(\Vec(G)\) for certain group \(G\), a \(T\)-gauge transformation on the gauge field reduces to a traditional gauge transformation:
\begin{itemize}
\item The gauge transformation of a tail in plaquette \(P\) is \(x \mapsto s^\ast xs\) if the plaquette operator in \(P\) is \(B_P^{p, s, p^\ast s, s^\ast ps}\).

\item If two adjacent plaquettes \(P, Q\) on the left and right sides of edge \(E\) labeled by a simple object $x$ are acted by plaquette operators \(B_P^{p,s_P,p^\ast s_P,s_P^\ast ps_P}\) and \(B_Q^{p,s_Q,p^\ast s_Q,s_Q^\ast ps_Q}\), then the gauge transformation on edge \(E\) should be \(x \mapsto s_P^\ast xs_Q\) for any \(x\in S_3\):
\end{itemize}
$$\TubeGaugeI \qquad\Longrightarrow\qquad \TubeGaugeJ\qquad =\qquad \TubeGaugeK \qquad\Longrightarrow\qquad \TubeGaugeL .$$
\end{example}

\begin{figure}
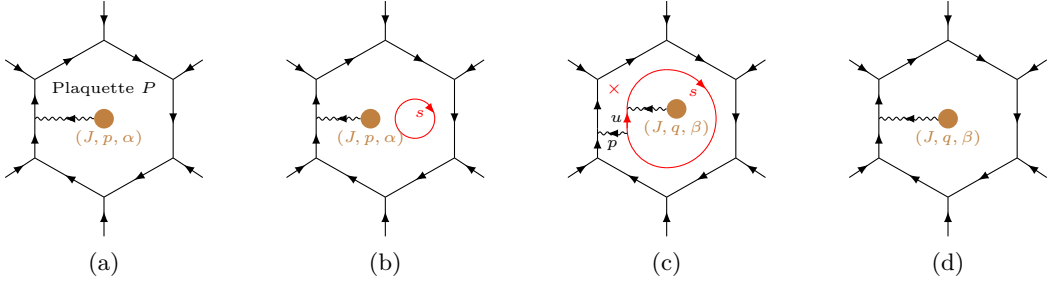
\centering
\subfloat[]{\TubeGaugeA}\hspace{20pt}
\subfloat[]{\TubeGaugeB}\hspace{20pt}
\subfloat[]{\TubeGaugeC}\hspace{20pt}
\subfloat[]{\TubeGaugeD}
\caption{Process of a tube gauge transformation \(B_P^{psuq}\) acting on the dyonic sector \(\ket{J, p, \alpha}_P\) in plaquette \(P\) as shown in (a), transforming it into sector \(\ket{J, q, \beta}_P\) as shown in (d). (b) Creating a loop edge carrying simple object \(s\) in plaquette \(P\). (c): Applying a dyonic hopping operator to move the anyon into the loop and contracts the outer plaquette.}
\label{fig:tubegauge}
\end{figure}

The set of all such operators \(\{B_P^{psuq}\}\) in any given plaquette \(P\) generates a \emph{tube algebra} under composition: 
\eqn[eq:tubecoeff]{B^{q'tvm}B^{pruq} = \delta_{qq'}\sum_{t, w\in L_\Fus}f_{qsvm, pruq}^{ptwm} B^{ptwm},}
where the structure constants \(f_{qsvm, pruq}^{ptwm}\in\CC\) are composed of products of \(6j\)-symbols; for simplicity, we do not list their explicit forms here. In the enlarged HGW model, the tube algebra for any given plaquette \(P\) depends only on the input UFC and is independent of the specific plaquette \(P\) where it is defined, while \(B_P^{psuq}\) operators in different plaquettes commute. Therefore, unless explicitly necessary, we omit the plaquette index \(P\) and refer to the operators simply as \(B^{psuq}\).

Now we move to the second part: How the gauge transformations act on the gauge space of a given anyon \(J\). Since the whole gauge theory shall be invariant under the gauge transformations, the anyon's gauge states should transform covariantly with the gauge transformation of the field configurations on edges and tails. For example, the flux of the anyon should always be equal to the dof on tail in the same plaquette due to the magnetic Gauss law \eqref{eq:magauss}. In other words, the anyon's internal gauge space and its transformation forms a \emph{representation} of the tube algebra of the plaquette operators. This property leads to another definition of anyon types equivalent to that due to Yang-Baxter equation \eqref{eq:yangbaxter}:

\begin{figure}
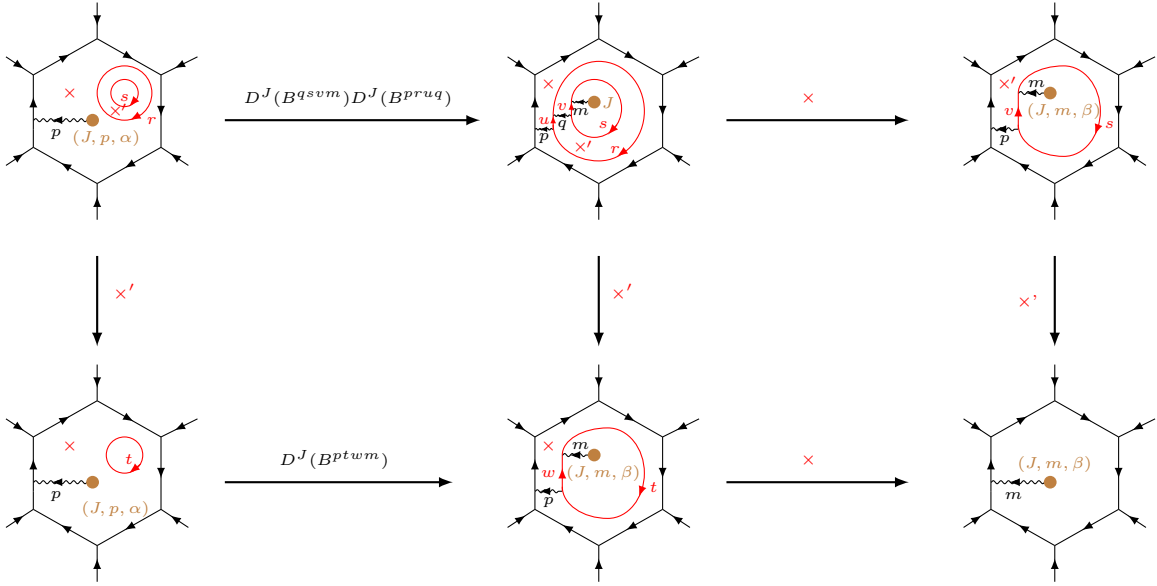
\centering
\Representation
\caption{Diagram depiction of Eq. \eqref{eq:representation}. The internal Hilbert space of an anyon \(J\) transforms under the \(T\)-gauge transformation implemented by the representation \(D^J\) tube algebra element \(D^J(B^{psuq})\). This transformation is compatible with the multiplication rules of the tube algebra and the fusion rules of the UFC \(\Fus\).}
\label{fig:representation}
\end{figure}

\begin{important}
Anyon types $J$ label the irreducible representations of the tube algebra \(\{B^{psuq}\}\), where the representation space is anyon \(J\)'s internal gauge space, and the representation matrix elements are the half-braiding tensor components
\eqn[eq:anyontrans]{ D^J(B^{psuq})\ket{J,p',\alpha} = \delta_{pp'}\sum_{\beta = 1}^{n_J^q}[z^J_{su}]_{p\alpha}^{q\beta}\ \ket{J,q,\beta}\ ,}
which forms a representations of the tube algebra. To be explicit, we have \cite{muger2003a, muger2003, muger2003b}
\eqn[eq:representation]{\sum_{q\in L_J}D^J(B^{qsvm})D^J(B^{pruq})f_{qsvr, pruq}^{ptwm}\  
=\ D^J(B^{ptwm}),}
where \(f_{qsvm, pruq}^{ptwm}\) \eqref{eq:tubecoeff} is the structure constants of the tueb algebra. These representations are not linear representations of the tube algebra. Instead, the representation equations ensure that the representation is compatible with the topological moves of the enlarged HGW lattice. See Figure \ref{fig:representation} for a diagrammatic interpretation.. 

The action of $B^{psuq}$ on the gauge space of an anyon type $J$ is through its representation matrix \(D^J(B^{psuq})\), which encodes a physical process in \((2+1)\)D where the gauge flux loop carrying dof \(s\) encircles the matter field located at the center of the plaquette, see Figure \ref{fig:tubegauge}. 

\end{important}

\begin{figure}
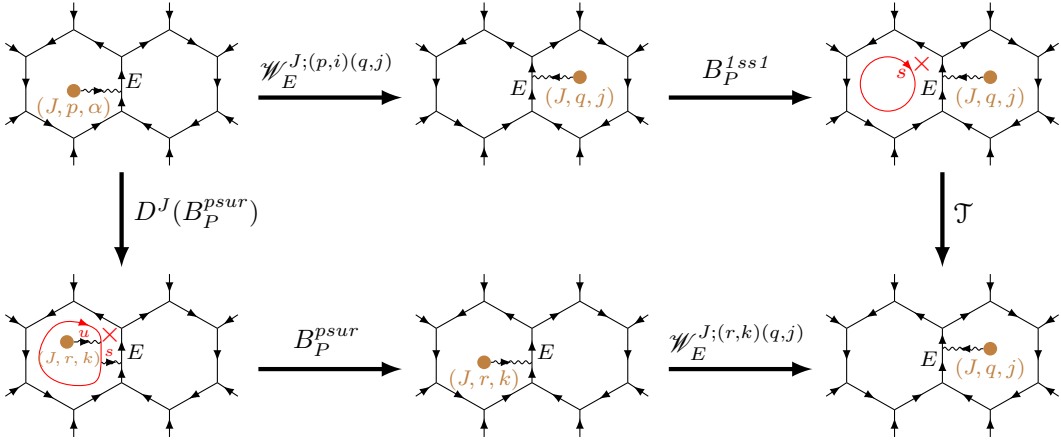
\centering
\Commutivity
\caption{Commuting graph of hopping and gauge transformations in \eqref{eq:commutativity}.}
\label{fig:commutivity}
\end{figure}

The transformation $B^{psuq}$ on the gauge field and the transformation \(D^J(B^{psuq})\) on an anyon's internal space precisely cancel each other to leave the theory gauge invariant. This cancellation is two-fold: First, as in Eq. \eqref{eq:anyontrans}, operator $B^{psuq}$ transforms the flux label on a tail from $p$ to $q$; commensurately, \(D^J(B^{psuq})\) transforms the flux \(p\) of an anyon \(J\) at the open end of the tail to \(q\), preserving the magnetic Gauss law \eqref{eq:magauss}. Second, the hopping operators commute with gauge transformations. Formally, this commutativity reads
\eqn[eq:commutativity]{
B_P^{\tob ss \tob} \hop_E^{J; (p, \alpha), (q, \beta)} = \hop_E^{J; (r, \gamma), (q, \beta)} \, B_P^{psur} D^J(B_P^{psur}),}
where \(P\) denotes directed edge $E$'s left plaquette. Operators \(B_P^{psuq}\) and \(D^J(B_P^{psuq})\) act on the gauge field and anyon's internal space respectively. This commutativity is depicted in Figure \eqref{fig:commutivity}. One can easily check that Eqs. \eqref{eq:tubecoeff}, \eqref{eq:representation}, and \eqref{eq:commutativity} reproduce the Yang-Baxter equation \eqref{eq:yangbaxter} for half-braiding tensors \(z\). This relation guarantees that the coupling between the gauge field and matter field---corresponding to a discrete covariant derivative and encoded by the hopping operators---is gauge invariant, ensuring the gauge invariance of the theory. 

\begin{example}
Let us return to the example of enlarged HGW model with input UFC \(\Vec(S_3)\). In this case, the tube algebra \(\{B^{p,s,p^\ast s,s^\ast ps}\}\) reduces to the group algebra \({\CC}[S_3]\):
\[B^{q,s,q^\ast s,s^\ast qs}B^{p,r,p^\ast r,r^\ast pr} = \delta_{rq, pr}B^{p,rs,p^\ast rs,s^\ast r^\ast prs}.\]

Anyon type \(C\) transforms as
\[ D^C(B^{e, g, g, e}) \ket{J, e, \alpha} =\sum_{\beta = 1}^{2} D^\pi_{\alpha\beta}(g)\ket{J, e, \beta},
\]
which is simply the \(\pi\)-representation matrix of the group element \(g \in S_3\). This reduces to the case of traditional lattice gauge theory, where matter fields transform under unitary irreducible representations of the gauge group.

In the enlarged HGW model with \(\Vec(G)\) as input UFC, anyons carry both flux and charge dofs. The charge corresponds to representations of the gauge group, while the flux is associated with a conjugacy class of the group, transformed under adjoint actions. As a result, the anyon spectrum is significantly richer than the spectrum of irreducible representations alone as in traditional lattice gauge theory.

For example, anyon \(D\) carries the trivial representation of the symmetric group \(S_3\), yet its nontrivial flux varies within the conjugacy class \(L_D = \{s, rs, rs^2\}\). The action of the plaquette operator on \(D\)'s gauge space
\[
D^D(B^{p, s, p^\ast s, s^\ast ps}) \ket{J, p'} = \delta_{p p'} \ket{J, s^\ast ps}
\]
conjugates the fluxes of anyon \(D\). Anyon \(E\) carries the same conjugacy class \(L_E = \{s, rs, rs^2\}\), but with a nontrivial charge---specifically, the sign representation \({\tt sgn}\) of \(S_3\). In this case, the plaquette operator acts as:
\[
D^E(B^{p, s, p^\ast s, s^\ast ps}) \ket{J, p'} = \delta_{p p'}\, {\tt sgn}(s)\, \ket{J, s^\ast ps},
\]
introducing a sign factor that depends on \(s\).

In fact, in the \(\Vec(G)\) model, each anyon type \(J\) is characterized by the flux component \(L_J = [p] = \{s^\ast ps| s\in G\}\) and a charge component \(\rho\)---an irreducible representation of the centralizer subgroup \(\mathcal{Z}(p) = \{s \in G \mid sp = ps\}\) of an arbitrary element \(p \in L_J\) in the conjugacy class. The \(T\)-gauge transformation described in Eq. \eqref{eq:anyontrans}, acting on the gauge spaces of anyon \(J\), forms an induced representation of the gauge group \(G\), induced from the irreducible representation of the subgroup \(\mathcal{Z}(p)\):
\[
D^{J = ([p], \rho)} \cong {\tt Ind}_{\mathcal{Z}(p)}^G(\rho).
\]

\end{example}

We refer to the gauge transformations of the enlarged HGW model both on the gauge field defined by Eqs. \eqref{eq:tube} and on the matter fields by \eqref{eq:anyontrans} as the $T$-gauge transformations for two reasons: First, they are pertaining to the tube algebra of the model. Second, they are analogous to and direct generalizations of the traditional lattice gauge transformations \eqref{eq:conventiongaugefield} and \eqref{eq:conventionmatterfield}.

\subsection{Anyon's Measurement Operator and Hamiltonian Mass-Term}\label{sec:Hamil}

In this section, we make use of the plaquette operators \(B^{psuq}\) to define the anyon measurement operators and the Hamiltonian of our enlarged HGW model.

The tube algebra generated by \(\{B^{psup}\}\) is a regular representation (free module) of its own via its algebra multiplication. According to the Peter–Weyl theorem, one can construct specific linear combinations of the plaquette operators \(B^{psup}\). Each such combination is a projector $\Pi_P^J$ labeled by an anyon type \(J\):
\eqn[eq:measure]{
\Pi_P^J := \frac{d_J}{D_\Fus} \sum_{p \in L_J} \sum_{s, u \in L_\Fus} \sqrt{\frac{d_s d_u}{d_p}} \left( \sum_{\alpha = 1}^{n_J^p} \overline{[z^J_{su}]_{p\alpha}^{p\alpha}} \right) B_P^{psup},}
where \(D_\Fus = \sum_{x \in L_\Fus} d_x^2\) is the total quantum dimension of the input UFC \(\Fus\), and \(\bar z\) denotes the complex conjugate. These projectors \(\Pi_P^J\) commute with all elements of the tube algebra and satisfy orthogonality and completeness:
\[
[\Pi^J, B^{qsur}] = 0,\quad\Pi^J \Pi^{J'} = \delta_{J J'} \Pi^J, \quad 
\sum_{J \in L_{\mathcal{Z}(\Fus)}} \Pi^J = \mathbb{I},
\]
where \(\mathbb{I}\) is the identity operator. This constitutes a Peter-Weyl-type decomposition of the tube algebra \(\{B^{psuq}\}\). Furthermore, the subalgebra of the tube algebra spanned by
\[
\{\Pi^J B^{qsur} \}
\]
is isomorphic to the representation algebra \(\{D^J(B^{psuq})\}\) acting on anyon \(J\)'s gauge space. 

Different irreducible representations of the tube algebra are orthogonal, and $\Pi^J_P$ is also a local gauge transformation, so the gauge transformation \(\Pi^J\) in the representation \(D^J\) ats on $J$'s dyonic sectors as
\[
D^J\left(\Pi^{J'}\right) \ket{J, p, \alpha} = \delta_{J, J'} \ket{J, p, \alpha}.
\]
Therefore, projector \(\Pi_P^J\) serves as the \emph{measurement operator} of anyon of type \(J\) in plaquette $P$: While the projectors \(\Pi^J\)---elements of the tube algebra \(\{B^{psuq}\}\)---act on the gauge-field Hilbert space and project it onto subspaces, the corresponding representations \(D^{J'}(\Pi^J) = \delta_{J J'}\) act on the anyon Hilbert spaces, selecting a specific anyon type \(J\). This commensurate action leads to the \emph{topological Gauss law}:

\begin{important}
In physical states, the gauge field configurations on edges and tails uniquely reflect the anyon type (and anyon's flux type) in each plaquette.
\end{important}

Conversely, due to the topological superselection rule, superpositions of different anyon types within a single plaquette are forbidden. This imposes a physical constraint on the gauge-field states: Not all states in the kinematic Hilbert space of the gauge field are physically allowed in the topological phase. Only those states bearing an anyon of definite (including trivial) type per plaquette are admissible. 

Nevertheless, if we did allow one or more plaquettes to contain a superposition of anyons of different types, the topological phase would have been locally broken in those plaquettes, rendering the perimeter of the plaquettes a domain wall separating the broken and unbroken regions. More extremely, such as in a process of anyon condensation, certain types of anyons would be proliferated in all plaquettes, breaking the entire topological phase globally \cite{zhao2024b}. We shall discuss anyon condensation in Section \ref{sec:cond}. 

Superposition of anyons of different types in a plaquette may also occur when the topological phase is endowed with a global symmetry, turning it into a symmetry-enriched topological phase \cite{Fu2025, Fu2025b}.

We can now express the Hamiltonian of the enlarged HGW model in terms of the measurment operators as follows. A ground state of the model bears no nontrivial anyon in any plaquette. An excited state must have at least one anyon of a nontrivial type $J$ in a certain plaquette. For simplicity, we assume that all nontrivial anyons carry the same energy \(M\), then the Hamiltonian of the enlarged HGW model can be expressed in the form:
\eqn[eq:HamilMass]{
H_{\text{mass}}\ \ =\ \ M\sum_{\text{plaquettes } P}\quad \sum_{\substack{J \in L_{\mathcal{Z}(\Fus)} \\ J \neq \idm}} \Pi_P^J\ \ =\ \ M\sum_{\text{plaquettes } P} \left( \mathbb{I} - \Pi_P^{\idm} \right).}
Note that for the trivial anyon \(\idm\), we have \(L_\idm = \{1\}\), \(d_\idm = 1\), and \([z^\idm_{su}]_{\tob}^{\tob} = \delta_{su}\). The measurement operator of trivial anyon in plaquette \(P\) is
\eqn{
\Pi_P^\idm = \sum_{s \in L_\Fus} d_s B_P^{1ss1},
}
which exactly matches the traditional plaquette operator of the string-net model defined in Ref. \cite{Levin2004, Hu2018, zhao2024b, zhao2025}, up to an energy shift and definition of energy unit.

\subsection{Dyon Braiding}\label{sec:braid}

\begin{figure}\centering
\subfloat[]{\BraidingA}\hspace{25pt}
\subfloat[]{\raisebox{-24pt}{\includegraphics[width=3.3cm]{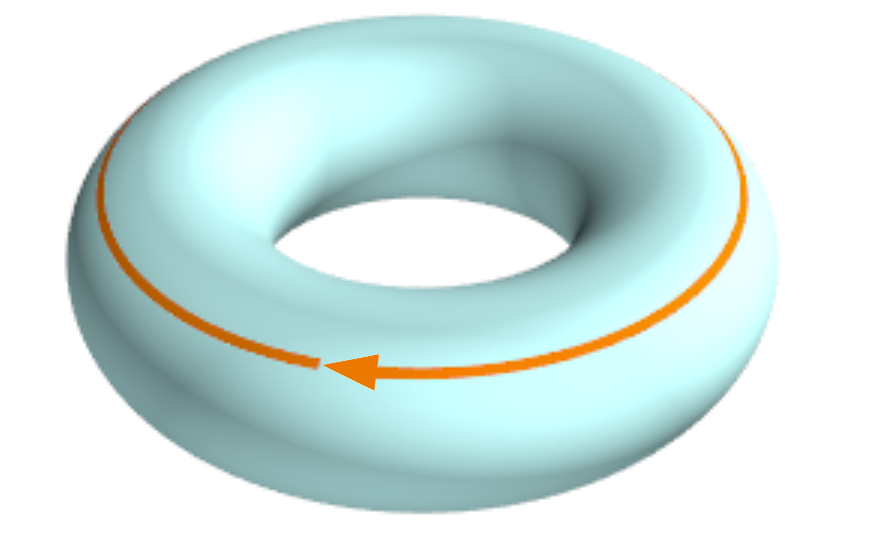}}}\hspace{25pt}
\subfloat[]{\BraidingC}\hspace{25pt}
\subfloat[]{\BraidingD}
\caption{The four types of loop operators. (a) A trivial loop operator. (b) A noncontractible loop operator on a torus. (c) A loop operator encircling a dyon. (d) A twisted loop operator.}
\label{fig:braid}
\end{figure}

In this section, we discuss the braiding of dyons in our enlarged HGW model. 

A special class of hopping operators are the \emph{loop operators}, which are defined along closed paths \(C\) whose initial and terminal plaquettes coincide, making it possible to compare the state before and after hopping along the loop. The braiding properties characteristic of topological phases are fundamentally rooted in the nontrivial action of these loop operators, which can be broadly classified into four distinct cases.

The first case concerns contractible loop paths---they enclose simply connected regions of plaquettes, i.e., free of anyons. Then, the flatness condition of topological gauge field theory renders these loop operators merely the measurement operators.

The second case concerns the situation where the entire lattice system is not simply connected (e.g, a torus or cylinder). The system then admits noncontractible loop operators that encloses no anyons at all. These operators act nontrivially on states in the Hilbert space, generating nontrivial transformations that preserve the anyon type in each plaquette. Namely, each Hilbert subspace with a fixed dyon type in each plaquette still exhibits a nontrivial topological degeneracy\footnote{For closed manifolds such as a torus, the degeneracy is $|L_\Cent{\Fus}|^{g}$, where \(g\) is the genus of the closed manifold (e.g., \(g = 1\) for a torus).}\cite{Kitaev2003a,Levin2004,Hung2012}. This topological degeneracy arises from the gauge-field Hilbert space rather than the matter fields. In this case, under the topological Gauss law, there exist multiple orthogonal gauge-field states corresponding to the same dyon distribution on the lattice.

The third case occurs when the loop \(C\) of a loop operator \(\hop_{C}^J\) encloses a dyon \((K, p, \alpha)\), which contributes a nontrivial gauge-field curvature and making the region within non-flat. In this scenario, the loop operator \(\hop_{C}^J\) acts as a transformation within the internal gauge space of anyon \(J\). If the single-anyon Hilbert space $\Hil_J^P$ is one-dimensional, this transformation yields simply a phase factor. Nevertheless, if $\Hil_J^P$ is multi-dimensional, the transformation becomes nonabelian and gives rise to richer structure; for instance, on Fibonacci anyons, such nonabelian unitary transformations can support universal quantum computation \cite{freedman2002, Bonesteel2005, Hormozi2007, Nayak2008, Li2023}.

The transformation of a dyonic sector \((J, p, \alpha)\) when it travels along a loop \(C\) encircling another dyonic sector \((K, r, \mu)\) can be described by a braiding matrix \(\mathcal{D}^J_\hop(K, r, \mu)\) acting on \((J, p, \alpha)\):
\eqn[eq:braiding]{
\hop_{C}^{J}\ket{J, p, \alpha} = \sum_{q \in L_J}\sum_{\beta = 1}^{n_J^q}\left[\mathcal{D}^J_\hop(K, r, \mu)\right]_{p\alpha}^{q\beta}\ket{J, q, \beta}.
}
This matrix \(\mathcal{D}^J_\hop(K, r, \mu)\), again, forms a representation of the loop operators \(\hop_C^{J}\) on anyon \(J\)'s internal Hilbert space \(\Hil_J^P\). 

We are particularly interested in the overlap between the initial and final states---namely, the amplitude in which the dyonic sector is preserved---when a dyon braids around another dyon through a loop path. Define the dyonic \(S\) matrix
\eqn{
S_{(J, p, \alpha), (K, q, \beta)} = \frac{d_p d_q}{\sqrt{D_{\Cent(\Fus)}}} \bra{J, p, \alpha} \mathcal{D}^J_{\mathrm{\hop}}(K, q, \beta) \ket{J, p, \alpha} = \frac{1}{D_{\Fus}}\sum_{r\in L_\Fus} d_r [z^J_{qr}]_{p\alpha}^{p\alpha}[z^K_{pr}]_{q\beta}^{q\beta}
}
that characterizes the braiding of two dyons, where \(D_{\Fus} = \sum_{p \in L_\Fus} d_p^2\) is the total quantum dimension of the output MTC \(\Cent(\Fus)\). This dyonic \(S\) matrix is in general not modular and not observable in a single topological order. Nevertheless, our previous work \cite{zhao2024} surpassed this limitation and demonstrates that (at least parts of) the entries of the dyonic \(S\) matrix can become physically observable during anyon-condensation-induced phase transitions or in a composite system of topological phases separated by gapped domain walls. As will be shown in the next section, after anyon condensation, the gauge invariance of the parent phase relating dyonic sectors is spontaneously broken to a global symmetry of the child phase, then several dyonic sectors split into topologically distinct anyons in the child phase, thereby revealing the values of the \(S\) matrix.

Two Dyons braid nontrivially because of two effects: (1) The Arharonov-Bohm effect, whereby the internal charges of anyons are transformed when encircling a nontrivial flux type \(p\) within the loop, and (2) the nontrivial braiding between two distinct fluxes \(p, q \in L_\Fus\). In general, these two effects are inseparable\footnote{It would be possible to disentangle the pure braiding of fluxes and AB effect only if the input UFC of the enlarged HGW model is a \emph{braided fusion category}\cite{Hu2018}.}. One notable exception arises when considering the braiding of dyons \((J, \tob, \alpha)\) with trivial flux type \(\tob\) around another dyon \((K, p, \mu)\). Since the trivial flux braids trivially with all other flux types, any nontrivial phase factors accumulated in the braiding process arise solely from the AB effect. In this simple scenario, upon being acted on by the loop operator, Eq. \eqref{eq:braiding} reduces to
\eqn[eq:braidingsimple]{
\hop_{C}^J\ket{J, \tob, \alpha}_Q = \sum_{q\in L_J}\sum_{\beta = 1}^{n_J^q}\ [z^J_{pp}]_{\tob\alpha}^{q\beta}\ket{J, q, \beta}_Q,
}
which depends only on the flux type \(p\) of the anyon \(K\) but not on the anyon type \(K\) per se or the \(T\)-charge sector associated with $p$.

\begin{example}
As an example, we consider braiding respectively the dyons \((\psi\bar\psi, 1)\) and \((\sigma\bar\sigma, 1)\) with other dyons in the doubled Ising phase.

\begin{itemize}
\item For dyon \((\psi\bar\psi, 1)\), the relevant half-braiding tensor components are:
\[
[z^{\psi\bar\psi}_{11}]_1^1 = [z^{\psi\bar\psi}_{\psi\psi}]_1^1 = 1,\qquad [z^{\psi\bar\psi}_{\sigma\sigma}]_1^1 = -1.
\]
By Eq. \eqref{eq:braidingsimple}, \((\psi\bar\psi, 1)\) braids trivially when encircling anyons of flux type \(1\) or \(\psi\): these are \(1\bar 1\), \(1\bar\psi\), \(\psi\bar 1\), and \(\psi\bar\psi\). When braiding around anyons with flux type \(\sigma\)---namely, \(\sigma\bar 1\), \(\sigma\bar\psi\), \(1\bar\sigma\), and \(\psi\bar\sigma\)---it acquires a phase factor of \(-1\). Hence, the dyonic \(S\)-matrix elements related to \((\psi\bar\psi, 1)\) are
\eq{&S_{(\psi\bar\psi, 1), (1\bar 1, 1)} = S_{(\psi\bar\psi, 1), (\psi\bar\psi, 1)} = S_{(\psi\bar\psi, 1), (\sigma\bar\sigma, 1)} = \frac{1}{4},\\
&S_{(\psi\bar\psi, 1), (\psi\bar 1, \psi)} = S_{(\psi\bar\psi, 1), (1\bar\psi, \psi)} = S_{(\psi\bar\psi, 1), (\sigma\bar\sigma, \psi)} = \frac{1}{4},\\
&S_{(\psi\bar\psi, 1), (\sigma\bar 1, \sigma)} = S_{(\psi\bar\psi, 1), (\sigma\bar\psi, \sigma)} = S_{(\psi\bar\psi, 1), (1\bar\sigma, \sigma)} = S_{(\psi\bar\psi, 1), (\psi\bar\sigma, \sigma)} = -\frac{\sqrt{2}}{4},}
where the total quantum dimension of the input Ising MTC is \(4\).
\item The half-braiding tensors related to dyon \((\sigma\bar\sigma, 1)\) are
\eq{[z^{\sigma\bar\sigma}_{11}]_1^1 = 1,\qquad [z^{\sigma\bar\sigma}_{\psi\psi}]_1^1 = -1,\qquad [z^{\sigma\bar\sigma}_{\sigma\sigma}]_1^1 = 0,\qquad [z^{\sigma\bar\sigma}_{\sigma\sigma}]_1^\psi = 1,}
so it braids trivially around dyons \((1\bar 1, 1)\), \((\psi\bar\psi, 1)\), and \((\sigma\bar\sigma, 1)\). When braiding around dyons with flux type \(\psi\), specifically \((1\bar\psi, \psi)\), \((\psi\bar 1, \psi)\), and \((\sigma\bar\sigma, \psi)\), it acquires a phase of \(-1\). Therefore, the dyonic \(S\)-matrix elements related to \((\sigma\bar\sigma, 1)\) are
\eq{&S_{(\sigma\bar\sigma, 1), (1\bar 1, 1)} = S_{(\sigma\bar\sigma, 1), (\psi\bar\psi, 1)} = S_{(\sigma\bar\sigma, 1), (\sigma\bar\sigma, 1)} = \frac{1}{4},\\
&S_{(\sigma\bar\sigma, 1), (\psi\bar 1, \psi)} = S_{(\sigma\bar\sigma, 1), (1\bar\psi, \psi)} = S_{(\sigma\bar\sigma, 1), (\sigma\bar\sigma, \psi)} = -\frac{1}{4},\\
&S_{(\sigma\bar\sigma, 1), (\sigma\bar 1, \sigma)} = S_{(\sigma\bar\sigma, 1), (\sigma\bar\psi, \sigma)} = S_{(\sigma\bar\sigma, 1), (1\bar\sigma, \sigma)} = S_{(\sigma\bar\sigma, 1), (\psi\bar\sigma, \sigma)} = 0.}

\item Notably, when dyon \((\sigma\bar\sigma, 1)\) braids around anyons \(\sigma\bar 1\), \(\sigma\bar\psi\), \(1\bar\sigma\), and \(\psi\bar\sigma\) with flux type \(\sigma\), it transforms into the dyon \((\sigma\bar\sigma, \psi)\). This demonstrates a nontrivial change in the internal dyonic sector induced by braiding.
\end{itemize}
\end{example}

The fourth case addresses \emph{twisting}. This occurs when a loop path is self-intersecting. Even if the loop of a dyon is contractible and encloses no anyons, the corresponding loop operator can yield a nontrivial phase---that is, the \emph{topological spin} of the dyon. Since such self-intersecting loops remain contractible, the resultant topological spins are always phase factors.

As we said before, the transformations of dyonic sectors given by braiding are generally unobservable within a single topological phase. Collecting the braidings of the \emph{dyonic sectors} of anyons results in the braidings of \emph{anyons}, which are one of the most celebrated observables of topological phases and a subject of extensive study over the past decades \cite{Wen1990a, Kitaev2006, Rowell2009}. The braidings of anyons are encoded in the modular \(S\) and \(T\) matrices of the topological phase. 

The entries of the \(S\) matrix, \(S_{JK}\), are defined as the sum of dyonic \(S\) matrix entries \(S_{(J, p, \alpha)(K, q, \beta)}\) over all possible dyonic sectors \(p, \alpha, q, \beta\) for the fixed anyon types \(J\) and \(K\). The \(T\) matrix is diagonal, with \(T_{JJ}\) defined as the topological spin of an arbitrary dyon belonging to anyon type \(J\)---all dyons of a given anyon type \(J\) have the same topological spin. The matrix elements of the \(S\) and \(T\) matrices can be directly computed from the half-braiding tensors:
\eqn{S_{JK} &= \frac{1}{\sqrt{D_{\Cent(\Fus)}}}\sum_{r\in L_\Fus}\sum_{p\in L_J}\sum_{q\in L_K}\sum_{\alpha = 1}^{n_J^p}\sum_{\beta = 1}^{n_K^q}S_{(J, p, \alpha)(K, q, \beta)} \\ &= \frac{1}{\sqrt{D_{\Cent(\Fus)}}}\sum_{r\in L_\Fus}\sum_{p\in L_J}\sum_{q\in L_K}\sum_{\alpha = 1}^{n_J^p}\sum_{\beta = 1}^{n_K^q}d_r [z^J_{qr}]_{p\alpha}^{p\alpha}[z^K_{pr}]_{q\beta}^{q\beta}\ ,}
\eqn{T_{JK} = \frac{\delta_{JK}}{d_p}\sum_{q\in L_\Fus}[z^J_{pq}]_{p\alpha}^{p\alpha}\qquad\forall p\in L_J,\ 1\le\alpha\le n_J^p\ .}

\begin{example}
The modular \(S\) matrix entries related to anyon \(\psi\bar\psi\) and \(\sigma\bar\sigma\) in the doubled Ising topological phase is
\eq{S_{\psi\bar\psi, 1\bar 1} = S_{\psi\bar\psi, 1\bar\psi} = S_{\psi\bar\psi, \psi\bar 1} = S_{\psi\bar\psi, \psi\bar\psi} = \frac{1}{4},\\
S_{\psi\bar\psi, \sigma\bar\sigma} = S_{(\psi\bar\psi, 1), (\sigma\bar\sigma, 1)} + S_{(\psi\bar\psi, 1), (\sigma\bar\sigma, \psi)} = \frac{1}{2},\\
S_{\psi\bar\psi, 1\bar\sigma} = S_{\psi\bar\psi, \sigma\bar\psi} = S_{\psi\bar\psi, \psi\bar\sigma} = S_{\psi\bar\psi, \sigma\bar 1} = -\frac{\sqrt{2}}{4},\\
S_{\sigma\bar\sigma, 1\bar 1} = S_{\sigma\bar\sigma, \psi\bar\psi} = \frac{1}{2},\quad S_{\psi\bar\psi, \psi\bar 1} = S_{\psi\bar\psi, 1\bar\psi} = -\frac{1}{4},\\
S_{\psi\bar\psi, 1\bar\sigma} = S_{\psi\bar\psi, \sigma\bar\psi} = S_{\psi\bar\psi, \psi\bar\sigma} = S_{\psi\bar\psi, \sigma\bar 1} = 0.}
One can check these modular \(S\) matrix elements coincide with the sums of the dyonic \(S\)-matrix entries with fixed corresponding anyon types.
\end{example}


\subsection{Dyon Fusion}\label{sec:fuse}

At the end of this section, and as a prelude to the next section, we discuss the fusion of two dyonic sectors.

Suppose we have two dyons, \(\ket{J_1, p, \alpha}_P\) and \(\ket{J_2, r, \gamma}_Q\), located in plaquettes \(P\) and \(Q\), respectively. If we force dyon \((J_1, p, \alpha)\) in plaquette \(P\) to hop into plaquette \(Q\) as dyon \((J_1, q, \beta)\), then---since at most one anyon can occupy a plaquette---the two dyons in plaquette \(Q\) must interact and subsequently \emph{fuse} into a single anyon:
\eqn[eq:fusion]{
&\quad\hop_E^{J_1;(p, \alpha)(q, \beta)}\quad \FusionLeft\\
&=\ \sum_{K\in L_{\Cent(\Fus)}}\sum_{s\in L_K}\sum_{\mu = 1}^{n_K^s}\ [z_{ux}^J]_{p\alpha}^{q\beta}\ G^{xqu^\ast}_{ry^\ast s}\sqrt{d_pd_s}\ \left[\begin{array}{cccc|cc}
J_1 && J_2 &&& K \\
q && r &&& s \\
\beta && \gamma &&& \mu
\end{array}\right]\quad \FusionRight\ .}
Here, the coefficients
\[
\left[\begin{array}{cccc|cc}
J && K &&& L \\
p && q &&& r \\
\alpha && \beta &&& \gamma
\end{array}\right]\quad\in\quad\CC
\]
are the Clebsch-Gordan coefficients relating the charge dofs of different anyons during anyon fusion, satisfying
\eq{
&\sum_{\substack{p\in L_{J_1}\\ q\in L_{J_2}}}\ \sum_{\alpha = 1}^{n_{J_1}^p}\ \sum_{\beta = 1}^{n_{J_1}^p}\ 
\left[\begin{array}{cccc|cc}
J_1 && J_2 &&& K \\
p && q &&& r \\
\alpha && \beta &&& \gamma
\end{array}\right]^\ast
\left[\begin{array}{cccc|cc}
J_1 && J_2 &&& K' \\
p && q &&& r' \\
\alpha && \beta &&& \gamma
\end{array}\right]\ =\ \delta_{KK'} \delta_{rr'}\delta_{\gamma\gamma'}\delta_{pqr^\ast}\delta_{JKL^\ast},}
for any \(J, K, L, p, q, r, \alpha, \beta\). Here, \(\delta_{JKL^\ast}\) denotes the fusion tensor of modular tensor category \(\Cent(\Fus)\), which is given by the \emph{Verlinde formula}\cite{verlinde1988}
\eqn{\delta_{J_1J_2K^\ast} = \sum_{M\in L_{\Cent(\Fus)}}\frac{S_{J_1M}S_{J_2M}S^{-1}_{MK}}{S_{\idm M}}\ .}
Throughout this article, we assume that for any triple \((J_1, J_2, K)\), there is at most one fusion channel fusing \(J\) and \(K\) into \(L\), namely, \(\delta_{J_1J_2K^\ast} \le 1\).

The modular square of the CG-type coefficients
\eqn[eq:fusedyon]{
&\delta_{(J_1, p, \alpha)(J_2, q, \beta)(J^\ast, r^\ast, \gamma)} := \left[\begin{array}{cccc|cc}
J_1 && J_2 &&& K \\
p && q &&& r \\
\alpha && \beta &&& \gamma
\end{array}\right]^\ast
\left[\begin{array}{cccc|cc}
J_1 && J_2 &&& K \\
p && q &&& r \\
\alpha && \beta &&& \gamma
\end{array}\right]}
yields the fusion coefficients for two \emph{dyons} \((J_1, p, \alpha)\) and \((J_2, q, \beta)\) to dyon \((K, r, \gamma)\). These dyonic fusion coefficients \(\delta_{(J_1, p, \alpha)(J_2, q, \beta)(J^\ast, r^\ast, \gamma)}\) also satisfy the Verlinde formula accompanied with the dyonic \(S\) matrix \cite{zhao2024}:
\eqn{\delta_{(J_1, p, \alpha)(J_2, q, \beta)(J^\ast, r^\ast, \gamma)} = \sum_{M\in L_{\Cent(\Fus)}}\sum_{s \in L_M} \sum_{\mu = 1}^{n_M^s} \frac{S_{(J_1, p, \alpha)(M, s, \mu)}S_{(J_2, q, \beta)(M, s, \mu)}S^{-1}_{(M, s, \mu)(J, r, \gamma)}}{S_{(\idm, \tob)(M, s, \mu)}}}
The dyonic fusion coefficients can be in general fractional and even irrational numbers \cite{zhao2024}. The sums of these coefficients reproduce the integer fusion rules for anyon types:
\eqn{\delta_{J_1J_2J_3} = \sum_{p_i\in L_{J_i}}\sum_{\alpha_i = 1}^{n_{J_i}^{p_i}}\delta_{(J_1, p_1, \alpha_1)(J_2, p_2, \alpha_2)(J_3, p_3, \alpha_3)}\in\NN.}

The matrix elements of creation operators when there have been anyons in the plquettes where the anyons are to be created can be defined likewise: 
\begin{equation}
\begin{aligned}
W_E^{J;(r, \mu)(s, \nu)}\ \FusionCreationLeft\ =\ \sum_{v\in L_\Fus}\ \sum_{m\in L_{K_1}}\ \sum_{n\in L_{K_2}}\ \sum_{\rho = 1}^{n_{K_1}^m}\ \sum_{\sigma = 1}^{n_{K_2}^n}\ \FusionCreationRight\ \times\\
[z_{uv}^J]_{r\mu}^{s\nu}\ G^{vsu^\ast}_{qy^\ast n}\ G^{v^\ast r^\ast u}_{p^\ast xm^\ast}\ \sqrt{d_ud_vd_md_n}\ \left[\begin{array}{cccc|cc}
J_1 && J &&& K_1 \\
p && r &&& m \\
\alpha && \mu &&& \rho
\end{array}\right]^\ast
\left[\begin{array}{cccc|cc}
J_2 && J &&& K_2 \\
q && s &&& n \\
\beta && \nu &&& \sigma 
\end{array}\right]\ .
\end{aligned}
\end{equation}

\subsection{Covariant Interaction Term in the Hamiltonian}\label{sec:cov}

Since hopping operators describe how anyonic matter fields couple to the gauge field and propagate on the lattice, they must be incorporated into the Hamiltonian in addition to the mass term:
\begin{equation}
H_\text{hop} = -g\sum_{\text{Edges }E}\sum_{\substack{J\in L_{\Cent(\Fus)}\\ J\ne\idm}}\hop_E^{J},
\label{eq:HamilHop}
\end{equation}
where \(g\) is the coupling constant. This interaction term not only captures the coupling between matter fields and gauge fields, but also encodes how different matter fields interact via the fusion rules (see Eq. \eqref{eq:fusion}). If two anyons are hopped into the same plaquette, the definition of the hopping operator (again, in analogy to Eq. \eqref{eq:fusion}) naturally specifies how they transform into superpositions of other anyons. 

Nevertheless, including the hopping term has two important consequences. First, it generally breaks the exact solvability of the Hamiltonian. Second, the fusion of anyons can locally break the topological phase---resulting in the formation of a defect within the plaquette or a domain wall around it. Therefore, if the coupling constant \(g\) is sufficiently small, we may neglect this term to maintain solvability.

It is also worth noting that the hopping term does not affect the ground states \(\ket{\Phi}\) of the exactly solvable Hamiltonian, since
\[
\hop_E^{J}\ket{\Phi} = \delta_{J, \idm}\ket{\Phi}
\]
as nothing can be hopped from the vacuum. So the interacting term will not break the topological order and can be neglected if \(g\) is small.

\section{Landau-Ginzburg Paradigm: Anyon Condensation in the Enlarged HGW String-Net Model as Higgs Mechanism}\label{sec:cond}

Topological phases and topological phase transitions have long been regarded as beyond the Landau-Ginzburg paradigm, in the sense that they do not involve apparent spontaneous symmetry breaking \cite{haldane1988,Wen1990a,WenTensorCat2004,senthil2004,ran2006,Wen2017}. As a gauge-invariant realization of the Landau-Ginzburg paradigm, the conventional Higgs mechanism is also believed to be unable to account for such topological phase transitions.

Nevertheless, our study of anyon condensation within the framework of the enlarged HGW model \cite{zhao2024b}---in particular, by reinterpreting the enlarged HGW model as a lattice gauge-theoretic description of topological phases---provides a powerful approach for examining topological phase transitions from a new perspective. We propose that topological phase transitions can, in fact, be understood as a generalization of the Landau-Ginzburg paradigm and the Higgs mechanism.

In this section, we formulate anyon condensation in topological phases as a Higgs mechanism, thereby incorporating topological phases transitions induced by anyon condensation back in the Landau-Ginzburg paradigm of symmetry breaking. We demonstrate that all the key mathematical structures and physical phenomena of the Higgs mechanism and Landau-Ginzburg paradigm find their counterparts in these topological phase transitions. This includes the proliferation of condensed particles in the ground state---leading to nonzero vacuum expectation values for condensed anyon field operators---apparent breaking of gauge invariance, and the emergence of Goldstone modes, which are then absorbed by the gapped gauge fields as new physical dofs. Conversely, phenomena usually thought to be unique to anyon condensation, such as the identification, splitting, and confinement of anyon types, also have natural correspondences in traditional Higgs mechanism. 

\subsection{Coherent States and Anyon Condensation}\label{sec:coherent}

During anyon-condensation-induced topological phase transitions, certain types of anyons are condensed \cite{Bais2009a,Kitaev2012,Hung2013,Eliens2013,Kong2013,Gu2014a,Burnell2018,Hu2021,zhao2022,zhao2024b}. That is, the Hamiltonian undergoes a specific evolution such that the new ground state becomes a \emph{coherent state} containing arbitrarily many condensed bosons from the parent order \cite{Hu2021,zhao2022,zhao2024b}, analogous to the Cooper-pair condensation in superconductivity phase transitions or to the Higgs condensation in the electroweak theory. Compared with the original topological superselection rules, this condensation process breaks the original topological phase, referred to as the \emph{parent phase}.

In general, proliferating a set of anyon types may break the topological invariance of the parent phase and lead to a phase with reduced symmetry, such as those described by conformal field theories \cite{Hung2025}. Nevertheless, under certain conditions---for example, when the condensed anyons braid trivially with each other—the topological invariance of the system is preserved, and the resulting phase remains a topological phase, referred to as the \emph{child phase}. As we will see in the following subsections, in the child topological phase, the set of anyon types is rearranged. The topological superselection rules are preserved with respect to these new anyon types.

In the traditional Landau-Ginzburg paradigm, such coherent states---the ground states of the child phase---are realized by adding a parameterized potential term \(V(\phi)\) for the condensed field to the Hamiltonian. Here,we do not write an explicit Higgs-like potential; instead we enforce condensation directly by adding to the Hamiltonian of the parent model with a sum of creation operators \(W_E^{J; (p, \alpha)(q, \beta)}\) for the condensed anyons \(J\) on specified edges \(E\). The coefficients are so chosen that the sum forms a set of commuting local projectors \(P_X\) over all positions \(X\):
\begin{align}
H_\Fus &\to H_\Fus -\lim_{\Lambda\to\infty}\Lambda\sum_{\text{Edges } E} P_E,\qquad\Lambda\to+\infty, \label{eq:EffHamil} \\
P_E &:= \sum_{J\in L_{\Cent(\Fus)}}\sum_{p, q\in L_J}\sum_{\alpha,\beta}\frac{\pi_J^{(p, \alpha)(q, \beta)}}{d_p d_q}\, W_E^{J; (p, \alpha)(q, \beta)}. \label{eq:CondenseHamil}
\end{align}
Here, \(\pi_J^{(p, \alpha)(q, \beta)}\in{\CC}\) are coefficients. The condensed anyons refer to those anyons \(J\) for which \(\pi_J^{(p, \alpha)(q, \beta)}\ne 0\) for some \(p,q,\alpha, \beta\), while the condensed dyons refer to those dyons \((J, p, \alpha)\) for which \(\pi_J^{(p, \alpha)(q, \beta)}\ne 0\) for some \(q, \beta\). Since the child phase is still a topological phase described by our enlarged HGW model, it is natural to require that after anyon condensation, the child model also takes a UFC \(\Sub\) as its input data. This child input UFC \(\Sub\) is a subcategory of the parent UFC \(\Fus\). This imposes the requirement that the coefficients \(\pi_J^{(p, \alpha)(q, \beta)}\) must render \(P_E\) a \emph{projector} that projects the original field configurations (valued in \(L_\Fus\)) onto the set \(L_\Sub\) of simple objects of the child UFC \(\Sub\), a \emph{subcategory} of \(\Fus\). 

For finite \(\Lambda\), Hamiltonian \eqref{eq:EffHamil} is not exactly solvable, representing a non-topological intermediate dynamical process and necessitating numerical calculation \cite{schulz2013,schulz2016,shi2024}. For \(\Lambda \to \infty\), projector \(P_E\) ensures that the new ground states are \(+1\) eigenstates of \(P_E\)---the sum of condensed anyons' creation operators \(W_E^{J;pq}\), making the new ground states coherent states with arbitrarily many condensed anyons \(J\) throughout the lattice. The resultant child order after anyon condensation is described by the child HGW model, with the child Hilbert space \(\Hil_{\rm Child}\) and child Hamiltonian \(H_{\rm Child}\) obtained by applying projector \(P_E\) to those in the parent model \cite{zhao2022}:
\eqn{
\Hil_{\rm Child} = \Bigg[\prod_{\text{Edges }E} P_E\Bigg]\Hil_\Fus,
\qquad H_{\rm Child} = \Bigg[\prod_{\text{Edges }E} P_E\Bigg]H_\Fus\Bigg[\prod_{\text{Edges }E}& P_E\Bigg],\\
\ket{\text{GS}}_\Sub = \Bigg[\prod_{\text{Edges }E}P_E\Bigg]\ket{\text{GS}}_\Fus,\qquad
\ket{J_\Sub, \psi'} = \Bigg[\prod_{\text{Edges }E}P_E\Bigg]\ket{J_\Fus, \psi}.
}
Hamiltonian \(H_{\rm Child}\) is exactly solvable up to an irrelevant global scalar factor arising from the projection. The spectrum of the child model is directly determined by projector \(P_E\) \eqref{eq:CondenseHamil}, which projects the ground states \(\ket{\text{GS}}_\Fus\) and excited states \(\ket{J, \psi}_\Fus\) of the parent model into those of the child model \cite{zhao2022}, where \(J_\Fus\) and \(J_\Sub\) denote the anyon types in the parent and child model, respectively, and \(\psi\) and \(\psi'\) corresponds to certain internal states of these two anyon type. 

Now we show how to calculate the projectors \(P_E\). In general, it is challenging to solve the projector condition \(P_E^2 = P_E\), where \(P_E\) is the sum of creation operators for all condensed anyons. The Hamiltonian for anyon condensation has been solved in certain cases before \cite{zhao2022, lin2023,christian2023,delcamp2024}. In Ref. \cite{zhao2024}, we devised a more efficient algorithm for determining these projectors. We showed that arbitrary bosonic dyon condensations are equivalent to certain \emph{fluxon condensation} in certain equivalent HGW model. So it is sufficient to study the condensations of dyon sectors with trivial flux type \(\tob\): For these dyons, each basis state of the gauge field is an eigenstate of the creation operator:
\eqn[eq:FluxCreate]{
&W_E^{J;(\tob, \alpha)(\tob, \beta)}\ \ExcitedFlux\ =\ &[z^J_{jj}]_{\tob\alpha}^{\tob\beta}\ \ExcitedFlux,}
where \(j\) is the dof on edge \(E\). Thus, solving for the projectors \(P_E\) reduces to considering creation operators that generate anyons with trivial flux types. Such kind of anyon condensation is called \emph{fluxon condensation}.

\begin{important}
We first select a subset \(L_\Sub \subset L_\Fus\) of simple objects that is closed under fusion: For any \(a, b \in L_\Sub\), \(\delta_{abc} \ne 0\) only if \(c \in L_\Sub\). The transition from the parent model to the child model involves gapping out those dofs not in \(L_\Sub\), such that the condensation projector \(P_E\) (Eq. \eqref{eq:CondenseHamil}) becomes
\begin{equation}
P_E^{\Sub|\Fus}\ket{\psi} = \delta_{j \in L_\Sub}\ket{\psi},
\label{eq:FluxonCondensation}
\end{equation}
where \(\delta\) is the Kronecker symbol, and \(j_E\) is the dof on edge \(E\) in the basis state \(\ket\psi\). The operator \(P_E^{\Sub|\Fus}\) is diagonal due to Eq. \eqref{eq:FluxonCondensation}. It can be shown that for any enlarged HGW model with input UFC \(\Fus\), the number of such creation operators \(W_E^{J; {\tob}{\tob}}\) always equals the number of simple objects. Thus, the system of linear equations indexed by simple objects \(j \in L_\Fus\),
\eqn{
\sum_{J\in L_{\Cent(\Fus)}}\ \sum_{\alpha, \beta = 1}^{n_J^\tob} \pi_J^{(\tob, \alpha)(\tob, \beta)} [z^J_{jj}]_{\tob\alpha}^{\tob\beta} = \delta_{j\in L_\Sub}
}
always has a unique solution for any subset \(L_\Sub\) and can be solved straightforwardly.
\end{important}

The anyon types \(J\) with \(\pi_J^{(\tob, \alpha)(\tob, \alpha)} \ne 0\) are those condensed during fluxon condensation. We denote the set of all condensed anyons as \(L_\text{Cond}\), a subset of \(L_{\Cent(\Fus)}\). All types of fluxon condensation are bosonic, since fluxons must be bosons with trivial self and mutual statistics. This is because the fluxon creation operators \(W_E^{J;{\tob\alpha}{\tob\beta}}\) are diagonal and hence commute. 

If the condensed anyon \(J\) with flux type \(\tob\) has a multidimensional \(T\)-charge space with dimension \(n_J^\tob > 1\), there may be multiple dyonic sectors of the same anyon type \(J\) condenses 	simultaneously, called condensed sectors of anyon type \(J\). The number \(n_{\text{Cond}}^J\) of condensed sectors of the same anyon type \(J\) is called the condensation \emph{multiplicity} of condensed anyon \(J\) during anyon condensation. Arrange the condensation coefficients \(\pi_J^{(\tob, \alpha)(\tob, \beta)}\) for fixed anyon type \(J\) into a matrix:
\[
\Pi^J = \begin{pmatrix}\pi_J^{(\tob, 1)(\tob, 1)} & \pi_J^{(\tob, 1)(\tob, 2)} & \cdots & \pi_J^{(\tob, 1)(\tob, n_J^\tob)} \\ \pi_J^{(\tob, 2)(\tob, 1)} & \pi_J^{(\tob, 2)(\tob, 2)} & \cdots & \pi_J^{(\tob, 2)(\tob, n_J^\tob)} \\ \vdots & \vdots & \ddots & \vdots \\ \pi_J^{(\tob, n_J^\tob)(\tob, 1)} & \pi_J^{(\tob, n_J^\tob)(\tob, 2)} & \cdots & \pi_J^{(\tob, n_J^\tob)(\tob, n_J^\tob)} \\ \end{pmatrix}.
\]
Then, we have
\eqn{n_\text{Cond}^J = {\tt rank}(\Pi^J).}
We can diagonalize the condensation matrix via a unitary transformation matrix \(U_J\):
\[U_J\Pi^JU_J^\dagger = {\tt diag}\{\pi_J^1, \pi_J^2, \cdots, \pi_J^{n_\text{Cond}^J}, 0, 0, \cdots, 0\},\]
where \(\pi_J^i\in\CC, 1\le i\le n_\text{Cond}^J\). Then, the \(i\)-th condensed sector of anyon \(J\) is
\eqn{\ket{J, i}_\text{Cond} = \sum_{\alpha = 1}^{n_J^\tob}[U_J]_{i\alpha}\ket{J, \tob, \alpha},}
where \(1\le i \le n_\text{Cond}^J \le n_J^\tob\).

After fluxon condensation, the set of simple objects \(L_\Fus\) of the parent UFC, which serves as the gauge-field dofs in the parent model, splits into two sets: \(L_\Sub\), the remaining gauge field dofs in the child model after anyon condensation, and \(L_\G = L_\Fus \setminus L_\Sub\), consisting of those dofs that become infinitely gapped. The basic dofs on the edges and tails of the child enlarged HGW model remain unchanged but are only restricted to values in \( L_\Sub \). Since \( L_\Sub \) is simply a subset of \( L_\Fus \), the simple objects in \( L_\Sub \) must retain the same \(\delta\), \(d\), and \(G\) functions as those in \( L_\Fus \). These simple objects in \( L_\Sub \), equipped with the same \(\delta\), \(d\), and \(G\) functions, define the input UFC \(\Sub\) of the child model. Mathematically, such a UFC \(\Sub\) is a \emph{full subcategory} of the parent input UFC \(\Fus\). The topological properties of the child order are encapsulated by the UMTC \(\Cent(\Sub)\). 

\begin{important}
After fluxon condensation, the gauge structure of the enlarged HGW lattice gauge theory undergoes spontaneous symmetry breaking, reducing the original (larger) gauge UFC \(\Fus\) to a smaller gauge UFC \(\Sub\).
\end{important}
\begin{example}
Let us return to the example of \(\Vec(S_3)\), where \(S_3 = \langle r, s \mid r^3 = s^2 = (rs)^2 = e \rangle\). The group \(S_3\) has five nontrivial subgroups: \(\{e\}\), \(\{e, s\}\), \(\{e, rs\}\), \(\{e, sr\}\), and \(\{e, r, r^2\}\), each of which corresponds to a full subcategory of \(\Vec(S_3)\).

There are three distinct nontrivial fluxon condensation scenarios in the \({\tt Vec}(S_3)\) model:
\begin{enumerate}
\item Gauge structure broken to \({\ZZ}_3 = \{e, r, r^2\}\). The condensation projector is
\[
P_E = \frac{W_E^{A;{e}{e}} + W_E^{B;{e}{e}}}{2}.
\]
Here, only the nontrivial dyon \(B\) condenses during anyon condensation.

\item Gauge structure broken to \({\ZZ}_2 = \{e, s\}\). The condensation projector is
\eq{\quad
P_E = &\frac{W_E^{A;ee} + W_E^{C;(e1)(e1)} + W_E^{C;(e2)(e2)} + W_E^{C;(e2)(e1)} +W_E^{C;(e2)(e1)}}{3}.}
Although both dyonic sectors \((C, e, 1)\) and \((C, e, 2)\) have nontrivial condensation coefficients, the condensation matrix
\[\Pi_C = \frac{1}{3}\begin{pmatrix}1 & & 1 \\ 1 & & 1\end{pmatrix}\]
is degenerate, so only one dyonic sector of \(C\)---namely, \(\ket{C, e, 1} + \ket{C, e, 2}\)---condensed, while the other sector \(\ket{C, e, 1} - \ket{C, e, 2}\) becomes a nontrivial anyonic excitation in the child model. We have \(n_\text{Cond}^C = 1\). 

The other two \({\ZZ}_2\) subgroups, \(\{1, sr\}\) and \(\{1, rs\}\) correspond to condensing other sectors of anyon \(C\).

\item Gauge structure broken to the trivial group. In this case, the condensation projector is
\eq{
P_E = &\frac{W_E^{A;ee} + W_E^{B;ee} + 2W_E^{C;(e1)(e1)} + 2W_E^{C;(e2)(e2)}}{6}.
}
The coefficient matrix \(\Pi_J = \tfrac{1}{3}{\tt diag}(1, 1)\) is full-rank, so both dyon sectors of anyon \(C\)---\(\ket{C, e, 1}\) and \(\ket{C, e, 2}\)---are condensed.
\end{enumerate}
\end{example}

\begin{example}
Let us consider another example: fluxon condensations in the enlarged HGW Ising model. The Ising UFC has two full subcategory, so the Ising model admits two fluxon condensations:
\begin{enumerate}
\item One full subcategory has two simple objects valued in \(\ZZ_2 = \{1, \psi\}\). The corresponding condensation projector is
\[
P_E = \frac{W_E^{1\bar{1};\,{\tob}{\tob}} + W_E^{\psi\bar{\psi};\,{\tob}{\tob}}}{2}.
\]
The resultant child model is the $\ZZ_2$ toric code HGW model.

\item Another full subcategory is the trivial UFC. The corresponding projector is
\[
P_E = \frac{W_E^{1\bar{1};\,{\tob}{\tob}} + W_E^{\psi\bar{\psi};\,{\tob}{\tob}} + 2W_E^{\sigma\bar{\sigma};\,{\tob}{\tob}}}{4}.
\]
It is known that anyon \(\sigma\bar{\sigma}\) possesses two flux types: \(1\) and \(\psi\). The dyonic sector with trivial flux type \(1\) is condensed, whereas the sector with nontrivial flux type \(\psi\) is, as we will see, confined after anyon condensation.
\end{enumerate}
\end{example}

\begin{example}
There is a unique fluxon condensation in the Fibonacci model, breaking the Fibonacci model to the trivial model. The subcategory is \(L_\Sub = \{1\}\), and the condensation projector is
\[P_E = \frac{W_E^{1\bar 1;(1, 1)(1, 1)} + \phi^2W_E^{\tau\bar\tau; (1, 1)(1, 1)}}{1 + \phi^2}.\]
\end{example}

\subsection{Rearrangement of Anyon Types}\label{sec:rearrangement}

\subsubsection{Splitting}\label{sec:splitting}

As discussed in the previous subsection, the proliferation of condensed anyons breaks the topological invariance. Analogous to the Higgs mechanism, the spontaneous breaking of gauge invariance from the parent gauge structure \(\Fus\) to the child gauge structure \(\Sub\) results in a \emph{rearrangement} of anyon types. After this rearrangement, the new set of anyon types---now labeled by \(L_{\Cent(\Sub)}\)---obey the topological superselection rule in the child model and thus preserve the new child topological phase.

The first effect of this rearrangement is the \emph{splitting} of anyon types. As discussed before, a physical state has a unique observable---the anyon type \(J\) in each plaquette---although it may possess a collection of internal dyonic sectors \(\ket{J, p, \alpha}_P\). These dyonic sectors are purely gauged: the coupling between the anyon and the gauge field does not preserve individual dyonic sectors. Namely, when a dyon \(\ket{J, p, \alpha}_P\) hops from plaquette \(P\) to a neighboring plaquette \(Q\) across some edge \(E\), it may be transformed into another dyonic sector \(\ket{J, q, \beta}_Q\). with nonzero hopping amplitude
\[
[z_{jk}^J]_{p\alpha}^{q\beta} \ne 0.
\]

This is no longer the case when the gauge invariance is broken, i.e., when certain gauge-field configurations in \(L_\G\) become gapped. In this case, there exist two dyonic sectors \(\ket{J, p, \alpha}\) and \(\ket{J, q, \beta}\) that can only be related via gapped gauge dofs in \(L_\G\) (even if \(p = q\)), but not by the child gauge dofs in \(L_\Sub\):
\[
\forall j, k \in L_\Sub,\qquad [z_{jk}^J]_{p\alpha}^{q\beta} = 0.
\]
Therefore, these two dyonic sectors can no long be regarded as  gauge sectors of a single anyon type and can never be transformed into each other via coupling to the child gauge field. Rather, they should be considered as distinct anyon types. This phenomenon is \emph{splitting}.

A similar splitting phenomenon occurs in the electroweak phase transition. For example, the left-handed neutrino \(\nu_L^e\) and the left-handed electron \(e_L\) are intrinsically two isospin components of the \({\tt SU}(2)_L\) gauge-covariant lepton doublet,
\[
\ell = \begin{pmatrix}\nu_L^e \\ e_L\end{pmatrix},\ \ 
\mathcal{L}_\ell = \ell^\dagger\Bigg(\ii\slashed\partial
- \frac{g'}{2}Y_W\slashed B
- \frac{g}{2}\sum_{a = 1, 2, 3}\slashed W^a\sigma^a
\Bigg)\ell,
\]
where \(Y_W\) is the weak hypercharge (\(-1\) for left-handed leptons and \(+\tfrac{1}{3}\) for left-handed quarks), \(g, g'\in\mathbb{R}\) are the coupling constants, \(\sigma^a\) are Pauli matrices, and \(B_\mu, W^a_\mu\) are the \({\tt U}(1)_Y\) and \({\tt SU}(2)_L\) gauge bosons. After spontaneous symmetry breaking, three components of the gauge fields are gapped, giving rise to the massive \(W^\pm_\mu\) and \(Z_\mu\) bosons, while one component remains massless---the photon field\footnote{\(W^\pm\) bosons are combinations \(W^\pm_\mu = (W^1_\mu \mp \ii W^2_\mu)/\sqrt{2}\), representing the particles with definite electric charge. The remaining neutral gauge bosons mix to form the massless photon \(A_\mu = \cos\theta_\text{W} B_\mu - \sin\theta_\text{W} W^3_\mu\) and the massive (gapped) \(Z_\mu = \cos\theta_\text{W} W^3_\mu + \sin\theta_\text{W} B_\mu\), where \(\theta_\text{W} = \arccos\frac{g}{\sqrt{g^2 + g'^2}}\) is the Weinberg angle.}. This symmetry breaking leads to the splitting of the two isospin components: the \(+\tfrac{1}{2}\) isospin component becomes the left-handed neutrino, while the \(-\tfrac{1}{2}\) isospin component becomes the left-handed electron. The latter can subsequently combine with the right-handed electron to form a Dirac lepton---the electron.

\begin{example}
In the anyon condensation from the \(\Vec(S_3)\)model to the \(\Vec(\ZZ_3)\) model, the gauge-field dofs \(s, sr, rs\) become gapped. This leads to the following splitting phenomena:
\begin{itemize}
\item The two dyonic sectors \(\ket{C, e, 1}\) and \(\ket{C, e, 2}\) of anyon \(C\) are split and become the two chargeons \(e\) and \(e^2\) in the \(\Vec(\ZZ_3)\) model, respectively. See Appendix for details.
\item The \(F\), \(G\), and \(H\) anyons are also split because their flux types \(r, r^2\) commutes with the residual child dofs \(e, r, r^2\).
\item Anyon \(A, B, D, E\) do not split.
\end{itemize}
\end{example}

\subsubsection{Confinement}\label{sec:confinement}

After fluxon condensation, the basic dofs on the tails of the child string-net model are restricted to values in \(L_\Sub\). By the magnetic Gauss law \eqref{eq:magauss}, the residual anyons in the child model can only have fluxes taking values in \(L_\Sub\); no additional constraints are imposed on the anyons. 
\begin{important}
A parent dyon sector \((J, p, \alpha)\) is \emph{confined} in the child model if the flux type \(p\) is not among the child simple objects in \(L_\Sub\).
\end{important}

An anyon may possess two dyonic sectors that behave differently under anyon condensation: one sector with flux type in \(L_\Sub\) can remain free in the child model, while another sector with a gapped flux type in \(L_\G\) becomes confined. These sectors necessarily split because \(L_\Sub\) is closed under fusion: A dyonic sector with flux in \(L_\Sub\) can never transform into one with flux in \(L_\G\) via hopping across edges carrying gauge connections in \(L_\Sub\).

\subsubsection{Condensation}\label{sec:condensation}

Mathematically, there are two complementary perspectives to describe a phase transition. On the one hand, one can focus on the gauge structure: namely, to which child gauge structure the parent gauge structure is broken. On the other hand, one can focus on the matter fields that couple to the gauge fields and are condensed. These two perspectives must be mathematically consistent, but physically, it is the condensation of matter fields that drives the spontaneous breaking of gauge invariance.

Since the new ground states are coherent states populated by arbitrarily many condensed anyons, the condensed anyons must braid trivially around any allowed curvature taking values in \(L_\Sub\). 
Conversely, the condensed anyons form the new vacuum of the child topological phase, which must be gauge invariant under the residual gauge structure described by the UFC \(\Sub\). That is, all condensed dyons \(\ket{J, \tob, \alpha}\) must be able to hop trivially across the entire lattice.  Therefore,
\begin{important}
Given a spontaneous breaking of gauge invariance from the parent gauge structure \(\Fus\) to the child gauge structure \(\Sub\), a parent dyonic sector \((J, \tob, \alpha)\) is \emph{condensed} in the child model if
\[\forall p\in L_\Sub,\quad [z^J_{pp}]_{\tob\alpha}^{\tob\alpha} = 1.\]
Conversely, given a set of condensed dyonic sectors \(\ket{J, \tob, \alpha}\) (when physically allowed), a parent gauge dof \(p\in L_\Fus\) remains in the child phase if all condensed \(\ket{J, \tob, \alpha}\) hops nontrivially across it:
\[\forall\text{ Condensed dyon }(J, \tob, \alpha),\quad [z^J_{pp}]_{\tob\alpha}^{\tob\alpha} = 1.\]
\end{important}

This phenomenon of condensation is analogous to the Higgs boson condensation in the electroweak phase transition. The Higgs boson \(\phi\) is a complex scalar doublet,
\[
\phi = \begin{pmatrix}\phi_1 + \ii\phi_2 \\ \phi_0 + \ii\phi_3\end{pmatrix}.
\]
The four components of \(\phi\) are gauged and couple to the \({\tt U}(1)_Y \times {\tt SU}(2)_L\) gauge fields:
\[
\mathcal{L}_\phi = \left|\left(\ii\slashed\partial - g' \slashed B - \frac{g}{2}\sum_{a=1}^3 \slashed W^a \sigma^a\right)\phi\right|^2 + V(\phi^\dagger \phi).
\]
After Higgs condensation, the vacuum expectation value \(\langle\phi\rangle\) of the Higgs field becomes nonzero. One can choose the gauge fixing
\[
\langle\phi_0\rangle = v, \qquad \langle\phi_1\rangle = \langle\phi_2\rangle = \langle\phi_3\rangle = 0.
\]
Closely in parallel with what happens in the enlarged HGW model, in this situation, three gauge bosons become gapped,
\eq{
W^+_\mu = &W^1_\mu - \ii W^-_\mu,\qquad W^-_\mu = W^+_\mu + \ii W^-_\mu,\qquad Z_\mu = \cos\theta_\text{W} W^3_\mu + \sin\theta_\text{W} B_\mu,
}
each corresponding to gauge field transformation generators \(\sigma^1 - \ii\sigma^2,\, \sigma^1 + \ii\sigma^2,\, \sigma^3 + \mathbb{I}\), all of which act nontrivially on the new vacuum with condensed Higgs field \(\langle\phi\rangle = (0\ v)^T\), where \(\theta_W = \arctan(g'/g)\). The remaining gauge field dof, \(A_\mu = \cos\theta_\text{W}B_\mu - \sin\theta_\text{W}W^3_\mu\), acts trivially on the condensed Higgs field under the gauge transformation generator \(\sigma^3 - \mathbb{I}\) and thus remains free in the child phase with \({\tt U}(1)_\text{EM}\) gauge invariant, whose excitations are known as photons.

Additionally, the Higgs field can still be excited around the new vacuum as the Higgs boson that was observed experimentally:
\[
h = \phi - \begin{pmatrix}0 \\ v\end{pmatrix}.
\]
This scalar excitation is a particle with mass approximately \(125.11\,\mathrm{GeV}\) in our low-energy universe. The counterpart of this Higgs particle in our model---defined as \emph{local excitations}---will be discussed in Section \ref{sec:local}.

\subsubsection{Identification}\label{sec:identification}

The last phenomenon is the \emph{identification} of anyon types. Since anyon types are coupled through \emph{fusion}, after anyon condensation, the condensed anyons proliferate to form the new vacuum of the child phase. Hence, two anyon types in the parent phase may become indistinguishable in the child phase, as they can transform into one another while propagating in the lattice and interacting with the vacuum background fulfilled with condensed anyons. Note that this is another form of transformation of anyon's internal Hilbert space, distinct from those induced by coupling to the gauge field. It is a \emph{nontopological} interaction from the perspective of the parent topological phase, as it alters anyon types.

\begin{important}
Two dyonic sectors $\ket{J_1, p, \alpha_1}$ and $\ket{J_2, p, \alpha_2}$ in the parent phase can be \emph{identified} as a single child dyonic sector $\ket{J_\text{Child}, p, \beta}$ if they are related by fusion with some parent condensed dyonic sector $\ket{J_\text{Cond}, \tob, \alpha_0}$.
\end{important}

As noted earlier, the parent topological phase is broken during anyon condensation because the original topological superselection rules are violated: the ground state (or excited states) of the new Hamiltonian \eqref{eq:EffHamil} becomes a superposition of states with definite parent anyon types in each plaquette. Nevertheless, condensation and identification processes rearrange the anyon types in such a way that these superposed parent anyon types are now regarded as the same child anyon type. Thus, the topological superselection rule is restored in terms of the anyon types of the child phase with input UFC $\Sub$.

This phenomenon also has an analogue in electroweak phase transition—specifically, in the formation of the Dirac electron. Three matter fields—the left-handed lepton doublet $\ell$, the Higgs boson $\phi$, and the right-handed electron $e_R$—interact via the Yukawa coupling:
\[
\mathcal{L}_\text{Yukawa} = Y_e\, \ell^\dagger \phi e_R + \text{c.c.}, \qquad
\ell = \begin{pmatrix}\nu_L^e \\ e_L\end{pmatrix},\qquad 
\phi = \begin{pmatrix}\phi_1 + \ii \phi_2 \\ \phi_0 + \ii \phi_3 \end{pmatrix},
\]
where $Y_e \in \mathbb{R}$ is the Yukawa coupling constant\footnote{In fact, in the standard model, there are three generations of particles and the Yukawa coupling constant is a \(3\times 3\) matrix.}. After Higgs condensation, we can choose the gauge fixing $\phi_0 = v$ and $\phi_1 = \phi_2 = \phi_3 = 0$, so that the Yukawa interaction reduces to
\[
\mathcal{L}_\text{Yukawa} \to \mathcal{L}_e = Y_e v\, \bar{e}_L e_R + + \text{c.c.}
\]
This is precisely the electron mass term, with $m_e = Y_e v$. In this process, the left-handed electron $e_L$ decouples from the left-handed neutrino $\nu_L^e$ (the splitting phenomenon), and subsequently combines with the right-handed electron $e_R$ to form the Dirac electron:
\[
e = \begin{pmatrix} e_L \\ e_R \end{pmatrix},\qquad \bar e = \begin{pmatrix} e_R^\dagger & e_L^\dagger\end{pmatrix},\qquad \mathcal{L}_e = m_e\bar ee.
\]

\subsection{Gapped Gauge-Field Dofs}\label{gapedof}

In the language of the Landau-Ginzburg paradigm and the Higgs mechanism, conventional gauge field theory describes a spontaneous breaking of the parent gauge Lie group \(G\) down to a subgroup \(H\). As a result, certain gauge bosons become gapped: they acquire finite mass and mediate only short-range (Yukawa-type) interactions. This mechanism underlies phenomena such as the short-ranged nature of weak interactions in our low-energy universe, as well as the free movement of electrons and the Meissner effect (expulsion of magnetic fields) in superconductors. 

Abstractly, the enlarged HGW model shows an analogous phenomenon: its input UFC can be spontaneously broken from a parent UFC \(\Fus\) to a child UFC \(\Sub\). Nevertheless, in our previous section \ref{sec:cond}, to ensure exact solvability in the lattice model, we assumed an infinite mass \(\Lambda\) of the gapped gauge-field configurations. This assumption, while convenient for calculation, is not strictly physical and requires a slight modification for reality.

Physically, we cannot truly take the limit \(\Lambda \to \infty\), and there are always sources of noise---such as quantum or thermal fluctuations---that can excite the system from the ground state to higher-energy states. Therefore, actually the gapped gauge-field dofs in \(L_\G\) can still manifest in the enlarged HGW lattice gauge theory. When the parameter \(\Lambda\) is large (compared with the mass constants \(M\) and coupling constants \(g\) of the parent Hamiltonian), each gapped dof on the edges in \(L_\G\) acquires a mass of order \(\Lambda\), but they can still locally exist. This situation is analogous to the Cooper-pair condensation, where the photon acquires a finite mass and becomes gapped, or to the electroweak phase transition, where the \(W\) and \(Z\) bosons acquire finite masses and become gapped excitations. As \(\Lambda\) increases, the population of gapped gauge-field dofs on edges decreases, and in the limit \(\Lambda \to \infty\), these excitations are completely suppressed. 

\begin{figure}
\subfloat[Allowed case.]{\GappedA}\hspace{30pt}
\subfloat[Forbidden case.]{\raisebox{-23pt}{\includegraphics[width=3.cm]{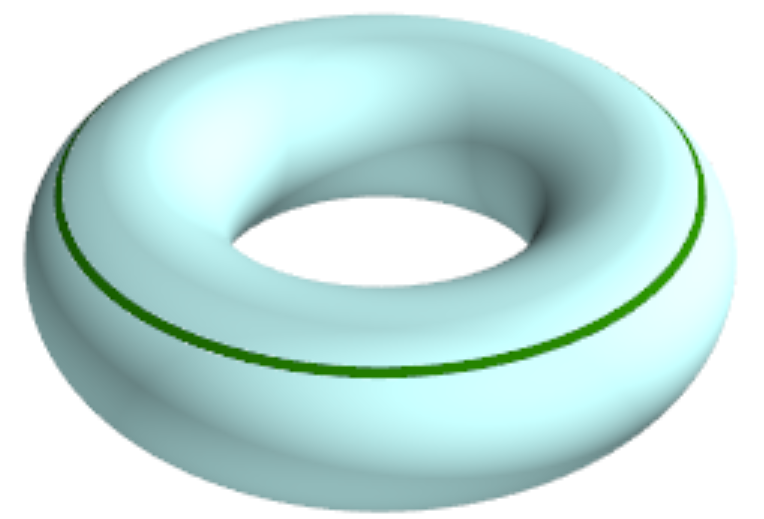}}}\hspace{30pt}
\subfloat[Forbidden case.]{\GappedC}\hspace{30pt}
\subfloat[]{\raisebox{-23pt}{\includegraphics[width=3.cm]{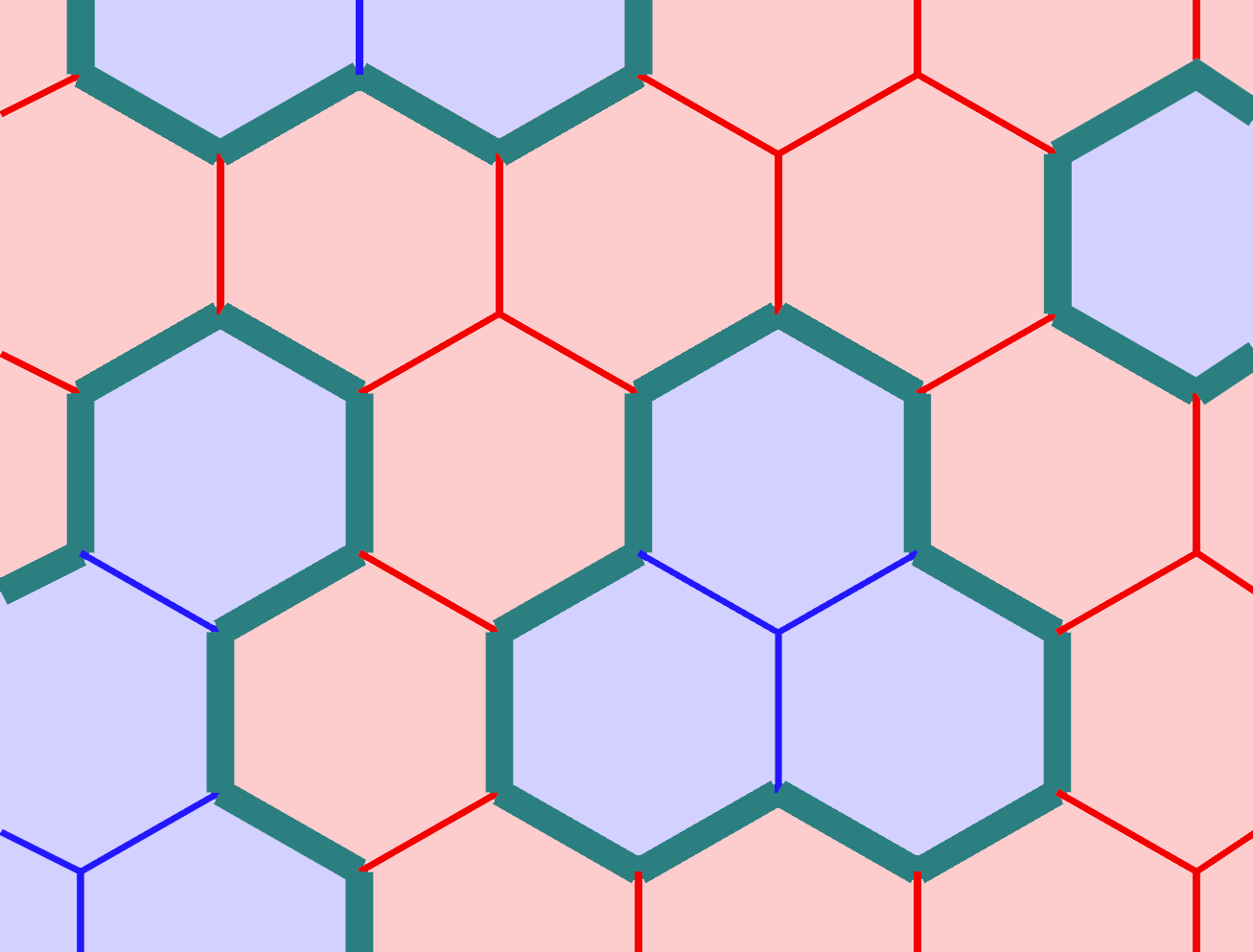}}}
\caption{Gapped dofs can only form contractible loops (a) on the lattice after anyon condensation. The green edges refer to gapped degrees of freedom. Noncontractible loops (b) or open lines (c) are forbidden after anyon condensation. This contractible gapped loops separate domains on the lattice, as depicted in (d).}
\label{fig:gapped}
\end{figure}

Hence, the gapped configurations manifest on the lattice under three constraints:
\begin{itemize}
\item Gapped gauge-field dofs cannot appear on tails because the gauge curvature within each plaquette must take values only in \(L_\Sub\). This restriction follows from that in the thermodynamic limit and beyond the critical point, the condensed anyons have been proliferated---even if \(\Lambda\) is finite.

\item On the other hand, since the residual gauge dofs in \(L_\Sub\) are closed under fusion, all gapped gauge-field configurations must form \emph{loops} along the edges of the lattice. This does not contradict our previous restriction: condensed anyons necessarily cross such loops an even number of times and return to the initial plaquette, which reduces to a trivial gauge transformation due to the flatness condition of the topological gauge field and that condensed anyons braid trivially with all unconfined anyons.

\item These loops of gapped gauge-field dofs in \(L_\G\) must be contractible. In the limit \(\Lambda \to \infty\), such loops are energetically progressively suppressed and effectively eliminated from the low-energy theory.  See Fig. \ref{fig:gapped}.
\end{itemize}

As a result,
\begin{important}
The gapped gauge-field configurations, forming contractible loops (called \emph{gapped loops}) on the lattice, serve as boundaries separating distinct \emph{domains}. Within each domain, the edges and tails have dofs restricted in \(L_\Sub\), and thus describe a pure topological phase given by the UMTC \(\Cent(\Sub)\), coupled to a gauge field described by the UFC \(\Sub\).
\end{important}

One of the most striking consequences of the Higgs mechanism is the emergence of Goldstone bosons. While the theory always retains gauge invariance under the full parent gauge group, the ground state of the child phase (partially) hides this gauge invariance. The dofs corresponding to gauge transformations that connect different ground states manifest as new scalar fields, \(\theta^a\), known as \emph{Goldstone bosons}. The covariant derivatives \(\partial_\mu\theta^a\) are then absorbed as longitudinal modes by the gauge fields, rendering them now massive. In the unbroken (massless) phase, these longitudinal modes are pure gauge.

As we will see, all of the above phenomena have analogues in our enlarged HGW model, albeit with certain modifications because our gauge structures are discrete unitary fusion categories (UFCs) \(\Fus\) rather than continuous Lie groups. On the other hand, a general UFC \(\Fus\) extends the notion of a gauge group \(G\), allowing for the emergence of novel structures—such as generalized symmetries—in the enlarged HGW model.
\begin{important}
\begin{itemize}
\item The broken gauge invariance becomes a global symmetry of the child model. 
\item Each domain carries a global symmetry charges that serve the same role analogous to the Goldstone bosons \(\theta^a\) in electroweak phase transition, which label different ground states at different positions. We refer to these symmetry charges as the \emph{Goldstone modes} within each domain.
\item The gapped loops separating different domains serve as carriers of global symmetry charge operators. 
\item The differences in Goldstone modes between adjacent domains are absorbed by the gapped dofs along the gapped loops, resulting in an enlarged set of dofs associated with these massive gauge-field configurations.
\end{itemize}
\end{important}

\subsection{Order Parameter and Goldstone Modes}\label{goldstone}

Let us define the goldstone modes in each plaquette in the enlarged HGW model after anyon condensation. 

It is straightforward to identify the order parameters that signal an anyon condensation process. Specifically, the order parameter is given by the creation operators of nontrivial condensed anyons. One subtlety is that the Goldstone mode is gauge covariant, so a choice of gauge fixing is necessary before defining the order parameter. We may fix a reference plaquette \(P_0\) and take its symmetry sector as the ``original point'' for the symmetry transformation. Accordingly, we define the order parameter at plaquette \(P\) as
\begin{equation}
\mathcal{O}_{P_0}(P, \Lambda) = \frac{1}{1 - \Big(\pi_{\idm}^{{\tob}{\tob}}\Big)^n}\Bigg\langle\Phi(\Lambda)\Bigg|\prod_{i=1}^n P_{E_i} - \Big(\pi_{\idm}^{{\tob}{\tob}}\Big)^n\mathbb{I}\Bigg|\Phi(\Lambda)\Bigg\rangle\ ,
\label{eq:orderpara}
\end{equation}
where \((E_1E_2\cdots E_n)\) is a path from \(P_0\) to the target plaquette \(P\), \(\ket{\Phi(\Lambda)}\) is the ground state of Hamiltonian \eqref{eq:EffHamil} with parameter \(\Lambda\), and
\[\pi_\idm^{\tob\tob} = \frac{1}{\sum_{J} n_\text{Cond}^Jd_J}\]
is the condensation coefficient of the ``trivial anyon'' \(\idm\). Since all gapped paths are contractible, the vacuum expectation values of the order parameter operator does not depend on the choices of the homotopic classes of paths \((E_1E_2\cdots E_n)\). Besides, we require \(P\ne P_0\) to preserve the topological superselection rules.

In the ground state of the parent phase, no condensed anyon is excited, hence, in the parent phase,
\eqn{
\mathcal{O}_{P_0}(P, \Lambda = 0) = 0.
}
In the ground state of the child phase in the limit \(\Lambda \to \infty\), which is the \((+1)\)-eigenvalued eigenstate of the projector \(P_E\), thus, in the child phase,
\eqn{
\mathcal{O}_{P_0}(P, \Lambda = +\infty) = 1.
}
This is because in the limit \(\Lambda\to\infty\), no gapped dofs are allowed on the lattice.

Things become interesting for finite \(\Lambda\) that is larger than the critical point. First, let's assume that \(P_0\) and \(P\) are in the same domain. Numerical calculations \cite{schulz2013,schulz2016,shi2024} show that the topological phase transition is first-order. In the thermodynamic limit, the vacuum expectation value of the operator \(\mathcal{O}\) exhibits a discontinuity at the critical point \(\Lambda_\text{critical} \in (0, +\infty)\), with the parent phase preserved for \(\Lambda < \Lambda_\text{critical}\) and the child phase realized for \(\Lambda > \Lambda_\text{critical}\). That is,
\eqn{\mathcal{O}_{P_0}(P, \Lambda) = \begin{cases}0,\qquad \Lambda < \Lambda_\text{critical},\\ 1,\qquad \Lambda > \Lambda_\text{critical}.\end{cases}}

\subsection{Local Excitations}\label{sec:local}

It is most effective to illustrate our ideas with an explicit example when discussing Goldstone modes in general. In this section, we use the case of \(\psi\bar\psi\) condensation in the Ising model to concretely demonstrate the concept of Goldstone modes and local excitations.

\begin{example}
Let's consider \(\psi\bar\psi\) anyon condensation in the doubled Ising phase. In this case, the order parameter is particularly simple:
\[
\mathcal{O}_{P_0}(P, \Lambda) = \langle \Phi(\Lambda)| W^{\psi\bar\psi; 11}_{E_0E_1\cdots E_n} |\Phi(\Lambda)\rangle.
\]
Note that
\[
W^{\psi\bar\psi; 11}_E\quad\Edge{1\text{ or }\psi} =\ \Edge{1\text{ or }\psi},\qquad\qquad
W^{\psi\bar\psi;11}_E\quad\Edge{\sigma} = -\ \ \Edge{\sigma}.
\]
Therefore,
\eqn{
\mathcal{O}_{P_0}(P, \Lambda) = (-1)^{\sum_{i=0}^n \delta_{\alpha_{E_i},\, \sigma}}.
}
If the path \(E_0E_1\cdots E_n\) crosses gapped $\sigma$-loops even (odd) times, the order parameter is \(+1\) (\(-1\)). 

This result shows that we can assign to each domain a \({\ZZ}_2\)-valued quantity, which we call the \emph{Goldstone mode}:
\begin{itemize}
\item The order parameter \(\mathcal{O}_{P_0}(P, \Lambda)\) takes the same value for all plaquettes \(P\) within a given domain, thus characterizing the Goldstone mode of that domain.
\item A gapped \(\sigma\)-loop acts as a domain wall separating two regions with different Goldstone modes.
\item By convention (a gauge choice), the domain containing the base plaquette \(P_0\) is assigned the Goldstone mode \(+1\).
\end{itemize}
\end{example}

From the example of the Ising model, it is evident that after anyon condensation in the enlarged HGW model, the parent gauge structure—the parent input UFC \(\Fus\)—is spontaneously broken to a smaller gauge structure, namely the child input UFC \(\Sub\), by gapping out certain gauge-field dofs in \(L_\G\).

In addition to coupling with the child \emph{gauge} fields \(L_\Sub\), the anyonic matter fields also couple to a \emph{global} symmetry generated by the gapped dofs in \(L_\G\). Different domains of the lattice gauge system are labeled by global-symmetry-graded charges, referred to as \emph{Goldstone modes}\cite{Hu2021}. An anyonic matter particle carries the Goldstone mode of the domain in which it resides, and its Goldstone mode changes whenever the anyon hops across a gapped loop from one domain to a neighboring domain.

\begin{example}
To substantiate the above statement, let us explicitly calculate how the unconfined anyons of the child toric code phase—namely, \(e\), \(m\), and \(\epsilon\)—transform when crossing a gapped \(\sigma\)-loop from one domain to another.
\begin{itemize}
\item The child anyons \(e\) and \(m\) result from the splitting of the parent anyon \(\sigma\bar\sigma\): the dyonic sector \((\sigma\bar\sigma, 1, 1)\) becomes \(e\), and the dyonic sector \((\sigma\bar\sigma, \psi, 1)\) becomes \(m\). Note that
\[
[z^{\sigma\bar\sigma}_{\sigma\sigma}]_{1}^{1} = [z^{\sigma\bar\sigma}_{\sigma\sigma}]_{\psi}^{\psi} = 0,\qquad
[z^{\sigma\bar\sigma}_{\sigma\sigma}]_{1}^{\psi} = [z^{\sigma\bar\sigma}_{\sigma\sigma}]_{\psi}^{1} = 1,
\]
so that when crossing the \(\sigma\)-loop, anyon \(e\) is transformed into anyon \(m\), and vice versa. Therefore, the \({\ZZ}_2\) global symmetry predicted earlier is precisely the celebrated \(e \leftrightarrow m\) exchange symmetry of the toric code phase.
\item The anyon \(\epsilon\) in the child phase results from the identification of the parent anyons \(\psi\bar{1}\) and \(1\bar{\psi}\). Noting that
\[
[z^{\psi\bar{1}}_{\sigma\sigma}]_{\psi}^{\psi} = \ii, \qquad [z^{1\bar{\psi}}_{\sigma\sigma}]_{\psi}^{\psi} = -\ii,
\]
we see that anyon \(\epsilon\) carries a \(\pm \frac{1}{4}\) global \({\ZZ}_2\) symmetry charge. Moreover, since \(1\bar{\psi} = \psi\bar{1} \otimes \psi\bar{\psi}\), the essential difference between them is the presence of a \(\psi\bar{\psi}\) anyon, which carries a \(\frac{1}{2}\) global \({\ZZ}_2\) symmetry charge.

\item The parent anyon \(\psi\bar{\psi}\) carries a \(\frac{1}{2}\) global \({\ZZ}_2\) symmetry charge in the child phase with gapped dofs. This demonstrates that the \(\psi\bar{\psi}\) anyon is not entirely absorbed as background in the coherent ground state of the child phase, but instead continues to exert a nontrivial influence when coupled with the gapped dofs.

The \(\tfrac{1}{4}\) symmetry charge of anyon \(\epsilon\) indicates that \(\epsilon\) is not a representation of the \({\ZZ}_2\) group. In fact, only the local excitations---\(1\bar{1}\) and \(\psi\bar{\psi}\)---constitute representations of the global \({\ZZ}_2\) symmetry after gauge invariance broken. The \(\tfrac{1}{4}\)-charge of the \(\epsilon\) anyon, which we refer to as \emph{fragmentation}, will be discussed in our subsequent work \cite{Fu2025, Fu2025b}.
\end{itemize}

\end{example}
The \(\psi\bar\psi\) excitation in the child toric code phase is referred to as a \emph{local excitation}, as it braids trivially with all other unconfined anyons. Moreover, a local excitation traversing a noncontractible loop operator on a non-simply-connected manifold does not generate any new degenerate topological sectors like anyons. This is because gapped loops \(\sigma\) always form contractible loops on the manifold in the condensed child model, so a local excitation necessarily crosses gapped loops an even number of times when traveling along any noncontractible path. Therefore, local excitations are nontopological, and thus it is necessary to extend the definition of an anyon type: 

\begin{important}
An anyon type should be understood as an equivalence class of all particles differing only by coupling to gapped dofs (gauge transformations) or local excitations (local excitations).
\end{important}
An analogue of these local excitations in the electroweak phase transition is the Higgs boson in the symmetry-broken phase—corresponding to fluctuations of the Higgs field around the new vacuum.

The above toric code example illustrates the two most common ways in which a child anyon couples to Goldstone modes:
\begin{itemize}
\item Condensed parent anyons become local excitations in the child model. Different local excitations behave like the trivial anyon within a single domain, but they undergo distinct transformations when crossing gapped loops from one domain to another.
\item Child anyons from identification may carry nontrivial global symmetry charges, which can even be nonabelian if the anyon is higher-dimensional. The internal state of such an anyon transforms when it crosses a gapped loop. Moreover, child anyons---regarded as equivalence classes of excitations differing only by local excitations---may undergo distinct transformations, depending on the local excitation they are coupled to, when traversing a gapped loop.
\item Child anyons arising from splitting may change their anyon type upon crossing gapped loops.
\end{itemize}

\subsection{The Gapped Gauge-Fields Absorbing Goldstone Modes}\label{sec:absorbing}

In this section, we continue to use the \(\psi\bar\psi\) condensation in the Ising model as an example to illustrate how gapped dofs absorb the Goldstone modes.

An edge is always directed. The \(\sigma\) dof on an edge of the parent Ising model, however, is self-dual, so the two directions of an edge carrying \(\sigma\) dof are completely equivalent.

Things change when anyon \(\psi\bar\psi\) anyon is condensed. The goldstone modes give a canonical classification of the two gapped edges carrying gapped \(\sigma\): There are direction edges whose left-side has Goldstone mode \(-1\) and right-side with Goldstone model \(+1\); there are also directed edges whose left-side has Goldstone mode \(+1\) and right-side with Goldstone model \(-1\). Reversing the direction of the edge exchanges these two types of gapped edges. That is, the gapped dof \(\sigma\) obtains an extra dof, that is the difference between the left-side of the directed edge and the right-side of the directed edge. We label these two dofs as
\eqn{\sigma_{-+},\qquad \sigma_{+-},}
where the first (second) subscript labels the Goldstone mode in the left (right) domain of the edge, on which gapped dof \(\sigma\) resides.

In other words, the gapped edge dof \(\sigma\) acquires an additional structure: it is sensitive not only to its local value but also to the difference in Goldstone mode across the directed edge. This extra label distinguishes the two types of gapped edges that appear after condensation. For clarity, we denote the dofs on such a directed edge as \(\sigma_{-+}\) and \(\sigma_{+-}\). These two are opposite:
\eqn{
\sigma_{-+}^* = \sigma_{+-}
}
because reversing the direction of an edge exchanges the left-side and right-side:
\[\GapEdgeUp{\sigma_{+-}}{+}{-}\qquad \Longrightarrow\qquad \GapEdgeDown{\sigma_{-+}}{+}{-}\qquad\equiv\qquad \GapEdgeUp{\sigma_{-+}}{-}{+},\]
where the third figure is simply rotating the second figure by \(180^\circ\). Here, we use red edges to indicate that the dofs on this edge is gapped. This phenomenon is analogous to the situation in the electroweak phase transition, where the gapped gauge bosons \(W^\pm_\mu\) and \(Z_\mu\) absorb the gradients of the Goldstone bosons, \(\partial_\mu\theta^a\), which become the longitudinal modes of these massive vector bosons.

\begin{figure}
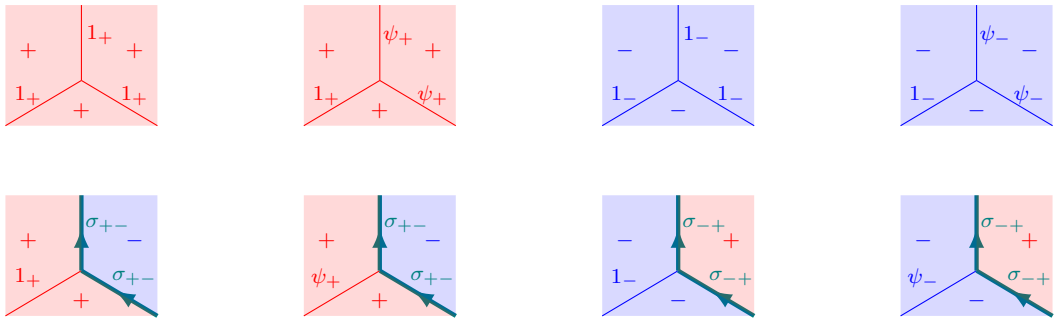
\centering
\subfloat{\YVertex{$1_+$}{$1_+$}{$1_+$}{$+$}{red}}\hspace{15pt}
\subfloat{\YVertex{$1_+$}{$\psi_+$}{$\psi_+$}{$+$}{red}}\hspace{15pt}
\subfloat{\YVertex{$1_-$}{$1_-$}{$1_-$}{$-$}{blue}}\hspace{15pt}
\subfloat{\YVertex{$1_-$}{$\psi_-$}{$\psi_-$}{$-$}{blue}}\\
\subfloat{\YDirect{$1_+$}{$\sigma_{+-}$}{$\sigma_{+-}$}{$+$}{$-$}{red}{blue}}\hspace{15pt}
\subfloat{\YDirect{$\psi_+$}{$\sigma_{+-}$}{$\sigma_{+-}$}{$+$}{$-$}{red}{blue}}\hspace{15pt}
\subfloat{\YDirect{$1_-$}{$\sigma_{-+}$}{$\sigma_{-+}$}{$-$}{$+$}{blue}{red}}\hspace{15pt}
\subfloat{\YDirect{$\psi_-$}{$\sigma_{-+}$}{$\sigma_{-+}$}{$-$}{$+$}{blue}{red}}
\caption{All possible vertex configurations in the child HGW model after \(\psi\bar\psi\) condensation in the parent Ising model.}
\end{figure}

We also assign Goldstone modes (\(+\) or \(-\)) to the residual gauge dofs in each domain. Thus, there are now four types of gauge dofs:
\eqn{
1_+, \qquad \psi_+, \qquad 1_-, \qquad \psi_-.
}
Because of the constraints imposed by the lattice geometry, the six dofs on the lattice cannot fuse arbitrarily. The nontrivial fusion rules are restricted as follows:
\eqn[eq:mulfus]{
&f_{1_+1_+1_+} = f_{1_+\psi_+\psi_+} = f_{1_-1_-1_-} = f_{1_-\psi_-\psi_-} = 1, \\
&f_{1_+\sigma_{+-}\sigma_{-+}} = f_{\psi_+\sigma_{+-}\sigma_{-+}} = 1\\
&f_{1_-\sigma_{-+}\sigma_{+-}} = f_{\psi_-\sigma_{-+}\sigma_{+-}} = 1.
}
Forbidden fusions---for example, the fusion of \(1_+\) and \(1_-\)---do not arise, since these dofs can never meet in the condensed model. This mathematical structure is a \emph{multifusion category}. We will study models with multifusion categories as input data in a separate work \cite{Fu2025, Fu2025b}.

\subsubsection{Gauge Invariance Breaking in Anyon Condensation}\label{sec:landau}

In the previous section, we have distinguished the child gauge-field dofs in the two types of domains by an additional label. In the \(+\) domain, the basic dofs are \(1_+\) and \(\psi_+\), while in the \(-\) domain, they are \(1_-\) and \(\psi_-\). A natural question arises: Is the distinction between \(1_+\) (\(\psi_+\)) and \(1_-\) (\(\psi_-\)) physically meaningful, or is it merely a mathematical convenience?

The answer is the former: These two sets of gauge-field dofs are indeed distinct. As we have shown before, the $\pm$ Goldstone modes of the enlarged HGW model after \(\psi\bar\psi\) condensation in the parent model characterize the two global symmetry sectors of the \(em\)-exchange symmetry in the child toric code phase. Our previous work \cite{zhao2025} has shown that while the gauge field in $+$ ($-$) domain takes value in the simple objects of $\Vec(\ZZ_2)$ ($\Rep(\ZZ_2)$, categorically Morita equivalent and isomorphic to but distinct UFC of \(\Vec(\ZZ_2)\)).

To illustrate, there may exist distinct UFCs that share isomorphic Drinfeld centers. These UFCs are \emph{categorically Morita equivalent} and are classified by the \emph{connected symmetric Frobenius algebras} \(\A\) in one UFC \(\Fus\) among them \cite{etingof2016}. Each such categorically Morita equivalent UFC is realized as the \emph{bimodule category} \(\Bimod_\Fus(\A, \A)\), consisting of all \(\A\)-\(\A\)-bimodules---that is, objects equipped with compatible left and right actions of \(\A\) with UFC \(\Fus\)\footnote{This is because the Frobenius algebra \(\A\) serves as the trivial object (trivial bimodule) in the categorically Morita equivalent UFC. An edge in the enlarged HGW model carrying the trivial dof \(A\) corresponds to a trivial gauge connection; that is, the flux type \(M\) of any dyonic sector \((J, M, \alpha)\) must remain invariant when crossing such a trivial edge. Mathematically, this condition means \([z^J_{\tob M}]_{M\alpha}^{N\beta} = \delta_{MN} \delta_{\alpha\beta}\). The flux types satisfying this property are precisely the \(\A\dash\A\)-bimodules, which are defined to half-braid trivially with the new ``trivial object'' \(\A\).}. The simple objects of \(\Bimod_\Fus(\A, \A)\) are the simple (irreducible) \(\A\)-\(\A\)-bimodules. Details of Frobenius algebras and related mathematics are reviewed in the Appendix.

The definition of categorical Morita equivalence implies that two enlarged HGW models with categorically Morita equivalent input UFCs are equivalent in the sense that they describe the same anyon spectrum---and hence the same topological phase---even though the same anyonic matter fields couple to different gauge fields. We have established in Ref. \cite{zhao2025} a duality map that transforms the gauge-field in one model to that in the other, while preserving the anyon types. The duality also transforms how the same anyon types couple with the gauge field, i.e., the flux types and charge spaces of anyons.

In certain cases, two Morita equivalent UFCs may even be isomorphic. Composing the duality map with the isomorphism leads to a map from one enlarged HGW model to itself. This map preserves the gauge-field while (possibly) permuting the anyon types in the model. This induces a genuine symmetry transformation of the topological phase described by the model, as the physical observables---the anyon types---are changed. In other words, in the same enlarged HGW model, there can exist different but equivalent ways in which the gauge field couple to anyon fields via the topological Gauss law. 

Moreover, there are instances in which the anyon types are invariant under the symmetry transformation, but an anyon's internal gauge space is transformed. In such a case, the symmetry transformations delineate a \emph{gauge redundancy} of the topological phase: There exist distinct but equivalent ways for the gauge field to couple with a given anyon type. The child HGW model after \(\psi\bar\psi\) condensation in the doubled Ising model is a very example. 

In the parent input Ising UFC, there are two Frobenius algebras \(\A_0 = 1\) and \(\A = 1 \oplus \psi\). The bimodule category over trivial algebra \(\A_0\) is the Ising UFC itself\footnote{Any UFC $\Fus$ has a trivial Frobenius algebra object $\A_0=1$, where $1$ is the trivial object. The bimodule category $\Bimod_\Fus(\A)$ is $\Fus$ itself.}, while \(\A = 1\oplus\psi\) admits three simple bimodules: \(M_1 = \A\) (Frobenius \(A\) itself as the trviial bimodule over \(\A\), using algebra multiplication as the action), \(M_\psi\), and \(M_\sigma\), which form a UFC that is isomorphic to the original UFC via
$$1 \leftrightarrow M_1 = \A,\qquad \psi \leftrightarrow M_\psi,\qquad \sigma\leftrightarrow M_\sigma.$$
The symmetry transformation of the enlarged HGW model with input Ising UFC, defined as the composition of the duality map and the above isomorphism, does not permute doubled Ising anyon types but acts merely within dyonic sectors of certain anyon types (for instance, exchanging \((\sigma\bar\sigma, 1)\) with \((\sigma\bar\sigma, \psi)\)). So, this symmetry transformation generates a \emph{gauge redundancy} of the parent doubled Ising HGW model.

After \(\psi\bar\psi\) condensation, the object \(1 \oplus \psi\) remains a Frobenius algebra in the child \(\Vec(\ZZ_2)\) UFC, which now admits only two simple bimodules \(M_1\) and \(M_\psi\) over \(\Vec(\ZZ_2)\). The bimodule category \(\Bimod_{\Vec(\ZZ_2)}(\A, \A)\) is known as \(\Rep(\ZZ_2)\) UFC, where the two simple objects (irreducible bimodules) \(M_1\) and \(M_\psi\) correspond to the two irreducible representations of the group \(\ZZ_2\). UFC \(\Rep(\ZZ_2)\) is isomorphic to \(\Vec(\ZZ_2)\):
\[
M_1 \leftrightarrow 1, \qquad M_\psi \leftrightarrow \psi.
\]
Nevertheless, the symmetry transformation of the child \(\Vec(\ZZ_2)\) UFC, defined by composing the duality map and the above isomorphism, does permute anyon types of the child \(\Vec(\ZZ_2)\) toric code HGW model: It exchanges anyon types \(e\) and \(m\). 

Now, \(\Vec(\ZZ_2)\) and \(\Rep(\ZZ_2)\) are respectively where the gauge fields in the two distinct domains in child enlarged HGW model take value:
\eqn{
1_+ = 1\in\Vec(\ZZ_2),\qquad \psi_+ = \psi\in\Vec(\ZZ_2),\\ 
1_- = M_1\in\Rep(\ZZ_2),\qquad \psi_-= M_\psi\in\Rep(\ZZ_2).
}
The Goldstone modes, denoted by \(+\) and \(-\), label the two distinct UFCs. Since UFCs can be realized as bimodule categories over Frobenius algebras, we can interpret the Goldstone modes as the two Frobenius algebras:
\eqn{+ = \A_0 = 1,\qquad - = \A = 1\oplus\psi.}

The two sides of an edge carrying \(\sigma_{+-}\) or \(\sigma_{-+}\) belong to distinct symmetry sectors. The gapped dofs \(\sigma_{+-}\) and \(\sigma_{-+}\) serve as \emph{generators} of the global symmetry in the child toric code model: when an anyon \(e\) (\(m\)) crosses such a gapped edge, it transforms into the anyon \(m\) (\(e\)).

Above discussion illustrates that the Landau-Ginzburg symmetry breaking paradigm can be generalized to the topological phase transition from the parent doubled Ising topological phase to the child toric code topological phase induced by condensing $\psi\bar\psi$:
\begin{important}
During \(\psi\bar\psi\) anyon condensation in the doubled Ising model, the \(\ZZ_2\) gauge redundancy of the parent Ising model, associated with Frobenius algebra \(1 \oplus \psi\), is broken to the \(\ZZ_2\) \(em\)-exchange symmetry of the child \(\ZZ_2\) toric code HGW model.
\end{important}

There are cases where the gauge transformation of the parent phase may act nontrivially on the condensed dyonic sectors, leading to global symmetries that permute condensed parent dyon sectors. In such situations, non-fluxon condensation must also be taken into account. One example is the \((\tau\bar\tau, 1)\) condensation in the Fibonacci model. This scenario was explored in Ref. \cite{zhao2024b}.

Not every global symmetry arising in a child phase is one that is due to Frobenius algebras. For instance, in the phase transition from \(\Vec(S_3)\) model by condensing dyon \((B, e)\) to the \(\Vec(\ZZ_3)\) model, the resultant global symmetry is the charge conjugation symmetry---the outer automorphism of the child \(\ZZ_3\) gauge group:
\[
e \mapsto e, \qquad r \mapsto r^2, \qquad r^2 \mapsto r,
\]
which is not caused by any Frobenius algebras of \(\Vec{S_3}\) but the global symmetry generator \(s, sr, rs\). Specifically, a \(\Vec(\ZZ_3)\) flux \(r\) is reversed when crossing a gapped dof due to non-Abelian structure:
\[
rs = sr^2, \qquad r^2s = sr.
\]
As a result, the corresponding Goldstone modes can be labeled by the cosets \([e] = \{e, r, r^2\}\) and \([s] = \{s, sr, rsr\}\) in the group quotient \(S_3/\ZZ_3\). This scenario will be discussed in detail in a forthcoming paper \cite{Fu2025, Fu2025b}.

\subsubsection{Gapped Dofs Carrying Gauge Charges}\label{sec:chargedgapedof}

In the anyon condensation from the \(\Vec(S_3)\) model to the \(\Vec(\ZZ_3)\) model, there are three gapped gauge dofs, \(L_\G = \{s, sr = r^2s, sr^2 = rs\}\). These are gapped excitations that carry the residual \({\ZZ}_3\) gauge charges \(1, r, r^2\), analogous to the situation in the electroweak phase transition where the \(W^\pm\) gauge bosons carry electric charges due to the residual ${\tt U(1)}_{\rm EM}$ gauge invariance.

\subsubsection{Algebraic Symmetry}\label{sec:algsymm}

The \((\tau\bar\tau, 1)\) condensation in the Fibonacci model is a lot more involved (see Ref. \cite{kawagoe2024} for a brief discussion). This condensation yields a topologically trivial phase, featuring only trivial child gauge field dofs \(1\) and trivial anyon \(\idm\) in the child model. Nevertheless, if \(\Lambda\) is large but finite, there still exist gapped dofs \(\tau\) that can form contractible loops, as well as local excitations \((\tau\bar\tau, 1)\). When a local excitation \((\tau\bar\tau, 1)\) crosses a gapped loop and enters from one domain to another, we have
\eqn{\hop_E^{\tau\bar\tau}\ \ \HoppingFiboLeft\quad 
&=\quad -\frac{1}{\phi^2}\ \ \HoppingFiboRight\ \ + \ \ \frac{\sqrt[4]{5}}{\phi}\ \ \HoppingFiboRightTau\quad \text{(Confined!)}\\
&\Longrightarrow -\frac{1}{\phi^2}\ \ \HoppingFiboRight\quad ,
}
where \(\phi = (1 + \sqrt{5}) / 2\) is the Golden ration. The ``$\Rightarrow $'' is because confined dyonic sector \((\tau\bar\tau, 1)\) is banned from the spectrum of the child model.
\begin{important}
The transformation of a local excitation \((\tau\bar\tau, 1)\) crossing a gapped loop is non-unitary in the perspective of the child model and thus cannot generate a group symmetry, resulting in an noninvertible global symmetry with infinitely many Goldstone modes. This symmetry structure is known as the algebraic symmetry \cite{polishchuk1998, Hung2013, HungWan2013a, Baratin2014, Gaiotto2014, Bhardwaj2017, Ji2019, ji2020, Kong2020, bartsch2023, bhardwaj2023, Ji2019, ji2020, bhardwaj2023, bhardwaj2022, choi2022, choi2023, choi2024, bartsch2024}. This is in contrast to the Ising and \(\Vec(S_3)\) cases.
\end{important}

Such cases of scenario may arise when an anyon to be condensed has more than one dyonic sectors, some of which would actually condense while some others confined. A further discussion of this involves Frobenius algebras of the input $\Fus$ of a parent model and will be reported elsewhere.

\subsection{Confined Anyons and Moduli Space}\label{sec:moduli}

Anyons that carry gapped dofs become confined in the child phase after condensation---for example, the anyons \(\sigma\bar 1\), \(\sigma\bar\psi\), \(1\bar\sigma\), and \(\psi\bar\sigma\) in the doubled Ising phase following \(\psi\bar\psi\) condensation. Nevertheless, confinement does not appear in the electroweak transition: No stable fluxes are confined after the Higgs condensation.

This difference is rooted in the structure of their \emph{moduli spaces}. In the electroweak phase transition, the gauge group is broken from \({\tt SU(2)}_L \times {\tt U(1)}_Y\) to \({\tt U(1)}_{\text{EM}}\), such that the moduli space (i.e., the parameter manifold of all new vacuum states) has the topology \(\mathbb{S}^3\), which is simply connected, which cannot support stable vortices, and hence no flux confinement.

In contrast, in our enlarged HGW model, the set of simple objects is always finite and endowed with the discrete topology. Therefore, the gapped dofs are disconnected from the residual, unbroken child dofs. This disconnect means that dyons carrying gapped fluxes cannot be smoothly deformed into dyons with trivial fluxes, resulting in genuine confinement.

\begin{example}
For example, the moduli space of the child toric code HGW model after \(\psi\bar\psi\) condensation in the parent Ising model is \(\ZZ_2\): there are exactly two child ground states, labeled by the two Frobenius algebras \(\mathcal{A}_0\) and \(\mathcal{A}_1\). As another example, the moduli space of the child \(\Vec({\ZZ}_3)\) HGW model after \(B\) condensation in the parent \(\Vec(S_3)\) model is also \(\ZZ_2\) formed by the Goldstone modes \([e]\) and \([s]\). The three gapped dofs, \(s\), \(sr\), and \(rs\), can freely deform into each other via gauge connections but cannot become the residual child dofs \(e\), \(r\), or \(r^2\); hence, they are confined.
\end{example}

In continuous Lie group gauge theories, confinement can arise if the moduli space is not simply connected. This occurs, for instance, in the case of Cooper-pair condensation that causes superconductivity: The moduli space is \(\mathbb{S}^1\), which is not simply connected. Consequently, together with energetics, magnetic-flux confinement can occur in Type-II superconductors.

\section{Discussion}\label{sec:dicussion}

This work presents a lattice gauge theory---the enlarged HGW string-net model---that reincorporates topologically ordered phases and their phase transitions in the Landau-Ginzburg symmetry breaking paradigm. In our enlarged HGW string-net model, gauge fields and matter fields are covariantly coupled. Within this framework, anyon condensation naturally emerges as a Higgs mechanism, complete with order parameters, Goldstone modes, and the breaking of certain gauge invariances.

Our enlarged HGW model treats the gauge and matter sectors as separate, interacting components, rather than merely realizing anyons as punctures in lattice-field excited states. Gauge variables---simple objects of an arbitrary UFC \(\Fus\)---reside on edges and tails, while plaquettes host anyonic matter particles whose internal gauge space (generally multidimensional) comprises both flux and \(T\)-charge dofs, together forming dyonic sectors. Different dyons may exhibit distinct braiding and fusion properties; however, two dyons of the same anyon type are related by gauge transformations, so only the anyon types are physically observable.

Within our enlarged HGW model, anyon condensation is realized as a Higgs transition. Specific dyon types are condensed by adding projector terms to the parent Hamiltonian, resulting in ground states that are coherent superpositions containing arbitrarily many condensed anyons. This process breaks the parent gauge category \(\Fus\) down to a subcategory \(\Sub\). The broken part of the gauge invariance turns to be the global symmetry of the child phase. Certain gauge-field dofs become gapped and serve as domain walls separating distinct condensed domains. The anyon condensation projector thus plays the role of the Landau order parameter, which may take different values in different domains and form a representation of the resultant global symmetry, leading to the emergence of Goldstone modes that indicate the global symmetry sector of each domain. In some cases, these Goldstone modes are labeled by Frobenius algebras; in others, they are identified with cosets in the quotient \(L_\Fus / L_\Sub\). Massive gauge loops constitute the gapped edge dofs that interpolate between domains and absorb the Goldstone modes, thereby enlarging their dofs and reproducing gauge-field mass generation.

This formalism enables a fine analysis of the splitting, identification, and confinement phenomena of anyon types during anyon condensation, elucidates how residual global or algebraic symmetries emerge from broken gauge redundancy, and connects confinement to the discrete---and thus non-simply-connected---moduli space of fusion-category gauge theories. All these phenomena have analogies in the traditional Higgs mechanism, including the splitting of left-handed isospin doublets and the formation of massive Dirac fermions during the electroweak phase transition, as well as the confinement of magnetic flux in type-II superconductors during Cooper-pair condensation. In conclusion, these results collectively demonstrate that topological phase transitions admit a fully fledged Landau–Ginzburg–Higgs description in the language of our enlarged HGW model.

The fine analysis of anyon condensation made possible by our enlarged HGW string-net model extends to the case of competing anyon condensates, which may lead to second order phase transitions, with critical conformal field theory description; hence it can be a key component of a systematic construction of 2D CFTs \cite{Hung2025}. Our enlarged HGW model also facilitates studying symmetry-enriched topological phases, allowing probing into the symmetry action on the internal spaces of non-Abelian anyons\cite{Fu2025, Fu2025b}.

Several important problems remain open in this paper and call for future investigation. Foremost among them are the \(F\)-gauge transformations \cite{zhao2024b, zhao2025}, which constitute a completely novel type of gauge transformation distinct from the conventional \(T\)-gauge transformations and have no analogy in conventional lattice gauge theories. These gauge transformations require a further enlargement of the single-anyon Hilbert space---the \(F\)-gauge space---for those nonabelian anyons. In particular, the interplay between \(T\)-gauge and \(F\)-gauge dofs---for instance, how \(T\)-gauge transformations act on the \(F\)-gauge space and vice versa---remains largely unexplored. In addition, the algebraic symmetries present in topological phases, such as those found in the doubled Fibonacci phase, appear rather unconventional and warrant deeper study. 

Moreover, now that we have explicitly defined the parameters of Hamiltonian during anyon condensation and the associated order parameters, more direct numerical analyses of topological phase transitions can be performed. Besides, we have modified the exactly solvable Hamiltonian by introducing a covariant derivative term:
\eqn{
H = 
M\sum_{\text{Plaquettes} P}
\sum_{\substack{J\in L_{\Cent(\Fus)}\\ J\ne\idm}}\Pi_P^J
- g\sum_{\text{Edge} E}
\sum_{\substack{J\in L_{\Cent(\Fus)}\\ J\ne\idm}} \hop_E^{J}.}
This Hamiltonian is no longer exactly solvable and may admit nontrivial dynamical evolutions of gauge field and matter fields, while topological gauge invariance is still preserved. New phenomena may appear when studying topological phases and their transitions under our modified Hamiltonian. This Hamiltonian deserves further and detailed studies.

\begin{acknowledgments}
The authors thank Hongguang Liu, Ling-Yan Hung, Yuting Hu, Nianrui Fu, Siyuan Wang, and Yifei Wang for inspiring and helpful discussions. YW is supported by NSFC Grant No. KRH1512711, the Shanghai Municipal Science and Technology Major Project (Grant No. 2019SHZDZX01), Science and Technology Commission of Shanghai Municipality (Grant No. 24LZ1400100), and the Innovation Program for Quantum Science and Technology (No. 2024ZD0300101). The authors are grateful for the hospitality of the Perimeter Institute during his visit, where the main part of this work is done. This research was supported in part by the Perimeter Institute for Theoretical Physics. Research at Perimeter Institute is supported by the Government of Canada through the Department of Innovation, Science and Economic Development and by the Province of Ontario through the Ministry of Research, Innovation and Science. 
\end{acknowledgments}

\appendix

\section{Review of the Unitary Fusion Category}\label{appen:ufc}

The input data of the string-net model is a \emph{unitary fusion category} \(\Fus\), described by a finite set \(L_\Fus\), whose elements are called \emph{simple objects}, equipped with three functions \(N: L_\Fus^3 \to \NN\), \(d: L_\Fus \to \RR^+\), and \(G: L^6_\Fus \to \C\). The function \(N\) sets the \emph{fusion rules} of the simple objects, satisfying
\eq{
\sum_{e \in L_\Fus} N_{ab}^e N_{ec}^d = \sum_{f \in L_\Fus} N_{af}^d N_{bc}^f, \qquad\qquad N_{ab}^c = N_{c^\ast a}^{b^\ast}.
}
There exists a special simple object \(1 \in L_\Fus\), called the \emph{trivial object}, such that for any \(a, b \in L_\Fus\),
\eq{
N_{1a}^b = N_{1b}^a = \delta_{ab},
}
where \(\delta\) is the Kronecker symbol. For each \(a \in L_\Fus\), there exists a unique simple object \(a^\ast \in L_\Fus\), called the \emph{opposite object} of \(a\), such that
\eq{
N_{ab}^1 = N_{ba}^1 = \delta_{ba^\ast}.
}
We only consider the case where for any \(a, b, c \in L_\Fus\), \(N_{ab}^c = 0\) or \(1\). In this case, we define
\eq{
\delta_{abc} = N_{ab}^{c^\ast} \in \{0, 1\}.
}
In this work we assume commutative fusion rules, \(\delta_{abc} = \delta_{bac}\); however, our construction extends readily to noncommutative fusion categories (see the Appendices of Ref. \cite{zhao2025}).

The fundamental setup of the string-net model is characterized by assigning a simple object from \(L_\Fus\) to each edge and tail, while enforcing a vertex constraint: \(\delta_{ijk} = 1\) for the trio of incident edges or tails converging at a vertex, all directed toward the vertex and sequentially labeled counterclockwise by \(i, j, k \in L_\Fus\). It is permissible to invert the direction of any edge or tail and simultaneously apply a conjugation to its label, expressed as \(j \to j^\ast\), without altering the configuration. The Hilbert space \(\Hil\) associated with the model is composed of the span of all conceivable configurations of these labels on the edges and tails.

The function \(d\) returns the \emph{quantum dimensions} of the simple objects in \(L_\Fus\). It is the largest eigenvalues of the fusion matrix and forms the \(1\)-dimensional representation of the fusion rule.
\eq{d_ad_b = \sum_{c\in L_\Fus}N_{ab}^cd_c.}
In particular, \(d_1 = 1\), and for any \(a\in L_\Fus, d_a = d_{a^\ast}\ge 1\). 

The function \(G\) defines the \(6j\)-\emph{symbols} of the fusion algebra. It satisfies
\eq{\sum_nd_nG^{pqn}_{v^*u^*a}G^{uvn}_{j^*i^*b}G^{ijn}_{q^*p^*c} = G^{abc}_{i^*pu^*}&G^{c^*b^*a^*}_{vq^*j},\qquad\sum_nd_nG^{ijp}_{kln}G^{j^*i^*q}_{l^*k^*n} = \frac{\delta_{pq^*}}{d_p}\delta_{ijp}\delta_{klq},\\
G^{ijm}_{kln} = G^{klm^*}_{ijn^*} = G^{jim}_{lkn^*}= G^{mij}_{nk^*l^*} &= \alpha_m\alpha_n\overline{G^{j^*i^*m^*}_{l^*k^*n^*}},\qquad \Big|G^{abc}_{1bc}\Big| = \frac{1}{\sqrt{d_bd_c}}\delta_{abc}.
}
where \(\alpha_m = G^{1mm^\ast}_{1m^\ast m}\in\{\pm 1\}\) is the Frobenius-Schur indicator of simple object \(m\).

\section{Review of the Topological Moves}\label{appen:pachner}

This appendix concisely reviews the topological characteristics of the ground-state subspace of the string-net model described in Ref. \cite{Hu2018}. Lattices possessing identical topologies can be interconverted via \emph{Pachner moves}. These moves correspond to unitary linear mappings between the Hilbert spaces of two string-net models that utilize the same input fusion category on distinct lattices, which are collectively denoted as \(\T\). The ground states exhibit invariance under these linear mappings. Three types of fundamental Pachner moves exist, with associated linear transformations specified as follows:
\eqn[eq:pachner]{
\T \quad \PachnerOne\ ,\\
\T \quad \PachnerTwo\ ,\\
\T \quad \PachnerThree\ .}
In this context, red ``\({\color{red}\times}\)'' symbols are utilized to denote the plaquettes that require contraction. All other Pachner moves and the associated linear transformations of Hilbert spaces can be decomposed into these three fundamental moves. Although several compositions of these elementary moves are possible for given initial and final lattices, they yield equivalent transformation matrices within the ground-state Hilbert space.

We have additionally observed that selecting different edges for tail attachment results in equivalent outcomes. These selections yield diverse lattice configurations and, in turn, distinct corresponding Hilbert spaces within the lattice model. The equivalence of states across these Hilbert spaces is achieved through the following linear transformation \(\T'\):
\eqn{
\T'\quad\PachnerFour\ .
}
In this way, states with tails attached to various edges can be derived iteratively.

For clarity and convenience, we will introduce auxiliary states that involve multiple tails on a single plaquette in specific situations. Although these states exhibit multiple tails on one plaquette, they remain equivalent to states found in the Hilbert space.
\eqn{
\PachnerFive\ .
}
\section{Frobenius Algebras, Bimodule Categories, and \eqs{F}-Gauge Transformations}\label{appen:frob}

We now briefly introduce the second type of gauge transformation—the \emph{\(F\)-gauge transformation}---which has no analog in traditional lattice gauge theory. The \(F\)-gauge transformation was originally introduced in our recent work \cite{zhao2024b}. As these transformations are entirely novel and require substantial mathematical background and technical detail, we divide this appendix into three parts. In the first, we introduce the necessary mathematical background; in the subsequent two, we describe the action of \(F\)-gauge transformations on gauge fields and matter fields, respectively.

We begin by summarizing the main ideas of the \(F\)-gauge transformation:
\begin{itemize}

\item \textbf{Morita Equivalent UFCs and Equivalent HGW Models:} Distinct UFCs can share isomorphic Drinfeld centers. Such UFCs are said to be \emph{categorically Morita equivalent} \cite{etingof2016}. This implies that two HGW models with categorically Morita equivalent input UFCs are equivalent in the sense that they describe the same anyonic matter spectrum—and hence the same topological phase—although they couple to different gauge fields.

\item \textbf{Classification of Morita Equivalent UFCs:} Given a UFC \(\Fus\), all UFCs that are categorically Morita equivalent to \(\Fus\) are classified by the \emph{connected symmetric Frobenius algebras} \(\A\) in \(\Fus\) \cite{etingof2016}. Each such UFC, denoted \(\Fus_\A\), is realized as the \emph{bimodule category} \(\Bimod_\Fus(\A, \A)\), consisting of all \(\A\)-\(\A\)-bimodules—a special class of coherent representations of \(\A\) in \(\Fus\). The simple objects of \(\Bimod_\Fus(\A, \A)\) are simple \(\A\)-\(\A\)-bimodules, i.e., bimodules that cannot be decomposed as a direct sum of two nontrivial bimodules. The trivial simple bimodule is the algebra \(\A\) itself, with bimodule action given by the algebra multiplication. The explicit definitions of Frobenius algebras and bimodules can be found in Appendix \ref{appen:frobdef}.

The bimodule category \(\Fus_\A = \Bimod_\Fus(\A, \A)\) can be defined as a subcategory of \(\Fus\) via a forgetful embedding functor \(\Delta_{\A, \A}: \Fus_\A \to \Fus\), which maps each simple object \(M\) of \(\Fus_\A\) to a (possibly composite) object in \(\Fus\), and each morphism in \(\Fus_\A\) to a morphism in \(\Fus\), preserving the categorical structure of \(\Fus_\A\). For simplicity, we will often not distinguish between a simple bimodule \(M\in\Fus_\A\) and its image \(\Delta_{\A, \A}(M)\in\Fus\).

\item \textbf{Duality Map Between Equivalent Enlarged HGW Models:} Based on the functor \(\Delta_{\A, \A}: \Fus_\A = \Bimod_\Fus(\A) \to \Fus\) associated with the Morita equivalent UFC \(\Fus_\A\), we construct a corresponding unitary duality map \(\D_\A: \Hil_\text{GF}^{\Fus_\A} \to \Hil_\text{GF}^{\Fus}\) that relates the gauge-field Hilbert spaces of the two equivalent HGW models with input UFCs \(\Fus_\A\) and \(\Fus\). This duality map changes the gauge field of the model by transforming simple objects \(M\in L_{\Fus_\A}\) on each edge and tail of the dual model into superposition states of the original model. It preserves the anyon types in each plaquette but redefines the gauge spaces of each anyon type according to the transformation of the gauge field:
\[
\D_\A: \Hil^{\Fus_\A}_{J, P} \to \Hil^{\Fus}_{J, P}.
\]

\item \textbf{Isomorphic UFCs and Symmetry Transformations of the Gauge Field:} It may occur that the Morita equivalent UFC \(\Fus_\A\) is \emph{isomorphic} to the original UFC \(\Fus\) via an isomorphism functor\footnote{In general, isomorphism functors between two isomorphic, categorically Morita equivalent UFCs are not unique; different isomorphisms yield different symmetry transformations.} \(\iota_\A: \Fus \to \Fus_\A\), which establishes a one-to-one correspondence between their simple objects \(a\in L_\Fus\) and \(M_a\in L_{\Fus_\A}\). By composing this isomorphism with the duality functor \(\D_\A\), we obtain an endofunctor \(\Gamma_\A = \Delta_\A \circ \iota_\A: \Fus \to \Fus\), which maps each simple object of \(\Fus\) to a simple \(\A\)-\(\A\)-bimodule object in \(\Fus\). This functor induces a unitary transformation \(\G_\A: \Hil^\Fus_\text{GF} \to \Hil^\Fus_\text{GF}\) on the gauge-field Hilbert space of the original enlarged HGW model. In contrast to the duality transformation \(\D_\A: \Hil_\Fus \to \Hil_{\Fus_\A}\), this transformation leaves the underlying UFC \(\Fus\) unchanged and thus defines a genuine \emph{symmetry} of the gauge field.

\item \textbf{Symmetry Transformation of Matter Fields and \(F\)-Gauge Redundancies:} Due to the topological Gauss law, when the symmetry transformation \(\G_\A\) alters the gauge-field states, it may correspondingly permute anyon types. In such cases, we interpret \(\G_\A\) as generating a \emph{global symmetry} of the topological phase.\footnote{All nontrivial Frobenius algebras in \(\Vec(G)\) for a group \(G\) are algebraically Morita inequivalent to \(\A_0 = e\), so they always induce global symmetries known as the \emph{(partial) electric-magnetic exchange symmetries}.} In contrast, if \(\A\) is algebraically Morita equivalent to the trivial algebra \(\A_0 = \tob\), \(\G_\A\) preserves the anyon types and acts only on an anyon's gauge space spanned by the flux and charge dofs. Since anyon types are the only physical observables in topological phases, such a transformation constitutes a \emph{gauge redundancy} and is referred to as an \(F\)-gauge transformation.

\item \textbf{Local Gauge Transformations:} What we have defined above is the uniform \(F\)-gauge transformation, which applies the same transformation everywhere on the lattice simultaneously. To define \emph{local} \(F\)-gauge transformations, we introduce the local dual model by labeling each plaquette \(P\) of the lattice with a Frobenius algebra \(\A_P\) in the original UFC \(\Fus\), where each \(\A_P\) is algebraically Morita equivalent to the trivial Frobenius algebra \(\A_0\); different plaquettes may be labeled by different algebras. In the resulting local dual model, an edge \(E\) separating two plaquettes \(P\) and \(Q\) carries an \(\A_P\)-\(\A_Q\)-bimodule, while a tail within plaquette \(P\) carries an \(\A_P\)-\(\A_P\)-bimodule.

We can still define a set of forgetful functors \(\Delta_{\A_P,\A_Q}\) that embed each simple \(\A_P\)-\(\A_Q\)-bimodule as a (possibly composite) object in \(\Fus\). This induces a \emph{local duality transformation} \(\D_{\{\A\}}\) between the Hilbert space of the local dual model (with a given algebra distribution \(\{\A\}\)) and that of the original model with input UFC \(\Fus\), by transforming each simple \(\A_P\)-\(\A_Q\)-bimodule on edges and tails to a superposition state in the original model. The anyon type \(J\) in each plaquette \(P\) is preserved, while its internal Hilbert space transforms as
\[
\D_{\{\A\}}|_{\Hil_{J, P}^\Fus} = \D_{\A_P}: \Hil_{J, P}^\Fus \to \Hil_{J, P}^{\Fus_{\A_P}}.
\]

An \(\A_P\)-\(\A_Q\)-bimodule can nontrivially fuse with an \(\A_{Q}\)-\(\A_{R}\)-bimodule if and only if \(\A_Q = \A_{P'}\). Therefore, when \(\A_P \neq \A_Q\), the set of all \(\A_P\)-\(\A_Q\)-bimodules does not form a category. Nevertheless, if both \(\A_P\) and \(\A_Q\) are algebraically Morita equivalent to the trivial Frobenius algebra \(\A_0 = \tob\), we can establish a collection of bijections \(\iota = \{\iota_{\A_P, \A_Q}\}\) between \(L_\Fus\) and the set of all simple \(\A_P\)-\(\A_Q\)-bimodules that is compatible with the fusion rules:
\[
\iota_{\A_P\text{-}\A_Q}(a) \otimes \iota_{\A_Q\text{-}\A_R}(b) = \iota_{\A_P\text{-}\A_R}(a \otimes b).
\]
By composing the local duality map \(\D_{\{\A\}}\) with the isomorphisms \(\iota\), we obtain a local gauge transformation \(\G_{\{\A\}} = \D_{\{\A\}} \circ \iota: \Hil^\Fus \to \Hil^\Fus\) on the original enlarged HGW model with input UFC \(\Fus\). Since all Frobenius algebras \(\A_P\) are algebraically Morita equivalent to \(\A_0\), this transformation does not alter the anyon types in each plaquette. Nevertheless, the single-anyon Hilbert spaces \(\Hil_{J, P}^\Fus\) of anyon type \(J\) in different plaquettes \(P\) undergo distinct local gauge transformations \(\G_{\{\A\}}|_{\Hil^\Fus_{J, P}} = \G_{\A_P}\), each depending only on the specific Frobenius algebra \(\A_P\) in plaquette \(P\).

\end{itemize}
\subsection{Frobenius Algebras and Bimodules}\label{appen:frobdef}

In this section, we briefly review the definitions of Frobenius algebras and their bimodules.

A Frobenius algebra \(\A\) is characterized by a pair of functions \((n_\A, f^\A)\). The function \(n_\A: L_\Fus \to \NN\) returns the \emph{multiplicity} \(n_\A^a\) of \(a\in L_\Fus\) appearing in the Frobenius algebra \(\A\), satisfying \(n_\A^\tob = 1\) and \(n_\A^a = n_{a^\ast}\). The basis dofs of \(\A\) are labeled by pairs \(a_\alpha\), where \(a \in L_\Fus\) with \(n_\A^a > 0\) and \(\alpha = 1, 2, \ldots, n_\A^a\) is the \emph{multiplicity index}. We denote the set of all basis dofs in \(\A\) as \(L_\A\). The function \(f^\A: L^3_\A \to \C\) gives the algebraic multiplication in \(\A\), satisfying:
\eqn[eq:frob]{
\sum_{t_\tau \in L_\A} f^\A_{r_\rho s_\sigma t_\tau} f^\A_{a_\alpha b_\beta t_\tau^\ast} G^{rst}_{abc} \sqrt{d_c d_t} &= \sum_{\gamma = 1}^{n_c} f^\A_{a_\alpha c_\gamma s_\sigma} f^\A_{r_\rho c_\gamma^\ast b_\beta}\ ,\\
\sum_{a_\alpha b_\beta \in L_\A} f^\A_{a_\alpha b_\beta c_\gamma} f^\A_{b_\beta^\ast a_\alpha^\ast c_\gamma^\ast} \sqrt{d_a d_b} &= d_\A \sqrt{d_c},\\
[f_\A]_{a_\alpha b_\beta c_\gamma} = f^\A_{b_\beta c_\gamma a_\alpha}, \qquad &f^\A_{\tob a_\alpha b_\beta} = \delta_{ab^\ast} \delta_{\alpha\beta},
}
where \(d_\A := \sum_{a \in L_\Fus} n_\A^a d_a\) is the \emph{quantum dimension} of the Frobenius algebra \(\A\).

An \(\A_1\)-\(\A_2\)-bimodule \(M\) in a fusion category \(\Fus\) is characterized by a pair of functions \((n_M, P_M)\). The function \(n_M: L_\Fus \to \NN\) returns the \emph{multiplicity} \(n_M^x\) of \(x \in L_\Fus\) appearing in \(M\), satisfying \(n_M^x = n_M^{x^\ast}\). The basis dofs of \(M\) are labeled by pairs \(x_\chi\), where \(x \in L_\Fus\) with \(n_M^x > 0\), and \(\chi = 1, 2, \ldots, n_M^x\) labels the multiplicity index. We denote the set of all basis dofs in \(M\) as \(L_M\). The action of the Frobenius algebras on \(M\) is characterized by the function \(P_M: L_{\A_1}\times L_{\A_2} \times L_M \times L_\Fus \times L_M \to \C\), satisfying the following defining equations:
\eqn[eq:bimod]{
&\sum_{uv\in L_\Fus}\ \sum_{y_\upsilon\in L_M}\ [P_M]^{a_\alpha r_\rho}_{x_\chi u y_\upsilon}\ [P_M]^{b_\beta s_\sigma}_{y_\upsilon v z_\zeta}\ G^{v^\ast by}_{urw}\ G^{w^\ast bu}_{axc}\ G^{sz^\ast v}_{wrt^\ast}\ \sqrt{d_ud_vd_wd_yd_cd_t}\\ 
=\ &\sum_{\gamma = 1}^{n_\A^c}\ \sum_{\tau = 1}^{n_\A^t}\ [P_M]^{c_\gamma t_\tau}_{x_\chi w z_\zeta}f^{\A_1}_{a_\alpha c_\gamma^\ast b_\beta}\ f^{\A_2}_{r_\rho s_\sigma t_\tau},\\
&\qquad\qquad [P_M]^{\tob\tob}_{x_\chi y z_\zeta} = \delta_{xy}\delta_{yz}\delta_{\chi\upsilon}\delta_{\upsilon\zeta},\qquad [P_M]^{a_\alpha b_\beta}_{x_\chi y z_\zeta} = [P_M]^{b_\beta a_\alpha}_{z_\zeta^\ast y^\ast x_\chi^\ast}.
}
A simple \(\A_1\)-\(\A_2\)-bimodule \(M\) is an \(\A_1\)-\(\A_2\)-bimodule that cannot be decomposed as a direct sum of two nontrivial bimodules. The forgetful functor \(\Delta_{\A_1, \A_2}\) maps an \(\A_1\)-\(\A_2\)-bimodule \(M\) to a (possibly composite) object in \(\Fus\):
\[
\Delta_{\A_1, \A_2}(M) = \bigoplus_{x \in L_\Fus} n_M^x\, x,
\]
where \(n_M^x\) denotes the multiplicity of the simple object \(x\) in \(M\).

\begin{example}
Every Frobenius algebra in the group category \(\Vec(G)\) is determined by a subgroup \(H \subseteq G\) and a \(2\)-cocycle \(\omega \in H^2(H, {\tt U(1)})\), with
\[
\A_H = \bigoplus_{g\in H} g,\qquad f_{g, h, h^\ast g^\ast} = \omega(g, h),\qquad \omega(g, h)\omega(gh, f) = \omega(g, hf)\omega(h, f).
\]
For \(\Vec(S_3)\), there are four distinct Frobenius algebras (up to \(T\)-gauge transformations), each generating a bimodule category whose Drinfeld center is the UMTC \(D(S_3)\). Let \(S_3 = \langle r, s \mid r^3 = s^2 = (rs)^2 = e\rangle\).
\begin{itemize}
\item The trivial subgroup \(\{e\}\) gives the trivial Frobenius algebra \(\A_0 = e\), whose bimodule category is simply \(\Vec(S_3)\) itself.

\item The largest subgroup of \(S_3\) is \(S_3\) itself, which corresponds to the Frobenius algebra \(\A_{S_3} = e\oplus r\oplus r^2\oplus s\oplus rs\oplus sr\), whose bimodule category is the representation category of \(S_3\), denoted \(\Rep(S_3)\). This UFC has three simple objects: \(\tob\) (the trivial representation), \(\tt sgn\) (the sign representation), and \(\pi\) (the standard \(2\)-dimensional representation):
\[
M_\tob = \bigoplus_{g \in S_3} g, \qquad 
M_{\tt sgn} = \bigoplus_{g \in S_3} g, \qquad 
M_\pi = \bigoplus_{g \in S_3} (g \oplus g),
\]
where \(M_\tob\) and \(M_{\tt sgn}\) correspond to the same object in \(\Vec(S_3)\) but have different bimodule actions, while \(M_\pi\) includes each simple object of \(\Vec(S_3)\) with multiplicity two, reflecting the dimension of \(\pi\).

\item \textbf{\(\ZZ_3\) algebra:} The subgroup \(\ZZ_3 = \{e, r, r^2\}\) gives the Frobenius algebra \(\A_3 = 1\oplus r\oplus r^2\), whose bimodule category has six simple objects. Three correspond to the object \(1\oplus r\oplus r^2\), forming a subcategory equivalent to \(\Rep(\ZZ_3)\), and three correspond to the object \(s\oplus sr\oplus rs\). This bimodule category is isomorphic to, but not equal to, \(\Vec(S_3)\).

\item \textbf{\(\ZZ_2\) algebras:} The three subgroups \(\{1, s\}, \{1, rs\}, \{1, sr\}\) of \(S_3\) yield three Frobenius algebras in \(\Vec(S_3)\). However, these algebras are related by \(T\)-gauge transformations and are thus indistinguishable up to gauge equivalence. The bimodule category for any of these Frobenius algebras is isomorphic to, but distinct from, the \(\Rep(S_3)\) category. The three simple bimodules over the algebra \(1\oplus s\) are \(1\oplus s\), \(1\oplus s\), and \(r\oplus r^2\oplus rs\oplus sr\).
\end{itemize}
\end{example}
\subsection{\eqs{F}-Gauge Transformations on Gauge Field}

As summarized previously, an \(F\)-gauge transformation consists of two steps: a local duality transformation \(\D_{\{\A\}}\) followed by an isomorphism. The local duality transformation \(\D_{\{\A\}}\) maps each state in the local dual model—where each plaquette \(P\) of the lattice is labeled by a Frobenius algebra \(\A_P\)—to a superposition of states in the original model with input UFC \(\Fus\). In the local dual model, the basic configuration is established by labeling each edge \(E\) separating two plaquettes \(P\) and \(Q\) with an \(\A_P\)-\(\A_Q\)-bimodule, and each tail within plaquette \(P\) with an \(\A_P\)-\(\A_P\)-bimodule. The gauge field Hilbert space \(\Hil_\text{GF}^{\{\A\}}\) of the local dual model is spanned by all such basic configurations. From now on, we always assume that all Frobenius algebras we introduce are algebraically Morita equivalent to the trivial Frobenius algebra \(\A_0 = \tob\).

The local duality transformation \(\D_{\{\A\}}: \Hil_\text{GF}^{\{\A\}} \to \Hil_\text{GF}^\Fus\) can be understood as a tensor product of transformations acting on each edge and tail in the local dual model:
\eqn[eq:FtransEdge]{
\D_{\{\A\}}\quad \Edge{M}\qquad =\qquad\sum_{x_\chi, y_\upsilon\in L_M}\ \ \sum_{a_\alpha\in L_{\A_P}}\ \ \sum_{b_\beta\in L_{\A_Q}}\ \ \sum_{u\in L_\Fus}\quad \BimoduleA,
}
followed by a topological move that removes all auxiliary tails (the red tails) in Eq. \eqref{eq:FtransEdge}. Here, \(M\) is a simple \(\A_P\)-\(\A_Q\)-bimodule on an edge \(E\) separating two plaquettes \(P\) and \(Q\), or a simple \(\A_P\)-\(\A_P\)-bimodule on a tail within plaquette \(P\).

This local duality transformation \(\D_{\{\A\}}\) can also be understood on a plaquette-by-plaquette basis, clarifying why Frobenius algebras must be assigned individually to each plaquette in the local dual model and illustrating the process by which auxiliary tails are removed:
\eqn[]{
\D_{\{\A\}}\quad\BimoduleB\quad = \quad \frac{1}{d_{\A_P}^9}\ \sum_{x_i, y_i\in L_{M_i}}\ \ \sum_{u_i\in L_\Fus}\ \ \sum_{a_i, b_i\in L_{\A_P}}\ \ \sum_{p_\alpha, q\in L_M}\ \ \sum_{w\in L_\Fus}\\ \mathcal{T}\ \ 
[P_M]^{b_7^\ast b_0}_{p_\alpha wq}\prod_{i = 0}^6[P_{M_i}]^{r_ia_i}_{x_iu_iy_i}\ \prod_{j = 0}^6f^{\A_P}_{a_i^\ast b_i^\ast b_{i+1}}\ \ \BimoduleC\quad .
}
Here, the left-hand side denotes a local configuration around plaquette \(P\) in the local dual model, where \(M_i\) are simple \(\A_{P_i}\)-\(\A_P\)-bimodules assigned to edges separating the central plaquette \(P\) from its neighboring plaquettes \(P_i\) (with the convention \(P_0 = P_6\)); \(N_i\) are simple \(\A_{P_i}\)-\(\A_{P_{i+1}}\)-bimodules on edges between adjacent neighboring plaquettes \(P_i\) and \(P_{i+1}\); and \(M\) labels a simple \(\A_P\)-\(\A_P\)-bimodule on the tail within plaquette \(P\). The right-hand side represents a superposition state in the original model with input UFC \(\Fus\). Note that we do not sum over \(e_i \in L_{N_i}\), as these are dofs associated with edges belonging to neighboring plaquettes. The symbol ``{\color{red}\(\times\)}'' marks plaquettes that are to be contracted, while ``{\color{red}\(\cdot\)}'' denotes the fusion of three Frobenius algebra objects. Although \(a_i, b_i, x_i, y_i, q\) represent basic dofs of the bimodules and may carry multiplicity indices, the topological moves will eliminate these multiplicity indices, leaving only the dofs on edges corresponding to simple objects in \(L_\Fus\). Nevertheless, the multiplicity index \(\alpha\) of the basic dof \(p_\alpha \in L_M\) at the endpoint of the tail cannot be removed by such moves. Finally, the transformation of plaquettes reduces to
\eqn[eq:FiboDualPlaq]{
\D_{\{A\}}\quad\BimoduleD\qquad\Longrightarrow\qquad\sum_{i_k, e_k\in L_\Fus}\ \ \sum_{p_\alpha\in L_M}\ \ \cdots\qquad\BimoduleE\ .
}
As we will show in the next subsection, the multiplicity indices \(\alpha\) are, in fact, the internal dofs of anyons rather than of the gauge field. Accordingly, we must enlarge the single-anyon Hilbert space to include the \(F\)\emph{-charge space}, in order to carry these multiplicity indices and faithfully represent the action of \(F\)-gauge transformations, analogous to how the \(T\)-gauge space captures the \(T\)-gauge transformations.

For each pair of algebraically Morita equivalent Frobenius algebras \(\A_P, \A_Q\), a bijection \(\iota_{\A_P, \A_Q}: L_\Fus \to L_{\A_P, \A_Q}\) establishes a one-to-one correspondence between the simple objects in \(L_\Fus\) and the set of simple \(\A_P\)-\(\A_Q\)-bimodules:
\[
\iota_{\A_P, \A_Q}: a \in L_\Fus \mapsto M_a \in L_{\A_P, \A_Q},
\]
where \(L_{\A_P, \A_Q}\) denotes the set of all simple \(\A_P\)-\(\A_Q\)-bimodules. The collection of bijections \(\{\iota_{\A_P, \A_Q}\}\) induces a linear transformation \(i_{\{\A\}}: \Hil_\text{GF}^\Fus \to \Hil_\text{GF}^{\{\A\}}\) between the Hilbert spaces of the original model and the local dual model, which can be understood edge-by-edge (or tail-by-tail) as
\eqn[eq:iso]{
i_{\{\A\}} \quad \Edge{\alpha_l} \qquad = \qquad \Edge{\iota_{\A_P, \A_Q}(\alpha_l)}\ ,
}
where \(P, Q\) are the two plaquettes separated by edge \(l\) (or \(P = Q\) if \(l\) is a tail within plaquette \(P\)), and \(\alpha_l \in L_\Fus\) is the dof on edge or tail \(l\) in the original model.

By composing the isomorphism \(i_{\{\A\}}: \Hil_\text{GF}^\Fus \to \Hil_\text{GF}^{\{\A\}}\) with the duality map \(\D_{\{\A\}}: \Hil_\text{GF}^{\{\A\}} \to \Hil_\text{GF}^\Fus\), we obtain a symmetry transformation
\begin{equation}
\G_{\{\A\}} = \D_{\{\A\}} \circ i_{\{\A\}}: \Hil_\text{GF}^\Fus \to \Hil_\text{GF}^\Fus,
\end{equation}
which acts within the original gauge field Hilbert space \(\Hil_\text{GF}^\Fus\). The enlargement of Hilbert spaces arising from the multiplicity indices of bimodules is manifested in the \(F\)-charge space of the matter fields, rather than as an enlargement of the gauge field Hilbert space \(\Hil_\text{GF}^\Fus\).

Due to the topological Gauss law, the gauge field configuration fully determines the anyon types \(J\) and flux types \(p\) of anyons residing in each plaquette \(P\). Since all \(\A_P\) are algebraically Morita equivalent to the trivial Frobenius algebra \(\A_0 = \tob\), anyon types remain invariant, but flux types and \(F\)-charge dofs may transform under the symmetry transformation \(\G_{\{\A\}}\). We emphasize again that the anyon types in plaquettes are the only physical observables of the topological phase. Thus, \(\G_{\{\A\}}\) constitutes a \emph{local gauge redundancy}, with different plaquettes undergoing different local gauge transformations. The action of these gauge transformations on each anyon's Hilbert space \(\Hil_{J, P}^\Fus\) will be discussed in the subsequent section.

\subsection{The \eqs{F}-Gauge Spaces}

A subtlety remains in our previous discussion of gauge transformations, as seen in Eq. \eqref{eq:FiboDualPlaq}. After performing the gauge transformation, each edge still carries basic dofs valued in \(L_\Fus\), but the tails may instead carry dofs \(p_\alpha\) in a bimodule \(M\), labeled by both a simple object \(p \in L_\Fus\) and a multiplicity index \(1 \leq \alpha \leq n_M^p\). This necessitates enlarging the Hilbert space of the model to include these multiplicity indices.

This enlargement is also physically justified.\footnote{As an analogy, recall that a massless photon has only two physical dofs---the transverse polarizations---at each momentum. In principle, just two independent functions would suffice to specify the electromagnetic field completely. However, to formulate gauge transformations, we introduce the full four-potential \(A_\mu(k)\), which contains unphysical longitudinal (and timelike) components. These extra modes are not observable and can always be removed by imposing an appropriate gauge fixing.} An anyon excitation resides in a plaquette, where the tail encodes the internal charge of the anyon, reflecting the action of the Frobenius algebra \(\A\). The precise action of \(\A\) is only discernible when different occurrences of \(p\) in the representation (bimodule) of \(\A\) are distinguished by their multiplicity indices, i.e., \(p_\alpha\). In contrast, the dofs on edges pertain to the ground state, as any path along edges forms a closed loop. At each vertex along such a loop, the fusion rules are automatically satisfied by construction, and the fusion rules treat all \(p_\alpha\) as equivalent.\footnote{For comparison: it makes no sense to ask about the electric charge within a closed electric flux loop, since Gauss's law (analogous to the fusion rules) is satisfied everywhere along the loop. Only when the loop is cut open into a path does it make sense to ask about the charges at its endpoints, where Gauss's law is violated; this corresponds to the tails in our string-net model.}

Thus, the Hilbert space of the model is given by the tensor product of the gauge field Hilbert space \(\Hil^\Fus_\text{GF}\) and the anyon Hilbert spaces \(\Hil_{J, P}^\Fus\). Each anyon contributes a flux at the endpoint of a tail, which in turn determines the curvature and coincides with the dofs assigned to edges. In our enlarged HGW model, tails serve as ``internal'' edges mediating between the edge and anyon sectors, just like ordinary edges in the lattice, and thus still carry dofs from \(L_\Fus\).

This reasoning demonstrates that the gauge field Hilbert space itself remains unchanged as \(\Hil_\text{GF}^\Fus\). The enlargement of the Hilbert space occurs solely in the anyon Hilbert spaces. Specifically, we introduce the \(F\)-gauge space for anyon \(J\) in plaquette \(P\) to enlarge its Hilbert space, thereby enabling a faithful representation of the \(F\)-gauge transformations. This is analogous to the multidimensional \(T\)-charge spaces that serve as representation spaces for the noncommutative \(T\)-gauge transformations (i.e., the tube algebras).

To elaborate, an anyon of type \(J\) with nonabelian flux type \(p\)---that is, \(d_p > 1\)---may carry an additional multidimensional \(F\)-gauge charge space of dimension
\eqn[]{
\nu_J^p = \max_{\substack{\text{all Frobenius algebras }\A\\\text{all simple }\A\dash\A\text{-bimodules }M}}\{n_M^p\}.
}
The single-anyon Hilbert space \(\bar\Hil_{J, P}^\Fus\) for an anyon of type \(J\) in plaquette \(P\) is thus enlarged, with anyonic sectors labeled by
\[
\ket{J, p, \alpha, \kappa},\qquad p \in L_J,\qquad 1 \leq \alpha \leq n_J^p,\qquad 1 \leq \kappa \leq \nu_J^p.
\]
Accordingly, the total Hilbert space of our extended model is
\eqn[]{
\Hil^\Fus = \Hil^\Fus_\text{GF} \otimes \bigotimes_{\text{plaquette } P} \bigoplus_{\text{anyon type } J} \bar\Hil_{J, P}^\Fus.
}
The duality transformation \(\D_{\{\A\}}\) then embeds each (un-enlarged) state in the local dual model into the total Hilbert space of the original model,
\eqn[]{
&\D_{\{\A\}}\Bigl(\BimoduleD \otimes \ket{J, M, \alpha}_P\Bigr)\\
&\qquad \Longrightarrow\ \sum_{i_k, e_k \in L_\Fus} \ \sum_{p \in L_M} \cdots\ A_{M}^{p, \kappa}\ \BimoduleG \otimes \sum_{\alpha = 1}^{n_M^p} \ket{J, p, \alpha, \kappa}.
}
The coefficients of this embedding are discussed in Ref. \cite{zhao2025}. Nevertheless, the precise structure of this enlargement is still unclear and requires further investigation.

\section{Data}\label{appen:data}

\subsection{Dyons in the \eqs{\Vec(\ZZ_2)} model}

The HGW model with input \(\Vec(\ZZ_2)\) UFC describes the simplest and most celebrated nontrivial topological phase—the \(\ZZ_2\) toric code phase. Let \(\ZZ_2 = \{1, \psi\}\) with fusion rules and \(6j\) symbols given by
\[
\delta_{111} = \delta_{1\psi\psi} = 1, \qquad G^{abm}_{cdn} = \delta_{abm}\,\delta_{bcn}\,\delta_{cdm}\,\delta_{dan}.
\]
The toric code topological phase possesses four anyon types, all of which do not have multidimensional \(T\)-charge spaces.
\begin{itemize}
\item \textbf{The trivial anyon} \(\idm\) with flux types \(L_\idm = \{1\}\):
\[
[z^\idm_{11}]_1^1 = [z^\idm_{\psi\psi}]_1^1 = 1.
\]

\item \textbf{The pure fluxon} \(m\) with nontrivial flux type \(L_m = \{\psi\}\), which has a unique dyonic sector \((m, \psi)\):
\[
[z^m_{1\psi}]_\psi^\psi = [z^m_{\psi 1}]_\psi^\psi = 1.
\]

\item \textbf{The pure chargeon} \(e\) with flux type \(L_e = \{1\}\), which has a unique dyonic sector \((e, 1)\):
\[
[z^e_{11}]_1^1 = 1, \qquad [z^e_{\psi\psi}]_1^1 = -1.
\]

\item \textbf{The composite} \(\epsilon\) (\(=e \times m\)), with flux type \(L_\epsilon = \{\psi\}\), which has a unique dyonic sector \((\epsilon, \psi)\):
\[
[z^\epsilon_{1\psi}]_\psi^\psi = 1, \qquad [z^\epsilon_{\psi 1}]_\psi^\psi = -1.
\]
\end{itemize}

\subsection{Dyons in the \eqs{\Vec(\ZZ_3)} model}

The HGW model with input \(\Vec(\ZZ_3)\) UFC describes the doubled \(\ZZ_2\) topological phase. Let \(\ZZ_2 = \{e, r, r^2\}\) with fusion rules and \(6j\) symbols given by
\[
\delta_{eee} = \delta_{err^2} = \delta_{er^2r} = 1, \qquad G^{abm}_{cdn} = \delta_{abm}\,\delta_{bcn}\,\delta_{cdm}\,\delta_{dan}.
\]
The topological phase possesses nine anyon types. All these anyons do not have multidimensional \(T\)-charge spaces.
\begin{center}
\begin{tabular}{|c|c|c|}\hline
Anyon Type & Flux Type & Half-Braiding Tensors \\ \hline
$\idm$ & $e$ & $[z^\idm_{e,e}]_e^e = [z^\idm_{r,r}]_e^e = [z^\idm_{r^2,r^2}]_e^e = 1$ \\ \hline
$e$ & $e$ & $[z^e_{e,e}]_e^e = 1,\ [z^e_{r,r}]_e^e = \omega,\ [z^e_{r^2,r^2}]_e^e = \omega^\ast$ \\ \hline
$e^2$ & $e$ & $[z^{e^2}_{e,e}]_e^e = 1,\ [z^\idm_{r,r}]_e^e = \omega^\ast,\ [z^\idm_{r^2,r^2}]_e^e = \omega$ \\ \hline
$m$ & $r$ & $[z^m_{e,r}]_r^r = [z^m_{r,r^2}]_r^r = [z^m_{r^2,e}]_r^r = 1$ \\ \hline
$me$ & $r$ & $[z^{me}_{e,r}]_r^r = 1,\ [z^{me}_{r,r^2}]_r^r = \omega,\ [z^{me}_{r^2,e}]_r^r = \omega^\ast$ \\ \hline
$me^2$ & $r$ & $[z^{me^2}_{e,r}]_r^r = 1,\ [z^{me^2}_{r,r^2}]_r^r = \omega^\ast,\ [z^{me^2}_{r^2,e}]_r^r = \omega$ \\ \hline
$m^2$ & $r^2$ & $[z^{m^2}_{e,r^2}]_{r^2}^{r^2} = [z^{m^2}_{r,e}]_{r^2}^{r^2} = [z^{m^2}_{r^2,r}]_{r^2}^{r^2} = 1$ \\ \hline
$m^2e$ & $r^2$ & $[z^{m^2e}_{e,r^2}]_{r^2}^{r^2} = 1,\ [z^{m^2e}_{r,e}]_{r^2}^{r^2} = \omega,\ [z^{m^2e}_{r^2,r}]_{r^2}^{r^2} = \omega^\ast$ \\ \hline
$m^2e^2$ & $r^2$ & $[z^{m^2e}_{e,r^2}]_{r^2}^{r^2} = 1,\ [z^{m^2e}_{r,e}]_{r^2}^{r^2} = \omega^\ast,\ [z^{m^2e}_{r^2,r}]_{r^2}^{r^2} = \omega$ \\ \hline
\end{tabular}
\end{center}
Here, \(\omega = (-1 + \sqrt{3}\ii)/2\) is the cubic root of unity.

\subsection{Dyons in the \eqs{\Vec(S_3)} model}

The HGW model with input \(\Vec(S_3)\) UFC describes the doubled \(S_3\) topological phase. Let 
\[S_3 = \{e, r, r^2, s, sr, rs \mid r^3 = s^2 = (sr)^2 = e\}\]
with the \(6j\) symbols given by
\[G^{abm}_{cdn} = \delta_{abm}\,\delta_{bcn^\ast}\,\delta_{cdm^\ast}\,\delta_{dan}.
\]
The topological phase possesses eight anyon types, labeled by \(A, B, C, D, E, F, G, H\). 
\begin{itemize}
\item Anyon \(A\) is the trivial anyon
\[[z^A_{xx}]_e^e = 1,\qquad x\in S_3.\]
\item Anyon \(B\) has only trivial flux type \(L_B = \{e\}\) with no multidimensional \(T\)-charge space:
\[[z^B_{xx}]_e^e = {\tt sgn}(x),\qquad x\in S_3,\]
where \(\tt sgn\) is the sign representation of \(S_3\) group:
\[{\tt sgn}(e) = {\tt sgn}(r) = {\tt sgn}(r^2) = 1,\qquad {\tt sgn}(s) = {\tt sgn}(rs) = {\tt sgn}(sr) = -1.\]
\item Anyon \(C\) has only trivial flux type \(L_C = \{e\}\). Nevertheless, it has a \(2\)-dimensional \(T\)-charge space. So there are two dyonic sectors \((C, e, 1)\) and \((C, e, 2)\) of anyon \(C\):
\[[z^C_{xx}]_{e\alpha}^{e\beta} = D^\pi_{\beta\alpha}(x),\qquad x\in S_3,\]
where \(D^\pi(x)\) is the \(2\)-dimensional irreducible representation matrix of \(S_3\) group. In this paper, we choose
\[D^\pi(r) = \begin{pmatrix}\omega & 0 \\ 0 & \omega^\ast\end{pmatrix},\qquad D^\pi(s) = \begin{pmatrix} 0 & 1 \\ 1 & 0\end{pmatrix}.\]
\item Anyon \(D, E\) both have three flux type \(L_D = L_E = \{s, sr, rs\}\) without multidimensional \(T\)-charge spaces.
\[[z^D_{xy}]_p^q = \delta_{p\in L_D}\delta_{pxy^\ast}\delta_{qx^\ast y},\qquad [z^E_{xy}]_p^q = {\tt sgn}(x)\delta_{p\in L_E}\delta_{pxy^\ast}\delta_{qx^\ast y},\]
\item Anyon \(F, G, H\) both have three flux type \(L_F = L_G = L_H = \{r, r^2\}\) without multidimensional \(T\)-charge spaces.
\begin{center}
\begin{tabular}{|c|c|c|c|c|c|}\hline
$J$ & $x$ & $y$ & $p$ & $q$ & $[z^J_{xy}]_p^q$ \\ \hline
$F$ & $r^i$ & $r^{i+1}$ & $r$ & $r$ & $1$\\ \hline
$F$ & $sr^i$ & $sr^{i+2}$ & $r$ & $r^2$ & $1$ \\ \hline
$F$ & $r^i$ & $r^{i+2}$ & $r^2$ & $r^2$ & $1$ \\ \hline
$F$ & $sr^i$ & $sr^{i+1}$ & $r^2$ & $r$ & $1$ \\ \hline
$G$ & $r^i$ & $r^{i+1}$ & $r$ & $r$ & $\omega^i$\\ \hline
$G$ & $r^i$ & $r^{i+2}$ & $r^2$ & $r^2$ & $\omega^{-i}$ \\ \hline
$G$ & $r^is$ & $r^{i+1}s$ & $r$ & $r^2$ & $\omega^{i}$ \\ \hline
$G$ & $r^is$ & $r^{i+2}s$ & $r^2$ & $r$ & $\omega^{-i}$ \\ \hline
$H$ & $r^i$ & $r^{i+1}$ & $r$ & $r$ & $\omega^{-i}$\\ \hline
$H$ & $r^i$ & $r^{i+2}$ & $r^2$ & $r^2$ & $\omega^{i}$ \\ \hline
$H$ & $r^is$ & $r^{i+1}s$ & $r$ & $r^2$ & $\omega^{-i}$ \\ \hline
$H$ & $r^is$ & $r^{i+2}s$ & $r^2$ & $r$ & $\omega^{i}$ \\ \hline
\end{tabular}
\end{center}
Here, \(\omega = (-1 + \sqrt{3}\ii)/2\) is the root of unity, and \(i = 0, 1, 2\). Direct calculation of the half-braiding tensors reveals gauge redundancies among their components; however, recent studies on gauging global symmetries~\cite{Fu2025, Fu2025b} demonstrate that these redundancies can be systematically fixed.
\end{itemize}

\subsection{Dyons in the Ising model}

The HGW model with input Ising UFC describes the doubled Ising topological phase.

There are three simple objects \(1, \psi\), and \(\sigma\) in the Ising UFC, with fusion rules and \(6j\) symbols
\[\delta_{111} = \delta_{1\psi\psi} = \delta_{1\sigma\sigma} = \delta_{\psi\sigma\sigma} = 1,\]
\[G^{111}_{111} = G^{1\psi\psi}_{1\psi\psi} = G^{111}_{\psi\psi\psi} = 1,\ G^{1\sigma\sigma}_{1\sigma\sigma} = G^{1\sigma\sigma}_{\psi\sigma\sigma} = \frac{1}{2},\ G^{111}_{\sigma\sigma\sigma} = G^{1\psi\psi }_{\sigma\sigma\sigma} = \frac{\sqrt{2}}{2},\ G^{\psi\sigma\sigma}_{\psi\sigma\sigma} = -\frac{1}{2}.\]
There are \(9\) types of doubled Ising anyons, all of which do not have multidimensional \(T\)-charge spaces.
\begin{itemize}
\item Anyon \(1\bar 1\) is the trivial anyon with trivial flux type \(L_{1\bar 1} = \{1\}\).
\[[z^{1\bar 1}_{11}]_1^1 = [z^{1\bar 1}_{\psi\psi}]_1^1 = [z^{1\bar 1}_{\sigma\sigma}]_1^1 = 1.\]
\item Anyon \(\psi\bar\psi\) has a unique trivial flux type \(L_{\psi\bar\psi} = \{1\}\).
\[[z^{\psi\bar\psi}_{11}]_1^1 = [z^{\psi\bar \psi}_{\psi\psi}]_1^1 = 1,\qquad [z^{\psi\bar\psi}_{\sigma\sigma}]_1^1 = -1.\]
\item Anyon \(\psi\bar 1\) and \(1\bar\psi\) both have a unique nontrivial flux type \(L_{\psi\bar 1} = L_{1\bar\psi} = \{\psi\}\).
\[[z^{\psi\bar 1}_{1\psi}]_\psi^\psi = [z^{1\bar \psi}_{1\psi}]_\psi^\psi = 1,\qquad [z^{\psi\bar 1}_{\psi 1}]_\psi^\psi = [z^{1\bar \psi}_{\psi 1}]_\psi^\psi = -1,\qquad [z^{\psi\bar 1}_{\sigma\sigma}]_\psi^\psi = \ii,\qquad [z^{1\bar \psi}_{\sigma\sigma}]_\psi^\psi = -\ii.\]
\item Anyon \(\sigma\bar 1, \sigma\bar\psi, 1\bar\sigma\), and \(\psi\bar\sigma\) all have a unique nontrivial flux type \(\sigma\).
\[[z^{\sigma\bar 1}_{1\sigma}]_\sigma^\sigma = [z^{\sigma\bar\psi}_{1\sigma}]_\sigma^\sigma = [z^{1\bar\sigma}_{1\sigma}]_\sigma^\sigma = [z^{\psi\bar\sigma}_{1\sigma}]_\sigma^\sigma = 1,\]
\[[z^{\sigma\bar 1}_{\psi\sigma}]_\sigma^\sigma = [z^{\sigma\bar\psi}_{\psi\sigma}]_\sigma^\sigma = \ii,\qquad [z^{1\bar\sigma}_{\psi\sigma}]_\sigma^\sigma = [z^{\psi\bar\sigma}_{\psi\sigma}]_\sigma^\sigma = -\ii,\]
\[[z^{\sigma\bar 1}_{\sigma 1}]_\sigma^\sigma = \ee^{\frac{\ii\pi}{8}},\qquad [z^{1\bar\sigma}_{\sigma \psi}]_\sigma^\sigma = \ee^{\frac{3\ii\pi}{8}},\qquad [z^{\sigma\bar\psi}_{\sigma\psi}]_\sigma^\sigma = \ee^{\frac{5\ii\pi}{8}},\qquad [z^{\psi\bar\sigma}_{\sigma 1}]_\sigma^\sigma = \ee^{\frac{7\ii\pi}{8}},\]
\[[z^{1\bar\sigma}_{\sigma 1}]_\sigma^\sigma = \ee^{-\frac{\ii\pi}{8}},\qquad [z^{\sigma\bar 1}_{\sigma \psi}]_\sigma^\sigma = \ee^{-\frac{3\ii\pi}{8}},\qquad [z^{\psi\bar\sigma}_{\sigma\psi}]_\sigma^\sigma = \ee^{-\frac{5\ii\pi}{8}},\qquad [z^{\sigma\bar\psi}_{\sigma 1}]_\sigma^\sigma = \ee^{-\frac{7\ii\pi}{8}}.\]

\item Anyon \(\sigma\bar\sigma\) has two flux types \(1\) and \(\psi\).
\[[z^{\sigma\bar\sigma}_{11}]_1^1 = 1,\qquad [z^{\sigma\bar\sigma}_{\psi\psi}]_1^1 = -1,\qquad [z^{\sigma\bar\sigma}_{1\psi}]_\psi^\psi = [z^{\sigma\bar\sigma}_{\psi 1}]_\psi^\psi = 1,\qquad [z^{\sigma\bar\sigma}_{\sigma\sigma}]_1^\psi = [z^{\sigma\bar\sigma}_{\sigma\sigma}]_\psi^1 = 1.\]
\end{itemize}

Here, we also list the modular \(S\) and \(T\) matrix elements of the doubled Ising topological phase.
\[S\quad =\quad \frac{1}{4}\quad\begin{pNiceMatrix}[first-row, first-col]
& 1\bar 1 & 1\bar\sigma & 1\bar\psi & \sigma\bar 1 & \sigma\bar\sigma & \sigma\bar\psi & \psi\bar 1 & \psi\bar\sigma & \psi\bar\psi \\
1\bar 1 & 1 & \sqrt{2} & 1 & \sqrt{2} & 2 & \sqrt{2} & 1 & \sqrt{2} & 1 \\
1\bar \sigma & \sqrt{2} & 0 & -\sqrt{2} & 2 & 0 & -2 & \sqrt{2} & 0 & -\sqrt{2} \\
1\bar \psi & 1 & -\sqrt{2} & 1 & \sqrt{2} & -2 & \sqrt{2} & 1 & -\sqrt{2} & 1 \\
\sigma\bar 1 & \sqrt{2} & 2 & \sqrt{2} & 0 & 0 & 0 & -\sqrt{2} & -2 & -\sqrt{2}\\
\sigma\bar \sigma & 2 & 0 & -2 & 0 & 0 & 0 & -2 & 0 & 2 \\
\sigma\bar \psi & \sqrt{2} & -2 & \sqrt{2} & 0 & 0 & 0 & -\sqrt{2} & 2 & -\sqrt{2}\\
\psi\bar 1 & 1 & \sqrt{2} & 1 & -\sqrt{2} & -2 & -\sqrt{2} & 1 & \sqrt{2} & 1 \\
\psi\bar \sigma & \sqrt{2} & 0 & -\sqrt{2} & -2 & 0 & 2 & \sqrt{2} & 0 & -\sqrt{2} \\
\psi\bar \psi & 1 & -\sqrt{2} & 1 & -\sqrt{2} & 2 & -\sqrt{2} & 1 & -\sqrt{2} & 1
\end{pNiceMatrix}.\]

\begin{center}
\begin{tabular}{|c|ccccccccc|}\hline
$J$ & $1\bar 1$ & $1\bar\sigma$ & $1\bar\psi$ & $\sigma\bar 1$ & $\sigma\bar\sigma$ & $\sigma\bar\psi$ & $\psi\bar 1$ & $\psi\bar\sigma$ & $\psi\bar\psi$ \\ \hline
$T_{JJ}$ & $1$ & $\ee^{\frac{\ii\pi}{8}}$ & $-1$ & $\ee^{-\frac{\ii\pi}{8}}$ & $1$ & $\ee^{\frac{7\ii\pi}{8}}$ & $-1$ & $\ee^{-\frac{7\ii\pi}{8}}$ & $-1$ \\ \hline
\end{tabular}
\end{center}

\subsection{Dyons in the Fobonacci model}

The Fibonacci fusion category has two simple objects, denoted as $1$ and $\tau$. The nonzero fusion rules are $\delta_{111} = \delta_{1\tau\tau} = \delta_{\tau\tau\tau} = 1$, and the quantum dimensions are 
$$d_1 = 1,\qquad d_\tau = \phi = \frac{\sqrt{5} + 1}{2}.$$
The nonzero $6j$ symbols are
$$G^{111}_{111} = 1,\qquad G^{111}_{\tau\tau\tau} = \frac{1}{\sqrt{\phi}},\qquad G^{1\tau\tau}_{1\tau\tau} = G^{1\tau\tau}_{\tau\tau\tau} = \frac{1}{\phi},\qquad G^{\tau\tau\tau}_{\tau\tau\tau} = -\frac{1}{\phi^2},$$
where \(\phi = (1 + \sqrt{5}) / 2\) is the Golden ration. 

The enlarged HGW model with Fibonacci input UFC has four anyon types:
\begin{itemize}
\item The trivial anyon \(1\bar 1\) with only trivial flux type \(1\):
\[[z^{1\bar 1}_{11}]_1^1 = [z^{1\bar 1}_{\tau\tau}]_1^1 = 1.\]
\item Anyon types \(\tau\bar 1\) and \(1\bar\tau\) both have a unique nontrivial flux type \(\tau\):
\[[z^{\tau\bar 1}_{1\tau}]_\tau^\tau = 1,\qquad [z^{\tau\bar 1}_{\tau 1}]_\tau^\tau = - \frac{\phi}{2} - \frac{\ii}{2}\sqrt{\frac{\sqrt{5}}{\phi}},\qquad [z^{\tau\bar 1}_{\tau\tau}]_\tau^\tau = -\frac{1}{2\phi} + \frac{\ii}{2}\sqrt{\sqrt{5}\phi}.\]
\[[z^{1\bar\tau}_{1\tau}]_\tau^\tau = 1,\qquad [z^{1\bar\tau}_{\tau 1}]_\tau^\tau = - \frac{\phi}{2} + \frac{\ii}{2}\sqrt{\frac{\sqrt{5}}{\phi}},\qquad [z^{1\bar\tau}_{\tau\tau}]_\tau^\tau = -\frac{1}{2\phi} - \frac{\ii}{2}\sqrt{\sqrt{5}\phi}.\]
\item Anyon \(\tau\bar\tau\) has two flux types \(1\) and \(\tau\):
\[[z^{\tau\bar\tau}_{11}]_1^1 = 1,\qquad [z^{\tau\bar\tau}_{\tau\tau}]_1^1 = -\frac{1}{\phi^2},\qquad [z^{\tau\bar\tau}_{1\tau}]_\tau^\tau = 1,\qquad [z^{\tau\bar\tau}_{\tau 1}]_\tau^\tau = 1,\]
\[[z^{\tau\bar\tau}_{\tau\tau}]_\tau^\tau = \frac{1}{\phi^2},\qquad [z^{\tau\bar\tau}_{\tau\tau}]_1^\tau = [z^{\tau\bar\tau}_{\tau\tau}]_\tau^1 = \frac{\sqrt[4]{5}}{\phi}.\]
\end{itemize}

\subsection{Dyon Type Rearrangement in Anyon Condensations}

During the anyon condensation from the Ising model to the \(\Vec(\ZZ_2)\) model, the dyon types transform as follows.

\begin{center}
\begin{tabular}{|c|c|c|c|c|}\hline
Parent Anyon & Parent Dyon & Child Anyon & Child Dyon & Phenomenon \\ \hline
$1\bar 1$ & $(1\bar 1, 1)$ & \multirow{2}{*}{$\idm$} & \multirow{2}{*}{$(\idm, 1)$} & \multirow{2}{*}{Condensation} \\ \cline{1-2}
$\psi\bar \psi$ & $(\psi\bar\psi, 1)$ & & & \\ \hline
$\psi\bar 1$ & $(\psi\bar 1, \psi)$ & \multirow{2}{*}{$\epsilon$} & \multirow{2}{*}{$(\epsilon, \psi)$} & \multirow{2}{*}{Condensation} \\ \cline{1-2}
$1\bar \psi$ & $(1\bar\psi, \psi)$ & & & \\ \hline
\multirow{2}{*}{$\sigma\bar\sigma$} & $(\sigma\bar\sigma, 1)$ & $e$ & \((e, 1)\) & \multirow{2}{*}{Splitting} \\ \cline{2-3}
& $(\sigma\bar\sigma, \psi)$ & $m$ & \((m, \psi)\) & \\ \hline
$\sigma\bar 1$ & $(\sigma\bar 1, \sigma)$ & \multicolumn{3}{c|}{} \\ \cline{1-2}
$\sigma\bar\psi$ & $(\sigma\bar\psi, \sigma)$ & \multicolumn{3}{c|}{\multirow{2}{*}{Confinement}} \\ \cline{1-2}
$1\bar\sigma$ & $(1\bar\sigma, \sigma)$ & \multicolumn{3}{c|}{} \\ \cline{1-2}
$\psi\bar\sigma$ & $(\psi\bar\sigma, \sigma)$ & \multicolumn{3}{c|}{} \\ \hline
\end{tabular}
\end{center}

During the anyon condensation from the \(\Vec(S_3)\) model to the \(\Vec(\ZZ_3)\) model, the dyon types transform as follows.

\begin{center}
\begin{tabular}{|c|c|c|c|c|}\hline
Parent Anyon & Parent Dyon & Child Anyon & Child Dyon & Phenomenon \\ \hline
$A$ & $(A, e)$ & \multirow{2}{*}{$\idm$} & \multirow{2}{*}{$(\idm, e)$} & \multirow{2}{*}{Condensation} \\ \cline{1-2}
$B$ & $(A, e)$ & & & \\ \hline 
\multirow{2}{*}{$C$} & $(C, e, 1)$ & $e$ & $(e, e)$ & \multirow{2}{*}{Splitting} \\ \cline{2-3}
& $(C, e, 2)$ & $e^2$ & $(e^2, e)$ & \\ \hline
\multirow{2}{*}{$F$} & $(F, r)$ & $m$ & $(m, r)$ & \multirow{2}{*}{Splitting} \\ \cline{2-4}
& $(F, r^2)$ & $m^2$ & $(m^2, r^2)$ & \\ \hline
\multirow{2}{*}{$G$} & $(G, r)$ & $em$ & $(em, r)$ & \multirow{2}{*}{Splitting} \\ \cline{2-4}
& $(G, r^2)$ & $em^2$ & $(em^2, r^2)$ & \\ \hline
\multirow{2}{*}{$H$} & $(H, r)$ & $e^2m$ & $(e^2m, r)$ & \multirow{2}{*}{Splitting} \\ \cline{2-4}
& $(H, r^2)$ & $e^2m^2$ & $(e^2m^2, r^2)$ & \\ \hline
$D, E$ & $(D, p), (E, p)$ & \multicolumn{3}{c|}{Confinement} \\ \hline
\end{tabular}
\end{center}

\subsection{Simple Bimodules over Nontrivial Frobenius in the Ising UFC}

In the Ising UFC, the nontrivial Frobenius algebra
\[\A = 1\oplus\psi\]
has three simple bimodules
\[M_1 = 1\oplus\psi,\qquad M_\psi = 1\oplus\psi,\qquad M_\sigma = \sigma\oplus\sigma.\]
The third bimodule \(M_\sigma\) contains simple object \(\sigma\) with multiplicity \(n_{M_\sigma}^\sigma = 2\). The corresponding bimodule tensor components are
\eq{
&[P_{M_0}]^{11}_{111}=[P_{M_0}]^{11}_{\psi\psi\psi}=[P_{M_0}]^{1\psi}_{11\psi}=[P_{M_0}]^{1\psi}_{\psi\psi1}=1,\\
&[P_{M_0}]^{\psi1}_{1\psi\psi}=[P_{M_0}]^{\psi1}_{\psi11}=[P_{M_0}]^{\psi\psi}_{1\psi1}=[P_{M_0}]^{\psi\psi}_{\psi1\psi}=1.\\
&[P_{M_1}]^{11}_{\sigma_0\sigma\sigma_0}=[P_{M_1}]^{11}_{\sigma_1\sigma\sigma_1}=1,\ [P_{M_1}]^{1\psi}_{\sigma_0\sigma\sigma_1}=[P_{M_1}]^{\psi1}_{\sigma_1\sigma\sigma_0}=\ee^{-\frac{\ii\pi}{4}},\\
&[P_{M_1}]^{1\psi}_{\sigma_1\sigma\sigma_0}=[P_{M_1}]^{\psi1}_{\sigma_0\sigma\sigma_1}=\ee^{\frac{\ii\pi}{4}},\ [P_{M_1}]^{\psi\psi}_{\sigma_0\sigma\sigma_0}=\ii,\ [P_{M_1}]^{\psi\psi}_{\sigma_1\sigma\sigma_1}=-\ii.\\
&[P_{M_2}]^{11}_{111}=[P_{M_2}]^{11}_{\psi\psi\psi}=1,\ [P_{M_2}]^{1\psi}_{11\psi}=[P_{M_2}]^{\psi1}_{\psi11}=\ii,\\
&[P_{M_2}]^{1\psi}_{\psi\psi1}=[P_{M_2}]^{\psi1}_{1\psi\psi}=-\ii,\ [P_{M_2}]^{\psi\psi}_{1\psi1}=[P_{M_2}]^{\psi\psi}_{\psi1\psi}=-1.\\
}

\bibliographystyle{apsrev4-1}
\bibliography{StringNet}
\end{document}

%% file: input.tex
\newcommand{\Lattice}{
\begin{tikzpicture}[baseline={([yshift=-.8ex]current bounding box.center)},scale=1]
\draw [decorate,decoration={snake,segment length=.12cm, amplitude=.02cm}] (1.7,6.8) -- (2.3,6.8);
\draw [->-] (1.7,6.4) -- (1.7,6.8);
\draw [->-] (1.7,6.8) -- (1.7,7.2);
\draw [->-] (1.7,7.2) -- (2.4,7.6);
\draw [->-] (2.4,7.6) -- (3.1,7.2);
\draw [->-] (2.4,6.) -- (1.7,6.4);
\draw [->-] (3.1,6.4) -- (2.4,6.);
\draw [->-] (2.4,5.2) -- (2.4,5.6);
\draw [->-] (2.4,5.6) -- (2.4,6.);
\draw [decorate,decoration={snake,segment length=.12cm, amplitude=.02cm}] (2.4,5.6) -- (3.,5.6);
\draw [decorate,decoration={snake,segment length=.12cm, amplitude=.02cm}] (3.1,6.8) -- (3.7,6.8);
\draw [->-] (3.1,6.4) -- (3.1,6.8);
\draw [->-] (3.1,6.8) -- (3.1,7.2);
\draw [->-] (3.1,7.2) -- (3.8,7.6);
\draw [->-] (3.8,7.6) -- (4.5,7.2);
\draw [->-] (3.8,6.) -- (3.1,6.4);
\draw [->-] (4.5,6.4) -- (3.8,6.);
\draw [->-] (3.8,5.2) -- (3.8,5.6);
\draw [->-] (3.8,5.6) -- (3.8,6.);
\draw [decorate,decoration={snake,segment length=.12cm, amplitude=.02cm}] (3.8,5.6) -- (4.4,5.6);
\draw [decorate,decoration={snake,segment length=.12cm, amplitude=.02cm}] (4.5,6.8) -- (5.1,6.8);
\draw [->-] (4.5,6.4) -- (4.5,6.8);
\draw [->-] (4.5,6.8) -- (4.5,7.2);
\draw [->-] (4.5,7.2) -- (5.2,7.6);
\draw [->-] (5.2,7.6) -- (5.9,7.2);
\draw [->-] (5.2,6.) -- (4.5,6.4);
\draw [->-] (5.9,6.4) -- (5.2,6.);
\draw [->-] (5.2,5.2) -- (5.2,5.6);
\draw [->-] (5.2,5.6) -- (5.2,6.);
\draw [decorate,decoration={snake,segment length=.12cm, amplitude=.02cm}] (5.2,5.6) -- (5.8,5.6);
\draw [decorate,decoration={snake,segment length=.12cm, amplitude=.02cm}] (5.9,6.8) -- (6.5,6.8);
\draw [->-] (5.9,6.4) -- (5.9,6.8);
\draw [->-] (5.9,6.8) -- (5.9,7.2);
\draw [->-] (5.9,7.2) -- (6.6,7.6);
\draw [->-] (6.6,7.6) -- (7.3,7.2);
\draw [->-] (6.6,6.) -- (5.9,6.4);
\draw [->-] (7.3,6.4) -- (6.6,6.);
\draw [->-] (6.6,5.2) -- (6.6,5.6);
\draw [->-] (6.6,5.6) -- (6.6,6.);
\draw [decorate,decoration={snake,segment length=.12cm, amplitude=.02cm}] (6.6,5.6) -- (7.2,5.6);
\draw [decorate,decoration={snake,segment length=.12cm, amplitude=.02cm}] (1.7,4.4) -- (2.3,4.4);
\draw [->-] (1.7,4.) -- (1.7,4.4);
\draw [->-] (1.7,4.4) -- (1.7,4.8);
\draw [->-] (1.7,4.8) -- (2.4,5.2);
\draw [->-] (2.4,5.2) -- (3.1,4.8);
\draw [->-] (2.4,3.6) -- (1.7,4.);
\draw [->-] (3.1,4.) -- (2.4,3.6);
\draw [->-] (2.4,3.2) -- (2.4,3.6);
\draw [decorate,decoration={snake,segment length=.12cm, amplitude=.02cm}] (3.1,4.4) -- (3.7,4.4);
\draw [->-] (3.1,4.) -- (3.1,4.4);
\draw [->-] (3.1,4.4) -- (3.1,4.8);
\draw [->-] (3.1,4.8) -- (3.8,5.2);
\draw [->-] (3.8,5.2) -- (4.5,4.8);
\draw [->-] (3.8,3.6) -- (3.1,4.);
\draw [->-] (4.5,4.) -- (3.8,3.6);
\draw [->-] (3.8,3.2) -- (3.8,3.6);
\draw [decorate,decoration={snake,segment length=.12cm, amplitude=.02cm}] (4.5,4.4) -- (5.1,4.4);
\draw [->-] (4.5,4.) -- (4.5,4.4);
\draw [->-] (4.5,4.4) -- (4.5,4.8);
\draw [->-] (4.5,4.8) -- (5.2,5.2);
\draw [->-] (5.2,5.2) -- (5.9,4.8);
\draw [->-] (5.2,3.6) -- (4.5,4.);
\draw [->-] (5.9,4.) -- (5.2,3.6);
\draw [->-] (5.2,3.2) -- (5.2,3.6);
\draw [decorate,decoration={snake,segment length=.12cm, amplitude=.02cm}] (5.9,4.4) -- (6.5,4.4);
\draw [->-] (5.9,4.) -- (5.9,4.4);
\draw [->-] (5.9,4.4) -- (5.9,4.8);
\draw [->-] (5.9,4.8) -- (6.6,5.2);
\draw [->-] (6.6,5.2) -- (7.3,4.8);
\draw [->-] (6.6,3.6) -- (5.9,4.);
\draw [->-] (7.3,4.) -- (6.6,3.6);
\draw [->-] (6.6,3.2) -- (6.6,3.6);
\draw [decorate,decoration={snake,segment length=.12cm, amplitude=.02cm}] (7.3,6.8) -- (7.9,6.8);
\draw [->-] (7.3,6.4) -- (7.3,6.8);
\draw [->-] (7.3,6.8) -- (7.3,7.2);
\draw [->-] (7.3,7.2) -- (7.6,7.4);
\draw [->-] (8.,6.) -- (7.3,6.4);
\draw [->-] (8.3,6.2) -- (8.,6.);
\draw [->-] (8.,5.2) -- (8.,5.6);
\draw [->-] (8.,5.6) -- (8.,6.);
\draw [decorate,decoration={snake,segment length=.12cm, amplitude=.02cm}] (8.,5.6) -- (8.6,5.6);
\draw [decorate,decoration={snake,segment length=.12cm, amplitude=.02cm}] (7.3,4.4) -- (7.9,4.4);
\draw [->-] (7.3,4.) -- (7.3,4.4);
\draw [->-] (7.3,4.4) -- (7.3,4.8);
\draw [->-] (7.3,4.8) -- (8.,5.2);
\draw [->-] (8.,5.2) -- (8.3,5.);
\draw [->-] (8.,3.6) -- (7.3,4.);
\draw [->-] (8.,3.2) -- (8.,3.6);
\draw [->-] (6.6,7.6) -- (6.6,8.);
\draw [->-] (5.2,7.6) -- (5.2,8.);
\draw [->-] (3.8,7.6) -- (3.8,8.);
\draw [->-] (2.4,7.6) -- (2.4,8.);
\draw [->-] (8.3,3.8) -- (8.,3.6);
\draw [->-] (1.7,4.) -- (1.3,3.8);
\draw [->-] (1.7,4.8) -- (1.3,5.);
\draw [->-] (1.7,6.4) -- (1.3,6.2);
\draw [->-] (1.7,7.2) -- (1.3,7.4);

\draw [<-] (2.6,5.6) -- (2.7,5.6);
\draw [<-] (4.,5.6) -- (4.1,5.6);
\draw [<-] (5.4,5.6) -- (5.5,5.6);
\draw [<-] (6.8,5.6) -- (6.9,5.6);
\draw [<-] (8.2,5.6) -- (8.3,5.6);
\draw [<-] (1.9,6.8) -- (2.,6.8);
\draw [<-] (3.3,6.8) -- (3.4,6.8);
\draw [<-] (4.7,6.8) -- (4.8,6.8);
\draw [<-] (6.1,6.8) -- (6.2,6.8);
\draw [<-] (7.5,6.8) -- (7.6,6.8);
\draw [<-] (1.9,4.4) -- (2.,4.4);
\draw [<-] (3.3,4.4) -- (3.4,4.4);
\draw [<-] (4.7,4.4) -- (4.8,4.4);
\draw [<-] (6.1,4.4) -- (6.2,4.4);
\draw [<-] (7.5,4.4) -- (7.6,4.4);
\end{tikzpicture}
}

\newcommand{\Interpret}{
\begin{tikzpicture}[baseline={([yshift=-.8ex]current bounding box.center)},scale=2.]
\draw [->-] (3.2,6.4) -- (3.2,6.8);
\draw [->-] (3.2,6.8) -- (3.2,7.2);
\draw [->-] (3.2,7.2) -- (3.9,7.6);
\draw [->-] (3.9,7.6) -- (4.6,7.2);
\draw [->-] (4.6,7.2) -- (4.6,6.4);
\draw [->-] (4.6,6.4) -- (3.9,6.);
\draw [->-] (3.9,6.) -- (3.2,6.4);
\draw [->-] (2.9,7.4) -- (3.2,7.2);
\draw [->-] (3.9,8.) -- (3.9,7.6);
\draw [->-] (5.,7.4) -- (4.6,7.2);
\draw [->-] (4.9,6.2) -- (4.6,6.4);
\draw [->-] (3.9,5.6) -- (3.9,6.);
\draw [->-] (3.2,6.4) -- (3.2,6.8);
\draw [->-] (2.9,6.2) -- (3.2,6.4);
\draw [snake it] (3.2,6.8) -- (3.8, 6.8);
\draw [<-] (3.45,6.8) -- (3.5, 6.8);
\node at (3.1, 7.) {\scriptsize $x$};
\node at (3.1, 6.6) {\scriptsize $x$};
\node at (2.8, 7.45) {\scriptsize $x$};
\node at (2.83, 6.15) {\scriptsize $x$};
\node at (3.5, 7.5) {\scriptsize $x$};
\node at (3.5, 6.1) {\scriptsize $x$};
\node at (3.5, 6.9) {\scriptsize $x_t$};
\node at (3.9, 7.2) {\scriptsize $({\tt Curl\ }x)_P = x_t$};

\fill [brown] (3.9, 6.8) circle (0.1);
\node [brown] at (3.95, 6.6) {\scriptsize ${\tt Flux} = ({\tt Curl\ }x)_P$};
\node [brown] at (4.25, 6.8) {\scriptsize Anyon};

\draw [red, very thin, <-] (3.8,6.8) -- (3.9, 7.1);
\draw [red, very thin, ->] (2.7,6.8) -- (3.2, 6.8);
\draw [red, very thin, ->] (2.7,7.05) -- (3.2, 7.2);
\draw [red, very thin, ->] (2.7,6.57) -- (3.2, 6.4);
\node at (2.43, 6.82) {\scriptsize $F = 0$};
\node at (2.43, 7.05) {\scriptsize $F = 0$};
\node at (2.43, 6.6) {\scriptsize $F = 0$};
\end{tikzpicture}
}

\newcommand{\HoppingLeft}{
\begin{tikzpicture}[baseline={([yshift=-.8ex]current bounding box.center)},scale=1]
\draw [->-] (4.,8.) -- (3.3,8.4);
\draw [->-] (3.3,8.4) -- (2.6,8.);
\draw [->-] (2.6,8.) -- (2.6,7.2);
\draw [->-] (2.6,7.2) -- (3.3,6.8);
\draw [->-] (3.3,6.8) -- (4.,7.2);
\draw [->-] (4.,8.) -- (4.7,8.4);
\draw [->-] (4.7,8.4) -- (5.4,8.);
\draw [->-] (5.4,8.) -- (5.4,7.2);
\draw [->-] (5.4,7.2) -- (4.7,6.8);
\draw [->-] (4.7,6.8) -- (4.,7.2);
\draw [->-] (4.,7.2) -- (4.,7.5);
\draw [->] (4.,7.2) -- (4.,7.7);
\draw [] (4.,7.5) -- (4.,7.7);
\draw [->-] (4.,7.7) -- (4.,8.);
\draw [->-] (3.3,8.4) -- (3.3,8.8);
\draw [->-] (4.7,8.4) -- (4.7,8.8);
\draw [->-] (3.3,6.4) -- (3.3,6.8);
\draw [->-] (4.7,6.4) -- (4.7,6.8);
\draw [->-] (5.7,7.) -- (5.4,7.2);
\draw [->-] (5.7,8.2) -- (5.4,8.);
\draw [->-] (2.3,8.2) -- (2.6,8.);
\draw [->-] (2.3,7.) -- (2.6,7.2);
\draw [snake it] (4.,7.5) -- (3.4,7.5);
\draw [->] (3.65,7.5) -- (3.75,7.5);
\draw [densely dotted] (4.,7.7) -- (4.4,7.7);
\node [brown] at (3.25, 7.8) {\tiny $(J, p, \alpha)$};
\node [brown] at (4.9, 7.7) {\tiny $(\idm, \tob)$};
\fill [brown] (3.3, 7.5) circle (0.1);
\node at (4.15, 7.3) {\tiny $x$};
\node at (4.15, 7.55) {\tiny $y$};
\node at (4.15, 7.9) {\tiny $y$};
\end{tikzpicture}
}

\newcommand{\HoppingRight}{
\begin{tikzpicture}[baseline={([yshift=-.8ex]current bounding box.center)},scale=1]
\draw [->-] (4.,8.) -- (3.3,8.4);
\draw [->-] (3.3,8.4) -- (2.6,8.);
\draw [->-] (2.6,8.) -- (2.6,7.2);
\draw [->-] (2.6,7.2) -- (3.3,6.8);
\draw [->-] (3.3,6.8) -- (4.,7.2);
\draw [->-] (4.,8.) -- (4.7,8.4);
\draw [->-] (4.7,8.4) -- (5.4,8.);
\draw [->-] (5.4,8.) -- (5.4,7.2);
\draw [->-] (5.4,7.2) -- (4.7,6.8);
\draw [->-] (4.7,6.8) -- (4.,7.2);
\draw [->-] (4.,7.2) -- (4.,7.5);
\draw [->] (4.,7.2) -- (4.,7.7);
\draw [] (4.,7.5) -- (4.,7.7);
\draw [->-] (4.,7.7) -- (4.,8.);
\draw [->-] (3.3,8.4) -- (3.3,8.8);
\draw [->-] (4.7,8.4) -- (4.7,8.8);
\draw [->-] (3.3,6.4) -- (3.3,6.8);
\draw [->-] (4.7,6.4) -- (4.7,6.8);
\draw [->-] (5.7,7.) -- (5.4,7.2);
\draw [->-] (5.7,8.2) -- (5.4,8.);
\draw [->-] (2.3,8.2) -- (2.6,8.);
\draw [->-] (2.3,7.) -- (2.6,7.2);
\draw [densely dotted] (4.,7.5) -- (3.6,7.5);
\draw [snake it] (4.,7.7) -- (4.6,7.7);
\draw [->] (4.4,7.7) -- (4.25,7.7);
\node [brown] at (3.1, 7.5) {\tiny $(\idm, \tob)$};
\fill [brown] (4.7, 7.7) circle (0.1);
\node [brown] at (4.8, 7.4) {\tiny $(J, q, \beta)$};
\node at (4.15, 7.3) {\tiny $x$};
\node at (4.15, 7.55) {\tiny $x$};
\node at (4.15, 7.9) {\tiny $y$};
\end{tikzpicture}
}

\newcommand{\DirectionA}{
\begin{tikzpicture}[baseline={([yshift=-.8ex]current bounding box.center)},scale=1.2]
	
\draw [ultra thick, teal] (4.,7.2) -- (4.,7.6);
\draw [ultra thick, teal] (4.,7.6) -- (4.,8.);
\draw [thick, teal, ->] (4.,7.2) -- (4.,7.5);
\draw [thick, teal, ->] (4.,7.2) -- (4.,7.95);
\draw [ultra thick, orange] (3.2,7.4) ..controls (3.5,7.275) and (3.6,7.25) .. (3.7,7.3)
			 ..controls (3.8,7.35) and (3.9,7.475) .. (4.,7.6)
			 ..controls (4.1,7.725) and (4.2,7.85) .. (4.3,7.9)
			 ..controls (4.4,7.95) and (4.5,7.925) .. (4.7,7.9);
\draw [thick, orange, ->] (4.7,7.9) -- (4.8,7.88);

\draw [->-] (4.,8.) -- (4.7,8.4);
\draw [->-] (4.7,8.4) -- (5.4,8.);
\draw [->-] (5.4,8.) -- (5.4,7.2);
\draw [->-] (5.4,7.2) -- (4.7,6.8);
\draw [->-] (4.7,6.8) -- (4.,7.2);
\draw [->-] (4.,8.) -- (3.3,8.4);
\draw [->-] (3.3,8.4) -- (2.6,8.);
\draw [->-] (2.6,8.) -- (2.6,7.2);
\draw [->-] (2.6,7.2) -- (3.3,6.8);
\draw [->-] (3.3,6.8) -- (4.,7.2);
\draw [->-] (3.3,8.8) -- (3.3,8.4);
\draw [->-] (4.7,8.8) -- (4.7,8.4);
\draw [->-] (5.7,8.2) -- (5.4,8.);
\draw [->-] (5.7,7.) -- (5.4,7.2);
\draw [->-] (4.7,6.4) -- (4.7,6.8);
\draw [->-] (3.3,6.4) -- (3.3,6.8);
\draw [->-] (2.3,7.) -- (2.6,7.2);
\draw [->-] (2.3,8.2) -- (2.6,8.);

\node at (3.3, 8) {\tiny Left-Side};
\node at (3.3, 7.8) {\tiny Plaquette};
\node at (4.7, 7.4) {\tiny Right-Side};
\node at (4.7, 7.2) {\tiny Plaquette};
\node [teal] at (4.4, 7.6) {\tiny Edge\hspace{1pt}$E$};
\node [orange] at (3.1, 7.5) {\tiny Hopping};
\node [orange] at (3., 7.3) {\tiny from};
\node [orange] at (4.8, 8.05) {\tiny Hopping};
\node [orange] at (4.95, 7.85) {\tiny to};
\end{tikzpicture}
}

\newcommand{\DirectionB}{
\begin{tikzpicture}[baseline={([yshift=-.8ex]current bounding box.center)},scale=1.2]
	
\draw [ultra thick, teal] (4.,7.2) -- (4.,7.6);
\draw [ultra thick, teal] (4.,7.6) -- (4.,8.);
\draw [thick, teal, <-] (4.,7.2) -- (4.,7.5);
\draw [thick, teal, -<] (4.,7.2) -- (4.,7.9);
\draw [ultra thick, orange] (3.2,7.4) ..controls (3.5,7.275) and (3.6,7.25) .. (3.7,7.3)
			 ..controls (3.8,7.35) and (3.9,7.475) .. (4.,7.6)
			 ..controls (4.1,7.725) and (4.2,7.85) .. (4.3,7.9)
			 ..controls (4.4,7.95) and (4.5,7.925) .. (4.7,7.9);
\draw [thick, orange, ->] (3.2,7.4) -- (3.1,7.44);

\draw [->-] (4.,8.) -- (4.7,8.4);
\draw [->-] (4.7,8.4) -- (5.4,8.);
\draw [->-] (5.4,8.) -- (5.4,7.2);
\draw [->-] (5.4,7.2) -- (4.7,6.8);
\draw [->-] (4.7,6.8) -- (4.,7.2);
\draw [->-] (4.,8.) -- (3.3,8.4);
\draw [->-] (3.3,8.4) -- (2.6,8.);
\draw [->-] (2.6,8.) -- (2.6,7.2);
\draw [->-] (2.6,7.2) -- (3.3,6.8);
\draw [->-] (3.3,6.8) -- (4.,7.2);
\draw [->-] (3.3,8.8) -- (3.3,8.4);
\draw [->-] (4.7,8.8) -- (4.7,8.4);
\draw [->-] (5.7,8.2) -- (5.4,8.);
\draw [->-] (5.7,7.) -- (5.4,7.2);
\draw [->-] (4.7,6.4) -- (4.7,6.8);
\draw [->-] (3.3,6.4) -- (3.3,6.8);
\draw [->-] (2.3,7.) -- (2.6,7.2);
\draw [->-] (2.3,8.2) -- (2.6,8.);

\node at (3.3, 8) {\tiny {\bf Right}-Side};
\node at (3.3, 7.8) {\tiny Plaquette};
\node at (4.7, 7.4) {\tiny {\bf Left}-Side};
\node at (4.7, 7.2) {\tiny Plaquette};
\node [teal] at (4.55, 7.6) {\tiny Edge\hspace{1pt}$(-E)$};
\node [orange] at (3.1, 7.55) {\tiny Hopping};
\node [orange] at (3., 7.35) {\tiny to};
\node [orange] at (4.8, 8.05) {\tiny Hopping};
\node [orange] at (4.95, 7.85) {\tiny from};
\end{tikzpicture}
}

\newcommand{\DirectionC}{
\begin{tikzpicture}[baseline={([yshift=-.8ex]current bounding box.center)},scale=1.2]
	
\draw [ultra thick, teal] (4.,7.2) -- (4.,8);
\draw [thick, teal, ->] (4.,7.2) -- (4.,7.5);
\draw [thick, teal, ->] (4.,7.2) -- (4.,7.95);
\draw [thick, teal, ->] (4.,7.2) -- (4.,7.74);
\draw [thick, snake it] (4.,7.5) -- (3.4,7.5);
\draw [thick, snake it] (4.,7.7) -- (4.6,7.7);
\fill [brown] (3.3, 7.5) circle (0.1);
\fill [brown] (4.7, 7.7) circle (0.1);
\draw [thick, ->] (3.7, 7.5) -- (3.75, 7.5);
\draw [thick, ->] (4.25, 7.7) -- (4.2, 7.7);

\draw [->-] (4.,8.) -- (4.7,8.4);
\draw [->-] (4.7,8.4) -- (5.4,8.);
\draw [->-] (5.4,8.) -- (5.4,7.2);
\draw [->-] (5.4,7.2) -- (4.7,6.8);
\draw [->-] (4.7,6.8) -- (4.,7.2);
\draw [->-] (4.,8.) -- (3.3,8.4);
\draw [->-] (3.3,8.4) -- (2.6,8.);
\draw [->-] (2.6,8.) -- (2.6,7.2);
\draw [->-] (2.6,7.2) -- (3.3,6.8);
\draw [->-] (3.3,6.8) -- (4.,7.2);
\draw [->-] (3.3,8.8) -- (3.3,8.4);
\draw [->-] (4.7,8.8) -- (4.7,8.4);
\draw [->-] (5.7,8.2) -- (5.4,8.);
\draw [->-] (5.7,7.) -- (5.4,7.2);
\draw [->-] (4.7,6.4) -- (4.7,6.8);
\draw [->-] (3.3,6.4) -- (3.3,6.8);
\draw [->-] (2.3,7.) -- (2.6,7.2);
\draw [->-] (2.3,8.2) -- (2.6,8.);

\node [teal] at (4.45, 7.4) {\tiny Edge\hspace{1pt}$E$};
\node [brown] at (3.25, 7.75) {\tiny $(J^\ast, p^\ast, \alpha)$};
\node [brown] at (4.75, 7.95) {\tiny $(J, q, \beta)$};
\node at (3.7, 7.35) {\tiny $p^\ast$};
\node at (4.3, 7.85) {\tiny $q$};

\end{tikzpicture}
}

\newcommand{\LongHopA}{
\begin{tikzpicture}[baseline={([yshift=-.8ex]current bounding box.center)},scale=.5]
\draw [] (2.1,7.2) -- (2.1,8.);
\draw [] (2.1,8.) -- (2.8,8.4);
\draw [] (2.8,8.4) -- (3.5,8.);
\draw [] (2.8,6.8) -- (2.1,7.2);
\draw [] (3.5,7.2) -- (2.8,6.8);
\draw [] (3.5,7.2) -- (3.5,8.);
\draw [] (3.5,8.) -- (4.2,8.4);
\draw [] (4.2,8.4) -- (4.9,8.);
\draw [] (4.2,6.8) -- (3.5,7.2);
\draw [] (4.9,7.2) -- (4.2,6.8);
\draw [] (4.9,7.2) -- (4.9,8.);
\draw [] (4.9,8.) -- (5.6,8.4);
\draw [] (5.6,8.4) -- (6.3,8.);
\draw [] (5.6,6.8) -- (4.9,7.2);
\draw [] (6.3,7.2) -- (5.6,6.8);
\draw [] (2.8,6.) -- (2.8,6.8);
\draw [] (4.2,6.) -- (4.2,6.8);
\draw [] (5.6,6.) -- (5.6,6.8);
\draw [ultra thick, teal] (6.3,7.2) -- (6.3,8.);
\draw [] (6.3,8.) -- (7.,8.4);
\draw [] (7.,8.4) -- (7.7,8.);
\draw [] (7.,6.8) -- (6.3,7.2);
\draw [ultra thick, teal] (7.7,7.2) -- (7.,6.8);
\draw [] (7.7,7.2) -- (7.7,8.);
\draw [] (7.7,8.) -- (8.,8.2);
\draw [] (8.4,6.8) -- (7.7,7.2);
\draw [] (7.,6.) -- (7.,6.8);
\draw [] (8.4,6.) -- (8.4,6.8);
\draw [] (2.1,4.8) -- (2.1,5.6);
\draw [] (2.1,5.6) -- (2.8,6.);
\draw [] (2.8,6.) -- (3.5,5.6);
\draw [] (2.8,4.4) -- (2.1,4.8);
\draw [] (3.5,4.8) -- (2.8,4.4);
\draw [] (3.5,4.8) -- (3.5,5.6);
\draw [] (3.5,5.6) -- (4.2,6.);
\draw [] (4.2,6.) -- (4.9,5.6);
\draw [] (4.2,4.4) -- (3.5,4.8);
\draw [] (4.9,4.8) -- (4.2,4.4);
\draw [ultra thick, teal] (4.9,4.8) -- (4.9,5.6);
\draw [] (4.9,5.6) -- (5.6,6.);
\draw [] (5.6,6.) -- (6.3,5.6);
\draw [] (5.6,4.4) -- (4.9,4.8);
\draw [] (6.3,4.8) -- (5.6,4.4);
\draw [] (2.8,4.) -- (2.8,4.4);
\draw [] (4.2,4.) -- (4.2,4.4);
\draw [] (5.6,4.) -- (5.6,4.4);
\draw [ultra thick, teal] (6.3,4.8) -- (6.3,5.6);
\draw [] (6.3,5.6) -- (7.,6.);
\draw [ultra thick, teal] (7.,6.) -- (7.7,5.6);
\draw [] (7.,4.4) -- (6.3,4.8);
\draw [] (7.7,4.8) -- (7.,4.4);
\draw [] (7.7,4.8) -- (7.7,5.6);
\draw [] (7.7,5.6) -- (8.4,6.);
\draw [] (8.4,4.4) -- (7.7,4.8);
\draw [] (7.,4.) -- (7.,4.4);
\draw [] (8.4,4.) -- (8.4,4.4);
\draw [] (8.4,6.8) -- (8.7,7.);
\draw [] (8.4,6.) -- (8.7,5.8);
\draw [] (8.4,4.4) -- (8.7,4.6);
\draw [] (1.4,8.4) -- (2.1,8.);
\draw [] (1.4,6.) -- (2.1,5.6);
\draw [] (1.4,8.4) -- (1.4,8.8);
\draw [] (2.8,8.4) -- (2.8,8.8);
\draw [] (4.2,8.4) -- (4.2,8.8);
\draw [] (5.6,8.4) -- (5.6,8.8);
\draw [] (7.,8.4) -- (7.,8.8);
\draw [] (0.7,8.) -- (1.4,8.4);
\draw [] (2.1,7.2) -- (1.4,6.8);
\draw [] (1.4,6.8) -- (0.7,7.2);
\draw [] (0.7,5.6) -- (1.4,6.);
\draw [] (1.4,6.) -- (1.4,6.8);
\draw [] (2.1,4.8) -- (1.4,4.4);
\draw [] (1.4,4.4) -- (0.7,4.8);
\draw [] (0.7,4.8) -- (0.7,5.6);
\draw [] (0.7,7.2) -- (0.7,8.);
\draw [] (1.4,4.) -- (1.4,4.4);
\draw [] (0.7,4.8) -- (0.4,4.6);
\draw [] (0.4,5.8) -- (0.7,5.6);
\draw [] (0.7,7.2) -- (0.4,7.);
\draw [] (0.4,8.2) -- (0.7,8.);

\draw [ultra thick, orange] (5.6, 5.) arc [x radius = 2, y radius = 1.3, start angle = -90, end angle = 85];
\draw [thick, orange, ->] (5.6, 5.) arc [x radius = 2, y radius = 1.3, start angle = -90, end angle = 95];
\draw [ultra thick, orange] (5.6, 5.) arc [x radius = 2.9, y radius = 1.3, start angle = -90, end angle = -120];
\node at (3.8, 5.2) {\tiny $P_0$};
\node at (5.2, 7.6) {\tiny $P$};
\node at (10, 6.4) {$=$};
\end{tikzpicture}
}

\newcommand{\LongHopB}{
\begin{tikzpicture}[baseline={([yshift=-.8ex]current bounding box.center)},scale=.5]
\draw [] (2.1,7.2) -- (2.1,8.);
\draw [] (2.1,8.) -- (2.8,8.4);
\draw [] (2.8,8.4) -- (3.5,8.);
\draw [] (2.8,6.8) -- (2.1,7.2);
\draw [] (3.5,7.2) -- (2.8,6.8);
\draw [] (3.5,7.2) -- (3.5,8.);
\draw [] (3.5,8.) -- (4.2,8.4);
\draw [] (4.2,8.4) -- (4.9,8.);
\draw [] (4.2,6.8) -- (3.5,7.2);
\draw [] (4.9,7.2) -- (4.2,6.8);
\draw [] (4.9,7.2) -- (4.9,8.);
\draw [] (4.9,8.) -- (5.6,8.4);
\draw [] (5.6,8.4) -- (6.3,8.);
\draw [ultra thick, teal] (5.6,6.8) -- (4.9,7.2);
\draw [] (6.3,7.2) -- (5.6,6.8);
\draw [] (2.8,6.) -- (2.8,6.8);
\draw [] (4.2,6.) -- (4.2,6.8);
\draw [] (5.6,6.) -- (5.6,6.8);
\draw [] (6.3,7.2) -- (6.3,8.);
\draw [] (6.3,8.) -- (7.,8.4);
\draw [] (7.,8.4) -- (7.7,8.);
\draw [] (7.,6.8) -- (6.3,7.2);
\draw [] (7.7,7.2) -- (7.,6.8);
\draw [] (7.7,7.2) -- (7.7,8.);
\draw [] (7.7,8.) -- (8.,8.2);
\draw [] (8.4,6.8) -- (7.7,7.2);
\draw [] (7.,6.) -- (7.,6.8);
\draw [] (8.4,6.) -- (8.4,6.8);
\draw [] (2.1,4.8) -- (2.1,5.6);
\draw [] (2.1,5.6) -- (2.8,6.);
\draw [] (2.8,6.) -- (3.5,5.6);
\draw [] (2.8,4.4) -- (2.1,4.8);
\draw [] (3.5,4.8) -- (2.8,4.4);
\draw [] (3.5,4.8) -- (3.5,5.6);
\draw [] (3.5,5.6) -- (4.2,6.);
\draw [ultra thick, teal] (4.2,6.) -- (4.9,5.6);
\draw [] (4.2,4.4) -- (3.5,4.8);
\draw [] (4.9,4.8) -- (4.2,4.4);
\draw [] (4.9,4.8) -- (4.9,5.6);
\draw [] (4.9,5.6) -- (5.6,6.);
\draw [] (5.6,6.) -- (6.3,5.6);
\draw [] (5.6,4.4) -- (4.9,4.8);
\draw [] (6.3,4.8) -- (5.6,4.4);
\draw [] (2.8,4.) -- (2.8,4.4);
\draw [] (4.2,4.) -- (4.2,4.4);
\draw [] (5.6,4.) -- (5.6,4.4);
\draw [] (6.3,4.8) -- (6.3,5.6);
\draw [] (6.3,5.6) -- (7.,6.);
\draw [] (7.,6.) -- (7.7,5.6);
\draw [] (7.,4.4) -- (6.3,4.8);
\draw [] (7.7,4.8) -- (7.,4.4);
\draw [] (7.7,4.8) -- (7.7,5.6);
\draw [] (7.7,5.6) -- (8.4,6.);
\draw [] (8.4,4.4) -- (7.7,4.8);
\draw [] (7.,4.) -- (7.,4.4);
\draw [] (8.4,4.) -- (8.4,4.4);
\draw [] (8.4,6.8) -- (8.7,7.);
\draw [] (8.4,6.) -- (8.7,5.8);
\draw [] (8.4,4.4) -- (8.7,4.6);
\draw [] (1.4,8.4) -- (2.1,8.);
\draw [] (1.4,6.) -- (2.1,5.6);
\draw [] (1.4,8.4) -- (1.4,8.8);
\draw [] (2.8,8.4) -- (2.8,8.8);
\draw [] (4.2,8.4) -- (4.2,8.8);
\draw [] (5.6,8.4) -- (5.6,8.8);
\draw [] (7.,8.4) -- (7.,8.8);
\draw [] (0.7,8.) -- (1.4,8.4);
\draw [] (2.1,7.2) -- (1.4,6.8);
\draw [] (1.4,6.8) -- (0.7,7.2);
\draw [] (0.7,5.6) -- (1.4,6.);
\draw [] (1.4,6.) -- (1.4,6.8);
\draw [] (2.1,4.8) -- (1.4,4.4);
\draw [] (1.4,4.4) -- (0.7,4.8);
\draw [] (0.7,4.8) -- (0.7,5.6);
\draw [] (0.7,7.2) -- (0.7,8.);
\draw [] (1.4,4.) -- (1.4,4.4);
\draw [] (0.7,4.8) -- (0.4,4.6);
\draw [] (0.4,5.8) -- (0.7,5.6);
\draw [] (0.7,7.2) -- (0.4,7.);
\draw [] (0.4,8.2) -- (0.7,8.);

\draw [ultra thick, orange] (4.2, 5.1) -- (5.55, 7.5);
\draw [thick, orange, ->] (4.2, 5.1) -- (5.67, 7.7);
\draw [snake it] (2.8, 6.4) -- (3.4, 6.4);
\fill [brown] (3.5, 6.4) circle (0.15);
\node at (10, 6.4) {$\ne$};
\end{tikzpicture}
}

\newcommand{\LongHopC}{
\begin{tikzpicture}[baseline={([yshift=-.8ex]current bounding box.center)},scale=.5]
\draw [] (2.1,7.2) -- (2.1,8.);
\draw [] (2.1,8.) -- (2.8,8.4);
\draw [] (2.8,8.4) -- (3.5,8.);
\draw [ultra thick, teal] (2.8,6.8) -- (2.1,7.2);
\draw [] (3.5,7.2) -- (2.8,6.8);
\draw [ultra thick, teal] (3.5,7.2) -- (3.5,8.);
\draw [] (3.5,8.) -- (4.2,8.4);
\draw [] (4.2,8.4) -- (4.9,8.);
\draw [] (4.2,6.8) -- (3.5,7.2);
\draw [] (4.9,7.2) -- (4.2,6.8);
\draw [ultra thick, teal] (4.9,7.2) -- (4.9,8.);
\draw [] (4.9,8.) -- (5.6,8.4);
\draw [] (5.6,8.4) -- (6.3,8.);
\draw [] (5.6,6.8) -- (4.9,7.2);
\draw [] (6.3,7.2) -- (5.6,6.8);
\draw [] (2.8,6.) -- (2.8,6.8);
\draw [] (4.2,6.) -- (4.2,6.8);
\draw [] (5.6,6.) -- (5.6,6.8);
\draw [] (6.3,7.2) -- (6.3,8.);
\draw [] (6.3,8.) -- (7.,8.4);
\draw [] (7.,8.4) -- (7.7,8.);
\draw [] (7.,6.8) -- (6.3,7.2);
\draw [] (7.7,7.2) -- (7.,6.8);
\draw [] (7.7,7.2) -- (7.7,8.);
\draw [] (7.7,8.) -- (8.,8.2);
\draw [] (8.4,6.8) -- (7.7,7.2);
\draw [] (7.,6.) -- (7.,6.8);
\draw [] (8.4,6.) -- (8.4,6.8);
\draw [] (2.1,4.8) -- (2.1,5.6);
\draw [ultra thick, teal] (2.1,5.6) -- (2.8,6.);
\draw [] (2.8,6.) -- (3.5,5.6);
\draw [] (2.8,4.4) -- (2.1,4.8);
\draw [] (3.5,4.8) -- (2.8,4.4);
\draw [ultra thick, teal] (3.5,4.8) -- (3.5,5.6);
\draw [] (3.5,5.6) -- (4.2,6.);
\draw [] (4.2,6.) -- (4.9,5.6);
\draw [] (4.2,4.4) -- (3.5,4.8);
\draw [] (4.9,4.8) -- (4.2,4.4);
\draw [] (4.9,4.8) -- (4.9,5.6);
\draw [] (4.9,5.6) -- (5.6,6.);
\draw [] (5.6,6.) -- (6.3,5.6);
\draw [] (5.6,4.4) -- (4.9,4.8);
\draw [] (6.3,4.8) -- (5.6,4.4);
\draw [] (2.8,4.) -- (2.8,4.4);
\draw [] (4.2,4.) -- (4.2,4.4);
\draw [] (5.6,4.) -- (5.6,4.4);
\draw [] (6.3,4.8) -- (6.3,5.6);
\draw [] (6.3,5.6) -- (7.,6.);
\draw [] (7.,6.) -- (7.7,5.6);
\draw [] (7.,4.4) -- (6.3,4.8);
\draw [] (7.7,4.8) -- (7.,4.4);
\draw [] (7.7,4.8) -- (7.7,5.6);
\draw [] (7.7,5.6) -- (8.4,6.);
\draw [] (8.4,4.4) -- (7.7,4.8);
\draw [] (7.,4.) -- (7.,4.4);
\draw [] (8.4,4.) -- (8.4,4.4);
\draw [] (8.4,6.8) -- (8.7,7.);
\draw [] (8.4,6.) -- (8.7,5.8);
\draw [] (8.4,4.4) -- (8.7,4.6);
\draw [] (1.4,8.4) -- (2.1,8.);
\draw [] (1.4,6.) -- (2.1,5.6);
\draw [] (1.4,8.4) -- (1.4,8.8);
\draw [] (2.8,8.4) -- (2.8,8.8);
\draw [] (4.2,8.4) -- (4.2,8.8);
\draw [] (5.6,8.4) -- (5.6,8.8);
\draw [] (7.,8.4) -- (7.,8.8);
\draw [] (0.7,8.) -- (1.4,8.4);
\draw [] (2.1,7.2) -- (1.4,6.8);
\draw [] (1.4,6.8) -- (0.7,7.2);
\draw [] (0.7,5.6) -- (1.4,6.);
\draw [] (1.4,6.) -- (1.4,6.8);
\draw [] (2.1,4.8) -- (1.4,4.4);
\draw [] (1.4,4.4) -- (0.7,4.8);
\draw [] (0.7,4.8) -- (0.7,5.6);
\draw [] (0.7,7.2) -- (0.7,8.);
\draw [] (1.4,4.) -- (1.4,4.4);
\draw [] (0.7,4.8) -- (0.4,4.6);
\draw [] (0.4,5.8) -- (0.7,5.6);
\draw [] (0.7,7.2) -- (0.4,7.);
\draw [] (0.4,8.2) -- (0.7,8.);

\draw [ultra thick, orange] (4.2, 5.1) arc [x radius = 2, y radius = 1.3, start angle = 270, end angle = 90];
\draw [ultra thick, orange] (4.2, 7.7) arc [x radius = 4, y radius = 1.3, start angle = 90, end angle = 70];
\draw [thick, orange, ->] (4.2, 7.7) arc [x radius = 4, y radius = 1.3, start angle = 90, end angle = 65];
\draw [snake it] (2.8, 6.4) -- (3.4, 6.4);
\fill [brown] (3.5, 6.4) circle (0.15);
\end{tikzpicture}
}

\newcommand{\FlatnessA}{
\begin{tikzpicture}[baseline={([yshift=-.8ex]current bounding box.center)},scale=.9]
\draw [->-] (3.2,7.2) -- (3.9,7.6);
\draw [->-] (3.9,7.6) -- (4.6,7.2);
\draw [->-] (4.6,7.2) -- (4.6,6.4);
\draw [->-] (4.6,6.4) -- (3.9,6.);
\draw [->-] (3.9,6.) -- (3.2,6.4);
\draw [->-] (3.2,6.4) -- (3.2,7.2);
\draw [->-] (3.2,7.2) -- (2.5,7.6);
\draw [->-] (2.5,7.6) -- (1.8,7.2);
\draw [->-] (1.8,7.2) -- (1.8,6.4);
\draw [->-] (1.8,6.4) -- (2.5,6.);
\draw [->-] (2.5,6.) -- (3.2,6.4);
\draw [->-] (3.9,7.6) -- (3.9,8.4);
\draw [->-] (3.9,8.4) -- (3.2,8.8);
\draw [->-] (3.2,8.8) -- (2.5,8.4);
\draw [->-] (2.5,8.4) -- (2.5,7.6);
\draw [->-] (2.2,8.6) -- (2.5,8.4);
\draw [->-] (3.2,9.2) -- (3.2,8.8);
\draw [->-] (4.2,8.6) -- (3.9,8.4);
\draw [->-] (4.9,7.4) -- (4.6,7.2);
\draw [->-] (4.9,6.2) -- (4.6,6.4);
\draw [->-] (3.9,5.6) -- (3.9,6.);
\draw [->-] (2.5,5.6) -- (2.5,6.);
\draw [->-] (1.5,6.2) -- (1.8,6.4);
\draw [->-] (1.5,7.4) -- (1.8,7.2);
\draw [snake it] (3.2, 6.6) -- (2.6, 6.6);
\draw [->] (2.85, 6.6) -- (2.92, 6.6);
\draw [snake it] (3.4, 7.3) -- (3.1, 7.8);
\draw [->] (3.26, 7.53) -- (3.195, 7.63);
\draw [decorate,decoration={snake,segment length=.12cm, amplitude=.02cm}] (3.2,6.95) ..controls (3.344,6.924) and (3.489,6.949) .. (3.6,7.05) ..controls (3.811,7.151) and (3.789,7.229) .. (3.8,7.3) ..controls (3.811,7.371) and (3.756,7.436) .. (3.7,7.5);
\draw [->] (3.67,7.06) -- (3.69,7.07);

\node [brown] at (2.5, 6.9) {\tiny $(J, p, \alpha)$};
\node [brown] at (4.05, 6.9) {\tiny $(J, q, \beta)$};
\node [brown] at (3., 8.15) {\tiny $(J, r, \gamma)$};
\node [red] at (3.5, 7.2) {\tiny $\times$};
\fill [brown] (2.5, 6.6) circle (0.1);
\fill [brown] (3.03, 7.88) circle (0.1);
\node at (3.4, 6.6) {\tiny $E_1$};
\node at (3.6, 7.65) {\tiny $E_2$};
\node at (2.7, 7.3) {\tiny $E_3$};
\end{tikzpicture}
}

\newcommand{\FlatnessB}{
\begin{tikzpicture}[baseline={([yshift=-.8ex]current bounding box.center)},scale=.9]
\draw [->-] (3.2,7.2) -- (3.9,7.6);
\draw [->-] (3.9,7.6) -- (4.6,7.2);
\draw [->-] (4.6,7.2) -- (4.6,6.4);
\draw [->-] (4.6,6.4) -- (3.9,6.);
\draw [->-] (3.9,6.) -- (3.2,6.4);
\draw [->-] (3.2,6.4) -- (3.2,7.2);
\draw [->-] (3.2,7.2) -- (2.5,7.6);
\draw [->-] (2.5,7.6) -- (1.8,7.2);
\draw [->-] (1.8,7.2) -- (1.8,6.4);
\draw [->-] (1.8,6.4) -- (2.5,6.);
\draw [->-] (2.5,6.) -- (3.2,6.4);
\draw [->-] (3.9,7.6) -- (3.9,8.4);
\draw [->-] (3.9,8.4) -- (3.2,8.8);
\draw [->-] (3.2,8.8) -- (2.5,8.4);
\draw [->-] (2.5,8.4) -- (2.5,7.6);
\draw [->-] (2.2,8.6) -- (2.5,8.4);
\draw [->-] (3.2,9.2) -- (3.2,8.8);
\draw [->-] (4.2,8.6) -- (3.9,8.4);
\draw [->-] (4.9,7.4) -- (4.6,7.2);
\draw [->-] (4.9,6.2) -- (4.6,6.4);
\draw [->-] (3.9,5.6) -- (3.9,6.);
\draw [->-] (2.5,5.6) -- (2.5,6.);
\draw [->-] (1.5,6.2) -- (1.8,6.4);
\draw [->-] (1.5,7.4) -- (1.8,7.2);
\draw [snake it] (3., 7.3) -- (2.7, 6.8);
\draw [->] (2.82, 7.) -- (2.88, 7.11);
\draw [snake it] (2.7, 7.5) -- (3., 8.);
\draw [<-] (2.82, 7.7) -- (2.88, 7.81);
\node [brown] at (2.45, 6.48) {\tiny $(J, p, \alpha)$};
\node [brown] at (3.2, 8.35) {\tiny $(J, r, \gamma)$};
\fill [brown] (2.67, 6.73) circle (0.1);
\fill [brown] (3.04, 8.08) circle (0.1);
\node at (3.4, 6.8) {\tiny $E_1$};
\node at (3.45, 7.55) {\tiny $E_2$};
\node at (2.7, 7.3) {\tiny $E_3$};
\end{tikzpicture}
}

\newcommand{\CreationLeft}{
\begin{tikzpicture}[baseline={([yshift=-.8ex]current bounding box.center)},scale=.9]
\draw [->-] (4.,8.) -- (3.3,8.4);
\draw [->-] (3.3,8.4) -- (2.6,8.);
\draw [->-] (2.6,8.) -- (2.6,7.2);
\draw [->-] (2.6,7.2) -- (3.3,6.8);
\draw [->-] (3.3,6.8) -- (4.,7.2);
\draw [->-] (4.,8.) -- (4.7,8.4);
\draw [->-] (4.7,8.4) -- (5.4,8.);
\draw [->-] (5.4,8.) -- (5.4,7.2);
\draw [->-] (5.4,7.2) -- (4.7,6.8);
\draw [->-] (4.7,6.8) -- (4.,7.2);
\draw [->-] (4.,7.2) -- (4.,7.5);
\draw [->] (4.,7.2) -- (4.,7.7);
\draw [] (4.,7.5) -- (4.,7.7);
\draw [->-] (4.,7.7) -- (4.,8.);
\draw [->-] (3.3,8.4) -- (3.3,8.8);
\draw [->-] (4.7,8.4) -- (4.7,8.8);
\draw [->-] (3.3,6.4) -- (3.3,6.8);
\draw [->-] (4.7,6.4) -- (4.7,6.8);
\draw [->-] (5.7,7.) -- (5.4,7.2);
\draw [->-] (5.7,8.2) -- (5.4,8.);
\draw [->-] (2.3,8.2) -- (2.6,8.);
\draw [->-] (2.3,7.) -- (2.6,7.2);
\draw [densely dotted] (4.,7.5) -- (3.6,7.5);
\draw [densely dotted] (4.,7.7) -- (4.4,7.7);
\node [brown] at (3.1, 7.5) {\tiny $(\idm, \tob)$};
\node [brown] at (4.9, 7.7) {\tiny $(\idm, \tob)$};
\node at (4.15, 7.3) {\tiny $x$};
\node at (4.15, 7.55 ) {\tiny $x$};
\node at (4.15, 7.9) {\tiny $x$};
\end{tikzpicture}
}

\newcommand{\CreationRight}{
\begin{tikzpicture}[baseline={([yshift=-.8ex]current bounding box.center)},scale=.9]
\draw [->-] (4.,8.) -- (3.3,8.4);
\draw [->-] (3.3,8.4) -- (2.6,8.);
\draw [->-] (2.6,8.) -- (2.6,7.2);
\draw [->-] (2.6,7.2) -- (3.3,6.8);
\draw [->-] (3.3,6.8) -- (4.,7.2);
\draw [->-] (4.,8.) -- (4.7,8.4);
\draw [->-] (4.7,8.4) -- (5.4,8.);
\draw [->-] (5.4,8.) -- (5.4,7.2);
\draw [->-] (5.4,7.2) -- (4.7,6.8);
\draw [->-] (4.7,6.8) -- (4.,7.2);
\draw [->-] (4.,7.2) -- (4.,7.5);
\draw [->] (4.,7.2) -- (4.,7.7);
\draw [] (4.,7.5) -- (4.,7.7);
\draw [->-] (4.,7.7) -- (4.,8.);
\draw [->-] (3.3,8.4) -- (3.3,8.8);
\draw [->-] (4.7,8.4) -- (4.7,8.8);
\draw [->-] (3.3,6.4) -- (3.3,6.8);
\draw [->-] (4.7,6.4) -- (4.7,6.8);
\draw [->-] (5.7,7.) -- (5.4,7.2);
\draw [->-] (5.7,8.2) -- (5.4,8.);
\draw [->-] (2.3,8.2) -- (2.6,8.);
\draw [->-] (2.3,7.) -- (2.6,7.2);
\node at (4.15, 7.3) {\tiny $x$};
\node at (4.15, 7.55) {\tiny $u$};
\node at (4.15, 7.9) {\tiny $x$};

\draw [snake it] (4.,7.5) -- (3.4,7.5);
\draw [->] (3.6,7.5) -- (3.7,7.5);
\draw [snake it] (4.,7.7) -- (4.6,7.7);
\draw [->] (4.4,7.7) -- (4.3,7.7);
\node [brown] at (3.2, 7.75) {\tiny $(J^\ast, p^\ast, \alpha)$};
\node [brown] at (4.8, 7.45) {\tiny $(J, q, \beta)$};
\fill [brown] (3.3, 7.5) circle (0.1);
\fill [brown] (4.7, 7.7) circle (0.1);
\end{tikzpicture}
}

\newcommand{\TubeGaugeA}{
\begin{tikzpicture}[baseline={([yshift=-.8ex]current bounding box.center)},scale=1.3]
\draw [->-] (3.2,7.6) -- (3.2,8.);
\draw [->-] (3.2,8.) -- (3.9,8.4);
\draw [->-] (3.9,8.4) -- (4.6,8.);
\draw [->-] (4.6,8.) -- (4.6,7.2);
\draw [->-] (4.6,7.2) -- (3.9,6.8);
\draw [->-] (3.9,6.8) -- (3.2,7.2);
\draw [->-] (3.2,7.2) -- (3.2,7.6);
\draw [->-] (3.9,8.8) -- (3.9,8.4);
\draw [->-] (3.9,6.4) -- (3.9,6.8);
\draw [->-] (4.9,7.) -- (4.6,7.2);
\draw [->-] (4.9,8.2) -- (4.6,8.);
\draw [->-] (2.9,8.2) -- (3.2,8.);
\draw [->-] (2.9,7.) -- (3.2,7.2);
\draw [snake it] (3.2,7.6) -- (3.8,7.6);
\draw [->] (3.6,7.6) -- (3.5,7.6);
\node at (3.9, 7.9) {\tiny Plaquette $P$};
\node [brown] at (3.95, 7.4) {\tiny $(J, p, \alpha)$};
\fill [brown] (3.9, 7.6) circle (0.1);
\end{tikzpicture}
}

\newcommand{\TubeGaugeB}{
\begin{tikzpicture}[baseline={([yshift=-.8ex]current bounding box.center)},scale=1.3]
\draw [->-] (3.2,7.6) -- (3.2,8.);
\draw [->-] (3.2,8.) -- (3.9,8.4);
\draw [->-] (3.9,8.4) -- (4.6,8.);
\draw [->-] (4.6,8.) -- (4.6,7.2);
\draw [->-] (4.6,7.2) -- (3.9,6.8);
\draw [->-] (3.9,6.8) -- (3.2,7.2);
\draw [->-] (3.2,7.2) -- (3.2,7.6);
\draw [->-] (3.9,8.8) -- (3.9,8.4);
\draw [->-] (3.9,6.4) -- (3.9,6.8);
\draw [->-] (4.9,7.) -- (4.6,7.2);
\draw [->-] (4.9,8.2) -- (4.6,8.);
\draw [->-] (2.9,8.2) -- (3.2,8.);
\draw [->-] (2.9,7.) -- (3.2,7.2);
\draw [snake it] (3.2,7.6) -- (3.65,7.6);
\draw [->] (3.5,7.6) -- (3.4,7.6);
\draw [red, draw, rotate around={0.:(4.2,7.6)}] (4.2,7.6) ++(0.:0.2 and 0.2) arc(0.:360.:0.2 and 0.2);
\draw [red, ->] (4.36,7.7) -- (4.4,7.65);
\node [red] at (4.25, 7.65) {\tiny $s$};
\node [brown] at (3.75, 7.4) {\tiny $(J, p, \alpha)$};
\fill [brown] (3.75, 7.6) circle (0.1);
\end{tikzpicture}
}

\newcommand{\TubeGaugeC}{
\begin{tikzpicture}[baseline={([yshift=-.8ex]current bounding box.center)},scale=1.3]
\draw [->-] (3.2,6.4) -- (3.2,6.7);
\draw [->-] (3.2,6.7) -- (3.2,7.2);
\draw [->-] (3.2,7.2) -- (3.9,7.6);
\draw [->-] (3.9,7.6) -- (4.6,7.2);
\draw [->-] (4.6,7.2) -- (4.6,6.4);
\draw [->-] (4.6,6.4) -- (3.9,6.);
\draw [->-] (3.9,6.) -- (3.2,6.4);
\draw [->-] (2.9,6.2) -- (3.2,6.4);
\draw [->-] (2.9,7.4) -- (3.2,7.2);
\draw [->-] (3.9,8.) -- (3.9,7.6);
\draw [->-] (4.9,7.4) -- (4.6,7.2);
\draw [->-] (4.9,6.2) -- (4.6,6.4);
\draw [->-] (3.9,5.6) -- (3.9,6.);

\draw [red] (3.5,6.9) arc (180:90:.4);
\draw [red] (3.9,7.3) arc (90:-90:.5);
\draw [red] (3.9,6.3) arc (-90:-180:.4);
\draw [red, ->] (4.25,7.15) -- (4.3,7.1);
\node [red] at (4.17, 7.07) {\tiny $s$};
\node [red] at (3.37, 7.1) {\tiny $\times$};

\draw [red] (3.5,6.7) -- (3.5,6.9);
\draw [red, ->] (3.5,6.7) -- (3.5,6.87);
\node at (3.39, 6.8) {\tiny $u$};

\draw [snake it] (3.2,6.65) -- (3.5,6.65);
\draw [<-] (3.3,6.65) -- (3.4,6.65);
\node at (3.35, 6.53) {\tiny $p$};

\draw [snake it] (3.5,6.9) -- (3.9,6.9);
\draw [->] (3.7,6.9) -- (3.65,6.9);
\fill [brown] (4., 6.9) circle (0.1);
\node [brown]at (4., 6.7) {\tiny $(J, q, \beta)$};
\end{tikzpicture}
}

\newcommand{\TubeGaugeD}{
\begin{tikzpicture}[baseline={([yshift=-.8ex]current bounding box.center)},scale=1.3]
\draw [->-] (3.2,7.6) -- (3.2,8.);
\draw [->-] (3.2,8.) -- (3.9,8.4);
\draw [->-] (3.9,8.4) -- (4.6,8.);
\draw [->-] (4.6,8.) -- (4.6,7.2);
\draw [->-] (4.6,7.2) -- (3.9,6.8);
\draw [->-] (3.9,6.8) -- (3.2,7.2);
\draw [->-] (3.2,7.2) -- (3.2,7.6);
\draw [->-] (3.9,8.8) -- (3.9,8.4);
\draw [->-] (3.9,6.4) -- (3.9,6.8);
\draw [->-] (4.9,7.) -- (4.6,7.2);
\draw [->-] (4.9,8.2) -- (4.6,8.);
\draw [->-] (2.9,8.2) -- (3.2,8.);
\draw [->-] (2.9,7.) -- (3.2,7.2);
\draw [snake it] (3.2,7.6) -- (3.9,7.6);
\draw [->] (3.55,7.6) -- (3.45, 7.6);
\node [brown] at (3.9, 7.4) {\tiny $(J, q, \beta)$};
\fill [brown] (3.9, 7.6) circle (0.1);
\end{tikzpicture}
}

\newcommand{\TubeGaugeE}{
\begin{tikzpicture}[baseline={([yshift=-.8ex]current bounding box.center)},scale=1.1]
\draw [->-] (3.2,7.6) -- (3.2,8.);
\draw [->-] (3.2,8.) -- (3.9,8.4);
\draw [->-] (3.9,8.4) -- (4.6,8.);
\draw [->-] (4.6,8.) -- (4.6,7.2);
\draw [->-] (4.6,7.2) -- (3.9,6.8);
\draw [->-] (3.9,6.8) -- (3.2,7.2);
\draw [->-] (3.2,7.2) -- (3.2,7.6);
\draw [->-] (3.9,8.8) -- (3.9,8.4);
\draw [->-] (3.9,6.4) -- (3.9,6.8);
\draw [->-] (4.9,7.) -- (4.6,7.2);
\draw [->-] (4.9,8.2) -- (4.6,8.);
\draw [->-] (2.9,8.2) -- (3.2,8.);
\draw [->-] (2.9,7.) -- (3.2,7.2);
\draw [snake it] (3.2,7.6) -- (3.9,7.6);
\draw [->] (3.6,7.6) -- (3.5,7.6);
\node at (3.58, 7.45) {\scriptsize $p'$};
\node at (3.05, 7.85) {\scriptsize $i_0$};
\node at (3.45, 8.35) {\scriptsize $i_1$};
\node at (4.3, 8.35) {\scriptsize $i_2$};
\node at (4.8, 7.6) {\scriptsize $i_3$};
\node at (4.35, 6.85) {\scriptsize $i_4$};
\node at (3.45, 6.9) {\scriptsize $i_5$};
\node at (3.05, 7.42) {\scriptsize $i_6$};
\node at (2.8, 8.27) {\scriptsize $e_0$};
\node at (4.08, 8.68) {\scriptsize $e_1$};
\node at (5.05, 8.25) {\scriptsize $e_2$};
\node at (5.03, 6.93) {\scriptsize $e_3$};
\node at (4.07, 6.46) {\scriptsize $e_4$};
\node at (2.83, 6.9) {\scriptsize $e_5$};
\end{tikzpicture}
}

\newcommand{\TubeGaugeF}{
\begin{tikzpicture}[baseline={([yshift=-.8ex]current bounding box.center)},scale=1.1]
\draw [->-] (3.2,6.4) -- (3.2,6.7);
\draw [->-] (3.2,6.7) -- (3.2,7.2);
\draw [->-] (3.2,7.2) -- (3.9,7.6);
\draw [->-] (3.9,7.6) -- (4.6,7.2);
\draw [->-] (4.6,7.2) -- (4.6,6.4);
\draw [->-] (4.6,6.4) -- (3.9,6.);
\draw [->-] (3.9,6.) -- (3.2,6.4);
\draw [->-] (2.9,6.2) -- (3.2,6.4);
\draw [->-] (2.9,7.4) -- (3.2,7.2);
\draw [->-] (3.9,8.) -- (3.9,7.6);
\draw [->-] (4.9,7.4) -- (4.6,7.2);
\draw [->-] (4.9,6.2) -- (4.6,6.4);
\draw [->-] (3.9,5.6) -- (3.9,6.);

\draw [red] (3.5,6.9) arc (180:90:.4);
\draw [red] (3.9,7.3) arc (90:-90:.5);
\draw [red] (3.9,6.3) arc (-90:-180:.4);
\draw [red,  ->] (4.25,7.15) -- (4.3,7.1);
\node [red]  at (4.17, 7.07) {\scriptsize $s$};
\node [red] at (3.37, 7.1) {$\times$};

\draw [red] (3.5,6.7) -- (3.5,6.9);
\draw [red,  ->] (3.5,6.7) -- (3.5,6.87);
\node [red] at (3.37, 6.8) {\scriptsize $u$};

\draw [snake it] (3.2,6.65) -- (3.5,6.65);
\draw [<-] (3.3,6.65) -- (3.4,6.65);
\node at (3.35, 6.5) {\scriptsize $p$};

\draw [snake it] (3.5,6.9) -- (4.,6.9);
\draw [->] (3.8,6.9) -- (3.7,6.9);
\node at (3.75, 6.78) {\scriptsize $q$};

\node at (3.05, 6.9) {\scriptsize $i_0$};
\node at (3.45, 7.55) {\scriptsize $i_1$};
\node at (4.3, 7.55) {\scriptsize $i_2$};
\node at (4.8, 6.8) {\scriptsize $i_3$};
\node at (4.35, 6.05) {\scriptsize $i_4$};
\node at (3.45, 6.1) {\scriptsize $i_5$};
\node at (3.05, 6.5) {\scriptsize $i_6$};
\node at (2.8, 7.47) {\scriptsize $e_0$};
\node at (4.08, 7.88) {\scriptsize $e_1$};
\node at (5.05, 7.45) {\scriptsize $e_2$};
\node at (5.03, 6.13) {\scriptsize $e_3$};
\node at (4.07, 5.66) {\scriptsize $e_4$};
\node at (2.83, 6.1) {\scriptsize $e_5$};
\end{tikzpicture}
}

\newcommand{\TubeGaugeG}{
\begin{tikzpicture}[baseline={([yshift=-.8ex]current bounding box.center)},scale=1.1]
\draw [->-] (3.2,7.6) -- (3.2,8.);
\draw [->-] (3.2,8.) -- (3.9,8.4);
\draw [->-] (3.9,8.4) -- (4.6,8.);
\draw [->-] (4.6,8.) -- (4.6,7.2);
\draw [->-] (4.6,7.2) -- (3.9,6.8);
\draw [->-] (3.9,6.8) -- (3.2,7.2);
\draw [->-] (3.2,7.2) -- (3.2,7.6);
\draw [->-] (3.9,8.8) -- (3.9,8.4);
\draw [->-] (3.9,6.4) -- (3.9,6.8);
\draw [->-] (4.9,7.) -- (4.6,7.2);
\draw [->-] (4.9,8.2) -- (4.6,8.);
\draw [->-] (2.9,8.2) -- (3.2,8.);
\draw [->-] (2.9,7.) -- (3.2,7.2);
\draw [snake it] (3.2,7.6) -- (3.9,7.6);
\draw [->] (3.6,7.6) -- (3.5,7.6);
\node at (3.55, 7.45) {\scriptsize $q$};
\node at (3.05, 7.85) {\scriptsize $j_0$};
\node at (3.45, 8.35) {\scriptsize $j_1$};
\node at (4.3, 8.35) {\scriptsize $j_2$};
\node at (4.8, 7.6) {\scriptsize $j_3$};
\node at (4.35, 6.85) {\scriptsize $j_4$};
\node at (3.45, 6.9) {\scriptsize $j_5$};
\node at (3.05, 7.42) {\scriptsize $j_6$};
\node at (2.8, 8.27) {\scriptsize $e_0$};
\node at (4.08, 8.68) {\scriptsize $e_1$};
\node at (5.05, 8.25) {\scriptsize $e_2$};
\node at (5.03, 6.93) {\scriptsize $e_3$};
\node at (4.07, 6.46) {\scriptsize $e_4$};
\node at (2.83, 6.9) {\scriptsize $e_5$};
\end{tikzpicture}
}

\newcommand{\TubeGaugeHLeft}{
\begin{tikzpicture}[baseline={([yshift=-.8ex]current bounding box.center)},scale=1.2]
\draw [->-] (3.2,7.6) -- (3.2,8.);
\draw [->-] (3.2,8.) -- (3.9,8.4);
\draw [->-] (3.9,8.4) -- (4.6,8.);
\draw [->-] (4.6,8.) -- (4.6,7.2);
\draw [->-] (4.6,7.2) -- (3.9,6.8);
\draw [->-] (3.9,6.8) -- (3.2,7.2);
\draw [->-] (3.2,7.2) -- (3.2,7.6);
\draw [->-] (3.9,8.8) -- (3.9,8.4);
\draw [->-] (3.9,6.4) -- (3.9,6.8);
\draw [->-] (4.9,7.) -- (4.6,7.2);
\draw [->-] (4.9,8.2) -- (4.6,8.);
\draw [->-] (2.9,8.2) -- (3.2,8.);
\draw [->-] (2.9,7.) -- (3.2,7.2);
\draw [snake it] (3.2,7.6) -- (3.9,7.6);
\draw [->] (3.6,7.6) -- (3.5,7.6);
\node at (3.58, 7.45) {\scriptsize $p'$};
\node at (3.05, 7.85) {\scriptsize $i_0$};
\node at (3.45, 8.35) {\scriptsize $i_1$};
\node at (4.3, 8.35) {\scriptsize $i_2$};
\node at (4.8, 7.6) {\scriptsize $i_3$};
\node at (4.35, 6.85) {\scriptsize $i_4$};
\node at (3.45, 6.9) {\scriptsize $i_5$};
\node at (3.05, 7.42) {\scriptsize $i_6$};
\node at (2.8, 8.27) {\scriptsize $e_0$};
\node at (4.08, 8.68) {\scriptsize $e_1$};
\node at (5.05, 8.25) {\scriptsize $e_2$};
\node at (5.03, 6.93) {\scriptsize $e_3$};
\node at (4.07, 6.46) {\scriptsize $e_4$};
\node at (2.83, 6.9) {\scriptsize $e_5$};
\end{tikzpicture}
}

\newcommand{\TubeGaugeHRight}{
\begin{tikzpicture}[baseline={([yshift=-.8ex]current bounding box.center)},scale=1.2]
\draw [->-] (3.2,7.6) -- (3.2,8.);
\draw [->-] (3.2,8.) -- (3.9,8.4);
\draw [->-] (3.9,8.4) -- (4.6,8.);
\draw [->-] (4.6,8.) -- (4.6,7.2);
\draw [->-] (4.6,7.2) -- (3.9,6.8);
\draw [->-] (3.9,6.8) -- (3.2,7.2);
\draw [->-] (3.2,7.2) -- (3.2,7.6);
\draw [->-] (3.9,8.8) -- (3.9,8.4);
\draw [->-] (3.9,6.4) -- (3.9,6.8);
\draw [->-] (4.9,7.) -- (4.6,7.2);
\draw [->-] (4.9,8.2) -- (4.6,8.);
\draw [->-] (2.9,8.2) -- (3.2,8.);
\draw [->-] (2.9,7.) -- (3.2,7.2);
\draw [snake it] (3.2,7.6) -- (3.9,7.6);
\draw [->] (3.6,7.6) -- (3.5,7.6);
\node at (3.63, 7.43) {\scriptsize $s^\ast ps$};
\node at (3., 7.85) {\scriptsize $i_0s$};
\node at (3.45, 8.35) {\scriptsize $i_1s$};
\node at (4.35, 8.35) {\scriptsize $i_2s$};
\node at (4.8, 7.6) {\scriptsize $i_3s$};
\node at (4.35, 6.85) {\scriptsize $i_4s$};
\node at (3.45, 6.85) {\scriptsize $i_5s$};
\node at (3., 7.4) {\scriptsize $i_6s$};
\node at (2.8, 8.27) {\scriptsize $e_0$};
\node at (4.08, 8.68) {\scriptsize $e_1$};
\node at (5.05, 8.25) {\scriptsize $e_2$};
\node at (5.03, 6.93) {\scriptsize $e_3$};
\node at (4.07, 6.46) {\scriptsize $e_4$};
\node at (2.83, 6.9) {\scriptsize $e_5$};
\end{tikzpicture}
}

\newcommand{\TubeGaugeI}{
\begin{tikzpicture}[baseline={([yshift=-.8ex]current bounding box.center)},scale=1.3]
\draw [->-] (1, 1) -- (1, 1.8);
\draw [->-] (0.65, 0.8) -- (1, 1);
\draw [->-] (1.35, 0.8) -- (1, 1);
\draw [->-] (1, 1.8) -- (0.65, 2.);
\draw [->-] (1, 1.8) -- (1.35, 2.);
\node at (1.1, 1.2) {\scriptsize $x$};
\draw [red] (.55, 1.) arc (-60:60:0.46188);
\draw [red, ->] (0.78, 1.4) -- (0.78, 1.34);
\draw [red, ->] (1.22, 1.35) -- (1.22, 1.43);
\draw [red] (1.45, 1.8) arc (120:240:0.46188);
\node [red] at (0.55, 1.4) {\scriptsize $s_P$};
\node [red] at (0.85, 1.65) {\small $\times$};
\node [red] at (1.45, 1.4) {\scriptsize $s_Q$};
\node [red] at (1.15, 1.65) {\small $\times$};
\draw [red] (1., 2.) arc (-90:-50:0.4);
\draw [red] (1., 2.) arc (-90:-130:0.4);
\draw [red] (1., 0.8) arc (90:50:0.4);
\draw [red] (1., 0.8) arc (90:130:0.4);
\draw [red, ->] (1., 2.) -- (.93, 2.);
\draw [red, ->] (1., 0.8) -- (1.07, 0.8);
\end{tikzpicture}
}

\newcommand{\TubeGaugeJ}{
\begin{tikzpicture}[baseline={([yshift=-.8ex]current bounding box.center)},scale=1.3]
\draw [->-] (1, 1) -- (1, 1.8);
\draw [->-] (0.65, 0.8) -- (1, 1);
\draw [->-] (1.35, 0.8) -- (1, 1);
\draw [->-] (1, 1.8) -- (0.65, 2.);
\draw [->-] (1, 1.8) -- (1.35, 2.);
\node at (1.3, 1.3) {\scriptsize $xs_Q$};
\node [red] at (0.85, 1.65) {\small $\times$};
\draw [red] (.55, 1.) arc (-60:60:0.46188);
\draw [red, ->] (0.78, 1.4) -- (0.78, 1.34);
\node [red] at (0.55, 1.4) {\scriptsize $s_P$};
\draw [red] (1., 2.) arc (-90:-50:0.4);
\draw [red] (1., 2.) arc (-90:-130:0.4);
\draw [red] (1., 0.8) arc (90:50:0.4);
\draw [red] (1., 0.8) arc (90:130:0.4);
\draw [red, ->] (1., 2.) -- (.93, 2.);
\draw [red, ->] (1., 0.8) -- (1.07, 0.8);
\end{tikzpicture}
}

\newcommand{\TubeGaugeK}{
\begin{tikzpicture}[baseline={([yshift=-.8ex]current bounding box.center)},scale=1.3]
\draw [->-] (1, 1) -- (1, 1.8);
\draw [->-] (0.65, 0.8) -- (1, 1);
\draw [->-] (1.35, 0.8) -- (1, 1);
\draw [->-] (1, 1.8) -- (0.65, 2.);
\draw [->-] (1, 1.8) -- (1.35, 2.);
\node at (1.3, 1.3) {\scriptsize $xs_Q$};
\draw [red] (.55, 1.) arc (-60:60:0.46188);
\draw [red, ->] (0.78, 1.4) -- (0.78, 1.44);
\node [red] at (0.55, 1.4) {\scriptsize $s_P^\ast$};
\node [red] at (0.85, 1.65) {\small $\times$};
\draw [red] (1., 2.) arc (-90:-50:0.4);
\draw [red] (1., 2.) arc (-90:-130:0.4);
\draw [red] (1., 0.8) arc (90:50:0.4);
\draw [red] (1., 0.8) arc (90:130:0.4);
\draw [red, ->] (1., 2.) -- (.93, 2.);
\draw [red, ->] (1., 0.8) -- (1.07, 0.8);
\end{tikzpicture}
}

\newcommand{\TubeGaugeL}{
\begin{tikzpicture}[baseline={([yshift=-.8ex]current bounding box.center)},scale=1.3]
\draw [->-] (1, 1) -- (1, 1.8);
\draw [->-] (0.65, 0.8) -- (1, 1);
\draw [->-] (1.35, 0.8) -- (1, 1);
\draw [->-] (1, 1.8) -- (0.65, 2.);
\draw [->-] (1, 1.8) -- (1.35, 2.);
\node at (1.5, 1.3) {\scriptsize $s_P^\ast xs_Q$};
\draw [red] (1., 2.) arc (-90:-50:0.4);
\draw [red] (1., 2.) arc (-90:-130:0.4);
\draw [red] (1., 0.8) arc (90:50:0.4);
\draw [red] (1., 0.8) arc (90:130:0.4);
\draw [red, ->] (1., 2.) -- (.93, 2.);
\draw [red, ->] (1., 0.8) -- (1.07, 0.8);
\end{tikzpicture}
}

\newcommand{\Commutivity}{
\begin{tikzpicture}[baseline={([yshift=-.8ex]current bounding box.center)},scale=.9]
\draw [->-] (2.,4.8) -- (2.,5.1);
\draw [->-] (2.,5.1) -- (2.,5.6);
\draw [->-] (2.,5.6) -- (1.3,6.);
\draw [->-] (1.3,6.) -- (0.6,5.6);
\draw [->-] (0.6,5.6) -- (0.6,4.8);
\draw [->-] (0.6,4.8) -- (1.3,4.4);
\draw [->-] (1.3,4.4) -- (2.,4.8);
\draw [->-] (2.,5.6) -- (2.7,6.);
\draw [->-] (2.7,6.) -- (3.4,5.6);
\draw [->-] (3.4,5.6) -- (3.4,4.8);
\draw [->-] (3.4,4.8) -- (2.7,4.4);
\draw [->-] (2.7,4.4) -- (2.,4.8);
\draw [->-] (1.3,6.4) -- (1.3,6.);
\draw [->-] (2.7,6.4) -- (2.7,6.);
\draw [->-] (3.7,5.8) -- (3.4,5.6);
\draw [->-] (3.7,4.6) -- (3.4,4.8);
\draw [->-] (2.7,4.) -- (2.7,4.4);
\draw [->-] (1.3,4.) -- (1.3,4.4);
\draw [->-] (0.3,4.6) -- (0.6,4.8);
\draw [->-] (0.3,5.8) -- (0.6,5.6);
\draw [snake it] (2.,5.1) -- (1.4,5.1);
\draw [->] (1.65,5.1) -- (1.75,5.1);
\fill [brown] (1.3, 5.1) circle (0.1);
\node [brown] at (1.3, 4.85) {\scriptsize $(J, p, \alpha)$};
\node at (2.18, 5.25) {\scriptsize $E$};

\draw [ultra thick, ->] (4., 5.) -- (6.1, 5.);
\node at (5., 5.4) {\small $\hop_E^{J;(p, i)(q,j)}$};
\draw [ultra thick, ->] (2., 3.9) -- (2., 2.5);
\node at (3.1, 3.3) {\small $D^J(B_P^{psur})$};

\draw [->-] (8.,4.8) -- (8.,5.3);
\draw [->-] (8.,5.3) -- (8.,5.6);
\draw [->-] (8.,5.6) -- (7.3,6.);
\draw [->-] (7.3,6.) -- (6.6,5.6);
\draw [->-] (6.6,5.6) -- (6.6,4.8);
\draw [->-] (6.6,4.8) -- (7.3,4.4);
\draw [->-] (7.3,4.4) -- (8.,4.8);
\draw [->-] (8.,5.6) -- (8.7,6.);
\draw [->-] (8.7,6.) -- (9.4,5.6);
\draw [->-] (9.4,5.6) -- (9.4,4.8);
\draw [->-] (9.4,4.8) -- (8.7,4.4);
\draw [->-] (8.7,4.4) -- (8.,4.8);
\draw [->-] (7.3,6.4) -- (7.3,6.);
\draw [->-] (8.7,6.4) -- (8.7,6.);
\draw [->-] (9.7,5.8) -- (9.4,5.6);
\draw [->-] (9.7,4.6) -- (9.4,4.8);
\draw [->-] (8.7,4.) -- (8.7,4.4);
\draw [->-] (7.3,4.) -- (7.3,4.4);
\draw [->-] (6.3,4.6) -- (6.6,4.8);
\draw [->-] (6.3,5.8) -- (6.6,5.6);
\draw [snake it] (8.,5.3) -- (8.6,5.3);
\draw [->] (8.35,5.3) -- (8.25,5.3);
\fill [brown] (8.7, 5.3) circle (0.1);
\node [brown] at (8.7, 5.) {\scriptsize $(J, q, j)$};
\node at (7.8, 5.15) {\scriptsize $E$};

\draw [ultra thick, ->] (10., 5.) -- (12.1, 5.);
\node at (11., 5.4) {\small $B_P^{\tob ss\tob}$};

\draw [->-] (14.,4.8) -- (14.,5.3);
\draw [->-] (14.,5.3) -- (14.,5.6);
\draw [->-] (14.,5.6) -- (13.3,6.);
\draw [->-] (13.3,6.) -- (12.6,5.6);
\draw [->-] (12.6,5.6) -- (12.6,4.8);
\draw [->-] (12.6,4.8) -- (13.3,4.4);
\draw [->-] (13.3,4.4) -- (14.,4.8);
\draw [->-] (14.,5.6) -- (14.7,6.);
\draw [->-] (14.7,6.) -- (15.4,5.6);
\draw [->-] (15.4,5.6) -- (15.4,4.8);
\draw [->-] (15.4,4.8) -- (14.7,4.4);
\draw [->-] (14.7,4.4) -- (14.,4.8);
\draw [->-] (13.3,6.4) -- (13.3,6.);
\draw [->-] (14.7,6.4) -- (14.7,6.);
\draw [->-] (15.7,5.8) -- (15.4,5.6);
\draw [->-] (15.7,4.6) -- (15.4,4.8);
\draw [->-] (14.7,4.) -- (14.7,4.4);
\draw [->-] (13.3,4.) -- (13.3,4.4);
\draw [->-] (12.3,4.6) -- (12.6,4.8);
\draw [->-] (12.3,5.8) -- (12.6,5.6);
\draw [snake it] (14.,5.3) -- (14.6,5.3);
\draw [->] (14.3,5.3) -- (14.2,5.3);
\fill [brown] (14.7, 5.3) circle (0.1);
\node [brown] at (14.7, 5.) {\scriptsize $(J, q, j)$};
\node at (13.8, 5.15) {\scriptsize $E$};
\draw [red] (13.2, 5.2) circle (.4);
\draw [red, ->] (13.47, 5.5) -- (13.56, 5.4);
\node [red] at (13.4, 5.35) {\scriptsize $s$};
\node [red] at (13.7, 5.5) {$\times$};

\draw [ultra thick, ->] (4., 1.) -- (6.1, 1.);
\node at (5., 1.4) {\small $B_P^{psur}$};

\draw [->-] (2.,0.8) -- (2.,1.1);
\draw [->-] (2.,1.1) -- (2.,1.6);
\draw [->-] (2.,1.6) -- (1.3,2.);
\draw [->-] (1.3,2.) -- (0.6,1.6);
\draw [->-] (0.6,1.6) -- (0.6,0.8);
\draw [->-] (0.6,0.8) -- (1.3,0.4);
\draw [->-] (1.3,0.4) -- (2.,0.8);
\draw [->-] (2.,1.6) -- (2.7,2.);
\draw [->-] (2.7,2.) -- (3.4,1.6);
\draw [->-] (3.4,1.6) -- (3.4,0.8);
\draw [->-] (3.4,0.8) -- (2.7,0.4);
\draw [->-] (2.7,0.4) -- (2.,0.8);
\draw [->-] (1.3,2.4) -- (1.3,2.);
\draw [->-] (2.7,2.4) -- (2.7,2.);
\draw [->-] (3.7,1.8) -- (3.4,1.6);
\draw [->-] (3.7,0.6) -- (3.4,0.8);
\draw [->-] (2.7,0.) -- (2.7,0.4);
\draw [->-] (1.3,0.) -- (1.3,0.4);
\draw [->-] (0.3,0.6) -- (0.6,0.8);
\draw [->-] (0.3,1.8) -- (0.6,1.6);
\draw [snake it] (2.,1.1) -- (1.7,1.1);
\draw [->] (1.85,1.1) -- (1.93,1.1);
\draw [red] (1.7,1.4) -- (1.7,1.1);
\draw [red] (1.7,1.4) ..controls (1.695,1.47) and (1.69,1.54) .. (1.6,1.6) ..controls (1.51,1.66) and (1.334,1.712) .. (1.2,1.7) ..controls (1.066,1.688) and (0.973,1.614) .. (0.9,1.5) ..controls (0.827,1.386) and (0.773,1.232) .. (0.8,1.1) ..controls (0.827,0.968) and (0.934,0.857) .. (1.1,0.8) ..controls (1.266,0.743) and (1.49,0.741) .. (1.6,0.8) ..controls (1.69,0.859) and (1.69,0.98) .. (1.7,1.1);
\draw [snake it] (1.7,1.4) -- (1.3,1.4);
\draw [->] (1.45,1.4) -- (1.55,1.4);
\fill [brown] (1.2, 1.4) circle (0.1);
\node [brown] at (1.23, 1.15) {\tiny $(J, r, k)$};
\draw [red, ->] (1.5, 1.65) -- (1.53, 1.639);
\node [red] at (1.45, 1.53) {\tiny $u$};
\node [red] at (1.82, 1.25) {\tiny $s$};
\node at (2.2, 1.25) {\scriptsize $E$};
\node [red] at (1.83, 1.5) {$\times$};

\draw [ultra thick, ->] (10., 1.) -- (12.1, 1.);
\node at (11., 1.4) {\small $\hop_E^{J;(r,k)(q,j)}$};

\draw [->-] (8.,0.8) -- (8.,1.1);
\draw [->-] (8.,1.1) -- (8.,1.6);
\draw [->-] (8.,1.6) -- (7.3,2.);
\draw [->-] (7.3,2.) -- (6.6,1.6);
\draw [->-] (6.6,1.6) -- (6.6,0.8);
\draw [->-] (6.6,0.8) -- (7.3,0.4);
\draw [->-] (7.3,0.4) -- (8.,0.8);
\draw [->-] (8.,1.6) -- (8.7,2.);
\draw [->-] (8.7,2.) -- (9.4,1.6);
\draw [->-] (9.4,1.6) -- (9.4,0.8);
\draw [->-] (9.4,0.8) -- (8.7,0.4);
\draw [->-] (8.7,0.4) -- (8.,0.8);
\draw [->-] (7.3,2.4) -- (7.3,2.);
\draw [->-] (8.7,2.4) -- (8.7,2.);
\draw [->-] (9.7,1.8) -- (9.4,1.6);
\draw [->-] (9.7,0.6) -- (9.4,0.8);
\draw [->-] (8.7,0.) -- (8.7,0.4);
\draw [->-] (7.3,0.) -- (7.3,0.4);
\draw [->-] (6.3,0.6) -- (6.6,0.8);
\draw [->-] (6.3,1.8) -- (6.6,1.6);
\draw [snake it] (8.,1.1) -- (7.4,1.1);
\draw [->] (7.6,1.1) -- (7.7,1.1);
\fill [brown] (7.3, 1.1) circle (0.1);
\node [brown] at (7.3, 0.85) {\scriptsize $(J, r, k)$};
\node at (8.2, 1.25) {\scriptsize $E$};

\draw [->-] (14.,.8) -- (14.,1.3);
\draw [->-] (14., 1.3) -- (14.,1.6);
\draw [->-] (14.,1.6) -- (13.3,2.);
\draw [->-] (13.3,2.) -- (12.6,1.6);
\draw [->-] (12.6,1.6) -- (12.6,0.8);
\draw [->-] (12.6,0.8) -- (13.3,0.4);
\draw [->-] (13.3,0.4) -- (14.,0.8);
\draw [->-] (14.,1.6) -- (14.7,2.);
\draw [->-] (14.7,2.) -- (15.4,1.6);
\draw [->-] (15.4,1.6) -- (15.4,0.8);
\draw [->-] (15.4,0.8) -- (14.7,0.4);
\draw [->-] (14.7,0.4) -- (14.,0.8);
\draw [->-] (13.3,2.4) -- (13.3,2.);
\draw [->-] (14.7,2.4) -- (14.7,2.);
\draw [->-] (15.7,1.8) -- (15.4,1.6);
\draw [->-] (15.7,0.6) -- (15.4,0.8);
\draw [->-] (14.7,0.) -- (14.7,0.4);
\draw [->-] (13.3,0.) -- (13.3,0.4);
\draw [->-] (12.3,0.6) -- (12.6,0.8);
\draw [->-] (12.3,1.8) -- (12.6,1.6);
\draw [snake it] (14.,1.3) -- (14.6,1.3);
\draw [->] (14.3,1.3) -- (14.25,1.3);
\fill [brown] (14.7, 1.3) circle (0.1);
\node [brown] at (14.7, 1.) {\scriptsize $(J, q, j)$};
\node at (13.8, 1.15) {\scriptsize $E$};

\draw [ultra thick, ->] (14., 3.9) -- (14., 2.5);
\node at (14.3, 3.3) {$\mathcal{T}$};
\end{tikzpicture}
}

\newcommand{\Representation}{
\begin{tikzpicture}[baseline={([yshift=-.8ex]current bounding box.center)},scale=1.2]
\draw [->-] (-3.1,8.) -- (-2.4,8.4);
\draw [->-] (-2.4,8.4) -- (-1.7,8.);
\draw [->-] (-1.7,8.) -- (-1.7,7.2);
\draw [->-] (-1.7,7.2) -- (-2.4,6.8);
\draw [->-] (-2.4,8.8) -- (-2.4,8.4);
\draw [->-] (-3.4,8.2) -- (-3.1,8.);
\draw [->-] (-2.4,6.4) -- (-2.4,6.8);
\draw [->-] (-1.4,7.) -- (-1.7,7.2);
\draw [->-] (-1.3,8.2) -- (-1.7,8.);
\draw [->-] (-3.4,7.) -- (-3.1,7.2);
\draw [->-] (-2.4,6.8) -- (-3.1,7.2);
\draw [] (-3.1,7.2) -- (-3.1,7.4);
\draw [->-] (-3.1,7.4) -- (-3.1,8.);

\draw [snake it] (-3.1,7.5) -- (-2.5,7.5);
\draw [->] (-2.85,7.5) -- (-2.9,7.5);
\fill [brown] (-2.45,7.5) circle (.07);
\draw [red] (-2.1, 7.8) circle (.3);
\draw [red] (-2.1, 7.8) circle (.15);
\draw [red, ->] (-2.005,7.7) -- (-2.05,7.65);
\draw [red, ->] (-1.93,7.56) -- (-2.04,7.5);
\node [red] at (-2.1,7.75) {\tiny $s$};
\node [red] at (-1.8,7.5) {\tiny $r$};

\node at (-2.85,7.35) {\tiny $p$};
\node [brown] at (-2.3,7.3) {\tiny $(J,p,\alpha)$};

\draw [->-] (-3.1,4.) -- (-2.4,4.4);
\draw [->-] (-2.4,4.4) -- (-1.7,4.);
\draw [->-] (-1.7,4.) -- (-1.7,3.2);
\draw [->-] (-1.7,3.2) -- (-2.4,2.8);
\draw [->-] (-2.4,4.8) -- (-2.4,4.4);
\draw [->-] (-3.4,4.2) -- (-3.1,4.);
\draw [->-] (-2.4,2.4) -- (-2.4,2.8);
\draw [->-] (-1.4,3.) -- (-1.7,3.2);
\draw [->-] (-1.3,4.2) -- (-1.7,4.);
\draw [->-] (-3.4,3.) -- (-3.1,3.2);
\draw [->-] (-2.4,2.8) -- (-3.1,3.2);
\draw [] (-3.1,3.2) -- (-3.1,3.4);
\draw [->-] (-3.1,3.4) -- (-3.1,4.);

\draw [snake it] (-3.1,3.5) -- (-2.5,3.5);
\draw [->] (-2.85,3.5) -- (-2.9,3.5);
\fill [brown] (-2.45,3.5) circle (.07);
\draw [red] (-2.1, 3.8) circle (.2);
\draw [red, ->] (-1.925,3.7) -- (-2.05,3.6);
\node [red] at (-2.05,3.75) {\tiny $t$};

\node at (-2.85,3.35) {\tiny $p$};
\node [brown] at (-2.2,3.2) {\tiny $(J,p,\alpha)$};

\draw [thick, ->] (-1., 7.5) -- (1.8, 7.5);
\node [] at (.35,7.75) {\tiny $D^J(B^{qsvm})D^J(B^{pruq})$};

\draw [->-] (2.4,8.) -- (3.1,8.4);
\draw [->-] (3.1,8.4) -- (3.8,8.);
\draw [->-] (3.8,8.) -- (3.8,7.2);
\draw [->-] (3.8,7.2) -- (3.1,6.8);
\draw [->-] (3.1,8.8) -- (3.1,8.4);
\draw [->-] (2.1,8.2) -- (2.4,8.);
\draw [->-] (3.1,6.4) -- (3.1,6.8);
\draw [->-] (4.1,7.) -- (3.8,7.2);
\draw [->-] (4.2,8.2) -- (3.8,8.);
\draw [->-] (2.1,7.) -- (2.4,7.2);
\draw [->-] (3.1,6.8) -- (2.4,7.2);
\draw [] (2.4,7.2) -- (2.4,7.4);
\draw [->-] (2.4,7.4) -- (2.4,8.);
\draw [decorate,decoration={snake,amplitude=0.1mm,segment length=.5mm}] (2.4,7.4) -- (2.6,7.4);
\draw [decoration={markings,mark=at position\pgfdecoratedpathlength with {\arrow[line width=0.01mm]{>}}},postaction={decorate}] (2.5,7.4) -- (2.45,7.4);
\draw [red] (2.6,7.4) -- (2.6,7.55);
\draw [decorate,decoration={markings,mark=at position\pgfdecoratedpathlength with {\arrow[line width=0.01mm,red]{>}}}] (2.6,7.4) -- (2.6,7.55);
\draw [decorate,decoration={snake,amplitude=0.1mm,segment length=.5mm}] (2.6,7.55) -- (2.8,7.55);
\draw [decoration={markings,mark=at position\pgfdecoratedpathlength with {\arrow[line width=0.01mm]{>}}},postaction={decorate}] (2.7,7.55) -- (2.65,7.55);
\draw [red] (2.8,7.55) -- (2.8,7.7);
\draw [decorate,decoration={markings,mark=at position\pgfdecoratedpathlength with {\arrow[line width=0.01mm,red]{>}}}] (2.8,7.5) -- (2.8,7.7);
\draw [decorate,decoration={snake,amplitude=0.1mm,segment length=.5mm}] (2.8,7.7) -- (3.,7.7);
\draw [decoration={markings,mark=at position\pgfdecoratedpathlength with {\arrow[line width=0.01mm]{>}}},postaction={decorate}] (2.9,7.7) -- (2.85,7.7);
\fill [brown] (3.05,7.7) circle (.07);

\draw [red] (2.6,7.55) arc [x radius = .5,y radius = .6,start angle = 180,end angle = 90] arc [x radius = .5,y radius = .55,start angle = 90,end angle = -90] arc [x radius = .5,y radius = .4,start angle = -90,end angle = -180];
\draw [red,->] (3.3,7.45) -- (3.2,7.34);

\draw [red] (2.8,7.7) arc [x radius = .25,y radius = .25,start angle = 180,end angle = 90] arc [x radius = .3,y radius = .32,start angle = 90,end angle = -90] arc [x radius = .25,y radius = .25,start angle = -90,end angle = -180];
\draw [red,->] (3.4,7.17) -- (3.3,7.1);

\node at (2.5,7.3) {\tiny $p$};
\node at (2.7,7.45) {\tiny $q$};
\node at (2.9,7.6) {\tiny $m$};
\node [red] at (2.5,7.5) {\tiny $u$};
\node [red] at (3.28,7.18) {\tiny $r$};
\node [red] at (2.7,7.65) {\tiny $v$};
\node [red] at (3.15,7.45) {\tiny $s$};
\node [brown] at (3.2,7.7) {\tiny $J$};

\draw [thick, ->] (4.5, 7.5) -- (6.5, 7.5);
\node [] at (5.4,7.75) {\tiny {\color{red} $\times$}};

\draw [->-] (7.4,8.) -- (8.1,8.4);
\draw [->-] (8.1,8.4) -- (8.8,8.);
\draw [->-] (8.8,8.) -- (8.8,7.2);
\draw [->-] (8.8,7.2) -- (8.1,6.8);
\draw [->-] (8.1,8.8) -- (8.1,8.4);
\draw [->-] (7.1,8.2) -- (7.4,8.);
\draw [->-] (8.1,6.4) -- (8.1,6.8);
\draw [->-] (9.1,7.) -- (8.8,7.2);
\draw [->-] (9.2,8.2) -- (8.8,8.);
\draw [->-] (7.1,7.) -- (7.4,7.2);
\draw [->-] (8.1,6.8) -- (7.4,7.2);
\draw [] (7.4,7.2) -- (7.4,7.4);
\draw [->-] (7.4,7.4) -- (7.4,8.);

\draw [snake it] (7.4,7.4) -- (7.7,7.4);
\draw [->] (7.5,7.4) -- (7.45,7.4);
\draw [red,->-] (7.7,7.4) -- (7.7,7.8);
\draw [snake it] (7.7,7.8) -- (8.,7.8);
\draw [->] (7.82,7.8) -- (7.78,7.8);
\fill [brown] (8.05,7.8) circle (.07);

\draw [red] (7.7,7.8) arc [x radius = .5,y radius = .3,start angle = 180,end angle = 90] arc [x radius = .4,y radius = .5,start angle = 90,end angle = -90] arc [x radius = .5,y radius = .35,start angle = -90,end angle = -180];
\draw [red,->] (8.58,7.45) -- (8.54,7.34);

\node at (7.55,7.28) {\tiny $p$};
\node at (7.9,7.9) {\tiny $m$};
\node [red] at (7.6,7.6) {\tiny $v$};
\node [red] at (8.7,7.45) {\tiny $s$};
\node [brown] at (8.2,7.6) {\tiny $(J, m, \beta)$};

\draw [->-] (2.4,4.) -- (3.1,4.4);
\draw [->-] (3.1,4.4) -- (3.8,4.);
\draw [->-] (3.8,4.) -- (3.8,3.2);
\draw [->-] (3.8,3.2) -- (3.1,2.8);
\draw [->-] (3.1,4.8) -- (3.1,4.4);
\draw [->-] (2.1,4.2) -- (2.4,4.);
\draw [->-] (3.1,2.4) -- (3.1,2.8);
\draw [->-] (4.1,3.) -- (3.8,3.2);
\draw [->-] (4.2,4.2) -- (3.8,4.);
\draw [->-] (2.1,3.) -- (2.4,3.2);
\draw [->-] (3.1,2.8) -- (2.4,3.2);
\draw [] (2.4,3.2) -- (2.4,3.4);
\draw [->-] (2.4,3.4) -- (2.4,4.);

\draw [snake it] (2.4,3.4) -- (2.7,3.4);
\draw [->] (2.5,3.4) -- (2.45,3.4);
\draw [red,->-] (2.7,3.4) -- (2.7,3.8);
\draw [snake it] (2.7,3.8) -- (3.,3.8);
\draw [->] (2.82,3.8) -- (2.78,3.8);
\fill [brown] (3.05,3.8) circle (.07);

\draw [red] (2.7,3.8) arc [x radius = .5,y radius = .3,start angle = 180,end angle = 90] arc [x radius = .4,y radius = .5,start angle = 90,end angle = -90] arc [x radius = .5,y radius = .35,start angle = -90,end angle = -180];
\draw [red,->] (3.58,3.45) -- (3.54,3.34);

\node at (2.55,3.28) {\tiny $p$};
\node at (2.9,3.9) {\tiny $m$};
\node [red] at (2.55,3.6) {\tiny $w$};
\node [red] at (3.7,3.45) {\tiny $t$};
\node [brown] at (3.15,3.6) {\tiny $(J, m, \beta)$};

\draw [->-] (7.4,4.) -- (8.1,4.4);
\draw [->-] (8.1,4.4) -- (8.8,4.);
\draw [->-] (8.8,4.) -- (8.8,3.2);
\draw [->-] (8.8,3.2) -- (8.1,2.8);
\draw [->-] (8.1,4.8) -- (8.1,4.4);
\draw [->-] (7.1,4.2) -- (7.4,4.);
\draw [->-] (8.1,2.4) -- (8.1,2.8);
\draw [->-] (9.1,3.) -- (8.8,3.2);
\draw [->-] (9.2,4.2) -- (8.8,4.);
\draw [->-] (7.1,3.) -- (7.4,3.2);
\draw [->-] (8.1,2.8) -- (7.4,3.2);
\draw [] (7.4,3.2) -- (7.4,3.4);
\draw [->-] (7.4,3.4) -- (7.4,4.);

\draw [snake it] (7.4,3.5) -- (8.,3.5);
\draw [->] (7.65,3.5) -- (7.6,3.5);
\fill [brown] (8.05,3.5) circle (.07);

\node at (7.65,3.35) {\tiny $m$};
\node [brown] at (8.1,3.7) {\tiny $(J,m,\beta)$};

\draw [thick, ->] (8.1, 6.) -- (8.1, 5.);
\node [] at (7.8,5.5) {\tiny {\color{red} $\times$'}};

\draw [thick, ->] (3.1, 6.) -- (3.1, 5.);
\node [right] at (3.12,5.6) {\tiny {\color{red} $\times'$}};

\draw [thick, ->] (4.5, 3.5) -- (6.5, 3.5);
\node [] at (5.4,3.75) {\tiny {\color{red} $\times$}};

\draw [thick, ->] (-2.4, 6.) -- (-2.4, 5.);
\node [] at (-2.1,5.6) {\tiny {\color{red} $\times'$}};

\draw [thick, ->] (-1., 3.5) -- (1.5, 3.5);
\node [] at (0.2,3.75) {\tiny $D^J(B^{ptwm})$};

\node [red] at (-2.7,7.8) {\tiny $\times$};
\node [red] at (-2.2,7.6) {\tiny $\times$};
\node [red] at (-2.1,7.6) {\tiny $'$};
\node [red] at (-2.7,3.9) {\tiny $\times$};
\node [red] at (2.55,3.9) {\tiny $\times$};
\node [red] at (2.55,7.9) {\tiny $\times$};
\node [red] at (2.9,7.2) {\tiny $\times$};
\node [red] at (3.,7.2) {\tiny $'$};
\node [red] at (7.55,7.9) {\tiny $\times$};
\node [red] at (7.65,7.9) {\tiny $'$};

\end{tikzpicture}
}

\newcommand{\Edge}[1]{
\begin{tikzpicture}[baseline={([yshift=-.8ex]current bounding box.center)},scale=1.]
\draw [->-] (0, 0) -- (0, 1.5);
\node [right, rotate=0.] at (0.,.75) {\scriptsize ${#1}$};
\end{tikzpicture}
}

\newcommand{\BimoduleA}{
\begin{tikzpicture}[baseline={([yshift=-.8ex]current bounding box.center)},scale=.8]
\draw [->-] (1, 0) -- (1, .6);
\draw [->-] (1, .6) -- (1, 1.3);
\draw [->-] (1, 1.3) -- (1, 2.);
\draw [snake it, red] (1, .6) -- (.3, .6);
\draw [->, red] (.6, .6) -- (.7, .6);
\draw [snake it, red] (1, 1.3) -- (1.7, 1.3);
\draw [->, red] (1.4, 1.3) -- (1.3, 1.3);
\node [red] at (0.05, .58) {\scriptsize $a_\alpha$};
\node [red] at (1.95, 1.28) {\scriptsize $b_\beta$};
\node at (1.3, .2) {\scriptsize $x_\chi$};
\node at (1.2,.9) {\scriptsize $u$};
\node at (1.3,1.8) {\scriptsize $y_\upsilon$};
\end{tikzpicture}
}

\newcommand{\BimoduleB}{
\begin{tikzpicture}[baseline={([yshift=-.8ex]current bounding box.center)},scale=.6]
\draw [->-] (3.2,5.6) -- (3.2,6.5);
\draw [->-] (3.2,6.5) -- (3.2,7.2);
\draw [->-] (3.2,7.2) -- (4.6,8.);
\draw [->-] (4.6,8.) -- (6.,7.2);
\draw [->-] (6.,7.2) -- (6.,5.6);
\draw [->-] (6.,5.6) -- (4.6,4.8);
\draw [->-] (4.6,4.8) -- (3.2,5.6);
\draw [snake it] (3.2,6.4) -- (4.2,6.4);
\draw [->] (3.7,6.4) -- (3.6,6.4);
\node [centered, rotate=0.] at (2.7,6.) {\scriptsize $M_6$};
\node [centered, rotate=0.] at (2.7,6.9) {\scriptsize $M_0$};
\node [centered, rotate=0.] at (3.6,7.9) {\scriptsize $M_1$};
\node [centered, rotate=0.] at (5.6,7.9) {\scriptsize $M_2$};
\node [centered, rotate=0.] at (6.4,6.4) {\scriptsize $M_3$};
\node [centered, rotate=0.] at (5.6,4.9) {\scriptsize $M_4$};
\node [centered, rotate=0.] at (3.6,4.9) {\scriptsize $M_5$};
\node [centered, rotate=0.] at (3.7,6.) {\scriptsize $M$};
\node [centered, rotate=0.] at (4.6,7.) {\tiny Plaquette $P$};
\draw [->-] (2.6,7.6) -- (3.2,7.2);
\draw [->-] (4.6,8.8) -- (4.6,8.);
\draw [->-] (6.6,7.6) -- (6.,7.2);
\draw [->-] (6.6,5.2) -- (6.,5.6);
\draw [->-] (4.6,4.) -- (4.6,4.8);
\draw [->-] (2.6,5.2) -- (3.2,5.6);
\node [centered, rotate=0.] at (2.3,7.8) {\scriptsize $N_1$};
\node [centered, rotate=0.] at (4.6,9.) {\scriptsize $N_2$};
\node [centered, rotate=0.] at (6.9,7.8) {\scriptsize $N_3$};
\node [centered, rotate=0.] at (6.9,5.) {\scriptsize $N_4$};
\node [centered, rotate=0.] at (4.6,3.7) {\scriptsize $N_5$};
\node [centered, rotate=0.] at (2.3,5.) {\scriptsize $N_6$};
\end{tikzpicture}
}

\newcommand{\BimoduleC}{
\begin{tikzpicture}[baseline={([yshift=-.8ex]current bounding box.center)},scale=1.5]
\draw [->-] (3.2,5.6) -- (3.2,5.8);
\draw [->-] (3.2,5.8) -- (3.2,6.1);
\draw [->-] (3.2,6.1) -- (3.2,6.4);
\node [centered, rotate=0.] at (3.05,5.7) {\scriptsize $x_6$};
\node [centered, rotate=0.] at (3.05,5.95) {\scriptsize $u_6$};
\node [centered, rotate=0.] at (3.05,6.25) {\scriptsize $y_6$};

\draw [red, -<-] (3.2, 5.8) -- (2.65, 5.8);
\draw [red, -<-] (3.2, 6.1) -- (4.1, 6.1);
\node [red] at (3.62,5.98) {\scriptsize $a_6$};
\node [red] at (2.5, 5.8) {\scriptsize $r_6$};
\fill [red] (4.05, 6.1) circle (0.05);
\draw [->-] (3.2,6.4) -- (3.2,6.7);
\draw [->-] (3.2,6.7) -- (3.2,7.);
\draw [->-] (3.2,7.) -- (3.2,7.2);
\node [centered, rotate=0.] at (3.05,6.55) {\scriptsize $x_0$};
\node [centered, rotate=0.] at (3.05,6.85) {\scriptsize $u_0$};
\node [centered, rotate=0.] at (3.05,7.1) {\scriptsize $y_0$};
\draw [red, -<-] (3.2, 6.7) -- (2.65, 6.7);
\draw [red, -<-] (3.2, 7.) -- (3.8, 7.);
\node [red] at (3.5,6.87) {\scriptsize $a_0$};
\node [red] at (2.5, 6.7) {\scriptsize $r_0$};
\fill [red] (3.8, 7.) circle (0.05);
\draw [->-] (3.2,7.2) -- (3.55,7.4);
\draw [->-] (3.55,7.4) -- (4.25,7.8);
\draw [->-] (4.25,7.8) -- (4.6,8.);
\node [centered, rotate=0.] at (3.47,7.18) {\scriptsize $x_1$};
\node [centered, rotate=0.] at (3.8,7.72) {\scriptsize $u_1$};
\node [centered, rotate=0.] at (4.35,8.) {\scriptsize $y_1$};
\draw [red, -<-] (3.55, 7.4) -- (3.25, 7.9);
\draw [red, -<-] (4.25, 7.8) -- (4.45, 7.45);
\node [red] at (4.25,7.55) {\scriptsize $a_1$};
\node [red] at (3.2, 8) {\scriptsize $r_1$};
\fill [red] (4.47, 7.4) circle (0.05);
\draw [->-] (4.6,8.) -- (4.95,7.8);
\draw [->-] (4.95,7.8) -- (5.65,7.4);
\draw [->-] (5.65,7.4) -- (6.,7.2);
\node [centered, rotate=0.] at (5.9,7.43) {\scriptsize $y_2$};
\node [centered, rotate=0.] at (5.4,7.7) {\scriptsize $u_2$};
\node [centered, rotate=0.] at (4.75,7.78) {\scriptsize $x_2$};
\draw [red, -<-] (4.95, 7.8) -- (5.25, 8.3);
\draw [red, -<-] (5.65, 7.4) -- (5.38, 7.);
\node [red] at (5.4,7.3) {\scriptsize $a_2$};
\node [red] at (5.35, 8.4) {\scriptsize $r_2$};
\fill [red] (5.39, 7.02) circle (0.05);
\draw [->-] (6.,7.2) -- (6.,6.7);
\draw [->-] (6.,6.7) -- (6.,6.1);
\draw [->-] (6.,6.1) -- (6.,5.6);
\draw [red, -<-] (6., 6.7) -- (6.5, 6.7);
\draw [red, ->-] (5.6, 6.1) -- (6., 6.1);
\fill [red] (5.56, 6.1) circle (0.05);
\node [centered, rotate=0.] at (5.85,6.9) {\scriptsize $x_3$};
\node [centered, rotate=0.] at (6.15,6.4) {\scriptsize $u_3$};
\node [centered, rotate=0.] at (6.15,5.85) {\scriptsize $y_3$};
\draw [->-] (6.,5.6) -- (5.65,5.4);
\draw [->-] (5.65,5.4) -- (4.95,5.);
\node [red] at (5.8,6.22) {\scriptsize $a_3$};
\node [red] at (6.65, 6.7) {\scriptsize $r_3$};
\draw [->-] (4.95,5.) -- (4.6,4.8);
\node [centered, rotate=0.] at (5.93,5.37) {\scriptsize $x_4$};
\node [centered, rotate=0.] at (5.4,5.05) {\scriptsize $u_4$};
\node [centered, rotate=0.] at (4.9,4.78) {\scriptsize $y_4$};
\draw [red, -<-] (5.65, 5.4) -- (5.95, 4.9);
\draw [red, -<-] (4.95, 5.) -- (4.57, 5.65);
\node [red] at (4.92,5.32) {\scriptsize $a_4$};
\node [red] at (6.05, 4.8) {\scriptsize $r_4$};
\fill [red] (4.58, 5.64) circle (0.05);
\draw [->-] (4.6,4.8) -- (4.25,5.);
\draw [->-] (4.25,5.) -- (3.55,5.4);
\draw [->-] (3.55,5.4) -- (3.2,5.6);
\node [centered, rotate=0.] at (3.48,5.6) {\scriptsize $y_5$};
\node [centered, rotate=0.] at (3.75,5.1) {\scriptsize $u_5$};
\node [centered, rotate=0.] at (4.5,5.03) {\scriptsize $x_5$};
\draw [red, -<-] (4.25, 5.) -- (3.95, 4.5);
\draw [red, -<-] (3.55, 5.4) -- (4.25, 5.8);
\node [red] at (4.,5.5) {\scriptsize $a_5$};
\node [red] at (3.9, 4.4) {\scriptsize $r_5$};
\fill [red] (4.25, 5.8) circle (0.05);
\draw [-<-] (3.2,6.4) -- (3.6,6.4);
\draw [-<-] (3.6,6.4) -- (4.,6.4);
\draw [-<-] (4.,6.4) -- (4.5,6.4);
\draw [red] (3.6, 6.4) arc (180:0:1.);
\draw [red] (4., 6.4) arc (-180:0:.8);
\draw [red, ->] (3.685, 6.8) -- (3.645, 6.7);
\draw [red, ->] (4.16, 7.296) -- (4.06, 7.242);
\draw [red, ->] (5.06, 7.29) -- (4.96, 7.337);
\draw [red, ->] (5.595, 6.5) -- (5.58, 6.6);
\draw [red, ->] (5.18, 5.695) -- (5.25, 5.74);
\draw [red, ->] (4.37, 5.73) -- (4.45, 5.68);
\draw [red, ->] (4.1395, 5.95) -- (4.175, 5.9);
\draw [red, ->] (4.005, 6.3) -- (4.025, 6.2);
\node [red] at (3.83,6.75) {\scriptsize $b_0$};
\node [red] at (4.15,7.15) {\scriptsize $b_1$};
\node [red] at (4.95,7.22) {\scriptsize $b_2$};
\node [red] at (5.46,6.55) {\scriptsize $b_3$};
\node [red] at (5.15,5.85) {\scriptsize $b_4$};
\node [red] at (4.45,5.83) {\scriptsize $b_5$};
\node [red] at (4.25,6.) {\scriptsize $b_6$};
\node [red] at (4.15,6.27) {\scriptsize $b_7$};

\node [centered, rotate=0.] at (3.4,6.53) {\scriptsize $q$};
\node [centered, rotate=0.] at (3.8,6.53) {\scriptsize $w$};
\node [centered, rotate=0.] at (4.3,6.53) {\scriptsize $p_\alpha$};
\node [centered, rotate=0.] at (3.4,6.7) {\color{red} $\times$};
\node [centered, rotate=0.] at (3.8,7.3) {\color{red} $\times$};
\node [centered, rotate=0.] at (4.9,7.55) {\color{red} $\times$};
\node [centered, rotate=0.] at (5.75,6.6) {\color{red} $\times$};
\node [centered, rotate=0.] at (5.4,5.6) {\color{red} $\times$};
\node [centered, rotate=0.] at (4.3,5.4) {\color{red} $\times$};
\node [centered, rotate=0.] at (3.6,5.8) {\color{red} $\times$};
\node [centered, rotate=0.] at (3.6,6.25) {\color{red} $\times$};

\draw [->-] (2.6,7.6) -- (2.9,7.4);
\draw [->-] (2.9,7.4) -- (3.2,7.2);
\draw [red, -<-] (2.9,7.4) -- (2.68,7.07);
\node [centered, rotate=0.] at (3.1,7.43) {\scriptsize $e_1$};
\draw [->-] (4.6,8.6) -- (4.6,8.3);
\draw [->-] (4.6,8.3) -- (4.6,8.);
\draw [red, -<-] (4.6,8.3) -- (4.2,8.3);
\node [centered, rotate=0.] at (4.74,8.1) {\scriptsize $e_2$};
\draw [->-] (6.6,7.6) -- (6.3,7.4);
\draw [->-] (6.3,7.4) -- (6.,7.2);
\draw [red, -<-] (6.3,7.4) -- (6.08,7.73);
\node [centered, rotate=0.] at (6.2,7.19) {\scriptsize $e_3$};
\draw [->-] (6.6,5.2) -- (6.3,5.4);
\draw [->-] (6.3,5.4) -- (6.,5.6);
\draw [red, -<-] (6.3,5.4) -- (6.7,5.73);
\node [centered, rotate=0.] at (6.25,5.6) {\scriptsize $e_4$};
\draw [->-] (4.6,4.2) -- (4.6,4.5);
\draw [->-] (4.6,4.5) -- (4.6,4.8);
\draw [red, -<-] (4.6,4.5) -- (5.,4.5);
\node [centered, rotate=0.] at (4.45,4.65) {\scriptsize $e_5$};
\draw [->-] (2.6,5.2) -- (2.9,5.4);
\draw [->-] (2.9,5.4) -- (3.2,5.6);
\draw [red, -<-] (2.9,5.4) -- (3.12,5.07);
\node [centered, rotate=0.] at (3.12,5.38) {\scriptsize $e_6$};
\end{tikzpicture}
}

\newcommand{\BimoduleD}{
\begin{tikzpicture}[baseline={([yshift=-.8ex]current bounding box.center)},scale=.5]
\draw [->-] (3.2,5.6) -- (3.2,6.5);
\draw [->-] (3.2,6.5) -- (3.2,7.2);
\draw [->-] (3.2,7.2) -- (4.6,8.);
\draw [->-] (4.6,8.) -- (6.,7.2);
\draw [->-] (6.,7.2) -- (6.,5.6);
\draw [->-] (6.,5.6) -- (4.6,4.8);
\draw [->-] (4.6,4.8) -- (3.2,5.6);
\draw [decorate,decoration={snake,segment length=.12cm, amplitude=.04cm}] (3.2,6.4) -- (4.4,6.4);
\draw [->] (3.8,6.4) -- (3.7,6.4);

\node [centered, rotate=0.] at (2.7,6.) {\scriptsize $M_6$};
\node [centered, rotate=0.] at (2.7,6.9) {\scriptsize $M_0$};
\node [centered, rotate=0.] at (3.6,7.9) {\scriptsize $M_1$};
\node [centered, rotate=0.] at (5.6,7.9) {\scriptsize $M_2$};
\node [centered, rotate=0.] at (6.5,6.4) {\scriptsize $M_3$};
\node [centered, rotate=0.] at (5.6,4.9) {\scriptsize $M_4$};
\node [centered, rotate=0.] at (3.7,4.9) {\scriptsize $M_5$};
\node [centered, rotate=0.] at (3.8,6.) {\scriptsize $M$};
\draw [->-] (2.6,7.6) -- (3.2,7.2);
\draw [->-] (4.6,8.8) -- (4.6,8.);
\draw [->-] (6.6,7.6) -- (6.,7.2);
\draw [->-] (6.6,5.2) -- (6.,5.6);
\draw [->-] (4.6,4.) -- (4.6,4.8);
\draw [->-] (2.6,5.2) -- (3.2,5.6);
\node [centered, rotate=0.] at (2.2,7.8) {\scriptsize $N_1$};
\node [centered, rotate=0.] at (4.7,9.) {\scriptsize $N_2$};
\node [centered, rotate=0.] at (7.,7.8) {\scriptsize $N_3$};
\node [centered, rotate=0.] at (7.,5.) {\scriptsize $N_4$};
\node [centered, rotate=0.] at (4.7,3.8) {\scriptsize $N_5$};
\node [centered, rotate=0.] at (2.2,5.) {\scriptsize $N_6$};
\end{tikzpicture}
}

\newcommand{\BimoduleE}{
\begin{tikzpicture}[baseline={([yshift=-.8ex]current bounding box.center)},scale=.5]
\draw [->-] (3.2,5.6) -- (3.2,6.5);
\draw [->-] (3.2,6.5) -- (3.2,7.2);
\draw [->-] (3.2,7.2) -- (4.6,8.);
\draw [->-] (4.6,8.) -- (6.,7.2);
\draw [->-] (6.,7.2) -- (6.,5.6);
\draw [->-] (6.,5.6) -- (4.6,4.8);
\draw [->-] (4.6,4.8) -- (3.2,5.6);
\draw [decorate,decoration={snake,segment length=.12cm, amplitude=.04cm}] (3.2,6.4) -- (4.4,6.4);
\draw [->] (3.8,6.4) -- (3.7,6.4);
\node [centered, rotate=0.] at (2.8,6.) {\scriptsize $i_6$};
\node [centered, rotate=0.] at (2.8,6.9) {\scriptsize $i_0$};
\node [centered, rotate=0.] at (3.6,7.9) {\scriptsize $i_1$};
\node [centered, rotate=0.] at (5.6,7.9) {\scriptsize $i_2$};
\node [centered, rotate=0.] at (6.4,6.4) {\scriptsize $i_3$};
\node [centered, rotate=0.] at (5.6,4.9) {\scriptsize $i_4$};
\node [centered, rotate=0.] at (3.7,4.9) {\scriptsize $i_5$};
\node [centered, rotate=0.] at (4.,6.) {\scriptsize $p_\alpha$};
\draw [->-] (2.6,7.6) -- (3.2,7.2);
\draw [->-] (4.6,8.8) -- (4.6,8.);
\draw [->-] (6.6,7.6) -- (6.,7.2);
\draw [->-] (6.6,5.2) -- (6.,5.6);
\draw [->-] (4.6,4.) -- (4.6,4.8);
\draw [->-] (2.6,5.2) -- (3.2,5.6);
\node [centered, rotate=0.] at (2.2,7.8) {\scriptsize $e_1$};
\node [centered, rotate=0.] at (4.7,9.) {\scriptsize $e_2$};
\node [centered, rotate=0.] at (7.,7.8) {\scriptsize $e_3$};
\node [centered, rotate=0.] at (7.,5.) {\scriptsize $e_4$};
\node [centered, rotate=0.] at (4.7,3.8) {\scriptsize $e_5$};
\node [centered, rotate=0.] at (2.2,5.) {\scriptsize $e_6$};
\end{tikzpicture}
}

\newcommand{\BimoduleG}{
\begin{tikzpicture}[baseline={([yshift=-.8ex]current bounding box.center)},scale=.5]
\draw [->-] (3.2,5.6) -- (3.2,6.5);
\draw [->-] (3.2,6.5) -- (3.2,7.2);
\draw [->-] (3.2,7.2) -- (4.6,8.);
\draw [->-] (4.6,8.) -- (6.,7.2);
\draw [->-] (6.,7.2) -- (6.,5.6);
\draw [->-] (6.,5.6) -- (4.6,4.8);
\draw [->-] (4.6,4.8) -- (3.2,5.6);
\draw [decorate,decoration={snake,segment length=.12cm, amplitude=.04cm}] (3.2,6.4) -- (4.4,6.4);
\draw [->] (3.8,6.4) -- (3.7,6.4);
\node [centered, rotate=0.] at (2.8,6.) {\scriptsize $i_6$};
\node [centered, rotate=0.] at (2.8,6.9) {\scriptsize $i_0$};
\node [centered, rotate=0.] at (3.6,7.9) {\scriptsize $i_1$};
\node [centered, rotate=0.] at (5.6,7.9) {\scriptsize $i_2$};
\node [centered, rotate=0.] at (6.4,6.4) {\scriptsize $i_3$};
\node [centered, rotate=0.] at (5.6,4.9) {\scriptsize $i_4$};
\node [centered, rotate=0.] at (3.7,4.9) {\scriptsize $i_5$};
\node [centered, rotate=0.] at (4.,6.) {\scriptsize $p$};
\draw [->-] (2.6,7.6) -- (3.2,7.2);
\draw [->-] (4.6,8.8) -- (4.6,8.);
\draw [->-] (6.6,7.6) -- (6.,7.2);
\draw [->-] (6.6,5.2) -- (6.,5.6);
\draw [->-] (4.6,4.) -- (4.6,4.8);
\draw [->-] (2.6,5.2) -- (3.2,5.6);
\node [centered, rotate=0.] at (2.2,7.8) {\scriptsize $e_1$};
\node [centered, rotate=0.] at (4.7,9.) {\scriptsize $e_2$};
\node [centered, rotate=0.] at (7.,7.8) {\scriptsize $e_3$};
\node [centered, rotate=0.] at (7.,5.) {\scriptsize $e_4$};
\node [centered, rotate=0.] at (4.7,3.8) {\scriptsize $e_5$};
\node [centered, rotate=0.] at (2.2,5.) {\scriptsize $e_6$};
\end{tikzpicture}
}

\newcommand{\ExcitedFlux}{
\begin{tikzpicture}[baseline={([yshift=-.8ex]current bounding box.center)},scale=1.1]
\draw [->-] (2.,2.9) -- (1.3,3.3);
\draw [->-] (1.3,3.3) -- (0.6,2.9);
\draw [->-] (0.6,2.9) -- (0.6,2.);
\draw [->-] (0.6,2.) -- (1.3,1.6);
\draw [->-] (2.,2.9) -- (2.7,3.3);
\draw [->-] (2.7,3.3) -- (3.4,2.9);
\draw [->-] (3.4,2.9) -- (3.4,2.);
\draw [->-] (3.4,2.) -- (2.7,1.6);
\draw [->-] (1.3,1.6) -- (2.,2.);
\draw [->-] (2.7,1.6) -- (2.,2.);
\draw [->-] (1.3,3.3) -- (1.3,3.7);
\draw [->-] (2.7,3.3) -- (2.7,3.7);
\draw [->-] (3.4,2.9) -- (3.7,3.1);
\draw [->-] (3.4,2.) -- (3.7,1.8);
\draw [->-] (2.7,1.6) -- (2.7,1.2);
\draw [->-] (1.3,1.6) -- (1.3,1.2);
\draw [->-] (0.6,2.) -- (0.3,1.8);
\draw [->-] (0.6,2.9) -- (0.3,3.1);
\draw [->-] (2, 2.1) -- (2, 2.9);
\draw [->-] (2, 2) -- (2, 2.3);
\draw [->-] (2, 2.7) -- (2, 2.9);
\draw [snake it] (2, 2.25) -- (1.2, 2.25);
\draw [->] (1.5, 2.25) -- (1.6, 2.25);
\draw [snake it] (2, 2.65) -- (2.8, 2.65);
\draw [->] (2.5, 2.65) -- (2.4, 2.65);
\node [centered, rotate=0.] at (2.15,2.42) {\scriptsize $j$};
\node [centered, rotate=0.] at (2.15,2.1) {\scriptsize $i$};
\node [centered, rotate=0.] at (2.15,2.8) {\scriptsize $k$};
\node [centered, rotate=0.] at (1.5,2.45) {\scriptsize $p$};
\node [centered, rotate=0.] at (2.5,2.45) {\scriptsize $q$};
\end{tikzpicture}
}

\newcommand{\FusionLeft}{
\begin{tikzpicture}[baseline={([yshift=-.8ex]current bounding box.center)},scale=1.]
\draw [->-] (4.,8.) -- (3.3,8.4);
\draw [->-] (3.3,8.4) -- (2.6,8.);
\draw [->-] (2.6,8.) -- (2.6,7.2);
\draw [->-] (2.6,7.2) -- (3.3,6.8);
\draw [->-] (3.3,6.8) -- (4.,7.2);
\draw [->-] (4.,8.) -- (4.7,8.4);
\draw [->-] (4.7,8.4) -- (5.4,8.);
\draw [->-] (5.4,8.) -- (5.4,7.2);
\draw [->-] (5.4,7.2) -- (4.7,6.8);
\draw [->-] (4.7,6.8) -- (4.,7.2);
\draw [->-] (4.,7.2) -- (4.,7.5);
\draw [->] (4.,7.2) -- (4.,7.7);
\draw [] (4.,7.5) -- (4.,7.7);
\draw [->-] (4.,7.7) -- (4.,8.);
\draw [->-] (3.3,8.4) -- (3.3,8.8);
\draw [->-] (4.7,8.4) -- (4.7,8.8);
\draw [->-] (3.3,6.4) -- (3.3,6.8);
\draw [->-] (4.7,6.4) -- (4.7,6.8);
\draw [->-] (5.7,7.) -- (5.4,7.2);
\draw [->-] (5.7,8.2) -- (5.4,8.);
\draw [->-] (2.3,8.2) -- (2.6,8.);
\draw [->-] (2.3,7.) -- (2.6,7.2);
\node at (4.15, 7.3) {\tiny $x$};
\node at (4.15, 7.55) {\tiny $u$};
\node at (4.15, 7.9) {\tiny $y$};

\draw [snake it] (4.,7.5) -- (3.4,7.5);
\draw [->] (3.6,7.5) -- (3.7,7.5);
\draw [snake it] (4.,7.7) -- (4.6,7.7);
\draw [->] (4.4,7.7) -- (4.3,7.7);
\node [brown] at (3.2, 7.75) {\tiny $(J_1, p, \alpha)$};
\node [brown] at (4.8, 7.45) {\tiny $(J_2, r, \gamma)$};
\fill [brown] (3.3, 7.5) circle (0.1);
\fill [brown] (4.7, 7.7) circle (0.1);
\end{tikzpicture}
}

\newcommand{\FusionRight}{
\begin{tikzpicture}[baseline={([yshift=-.8ex]current bounding box.center)},scale=1.]
\draw [->-] (4.,8.) -- (3.3,8.4);
\draw [->-] (3.3,8.4) -- (2.6,8.);
\draw [->-] (2.6,8.) -- (2.6,7.2);
\draw [->-] (2.6,7.2) -- (3.3,6.8);
\draw [->-] (3.3,6.8) -- (4.,7.2);
\draw [->-] (4.,8.) -- (4.7,8.4);
\draw [->-] (4.7,8.4) -- (5.4,8.);
\draw [->-] (5.4,8.) -- (5.4,7.2);
\draw [->-] (5.4,7.2) -- (4.7,6.8);
\draw [->-] (4.7,6.8) -- (4.,7.2);
\draw [->-] (4.,7.2) -- (4.,7.5);
\draw [->] (4.,7.2) -- (4.,7.7);
\draw [] (4.,7.5) -- (4.,7.7);
\draw [->-] (4.,7.7) -- (4.,8.);
\draw [->-] (3.3,8.4) -- (3.3,8.8);
\draw [->-] (4.7,8.4) -- (4.7,8.8);
\draw [->-] (3.3,6.4) -- (3.3,6.8);
\draw [->-] (4.7,6.4) -- (4.7,6.8);
\draw [->-] (5.7,7.) -- (5.4,7.2);
\draw [->-] (5.7,8.2) -- (5.4,8.);
\draw [->-] (2.3,8.2) -- (2.6,8.);
\draw [->-] (2.3,7.) -- (2.6,7.2);
\node at (4.15, 7.3) {\tiny $x$};
\node at (4.15, 7.55) {\tiny $u$};
\node at (4.15, 7.9) {\tiny $y$};

\draw [densely dashed] (4.,7.5) -- (3.4,7.5);
\draw [snake it] (4.,7.7) -- (4.6,7.7);
\draw [->] (4.4,7.7) -- (4.3,7.7);
\node [brown] at (3.2, 7.75) {\tiny $(\idm, \tob)$};
\node [brown] at (4.8, 7.45) {\tiny $(K, s, \mu)$};
\fill [brown] (3.3, 7.5) circle (0.1);
\fill [brown] (4.7, 7.7) circle (0.1);
\end{tikzpicture}
}

\newcommand{\FusionCreationLeft}{
\begin{tikzpicture}[baseline={([yshift=-.8ex]current bounding box.center)},scale=1.]
\draw [->-] (4.,8.) -- (3.3,8.4);
\draw [->-] (3.3,8.4) -- (2.6,8.);
\draw [->-] (2.6,8.) -- (2.6,7.2);
\draw [->-] (2.6,7.2) -- (3.3,6.8);
\draw [->-] (3.3,6.8) -- (4.,7.2);
\draw [->-] (4.,8.) -- (4.7,8.4);
\draw [->-] (4.7,8.4) -- (5.4,8.);
\draw [->-] (5.4,8.) -- (5.4,7.2);
\draw [->-] (5.4,7.2) -- (4.7,6.8);
\draw [->-] (4.7,6.8) -- (4.,7.2);
\draw [->-] (4.,7.2) -- (4.,7.5);
\draw [->] (4.,7.2) -- (4.,7.7);
\draw [] (4.,7.5) -- (4.,7.7);
\draw [->-] (4.,7.7) -- (4.,8.);
\draw [->-] (3.3,8.4) -- (3.3,8.8);
\draw [->-] (4.7,8.4) -- (4.7,8.8);
\draw [->-] (3.3,6.4) -- (3.3,6.8);
\draw [->-] (4.7,6.4) -- (4.7,6.8);
\draw [->-] (5.7,7.) -- (5.4,7.2);
\draw [->-] (5.7,8.2) -- (5.4,8.);
\draw [->-] (2.3,8.2) -- (2.6,8.);
\draw [->-] (2.3,7.) -- (2.6,7.2);
\node at (4.15, 7.3) {\tiny $x$};
\node at (4.15, 7.55) {\tiny $u$};
\node at (4.15, 7.9) {\tiny $y$};

\draw [snake it] (4.,7.5) -- (3.4,7.5);
\draw [->] (3.6,7.5) -- (3.7,7.5);
\draw [snake it] (4.,7.7) -- (4.6,7.7);
\draw [->] (4.4,7.7) -- (4.3,7.7);
\node [brown] at (3.2, 7.75) {\tiny $(J_1^\ast, p^\ast, \alpha)$};
\node [brown] at (4.8, 7.45) {\tiny $(J_2, q, \gamma)$};
\fill [brown] (3.3, 7.5) circle (0.1);
\fill [brown] (4.7, 7.7) circle (0.1);
\end{tikzpicture}
}

\newcommand{\FusionCreationRight}{
\begin{tikzpicture}[baseline={([yshift=-.8ex]current bounding box.center)},scale=1.]
\draw [->-] (4.,8.) -- (3.3,8.4);
\draw [->-] (3.3,8.4) -- (2.6,8.);
\draw [->-] (2.6,8.) -- (2.6,7.2);
\draw [->-] (2.6,7.2) -- (3.3,6.8);
\draw [->-] (3.3,6.8) -- (4.,7.2);
\draw [->-] (4.,8.) -- (4.7,8.4);
\draw [->-] (4.7,8.4) -- (5.4,8.);
\draw [->-] (5.4,8.) -- (5.4,7.2);
\draw [->-] (5.4,7.2) -- (4.7,6.8);
\draw [->-] (4.7,6.8) -- (4.,7.2);
\draw [->-] (4.,7.2) -- (4.,7.5);
\draw [->] (4.,7.2) -- (4.,7.7);
\draw [] (4.,7.5) -- (4.,7.7);
\draw [->-] (4.,7.7) -- (4.,8.);
\draw [->-] (3.3,8.4) -- (3.3,8.8);
\draw [->-] (4.7,8.4) -- (4.7,8.8);
\draw [->-] (3.3,6.4) -- (3.3,6.8);
\draw [->-] (4.7,6.4) -- (4.7,6.8);
\draw [->-] (5.7,7.) -- (5.4,7.2);
\draw [->-] (5.7,8.2) -- (5.4,8.);
\draw [->-] (2.3,8.2) -- (2.6,8.);
\draw [->-] (2.3,7.) -- (2.6,7.2);
\node at (4.15, 7.3) {\tiny $x$};
\node at (4.15, 7.55) {\tiny $v$};
\node at (4.15, 7.9) {\tiny $y$};

\draw [snake it] (4.,7.5) -- (3.4,7.5);
\draw [->] (3.6,7.5) -- (3.7,7.5);
\draw [snake it] (4.,7.7) -- (4.6,7.7);
\draw [->] (4.4,7.7) -- (4.3,7.7);
\node [brown] at (3.3, 7.75) {\tiny $(K_1^\ast, m^\ast, \rho)$};
\node [brown] at (4.8, 7.45) {\tiny $(K_2, n, \sigma)$};
\fill [brown] (3.3, 7.5) circle (0.1);
\fill [brown] (4.7, 7.7) circle (0.1);
\end{tikzpicture}
}

\newcommand{\PachnerOne}{\begin{tikzpicture}[baseline={([yshift=-.8ex]current bounding box.center)},scale=.6]
\draw [->-] (0.,0.2) -- (.5,1.);
\draw [->-] (0.,1.8) -- (.5,1.);
\draw [->-] (2.,1.) -- (.5,1.);
\draw [->-] (2.5,1.8) -- (2.,1.);
\draw [->-] (2.5,0.2) -- (2.,1.);
\node at (1.25, 1.3) {\scriptsize $m$};
\node at (-.15, 2.) {\scriptsize $a$};
\node at (-.15, 0.) {\scriptsize $b$};
\node at (2.65, 2.) {\scriptsize $d$};
\node at (2.65, 0.) {\scriptsize $c$};
\end{tikzpicture}\qquad = \qquad \sum_{n\in L_{\Fus}}\ \sqrt{d_md_n}\ G^{abm}_{cdn}\quad \begin{tikzpicture}[baseline={([yshift=-.8ex]current bounding box.center)},scale=.6]
\draw [->-] (0.,0.) -- (.8,.5);
\draw [->-] (1.6,0.) -- (.8,.5);
\draw [->-] (.8,.5) -- (.8,2.);
\draw [->-] (0.,2.5) -- (.8,2.);
\draw [->-] (1.6,2.5) -- (.8,2.);
\node at (1.05, 1.3) {\scriptsize $n$};
\node at (-.15, 2.6) {\scriptsize $a$};
\node at (-.15, -.1) {\scriptsize $b$};
\node at (1.75, 2.6) {\scriptsize $d$};
\node at (1.75, -0.1) {\scriptsize $c$};
\end{tikzpicture}}

\newcommand{\PachnerTwo}{\begin{tikzpicture}[baseline={([yshift=-.8ex]current bounding box.center)},scale=.6]
\draw [->-] (1, 0) -- (1, 1);
\draw [->-] (1, 1) arc (270:90:.5);
\draw [->-] (1, 1) arc (-90:90:.5);
\draw [->-] (1, 2) -- (1, 3);
\node at (1.25, .3) {\scriptsize $a$};
\node at (1., 1.5) {\scriptsize \color{red}$\times$};
\node at (.2, 1.5) {\scriptsize $x$};
\node at (1.8, 1.5) {\scriptsize $y$};
\node at (1.25, 2.7) {\scriptsize $b$};
\end{tikzpicture}\qquad = \qquad \sqrt{\frac{d_xd_y}{d_a}}\ \delta_{ab}\ \delta_{xya^\ast}\quad \begin{tikzpicture}[baseline={([yshift=-.8ex]current bounding box.center)},scale=.5]
\draw [->-] (1, 0) -- (1, 3);
\node at (1.25, 1.5) {\scriptsize $a$};
\end{tikzpicture}}

\newcommand{\PachnerThree}{\begin{tikzpicture}[baseline={([yshift=-.8ex]current bounding box.center)},scale=.5]
\draw [->-] (1, 0) -- (1, 3);
\node at (1.25, 1.5) {\scriptsize $a$};
\end{tikzpicture}\quad = \quad \frac{1}{D}\sum_{xy\in L_{\Fus}} \sqrt{\frac{d_xd_y}{d_a}}\ \delta_{xya^\ast} \quad \begin{tikzpicture}[baseline={([yshift=-.8ex]current bounding box.center)},scale=.6]
\draw [->-] (1, 0) -- (1, 1);
\draw [->-] (1, 1) arc (270:90:.5);
\draw [->-] (1, 1) arc (-90:90:.5);
\draw [->-] (1, 2) -- (1, 3);
\node at (1.25, .3) {\scriptsize $a$};
\node at (.2, 1.5) {\scriptsize $x$};
\node at (1.8, 1.5) {\scriptsize $y$};
\node at (1.25, 2.7) {\scriptsize $a$};
\end{tikzpicture}}

\newcommand{\PachnerFour}{
\begin{tikzpicture}[baseline={([yshift=-.8ex]current bounding box.center)},scale=.6]
\draw [->-] (3.2,5.6) -- (3.2,6.8);
\draw [->-] (3.2,7.2) -- (4.6,8.);
\draw [->-] (4.6,8.) -- (6.,7.2);
\draw [->-] (6.,5.6) -- (4.6,4.8);
\draw [->-] (4.6,4.8) -- (3.2,5.6);
\draw [->-] (6.,7.2) -- (6.,5.6);
\draw [decorate,decoration={snake,segment length=.12cm, amplitude=.04cm}, ->] (3.2,6.6) -- (4.4,6.6);
\draw [->-] (2.6,7.6) -- (3.2,7.2);
\draw [->-] (4.6,8.8) -- (4.6,8.);
\draw [->-] (6.6,7.6) -- (6.,7.2);
\draw [->-] (6.6,5.2) -- (6.,5.6);
\draw [->-] (4.6,4.) -- (4.6,4.8);
\draw [->-] (2.6,5.2) -- (3.2,5.6);
\draw [->-] (3.2,6.6) -- (3.2,7.2);
\node [centered, rotate=0.] at (2.5,7.8) {\scriptsize $e_1$};
\node [centered, rotate=0.] at (2.9,6.) {\scriptsize $i_0$};
\node [centered, rotate=0.] at (2.9,6.9) {\scriptsize $i_1$};
\node [centered, rotate=0.] at (3.6,7.8) {\scriptsize $i_2$};
\node [centered, rotate=0.] at (4.7,6.6) {\scriptsize $p$};
\end{tikzpicture}\ =\quad\sum_{j\in L_\Fus}\sqrt{d_{i_1}d_j}\ G^{i_2^\ast e_1i_1}_{i_0 p^\ast j}\ 
\begin{tikzpicture}[baseline={([yshift=-.8ex]current bounding box.center)},scale=.6]
\draw [->-] (3.2,5.6) -- (3.2,7.2);
\draw [->-] (3.2,7.2) -- (3.55,7.4);
\draw [->-] (3.55,7.4) -- (4.6,8.);
\draw [->-] (4.6,8.) -- (6.,7.2);
\draw [->-] (6.,5.6) -- (4.6,4.8);
\draw [->-] (4.6,4.8) -- (3.2,5.6);
\draw [->-] (6.,7.2) -- (6.,5.6);
\draw [->-] (2.6,7.6) -- (3.2,7.2);
\draw [->-] (4.6,8.8) -- (4.6,8.);
\draw [->-] (6.6,7.6) -- (6.,7.2);
\draw [->-] (6.6,5.2) -- (6.,5.6);
\draw [->-] (4.6,4.) -- (4.6,4.8);
\draw [->-] (2.6,5.2) -- (3.2,5.6);
\draw [decorate,decoration={snake,segment length=.12cm, amplitude=.04cm}, ->] (3.59,7.42) -- (4.1,6.5);
\node [centered, rotate=0.] at (2.5,5.) {\scriptsize $e_n$};
\node [centered, rotate=0.] at (2.5,7.8) {\scriptsize $e_1$};
\node [centered, rotate=0.] at (2.9,6.5) {\scriptsize $i_0$};
\node [centered, rotate=0.] at (3.3,7.6) {\scriptsize $j$};
\node [centered, rotate=0.] at (3.9,8.) {\scriptsize $i_2$};
\node [centered, rotate=0.] at (4.3,6.3) {\scriptsize $p$};
\node [centered, rotate=0.] at (3.7,5.1) {\scriptsize $i_n$};
\end{tikzpicture}
}

\newcommand{\PachnerFive}{
\begin{tikzpicture}[baseline={([yshift=-.8ex]current bounding box.center)},scale=.6]
\draw [->-] (3.2,5.6) -- (3.2,6.1);
\draw [->-] (3.2,6.1) -- (3.2,6.7);
\draw [->-] (3.2,6.7) -- (3.2,7.2);
\draw [->-] (3.2,7.2) -- (4.6,8.);
\draw [->-] (4.6,8.) -- (6.,7.2);
\draw [->-] (6.,5.6) -- (4.6,4.8);
\draw [->-] (4.6,4.8) -- (3.2,5.6);
\draw [->-] (6.,7.2) -- (6.,5.6);
\draw [decorate,decoration={snake,segment length=.12cm, amplitude=.04cm}, ->] (3.2,6.1) -- (4.4,6.1);
\draw [decorate,decoration={snake,segment length=.12cm, amplitude=.04cm}, ->] (3.2,6.7) -- (4.4,6.7);
\draw [->-] (2.6,7.6) -- (3.2,7.2);
\draw [->-] (4.6,8.8) -- (4.6,8.);
\draw [->-] (6.6,7.6) -- (6.,7.2);
\draw [->-] (6.6,5.2) -- (6.,5.6);
\draw [->-] (4.6,4.) -- (4.6,4.8);
\draw [->-] (2.6,5.2) -- (3.2,5.6);
\node [centered, rotate=0.] at (2.9,5.8) {\scriptsize $i_0$};
\node [centered, rotate=0.] at (2.9,7.1) {\scriptsize $i_1$};
\node [centered, rotate=0.] at (4.7,6.7) {\scriptsize $r$};
\node [centered, rotate=0.] at (4.7,6.1) {\scriptsize $s$};
\node [centered, rotate=0.] at (2.9,6.4) {\scriptsize $j$};
\end{tikzpicture}\quad = \quad \sum_{u\in L_\Fus}\sqrt{d_jd_p}\ G^{r^\ast i_1^\ast j}_{i_0s^\ast p}\quad 
\begin{tikzpicture}[baseline={([yshift=-.8ex]current bounding box.center)},scale=.6]
\draw [->-] (3.2,5.6) -- (3.2,6.8);
\draw [->-] (3.2,7.2) -- (4.6,8.);
\draw [->-] (4.6,8.) -- (6.,7.2);
\draw [->-] (6.,5.6) -- (4.6,4.8);
\draw [->-] (4.6,4.8) -- (3.2,5.6);
\draw [->-] (6.,7.2) -- (6.,5.6);
\draw [decorate,decoration={snake,segment length=.12cm, amplitude=.04cm}, ->] (3.2,6.6) -- (4.4,6.6);
\draw [->-] (2.6,7.6) -- (3.2,7.2);
\draw [->-] (4.6,8.8) -- (4.6,8.);
\draw [->-] (6.6,7.6) -- (6.,7.2);
\draw [->-] (6.6,5.2) -- (6.,5.6);
\draw [->-] (4.6,4.) -- (4.6,4.8);
\draw [->-] (2.6,5.2) -- (3.2,5.6);
\draw [->-] (3.2,6.6) -- (3.2,7.2);
\node [centered, rotate=0.] at (2.9,6.) {\scriptsize $i_0$};
\node [centered, rotate=0.] at (2.9,6.9) {\scriptsize $i_1$};
\node [centered, rotate=0.] at (4.7,6.6) {\scriptsize $p$};
\end{tikzpicture}
}

\newcommand{\YVertex}[5]{
\begin{tikzpicture}
\draw [#5] (0, 0) -- (1, .6);
\draw [#5] (2, 0) -- (1, .6);
\draw [#5] (1, .6) -- (1, 1.6);
\fill [#5, opacity = 0.15] (0, 0) -- (1, .6) -- (1, 1.6) -- (0, 1.6);
\fill [#5, opacity = 0.15] (2, 0) -- (1, .6) -- (1, 1.6) -- (2, 1.6);
\fill [#5, opacity = 0.15] (0, 0) -- (1, .6) -- (2,  0);
\node[#5] at (0.3, .4) {\scriptsize #1};
\node[#5] at (1.7, .4) {\scriptsize #2};
\node[#5] at (1.25, 1.2) {\scriptsize #3};
\node[#5] at (.3, 1.) {\scriptsize #4};
\node[#5] at (1.7, 1.) {\scriptsize #4};
\node[#5] at (1., 0.2) {\scriptsize #4};
\draw[white] (-.5, -.5) -- (2.5,-.5);
\end{tikzpicture}
}

\newcommand{\YDirect}[7]{
\begin{tikzpicture}
\draw[#6] (0, 0) -- (1, .6);
\draw [ultra thick, teal, ->-] (2, 0) -- (1, .6);
\draw [ultra thick, teal, ->-] (1, .6) -- (1, 1.6);
\fill [#6, opacity = 0.15] (0, 0) -- (1, .6) -- (1, 1.6) -- (0, 1.6);
\fill [#6, opacity = 0.15] (0, 0) -- (1, .6) -- (2,  0);
\fill [#7, opacity = 0.15] (2, 0) -- (1, .6) -- (1, 1.6) -- (2, 1.6);
\node[#6] at (0.3, .5) {\scriptsize #1};
\node[teal] at (1.7, .5) {\scriptsize #2};
\node[teal] at (1.35, 1.2) {\scriptsize #3};
\node[#6] at (.3, 1.) {\scriptsize #4};
\node[#6] at (1., 0.2) {\scriptsize #4};
\node[#7] at (1.7, 1.) {\scriptsize #5};
\draw[white] (-.5, -.5) -- (2.5,-.5);
\end{tikzpicture}
}

\newcommand{\GapEdgeUp}[3]{
\begin{tikzpicture}[baseline={([yshift=-.8ex]current bounding box.center)},scale=1.]
\draw [ultra thick, teal, ->-] (0, 0) -- (0, 1.5);
\node [right, rotate=0.] at (0.,.35) {\scriptsize ${#1}$};
\node [red, right, rotate=0.] at (-1.,.75) {\scriptsize ${#2}$};
\node [blue, right, rotate=0.] at (.5,.75) {\scriptsize ${#3}$};

\end{tikzpicture}
}

\newcommand{\GapEdgeDown}[3]{
\begin{tikzpicture}[baseline={([yshift=-.8ex]current bounding box.center)},scale=1.]
\draw [ultra thick, teal, -<-] (0, 0) -- (0, 1.5);
\node [right, rotate=0.] at (0.,.35) {\scriptsize ${#1}$};
\node [right, rotate=0.] at (-1.,.75) {\scriptsize ${#2}$};
\node [right, rotate=0.] at (.5,.75) {\scriptsize ${#3}$};

\end{tikzpicture}
}

\newcommand{\HoppingFiboLeft}{
\begin{tikzpicture}[baseline={([yshift=-.8ex]current bounding box.center)},scale=.7]
\draw [teal, ultra thick] (4.,8.) -- (3.3,8.4);
\draw [] (3.3,8.4) -- (2.6,8.);
\draw [] (2.6,8.) -- (2.6,7.2);
\draw [] (2.6,7.2) -- (3.3,6.8);
\draw [] (3.3,6.8) -- (4.,7.2);
\draw [] (4.,8.) -- (4.7,8.4);
\draw [] (4.7,8.4) -- (5.4,8.);
\draw [] (5.4,8.) -- (5.4,7.2);
\draw [] (5.4,7.2) -- (4.7,6.8);
\draw [teal, ultra thick] (4.7,6.8) -- (4.,7.2);
\draw [teal, ultra thick] (4.,7.2) -- (4.,8.);
\draw [teal, ultra thick] (3.3,8.4) -- (3.3,8.8);
\draw [] (4.7,8.4) -- (4.7,8.8);
\draw [] (3.3,6.4) -- (3.3,6.8);
\draw [teal, ultra thick] (4.7,6.4) -- (4.7,6.8);
\draw [] (5.7,7.) -- (5.4,7.2);
\draw [] (5.7,8.2) -- (5.4,8.);
\draw [] (2.3,8.2) -- (2.6,8.);
\draw [] (2.3,7.) -- (2.6,7.2);
\draw [snake it] (4.,7.5) -- (3.4,7.5);
\node [brown] at (3.25, 7.8) {\tiny $(\tau\bar\tau, 1)$};
\fill [brown] (3.3, 7.5) circle (0.1);
\node [teal] at (3.45, 8.6) {\tiny $\tau$};
\node [teal] at (3.5, 8.1) {\tiny $\tau$};
\node [teal] at (4.15, 7.35) {\tiny $\tau$};
\node [teal] at (4.15, 7.8) {\tiny $\tau$};
\node [teal] at (4.25, 6.85) {\tiny $\tau$};
\node [teal] at (4.5, 6.6) {\tiny $\tau$};
\end{tikzpicture}
}

\newcommand{\HoppingFiboRight}{
\begin{tikzpicture}[baseline={([yshift=-.8ex]current bounding box.center)},scale=.7]
\draw [teal, ultra thick] (4.,8.) -- (3.3,8.4);
\draw [] (3.3,8.4) -- (2.6,8.);
\draw [] (2.6,8.) -- (2.6,7.2);
\draw [] (2.6,7.2) -- (3.3,6.8);
\draw [] (3.3,6.8) -- (4.,7.2);
\draw [] (4.,8.) -- (4.7,8.4);
\draw [] (4.7,8.4) -- (5.4,8.);
\draw [] (5.4,8.) -- (5.4,7.2);
\draw [] (5.4,7.2) -- (4.7,6.8);
\draw [teal, ultra thick] (4.7,6.8) -- (4.,7.2);
\draw [teal, ultra thick] (4.,7.2) -- (4.,8.);
\draw [teal, ultra thick] (3.3,8.4) -- (3.3,8.8);
\draw [] (4.7,8.4) -- (4.7,8.8);
\draw [] (3.3,6.4) -- (3.3,6.8);
\draw [teal, ultra thick] (4.7,6.4) -- (4.7,6.8);
\draw [] (5.7,7.) -- (5.4,7.2);
\draw [] (5.7,8.2) -- (5.4,8.);
\draw [] (2.3,8.2) -- (2.6,8.);
\draw [] (2.3,7.) -- (2.6,7.2);
\draw [snake it] (4.,7.7) -- (4.6,7.7);
\fill [brown] (4.7, 7.7) circle (0.1);
\node [] at (4.8, 7.4) {\tiny $(\tau\bar\tau, 1)$};
\node [teal] at (3.45, 8.6) {\tiny $\tau$};
\node [teal] at (3.5, 8.1) {\tiny $\tau$};
\node [teal] at (3.85, 7.4) {\tiny $\tau$};
\node [teal] at (3.85, 7.83) {\tiny $\tau$};
\node [teal] at (4.25, 6.85) {\tiny $\tau$};
\node [teal] at (4.5, 6.6) {\tiny $\tau$};
\end{tikzpicture}
}

\newcommand{\HoppingFiboRightTau}{
\begin{tikzpicture}[baseline={([yshift=-.8ex]current bounding box.center)},scale=.7]
\draw [teal, ultra thick] (4.,8.) -- (3.3,8.4);
\draw [] (3.3,8.4) -- (2.6,8.);
\draw [] (2.6,8.) -- (2.6,7.2);
\draw [] (2.6,7.2) -- (3.3,6.8);
\draw [] (3.3,6.8) -- (4.,7.2);
\draw [] (4.,8.) -- (4.7,8.4);
\draw [] (4.7,8.4) -- (5.4,8.);
\draw [] (5.4,8.) -- (5.4,7.2);
\draw [] (5.4,7.2) -- (4.7,6.8);
\draw [teal, ultra thick] (4.7,6.8) -- (4.,7.2);
\draw [teal, ultra thick] (4.,7.2) -- (4.,8.);
\draw [teal, ultra thick] (3.3,8.4) -- (3.3,8.8);
\draw [] (4.7,8.4) -- (4.7,8.8);
\draw [] (3.3,6.4) -- (3.3,6.8);
\draw [teal, ultra thick] (4.7,6.4) -- (4.7,6.8);
\draw [] (5.7,7.) -- (5.4,7.2);
\draw [] (5.7,8.2) -- (5.4,8.);
\draw [] (2.3,8.2) -- (2.6,8.);
\draw [] (2.3,7.) -- (2.6,7.2);
\draw [teal, snake it] (4.,7.7) -- (4.6,7.7);
\fill [brown] (4.7, 7.7) circle (0.1);
\node [brown] at (4.8, 7.4) {\tiny $(\tau\bar\tau, \tau)$};
\node [teal] at (3.45, 8.6) {\tiny $\tau$};
\node [teal] at (3.5, 8.1) {\tiny $\tau$};
\node [teal] at (3.85, 7.4) {\tiny $\tau$};
\node [teal] at (3.85, 7.83) {\tiny $\tau$};
\node [teal] at (4.25, 6.85) {\tiny $\tau$};
\node [teal] at (4.5, 6.6) {\tiny $\tau$};
\end{tikzpicture}
}

\newcommand{\BraidingA}{
\begin{tikzpicture}[baseline={([yshift=-.8ex]current bounding box.center)},scale=.4]
\draw [] (2.4,8.2) -- (2.8,8.4);
\draw [] (2.8,6.8) -- (2.4,7.);
\draw [] (2.4,5.8) -- (2.8,6.);
\draw [] (2.8,4.4) -- (2.4,4.6);
\draw [] (2.8,8.4) -- (3.5,8.);
\draw [] (3.5,7.2) -- (2.8,6.8);
\draw [] (3.5,7.2) -- (3.5,8.);
\draw [] (3.5,8.) -- (4.2,8.4);
\draw [] (4.2,8.4) -- (4.9,8.);
\draw [] (4.2,6.8) -- (3.5,7.2);
\draw [] (4.9,7.2) -- (4.2,6.8);
\draw [] (4.9,7.2) -- (4.9,8.);
\draw [] (4.9,8.) -- (5.6,8.4);
\draw [] (5.6,8.4) -- (6.3,8.);
\draw [] (5.6,6.8) -- (4.9,7.2);
\draw [] (6.3,7.2) -- (5.6,6.8);
\draw [] (2.8,6.) -- (2.8,6.8);
\draw [] (4.2,6.) -- (4.2,6.8);
\draw [] (5.6,6.) -- (5.6,6.8);
\draw [] (6.3,7.2) -- (6.3,8.);
\draw [] (6.3,8.) -- (7.,8.4);
\draw [] (7.,8.4) -- (7.7,8.);
\draw [] (7.,6.8) -- (6.3,7.2);
\draw [] (7.7,7.2) -- (7.,6.8);
\draw [] (7.7,7.2) -- (7.7,8.);
\draw [] (7.7,8.) -- (8.,8.2);
\draw [] (8.4,6.8) -- (7.7,7.2);
\draw [] (7.,6.) -- (7.,6.8);
\draw [] (8.4,6.) -- (8.4,6.8);
\draw [] (2.8,6.) -- (3.5,5.6);
\draw [] (3.5,4.8) -- (2.8,4.4);
\draw [] (3.5,4.8) -- (3.5,5.6);
\draw [] (3.5,5.6) -- (4.2,6.);
\draw [] (4.2,6.) -- (4.9,5.6);
\draw [] (4.2,4.4) -- (3.5,4.8);
\draw [] (4.9,4.8) -- (4.2,4.4);
\draw [] (4.9,4.8) -- (4.9,5.6);
\draw [] (4.9,5.6) -- (5.6,6.);
\draw [] (5.6,6.) -- (6.3,5.6);
\draw [] (5.6,4.4) -- (4.9,4.8);
\draw [] (6.3,4.8) -- (5.6,4.4);
\draw [] (2.8,4.) -- (2.8,4.4);
\draw [] (4.2,4.) -- (4.2,4.4);
\draw [] (5.6,4.) -- (5.6,4.4);
\draw [] (6.3,4.8) -- (6.3,5.6);
\draw [] (6.3,5.6) -- (7.,6.);
\draw [] (7.,6.) -- (7.7,5.6);
\draw [] (7.,4.4) -- (6.3,4.8);
\draw [] (7.7,4.8) -- (7.,4.4);
\draw [] (7.7,4.8) -- (7.7,5.6);
\draw [] (7.7,5.6) -- (8.4,6.);
\draw [] (8.4,4.4) -- (7.7,4.8);
\draw [] (7.,4.) -- (7.,4.4);
\draw [] (8.4,4.) -- (8.4,4.4);
\draw [] (8.4,6.8) -- (8.7,7.);
\draw [] (8.4,6.) -- (8.7,5.8);
\draw [] (8.4,4.4) -- (8.7,4.6);
\draw [] (2.8,8.4) -- (2.8,8.8);
\draw [] (4.2,8.4) -- (4.2,8.8);
\draw [] (5.6,8.4) -- (5.6,8.8);
\draw [] (7.,8.4) -- (7.,8.8);

\draw [ultra thick, orange, ->] (5.75, 5.1) arc [x radius = 2, y radius = 1.3, start angle = -85, end angle = 270];
\end{tikzpicture}
}

\newcommand{\BraidingC}{
\begin{tikzpicture}[baseline={([yshift=-.8ex]current bounding box.center)},scale=.4]
\draw [] (2.4,8.2) -- (2.8,8.4);
\draw [] (2.8,6.8) -- (2.4,7.);
\draw [] (2.4,5.8) -- (2.8,6.);
\draw [] (2.8,4.4) -- (2.4,4.6);
\draw [] (2.8,8.4) -- (3.5,8.);
\draw [] (3.5,7.2) -- (2.8,6.8);
\draw [] (3.5,7.2) -- (3.5,8.);
\draw [] (3.5,8.) -- (4.2,8.4);
\draw [] (4.2,8.4) -- (4.9,8.);
\draw [] (4.2,6.8) -- (3.5,7.2);
\draw [] (4.9,7.2) -- (4.2,6.8);
\draw [] (4.9,7.2) -- (4.9,8.);
\draw [] (4.9,8.) -- (5.6,8.4);
\draw [] (5.6,8.4) -- (6.3,8.);
\draw [] (5.6,6.8) -- (4.9,7.2);
\draw [] (6.3,7.2) -- (5.6,6.8);
\draw [] (2.8,6.) -- (2.8,6.8);
\draw [] (4.2,6.) -- (4.2,6.8);
\draw [] (5.6,6.) -- (5.6,6.8);
\draw [] (6.3,7.2) -- (6.3,8.);
\draw [] (6.3,8.) -- (7.,8.4);
\draw [] (7.,8.4) -- (7.7,8.);
\draw [] (7.,6.8) -- (6.3,7.2);
\draw [] (7.7,7.2) -- (7.,6.8);
\draw [] (7.7,7.2) -- (7.7,8.);
\draw [] (7.7,8.) -- (8.,8.2);
\draw [] (8.4,6.8) -- (7.7,7.2);
\draw [] (7.,6.) -- (7.,6.8);
\draw [] (8.4,6.) -- (8.4,6.8);
\draw [] (2.8,6.) -- (3.5,5.6);
\draw [] (3.5,4.8) -- (2.8,4.4);
\draw [] (3.5,4.8) -- (3.5,5.6);
\draw [] (3.5,5.6) -- (4.2,6.);
\draw [] (4.2,6.) -- (4.9,5.6);
\draw [] (4.2,4.4) -- (3.5,4.8);
\draw [] (4.9,4.8) -- (4.2,4.4);
\draw [] (4.9,4.8) -- (4.9,5.6);
\draw [] (4.9,5.6) -- (5.6,6.);
\draw [] (5.6,6.) -- (6.3,5.6);
\draw [] (5.6,4.4) -- (4.9,4.8);
\draw [] (6.3,4.8) -- (5.6,4.4);
\draw [] (2.8,4.) -- (2.8,4.4);
\draw [] (4.2,4.) -- (4.2,4.4);
\draw [] (5.6,4.) -- (5.6,4.4);
\draw [] (6.3,4.8) -- (6.3,5.6);
\draw [] (6.3,5.6) -- (7.,6.);
\draw [] (7.,6.) -- (7.7,5.6);
\draw [] (7.,4.4) -- (6.3,4.8);
\draw [] (7.7,4.8) -- (7.,4.4);
\draw [] (7.7,4.8) -- (7.7,5.6);
\draw [] (7.7,5.6) -- (8.4,6.);
\draw [] (8.4,4.4) -- (7.7,4.8);
\draw [] (7.,4.) -- (7.,4.4);
\draw [] (8.4,4.) -- (8.4,4.4);
\draw [] (8.4,6.8) -- (8.7,7.);
\draw [] (8.4,6.) -- (8.7,5.8);
\draw [] (8.4,4.4) -- (8.7,4.6);
\draw [] (2.8,8.4) -- (2.8,8.8);
\draw [] (4.2,8.4) -- (4.2,8.8);
\draw [] (5.6,8.4) -- (5.6,8.8);
\draw [] (7.,8.4) -- (7.,8.8);

\draw [snake it](4.2, 6.4) -- (4.7, 6.4);
\fill [brown](4.9, 6.4) circle (.2);

\draw [ultra thick, orange, ->] (5.75, 5.1) arc [x radius = 2, y radius = 1.3, start angle = -85, end angle = 270];
\end{tikzpicture}
}

\newcommand{\BraidingD}{
\begin{tikzpicture}[baseline={([yshift=-.8ex]current bounding box.center)},scale=.4]
\draw [] (2.4,8.2) -- (2.8,8.4);
\draw [] (2.8,6.8) -- (2.4,7.);
\draw [] (2.4,5.8) -- (2.8,6.);
\draw [] (2.8,4.4) -- (2.4,4.6);
\draw [] (2.8,8.4) -- (3.5,8.);
\draw [] (3.5,7.2) -- (2.8,6.8);
\draw [] (3.5,7.2) -- (3.5,8.);
\draw [] (3.5,8.) -- (4.2,8.4);
\draw [] (4.2,8.4) -- (4.9,8.);
\draw [] (4.2,6.8) -- (3.5,7.2);
\draw [] (4.9,7.2) -- (4.2,6.8);
\draw [] (4.9,7.2) -- (4.9,8.);
\draw [] (4.9,8.) -- (5.6,8.4);
\draw [] (5.6,8.4) -- (6.3,8.);
\draw [] (5.6,6.8) -- (4.9,7.2);
\draw [] (6.3,7.2) -- (5.6,6.8);
\draw [] (2.8,6.) -- (2.8,6.8);
\draw [] (4.2,6.) -- (4.2,6.8);
\draw [] (5.6,6.) -- (5.6,6.8);
\draw [] (6.3,7.2) -- (6.3,8.);
\draw [] (6.3,8.) -- (7.,8.4);
\draw [] (7.,8.4) -- (7.7,8.);
\draw [] (7.,6.8) -- (6.3,7.2);
\draw [] (7.7,7.2) -- (7.,6.8);
\draw [] (7.7,7.2) -- (7.7,8.);
\draw [] (7.7,8.) -- (8.,8.2);
\draw [] (8.4,6.8) -- (7.7,7.2);
\draw [] (7.,6.) -- (7.,6.8);
\draw [] (8.4,6.) -- (8.4,6.8);
\draw [] (2.8,6.) -- (3.5,5.6);
\draw [] (3.5,4.8) -- (2.8,4.4);
\draw [] (3.5,4.8) -- (3.5,5.6);
\draw [] (3.5,5.6) -- (4.2,6.);
\draw [] (4.2,6.) -- (4.9,5.6);
\draw [] (4.2,4.4) -- (3.5,4.8);
\draw [] (4.9,4.8) -- (4.2,4.4);
\draw [] (4.9,4.8) -- (4.9,5.6);
\draw [] (4.9,5.6) -- (5.6,6.);
\draw [] (5.6,6.) -- (6.3,5.6);
\draw [] (5.6,4.4) -- (4.9,4.8);
\draw [] (6.3,4.8) -- (5.6,4.4);
\draw [] (2.8,4.) -- (2.8,4.4);
\draw [] (4.2,4.) -- (4.2,4.4);
\draw [] (5.6,4.) -- (5.6,4.4);
\draw [] (6.3,4.8) -- (6.3,5.6);
\draw [] (6.3,5.6) -- (7.,6.);
\draw [] (7.,6.) -- (7.7,5.6);
\draw [] (7.,4.4) -- (6.3,4.8);
\draw [] (7.7,4.8) -- (7.,4.4);
\draw [] (7.7,4.8) -- (7.7,5.6);
\draw [] (7.7,5.6) -- (8.4,6.);
\draw [] (8.4,4.4) -- (7.7,4.8);
\draw [] (7.,4.) -- (7.,4.4);
\draw [] (8.4,4.) -- (8.4,4.4);
\draw [] (8.4,6.8) -- (8.7,7.);
\draw [] (8.4,6.) -- (8.7,5.8);
\draw [] (8.4,4.4) -- (8.7,4.6);
\draw [] (2.8,8.4) -- (2.8,8.8);
\draw [] (4.2,8.4) -- (4.2,8.8);
\draw [] (5.6,8.4) -- (5.6,8.8);
\draw [] (7.,8.4) -- (7.,8.8);

\draw [ultra thick, orange] (4.35, 5.1) arc [x radius = 2, y radius = 1.3, start angle = -85, end angle = -25] arc [x radius = .4, y radius = .5, start angle = -25, end angle = 210] arc [x radius = 1., y radius = 1.2, start angle = 210, end angle = 270] arc [x radius = 1.4, y radius = 1.2, start angle = 270, end angle = 440] arc [x radius = 5., y radius = 6., start angle = 440, end angle = 460] arc [x radius = 1., y radius = 1.3, start angle = 460, end angle = 570];
\draw [ultra thick, orange, <-] (4.1, 5.25) arc [x radius = 1., y radius = 1.3, start angle = 570, end angle = 550];
\end{tikzpicture}
}

\newcommand{\GappedA}{
\begin{tikzpicture}[baseline={([yshift=-.8ex]current bounding box.center)},scale=.4]
\draw [] (2.4,8.2) -- (2.8,8.4);
\draw [] (2.8,6.8) -- (2.4,7.);
\draw [] (2.4,5.8) -- (2.8,6.);
\draw [] (2.8,4.4) -- (2.4,4.6);
\draw [] (2.8,8.4) -- (3.5,8.);
\draw [] (3.5,7.2) -- (2.8,6.8);
\draw [] (3.5,7.2) -- (3.5,8.);
\draw [] (3.5,8.) -- (4.2,8.4);
\draw [] (4.2,8.4) -- (4.9,8.);
\draw [] (4.2,6.8) -- (3.5,7.2);
\draw [] (4.9,7.2) -- (4.2,6.8);
\draw [] (4.9,7.2) -- (4.9,8.);
\draw [] (4.9,8.) -- (5.6,8.4);
\draw [] (5.6,8.4) -- (6.3,8.);
\draw [] (5.6,6.8) -- (4.9,7.2);
\draw [ultra thick, teal] (6.3,7.2) -- (5.6,6.8);
\draw [] (2.8,6.) -- (2.8,6.8);
\draw [] (4.2,6.) -- (4.2,6.8);
\draw [ultra thick, teal] (5.6,6.) -- (5.6,6.8);
\draw [] (6.3,7.2) -- (6.3,8.);
\draw [] (6.3,8.) -- (7.,8.4);
\draw [] (7.,8.4) -- (7.7,8.);
\draw [ultra thick, teal] (7.,6.8) -- (6.3,7.2);
\draw [] (7.7,7.2) -- (7.,6.8);
\draw [] (7.7,7.2) -- (7.7,8.);
\draw [] (7.7,8.) -- (8.,8.2);
\draw [] (8.4,6.8) -- (7.7,7.2);
\draw [ultra thick, teal] (7.,6.) -- (7.,6.8);
\draw [] (8.4,6.) -- (8.4,6.8);
\draw [] (2.8,6.) -- (3.5,5.6);
\draw [] (3.5,4.8) -- (2.8,4.4);
\draw [] (3.5,4.8) -- (3.5,5.6);
\draw [] (3.5,5.6) -- (4.2,6.);
\draw [] (4.2,6.) -- (4.9,5.6);
\draw [] (4.2,4.4) -- (3.5,4.8);
\draw [] (4.9,4.8) -- (4.2,4.4);
\draw [ultra thick, teal] (4.9,4.8) -- (4.9,5.6);
\draw [ultra thick, teal] (4.9,5.6) -- (5.6,6.);
\draw [] (5.6,6.) -- (6.3,5.6);
\draw [ultra thick, teal] (5.6,4.4) -- (4.9,4.8);
\draw [ultra thick, teal] (6.3,4.8) -- (5.6,4.4);
\draw [] (2.8,4.) -- (2.8,4.4);
\draw [] (4.2,4.) -- (4.2,4.4);
\draw [] (5.6,4.) -- (5.6,4.4);
\draw [] (6.3,4.8) -- (6.3,5.6);
\draw [] (6.3,5.6) -- (7.,6.);
\draw [ultra thick, teal] (7.,6.) -- (7.7,5.6);
\draw [ultra thick, teal] (7.,4.4) -- (6.3,4.8);
\draw [ultra thick, teal] (7.7,4.8) -- (7.,4.4);
\draw [ultra thick, teal] (7.7,4.8) -- (7.7,5.6);
\draw [] (7.7,5.6) -- (8.4,6.);
\draw [] (8.4,4.4) -- (7.7,4.8);
\draw [] (7.,4.) -- (7.,4.4);
\draw [] (8.4,4.) -- (8.4,4.4);
\draw [] (8.4,6.8) -- (8.7,7.);
\draw [] (8.4,6.) -- (8.7,5.8);
\draw [] (8.4,4.4) -- (8.7,4.6);
\draw [] (2.8,8.4) -- (2.8,8.8);
\draw [] (4.2,8.4) -- (4.2,8.8);
\draw [] (5.6,8.4) -- (5.6,8.8);
\draw [] (7.,8.4) -- (7.,8.8);
\end{tikzpicture}
}

\newcommand{\GappedC}{
\begin{tikzpicture}[baseline={([yshift=-.8ex]current bounding box.center)},scale=.4]
\draw [] (2.4,8.2) -- (2.8,8.4);
\draw [] (2.8,6.8) -- (2.4,7.);
\draw [] (2.4,5.8) -- (2.8,6.);
\draw [] (2.8,4.4) -- (2.4,4.6);
\draw [] (2.8,8.4) -- (3.5,8.);
\draw [] (3.5,7.2) -- (2.8,6.8);
\draw [] (3.5,7.2) -- (3.5,8.);
\draw [] (3.5,8.) -- (4.2,8.4);
\draw [] (4.2,8.4) -- (4.9,8.);
\draw [] (4.2,6.8) -- (3.5,7.2);
\draw [ultra thick, teal] (4.9,7.2) -- (4.2,6.8);
\draw [] (4.9,7.2) -- (4.9,8.);
\draw [] (4.9,8.) -- (5.6,8.4);
\draw [] (5.6,8.4) -- (6.3,8.);
\draw [ultra thick, teal] (5.6,6.8) -- (4.9,7.2);
\draw [ultra thick, teal] (6.3,7.2) -- (5.6,6.8);
\draw [] (2.8,6.) -- (2.8,6.8);
\draw [] (4.2,6.) -- (4.2,6.8);
\draw [ultra thick, teal] (4.2,6.4) -- (4.2,6.8);
\draw [teal, snake it] (4.2,6.4) -- (3.7,6.4);
\fill [brown] (3.5,6.4) circle (.2);
\draw [] (5.6,6.) -- (5.6,6.8);
\draw [] (6.3,7.2) -- (6.3,8.);
\draw [ultra thick, teal] (6.3,7.2) -- (6.3,7.6);
\draw [teal, snake it] (6.3,7.6) -- (6.8,7.6);
\fill [brown] (7.,7.6) circle (.2);
\draw [] (6.3,8.) -- (7.,8.4);
\draw [] (7.,8.4) -- (7.7,8.);
\draw [] (7.,6.8) -- (6.3,7.2);
\draw [] (7.7,7.2) -- (7.,6.8);
\draw [] (7.7,7.2) -- (7.7,8.);
\draw [] (8.4,6.8) -- (7.7,7.2);
\draw [] (7.,6.) -- (7.,6.8);
\draw [] (8.4,6.) -- (8.4,6.8);
\draw [] (2.8,6.) -- (3.5,5.6);
\draw [] (3.5,4.8) -- (2.8,4.4);
\draw [] (3.5,4.8) -- (3.5,5.6);
\draw [] (3.5,5.6) -- (4.2,6.);
\draw [] (4.2,6.) -- (4.9,5.6);
\draw [] (4.2,4.4) -- (3.5,4.8);
\draw [] (4.9,4.8) -- (4.2,4.4);
\draw [] (4.9,4.8) -- (4.9,5.6);
\draw [] (4.9,5.6) -- (5.6,6.);
\draw [] (5.6,6.) -- (6.3,5.6);
\draw [] (5.6,4.4) -- (4.9,4.8);
\draw [] (6.3,4.8) -- (5.6,4.4);
\draw [] (2.8,4.) -- (2.8,4.4);
\draw [] (4.2,4.) -- (4.2,4.4);
\draw [] (5.6,4.) -- (5.6,4.4);
\draw [] (6.3,4.8) -- (6.3,5.6);
\draw [] (6.3,5.6) -- (7.,6.);
\draw [] (7.,6.) -- (7.7,5.6);
\draw [] (7.,4.4) -- (6.3,4.8);
\draw [] (7.7,4.8) -- (7.,4.4);
\draw [] (7.7,4.8) -- (7.7,5.6);
\draw [] (7.7,5.6) -- (8.4,6.);
\draw [] (8.4,4.4) -- (7.7,4.8);
\draw [] (7.,4.) -- (7.,4.4);
\draw [] (8.4,4.) -- (8.4,4.4);
\draw [] (8.4,6.8) -- (8.7,7.);
\draw [] (8.4,6.) -- (8.7,5.8);
\draw [] (8.4,4.4) -- (8.7,4.6);
\draw [] (2.8,8.4) -- (2.8,8.8);
\draw [] (4.2,8.4) -- (4.2,8.8);
\draw [] (5.6,8.4) -- (5.6,8.8);
\draw [] (7.,8.4) -- (7.,8.8);
\end{tikzpicture}
}